\documentclass[usenatbib,usegraphicx]{mn2e}
\pdfoutput=1
\usepackage{txfonts}
\usepackage{multirow}
\usepackage{lscape}
\usepackage{longtable}
\usepackage{afterpage}
\usepackage{grffile}

\bibpunct[, ]{(}{)}{;}{a}{}{,}

\def\llm{\textsc{LLmodels}}

\def\kms{km~s$^{-1}$}

\def\synth3{\textsc{Synth3}}

\def\vsini{\upsilon\sin i}

\def\Rsun{R_{\odot}}

\def\mum{$\mu$m}

\ifdefined\ion
\renewcommand{\ion}[2]{\textup{#1\,\textsc{\lowercase{#2}}}}
\else
\newcommand{\ion}[2]{\textup{#1\,\textsc{\lowercase{#2}}}}
\fi

\def\araa{ARA\&A}%
\def\apj{ApJ}%
\def\apjl{ApJ}%
\def\apjs{ApJS}%
\def\aap{A\&A}%
\def\aaps{A\&AS}%
\def\mnras{MNRAS}%
\def\pasp{PASP}%
\def\procspie{Proc.~SPIE}%

\def\uma{$\varepsilon$~UMa}

\begin{document}

\title[Interferometry of chemically peculiar stars]{Interferometry of chemically peculiar stars:\\theoretical predictions vs. modern observing facilities}

\author[D. Shulyak et al.]{D. Shulyak,$^{1}$  C. Paladini,$^{2}$ G. Li Causi,$^{3}$ K. Perraut,$^{4}$ O. Kochukhov$^{5}$\\
$^1$~Institute of Astrophysics, Georg-August University, Friedrich-Hund-Platz 1, D-37077 G\"ottingen, Germany\\
$^2$~Institut of Astronomy and Astrophysics, Universit\'e Libre de Bruxelles, CP226, Boulevard du Triomphe, B-1050 Brussels, Belgium\\
$^3$~National Institute for Astrophysics, Rome Astronomical Observatory, Via Frascati 33, 00040 Monteporzio Catone (RM), Italy\\
$^4$~Univ. Grenoble Alpes, IPAG, F-38000 Grenoble, France \\
     CNRS, IPAG, F-38000 Grenoble, France\\
$^5$~Department of Physics and Astronomy, Uppsala University, Box 516, 751 20, Uppsala, Sweden
}


\maketitle

\begin{abstract}
{
{By means of numerical experiments we explore the application of interferometry to the detection and characterization of abundance spots
in chemically peculiar (CP) stars using the brightest star \uma\ as a case study.}
{We find that the best spectral regions to search for spots and stellar rotation signatures are in the visual domain. 
The spots can clearly be detected already at a first visibility lobe and 
their signatures can be uniquely disentangled from that of rotation.
The spots and rotation signatures {can also be detected} in NIR at low spectral resolution but
baselines longer than $180$~m are needed for all potential CP candidates.
According to our simulations,
an instrument like VEGA (or its successor e.g., FRIEND) should be able {to detect, in the visual, the effect} of spots and spots+rotation, provided
that the instrument is able to measure $V^2\approx10^{-3}$, and/or closure phase. 
In infrared, an instrument like AMBER but with longer baselines than the ones available so far
would be able to measure rotation and spots. Our study provides necessary details about strategies of spot detections and 
the requirements for modern and planned interferometric facilities essential for CP star research.}}
\end{abstract}

\section{Introduction}

Optical and infrared interferometry is a powerful observational technique capable of reaching a very high angular resolution.
Potentially, interferometry allows one to derive not only the sizes of stellar objects, but, under the conditions
of sufficient spatial coverage, even to reconstruct the details on the stellar surfaces.
{First} successful applications of interferometry to imaging the stellar surfaces
were carried out for supergiants \citep{2011ApJ...732...68C,2009ApJ...701..209Z,2007Sci...317..342M}
and giants \citep{2010A&A...511A..51C,2009A&A...496L...1L} that have large angular diameters because of their large sizes.

The application of interferometry to  main-sequence (MS) stars, unfortunately, is still limited to only brightest objects.
Modern facilities are capable of resolving the closest and/or largest MS stars and measure their diameters already on a regular
basis \citep[e.g.][]{2013ApJ...771...40B,2013MNRAS.434.1321M}. But the detailed study of surface morphology is still a challenging task for most of them.

Among MS stars there is one class of objects that obviously deserves interferometric attention~--~chemically peculiar (CP) stars.
These stars possess strong abundance inhomogeneities in their atmospheres where atoms and ions of certain elements tend to accumulate 
at different regions on the stellar surface driven by diffusion processes \citep{1970ApJ...160..641M}. 
The abundance inhomogeneities on CP stars are
usually obtained by a mapping technique known as Doppler Imaging (DI) 
\citep[e.g.][]{1958IAUS....6..209D,1977SvAL....3..147G}
which restores the information about surface abundance
inhomogeneities (spots) from the rotational modulated profiles of spectral lines 
\citep[e.g.][for details and some practical applications of the method]{1993PASP..105.1415P,2010A&A...509A..43L,2004A&A...424..935K}.
Note that the details of inversion problem used to recover spot and rotation information is out of scope of this paper.

Recently, interferometry has been successfully applied to CP stars resulting in a first estimate of the radii
of a few stars: $\alpha$~Cir \citep[HD~128898,][]{2008MNRAS.386.2039B}, $\beta$~CrB  \citep[HD~137909,][]{2010A&A...512A..55B},
$\gamma$~Equ \citep[HD~201601,][]{2011A&A...526A..89P}, and $10$~Aql \citep[HD~176232,][]{2013A&A...559A..21P}.
Detailed studies of surface morphology of CP stars, however, remain challenging and have never been done so far.

{It should be stressed that, in spite of the recent success of DI technique in {recovering} surface 
structures of stars, its application is limited to stars rotating fast enough so that rotation dominates
the broadening of spectroscopic lines. In other words, the faster the rotation is, the more spectroscopic features
can be resolved for a given resolving power of the spectrograph and thus more details on the stellar surface
can be studied. However, there are CP stars that rotate very slowly or have small projected rotational velocities, 
but still show significant spectral variability indicating existence of spots in their atmospheres 
\citep[e.g.,][]{2001A&A...377L..22K,2008MNRAS.389..441F}. 
For those stars no DI {is} possible, and interferometry thus appears to be a promising technique to study their 
surface morphology. Interferometry also allows one to derive 
both the inclination and position angle of stellar rotational axis
if sufficient spectral resolution and baseline configurations are provided.}

Besides the very few observations available, some authors explored the applications of interferometry to CP stars 
by means of numerical experiments. In the most recent study by \citet{2004A&A...422..193R}  authors used two
cases of well-known  CP stars $\alpha^2$~CVn and $\beta$~CrB to explore the possibility to derive the abundance and magnetic 
maps with interferometric instruments based on fringe phase signals. They concluded that
signals from abundance inhomogeneities and magnetic field could be in principle
 detected in visual and infrared already with modern instruments.

In this work we follow similar ideas outlined in \citet{2004A&A...422..193R}, but concentrating on the detection of abundance spots
in the one of the brightest CP star \uma. This star has a radii of about $4.2\Rsun$ \citep{2003A&A...406.1033L} 
and distance of $25.31$~pc, which results in one
of the largest angular diameter $\theta=1.54$~mas among closest CP stars, and which is easily resolved with modern interferometric instruments.
However, contrary to the previous work, we aim to explore
the behaviour of wavelength dispersed visibility and closure phase signals, while \citet{2004A&A...422..193R} 
simulated the differential fringe phase.

Our research is based on accurate model atmospheres that predict intensities from abundance maps  published
in \citet{2003A&A...406.1033L}. The same model atmospheres were successfully used in \citet{2010A&A...524A..66S} 
to predict the observed light variability of \uma\ in
narrow and broad-band photometric filters.


\section{Overview of modern and planned interferometric facilities}

It is essential to compare theoretical predictions that we make in this investigation against
available and planned interferometric facilities around the world. Important characteristics
of these instruments are the angular resolution, wavelength domain 
where instrument operate, spectral resolution provided, 
as well as the sensitivity of detectors in terms of limiting stellar magnitude. 
Table~\ref{tab:inters}
lists the major facilities and instruments already available and ones that will become available 
in nearest future to the scientific community. This table does not include instruments that operate
in mid-infrared range because CP stars are very faint in there.
Most of the information have been adopted from links listed in the OLBIN
(Optical Long Baseline Interferometry News) website\footnote{http://olbin.jpl.nasa.gov}
and references provided therein. In the table, we list four major interferometers
that provide long (on the order of tens of meters and longer) baselines:
VLTI (Very Large Telescope Interferometer,  Cerro Paranal, Chile), CHARA (Center
for High Angular Resolution,  Mount Wilson, California, USA),
SUSI (Sydney University Stellar Interferometer, Australia), and
NPOI (Navy Precision Optical Interferometer, Lowell Observatory, Arizona, USA).

\begin{table*}

\begin{minipage}{\textwidth}

\renewcommand{\thefootnote}{\alph{footnote}}

\caption{Interferometric facilities around the world.}

\label{tab:inters}

\begin{center}

\begin{tabular}{llclllllc}
\hline

Facility  & Instrument & Apertures & Sensitivity\footnotemark[1], $V^2_{\rm min}$ & Baselines range, m                             & Wavelength        & Resolution, $\lambda/\Delta\lambda$ & Limiting magnitude & ref.\\

\hline

\multirow{3}{*}{VLTI}  & PIONIER  & 4    & $10^{-5}$  (11)   & \multirow{3}{*}{$11$~--~$140$\footnotemark[2] }         & $H$-band          & 40                      & 7.5 (ATs) & 1\\

                       & AMBER    & 3    & $10^{-4}$   (12)  &                                       & $HK$-bands        & $30$, $1500$, $12\,000$ & 9.0 (UTs)\footnotemark[3] & 2\\


                       
                       & \textbf{GRAVITY}  &  4         & $10^{-5}$  &                                    &  $K$-band         & 22, 500, 4\,000    & 10 (UTs) & 3\\

\hline

\multirow{7}{*}{CHARA} & JouFLU    & 2   & $10^{-2}$ (13)    & \multirow{7}{*}{$34$~--~$331$}        & $K$-band         & $6$                  & 6 & 4\\

                       & VEGA     & 4    & $10^{-2}$ (14) &                                   & $480$~--~$850$~nm & 6\,000, 30\,000        & 7.5, 4.5 & 5,6\\
                       
                       & \textbf{FRIEND}\footnotemark[4] & 4    & $10^{-3}$ &                                   & $480$~--~$850$~nm & 6\,000, 30\,000        & 7.5, 4.5 & 14\\
                       & MIRC     & 6    & $10^{-4}$ (15) &                                      & $H$-band        & 40                     & 4.5 & 7\\

                       & CLASSIC   & 2   & $>10^{-3}$\footnotemark[6]  &                                   & $HK$-bands        & Broad band      & 8.5 & 8\\
                       
                       & CLIMB    & 3    & $10^{-3}$  (16) &                                      & $K$-band        & Broad band      & 6.5 & 8\\
                       
                       & PAVO     & 3    & $10^{-3}$  (17) &                                     & $650$~--~$800$~nm & 30                     & 8.0 & 9\\



\hline

SUSI                  &  PAVO        & 2     & $10^{-3}$  (17) & $5$~--~$160$, \textbf{up to} $\mathbf{640}$                          & $550$~--~$800$~mn & Broad band             & 7.0 & 9\\

\hline

NPOI                   &  \textbf{VISION}        & 6\footnotemark[5]       & $\sim$\footnotemark[7] &  $17$~--~$437$                         & $570$~--~$850$~mn & $200$~--~$1000$            & 5 & 10\\

\hline

\end{tabular}

\end{center}

\begin{flushleft}
\par\medskip
col.~1~--~name of the facility; col.~2~--acronym of the instrument;
col.~3~--~number of apertures; col.~4~--~baseline range; col.~5~--~spectral region;
col.~6~--~spectral resolution;
col.~7~--~limiting magnitude at the lowest spectral resolution
col.~8~--~reference to the article describing the instrument
\par\medskip
Future new {or expected upgrades of the available present} interferometric facilities are marked by bold font.
\par\medskip
\textbf{References:}
(1)~\citet{2011A&A...535A..67L}; (2)~\citet{2007A&A...464....1P}; (3)~\citet{2008SPIE.7013E..69E};
(4)~\citet{2013JAI.....240005S}; (5)~\citet{2009A&A...508.1073M}; (6)~\citet{2011A&A...531A.110M};
(7)~\citet{2004SPIE.5491.1370M};
(8)~\citet{2013JAI.....240004T}; (9)~\citet{2008SPIE.7013E..63I};
(10)~\citet{2012AAS...21944613G}; (11)~\citet{Montarges};
(12)~\citet{2009A&A...503..183O}; (13)~\citet{2009A&A...503..521M};
(14)~\citet{berio2014}; (15)~\citet{2012ApJ...761L...3M};
(16)~\citet{2011ApJ...728..111O}; (17)~\citet{2013MNRAS.434.1321M}

\footnotetext[1]{f{supplementary references to the instrument sensitivity values are given in brackets}}
\footnotetext[2]{baselines span from $11$~m to $140$~m for ATs (but the limiting magnitudes are brighter). Baselines span from $47$~m to $130$~m for UTs}
\footnotetext[3]{limiting magnitude is given for the seeing smaller than 0.8$\arcsec$}
\footnotetext[4]{{FRIEND is a successor of VEGA}}
\footnotetext[5]{first fringes only with $4$ apertures}
\footnotetext[6]{{CLASSIC is worse than CLIMB because the measurement principle is the same but it has only $2$ telescopes and cannot perform boot-strapping}}
\footnotetext[7]{{there are no published astrophysical results with VISION so the sensitivity has not been quantified (note that $V^2_{\rm min}\approx10^{-3}$ 
is routinely reached in modern optical interferometry)}}
\end{flushleft}

\end{minipage}

\end{table*}

\section{Methods and simulation setup}

The simulations provided in this paper are based on custom numerical routines {we have} coded in IDL language that
compute the complex visibility from a 2D intensity image of the source for the desired set of (u,v) points \citep{2008SPIE.7013E.134L}.

The images of the star were constructed employing abundance maps of elements Ca, Cr, Fe, Mg, Mn, Ti, and Sr 
obtained with the help of DI technique in \citet{2003A&A...406.1033L}. 
{The \llm\ stellar model atmosphere code \citep{2004A&A...428..993S} was used to compute local model atmospheres.
The code computes 1D, LTE model atmospheres and can account for individual abundances of chemical elements.
The Stark-broadenined profiles of hydrogen lines were
computed using tables of \citet{1997A&AS..122..285L}  based on the VCS theory by \citet{1973ApJS...25...37V}.}

Local model atmospheres were calculated for each 
of $2244$ surface elements of original DI maps ($68$ longitudes and $33$ latitudes). Thus each model atmosphere 
takes into account local abundance pattern of all mapped elements, and the solar abundance from \citet{2009ARA&A..47..481A} was assumed for all other
chemical elements. 
The up-to-date version of VALD database \citep{1995A&AS..112..525P,1999A&AS..138..119K}
was used as a source of atomic line transition parameters (including transitions
originating from predicted energy levels).
Theoretical fluxes were computed in a wide wavelength range from visual and IR wavelength domains with a characteristic 
resolution of $R\approx500\,000$ in V-band and $R\approx20\,000$ in K-band respectively.
{Note that the star's  surface magnetic field is weak, on the order of a few hundred Gauss \citep{2000MNRAS.313..851W}, 
and therefore has negligible influence on the spectrum and interferometric observables that we modeled in this study.}

\begin{figure}
\centering
\includegraphics[width=\hsize]{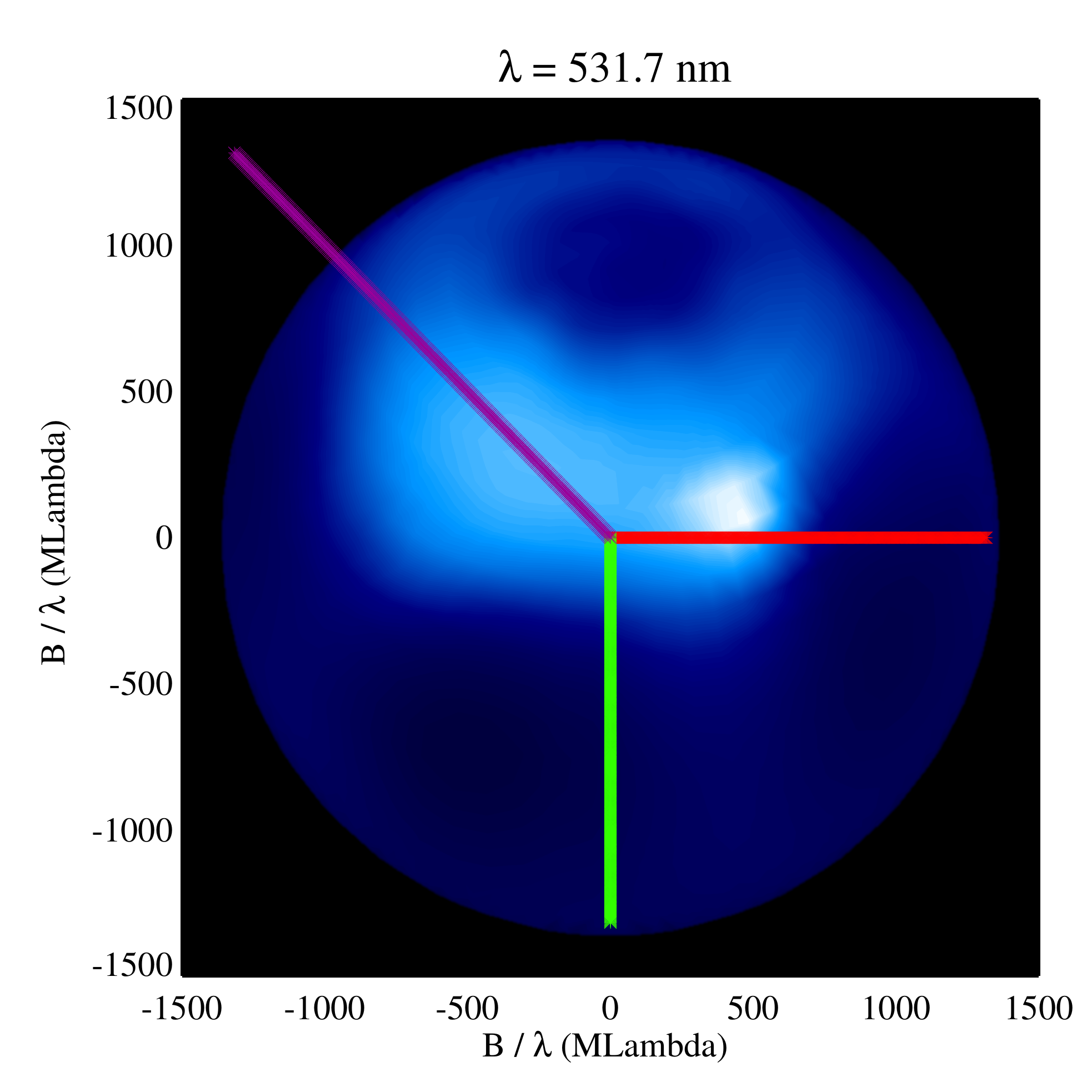}
\caption{Simulated intensity images of \uma\ at $\lambda=531.7$~nm and rotation phase $\phi = 0$.
Also shown are the three projected baselines which correspond to position angles of $0^\circ$ (red), $135^\circ$ (violet), and $270^\circ$ (green).}
\label{fig:looks-like}
\end{figure}

In order to obtain accurate interferometric predictions from a 2D image some caution must be taken.
An image made of pixels is a discrete sampling of the true source intensity distribution on the sky.
Such a discrete image necessarily has a maximum spatial frequency (Nyquist frequency) defined by twice the pixel size, 
while the true image of a circular object has frequencies up to infinity. 
Thus, if the pixel values represent the intensity at pixel center, its Fourier Transform at each frequency, and thus the visibility, 
will not exactly match the true value measured by the instrument on sky: the power in the over-Nyquist frequency 
will be folded in the sampled frequency range, a phenomenon called ``aliasing''. 
This problem can be mitigated by using two solutions (both of which we used): a very high resolution for the pixel image,
yielding very high computation time, or by smoothing the image by a suitable anti-aliasing filter, e.g. the Hanning one
which slightly changes the pixel values so that they represent much likely the integral of source flux over the pixel area,
rather than the intensity at center only. The residual aliasing error on the computed visibility can be roughly estimated 
by assuming that the true model does not have high contrast features  at a scale smaller than the chosen pixelization, 
which means that, on average, the visibility above Nyquist frequency  is decreasing with increasing frequency. 
With this assumption we can say that the average aliasing error on the visibility is of the order of the visibility value at Nyquist frequency, or less. 
We thus used this value as a check on our image resolution to finally choose a $680\times680$  surface elements for the surface modeling,
i.e. $\approx20$ times finer pixel sampling than the original $33\times33$ image.
The average error on $V^2$ was found to be of the order of $10^{-8}$, well below the effects we want to measure.


In all simulations we assumed a maximum baseline of $700$~m with the sampling every two meters. This baseline is projected onto the sky
by employing the three position angles of $0^\circ$, $135^\circ$, and $270^\circ$. 
As an example, the stellar intensity image
at the center of the strong \ion{Fe}{ii} line, as well as positions of the three projected baselines, 
are illustrated in Fig.~\ref{fig:looks-like}. One can clearly see the patchy distribution of surface brightness caused by 
inhomogeneously distributed abundances.

The calculation of a complex visibility at a given spatial frequency requires the application of the Van~Citter-Zernike theorem to the original
intensity image. This must be done for each monochromatic wavelength, position angle, and baseline sampling point, and thus
may easily result in enormous computing time required. In order to keep the latter within affordable limits we decided to restrict ourselves 
to a few characteristic spectral intervals containing pronounced spectroscopic features of metallic lines.
Also, the spectral resolution was degraded to match that of the modern interferometric instruments 
which in all cases is much lower than the resolution provided by our synthetic spectra.
Specifically, in the visual wavelength domain the following spectral windows and resolution modes have been chosen: 
$454.0$--$464.0$~nm, $524.0$--$534.0$~nm, and $605.0$--$613.5$~nm with $R=6\,000$ and $30\,000$; and $1000.0$--$2600.0$~nm covering the range of $J$, $H$, and $K$
infrared bands with $R=30$ plus a region $1000.0$--$1200.0$~nm with $R=6\,000$. 
The choice of spectral windows in visual wavelength domain is such because of a number of pronounced spectroscopic
features of Fe and Cr located in there. These features provide highest contrast between the stellar surface regions 
of enhanced and depleted abundances of these elements at corresponding wavelengths.

\section{Results}

According to theoretical predictions, the light variability in CP stars is controlled by the radiative flux redistribution from UV to visual and IR regions
caused by inhomogeneous abundance contents. As star rotates, regions of enhanced or depleted abundances move in and out of view, 
and this produces a characteristic light variations seen in different photometric
filters. The amplitude of this variability is wavelength dependent and, in general, decreases from visual to IR 
\citep[e.g.][]{2010A&A...524A..66S}.
Consequently, the spot contrast
is larger in UV and visual wavelength domains and {becomes} dimmer towards longer wavelengths.
Therefore UV and visual are the preferred regions to search for spot signatures. 
In addition, CP stars are main-sequence stars of types from late F to B, i.e. they radiate maximum of their flux at visual wavelengths 
thus favoring interferometric detectors operating in there. 
On the other hand, even if the flux contrast is weaker in IR region, it can still be large enough in individual 
spectral lines and potentially be detectable if sufficiently high spectral resolution is available.
Therefore below we test both these wavelength domains.

It is important to understand that the spot detection with interferometry does not necessary 
require observations of the star at different rotation phases, unlike, say, photometry.
It is still possible to detect spots from spectroscopy from a single spectrum because,
as discussed in e.g., \citet{2005A&A...439.1093K} and \citet{1999A&A...351..963R}, a spotted star would demonstrate a characteristic
deviation of the spectral line shapes from a pure rotational profile. 
However, the complete and unique characterization of spots (e.g., positions, numbers, shapes) 
requires time-series observations.
Considering that the abundance spots in CP stars are large-scale structures and spots of different elements are often found at 
different locations, a set of position angles would already be enough to look for characteristic changes 
in the interferometric visibility profiles at least in one particularly chosen rotation phase.
Therefore, in the case study of \uma, we will concentrate on the predictions made for a single rotational phase $\phi=0$ where the star
has regions of strong overabundance of most of mapped elements. 

\subsection{Visual wavelength domain}

\subsubsection{Visibility vs. baseline}

\begin{figure*}
\rotatebox{90}{
\begin{minipage}{\textheight}
\centerline{
\includegraphics[width=0.1\textwidth]{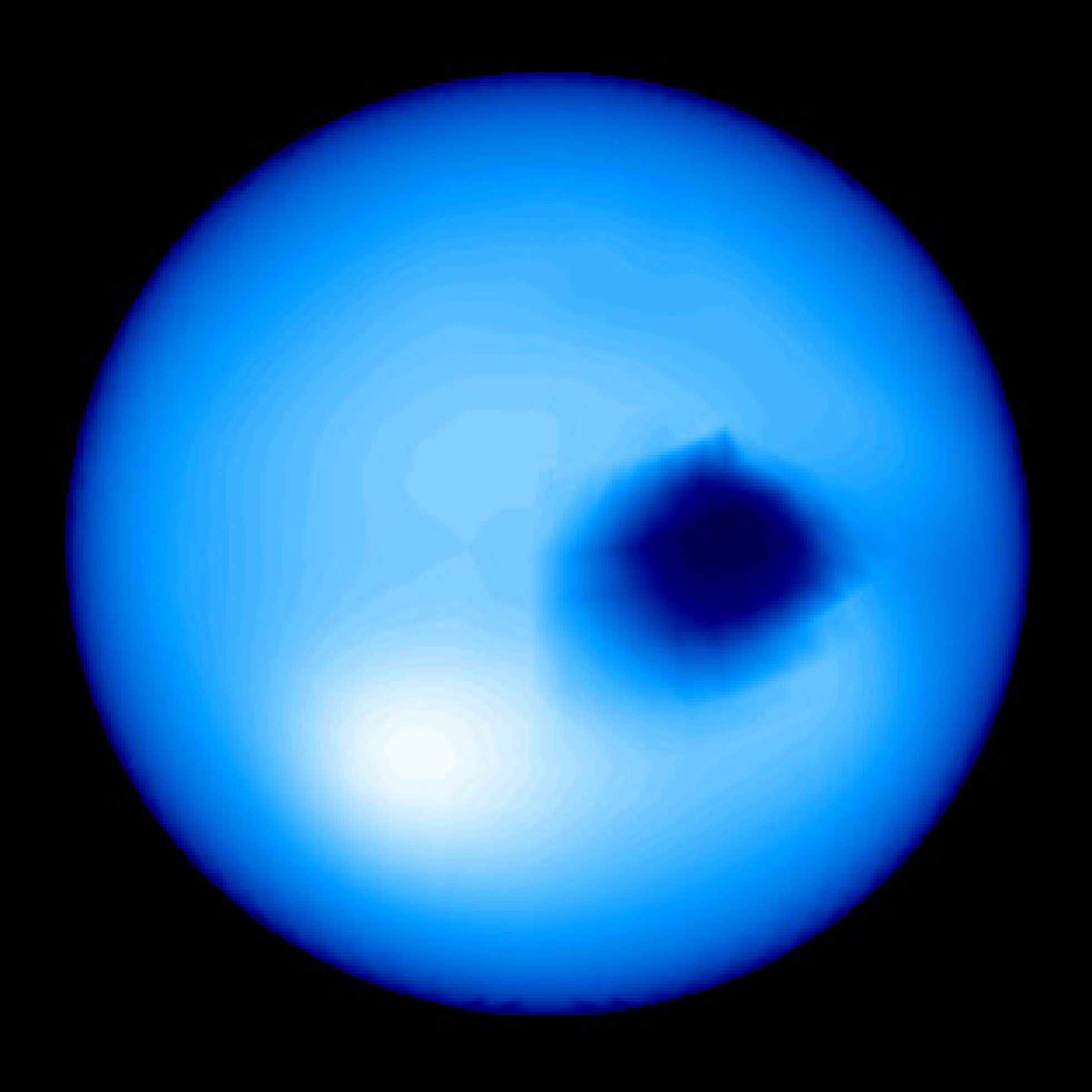}
\includegraphics[width=0.1\textwidth]{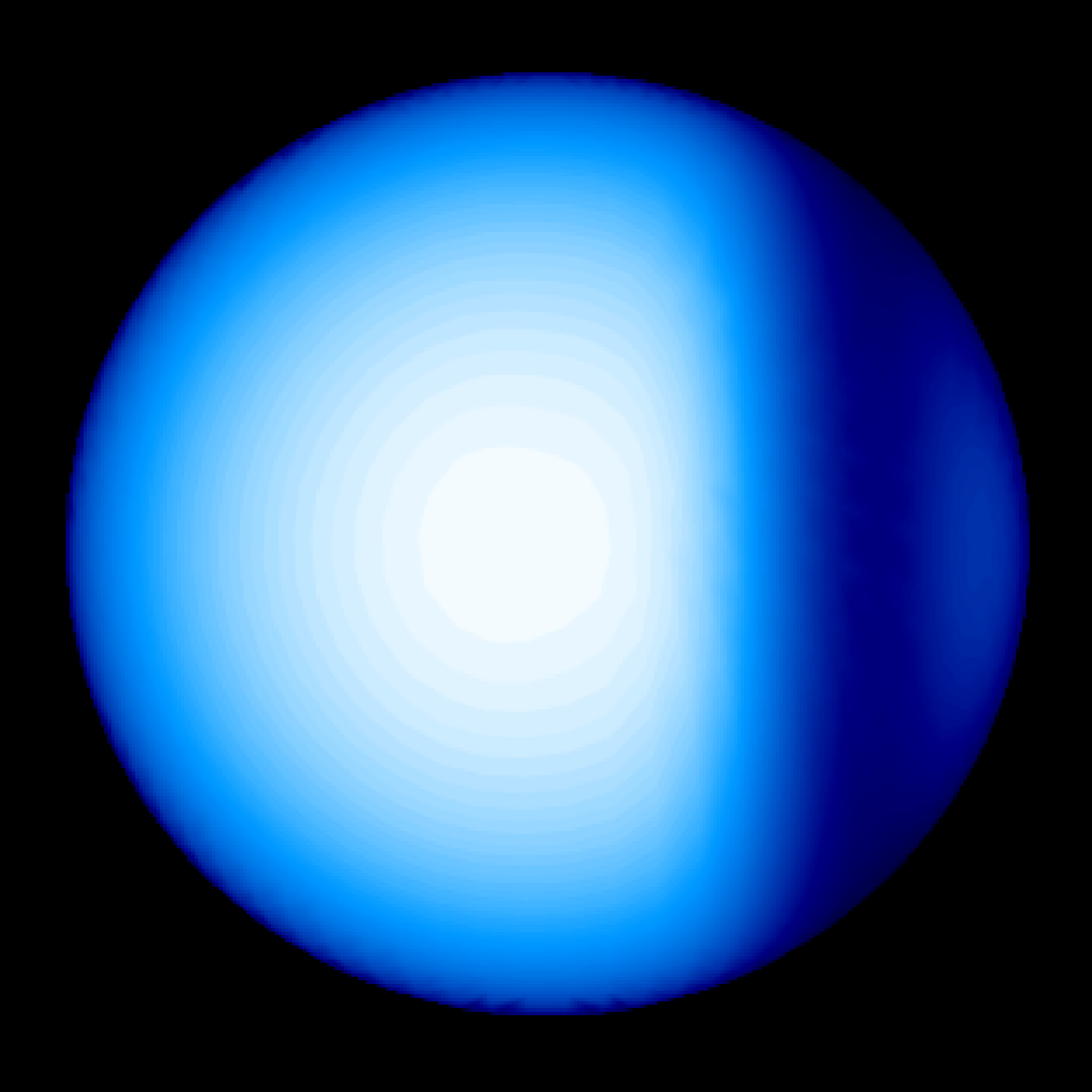}
\includegraphics[width=0.1\textwidth]{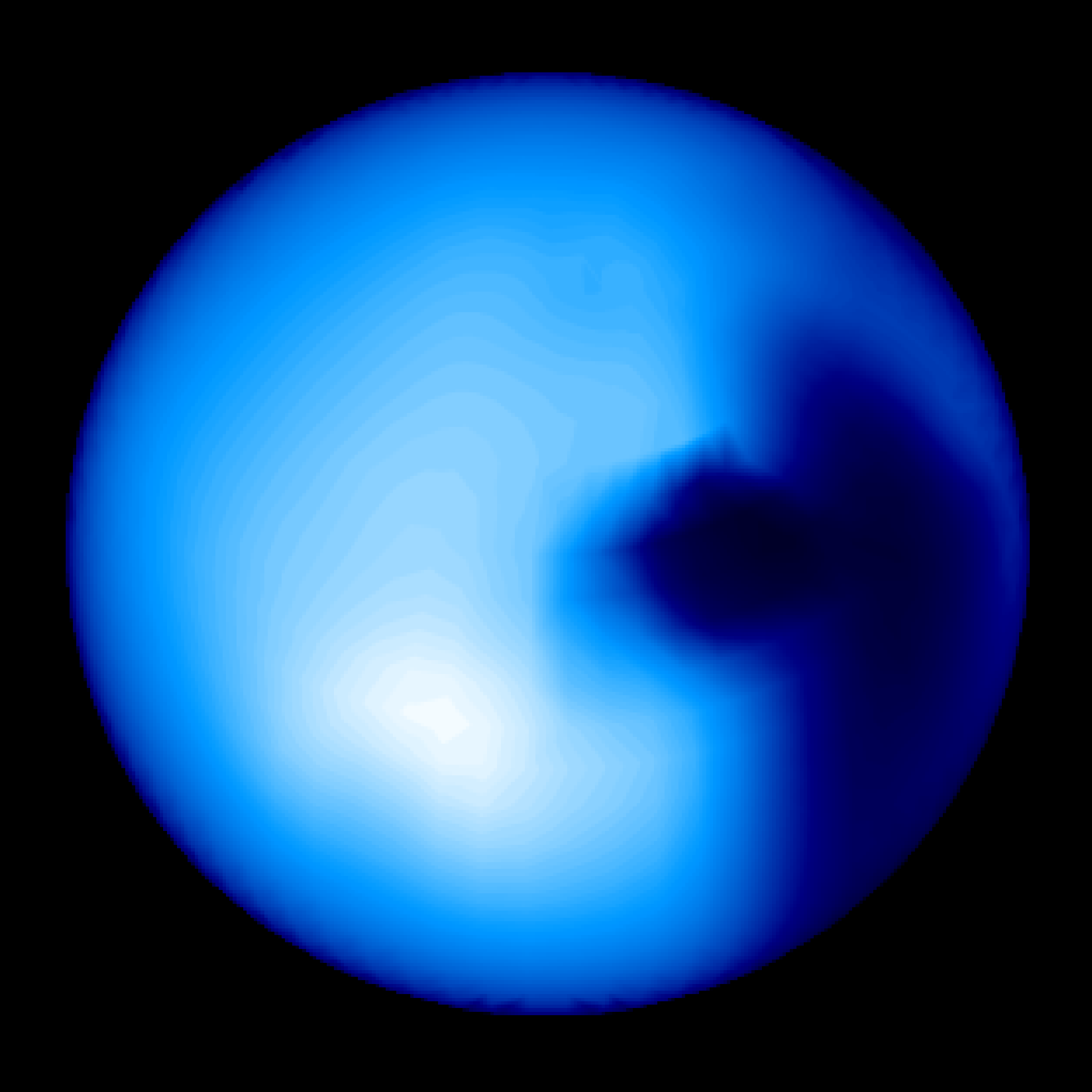}\hfill
\includegraphics[width=0.1\textwidth]{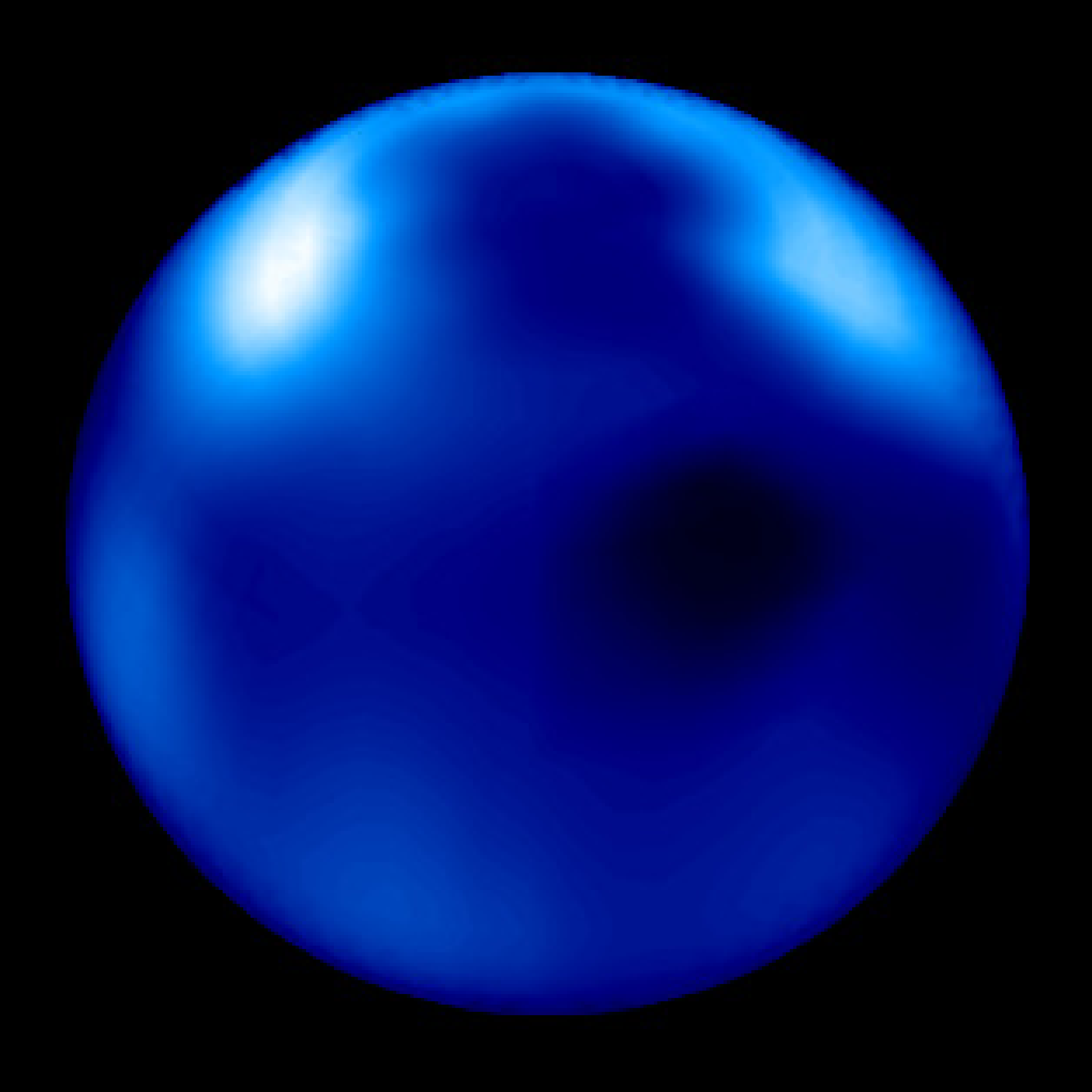}
\includegraphics[width=0.1\textwidth]{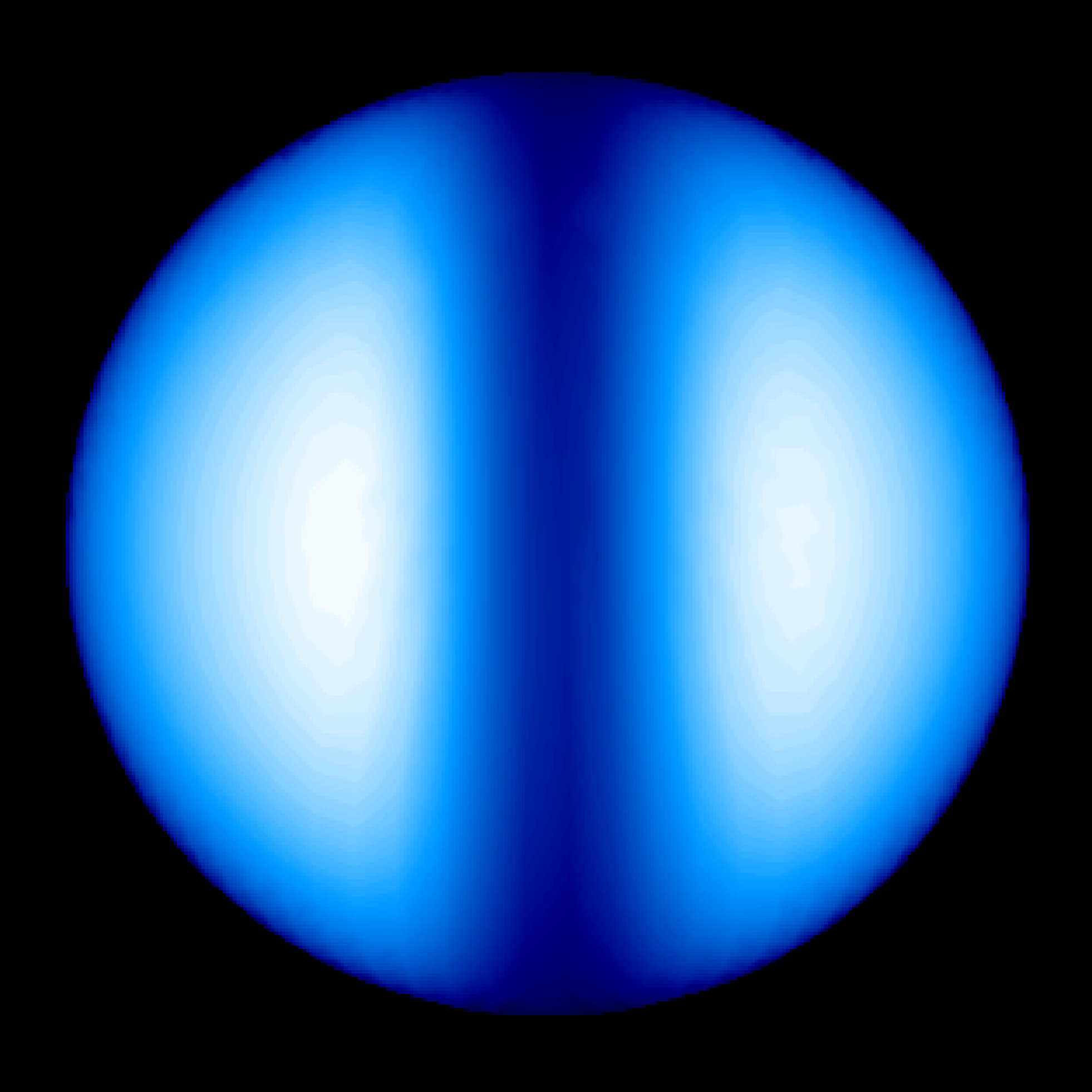}
\includegraphics[width=0.1\textwidth]{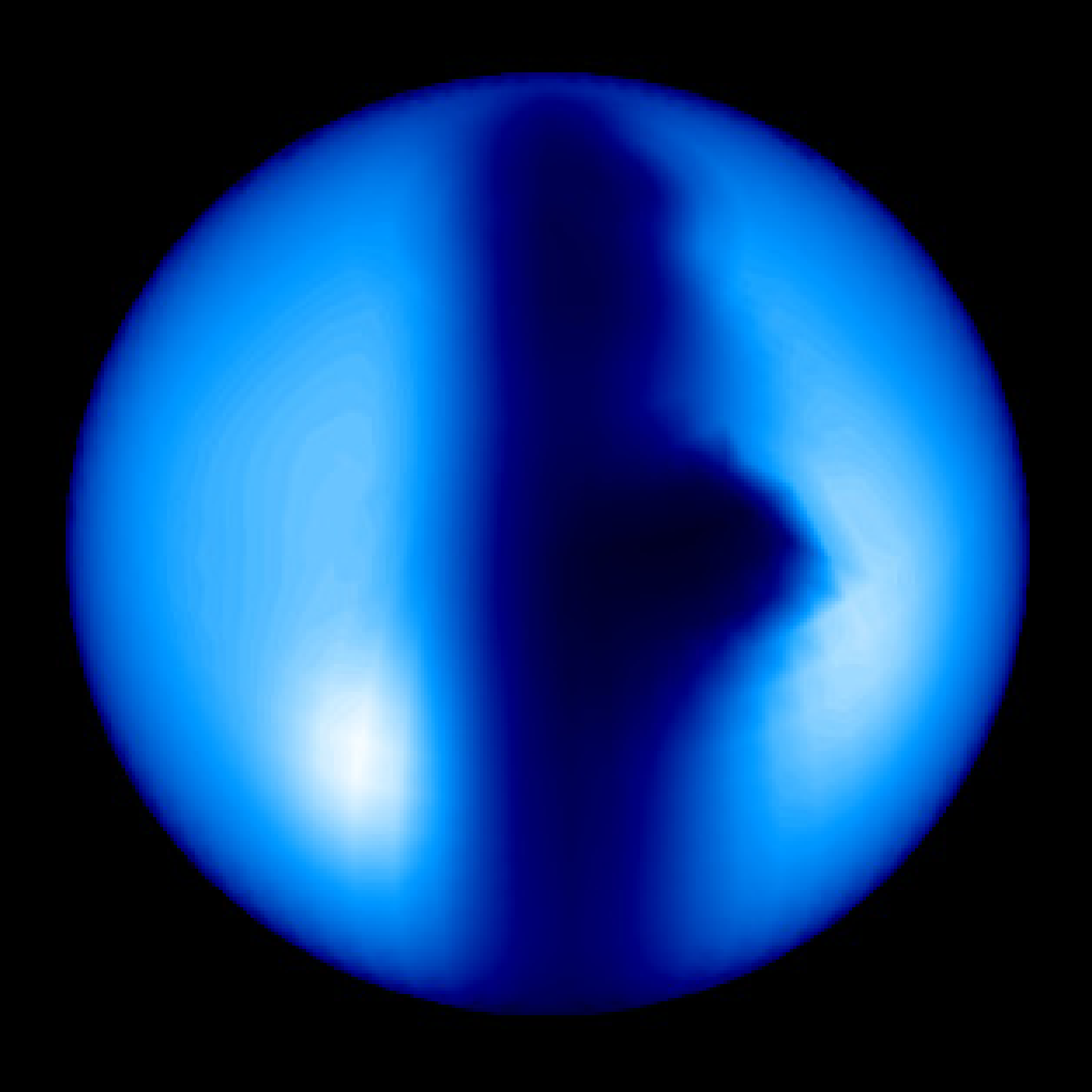}\hfill
\includegraphics[width=0.1\textwidth]{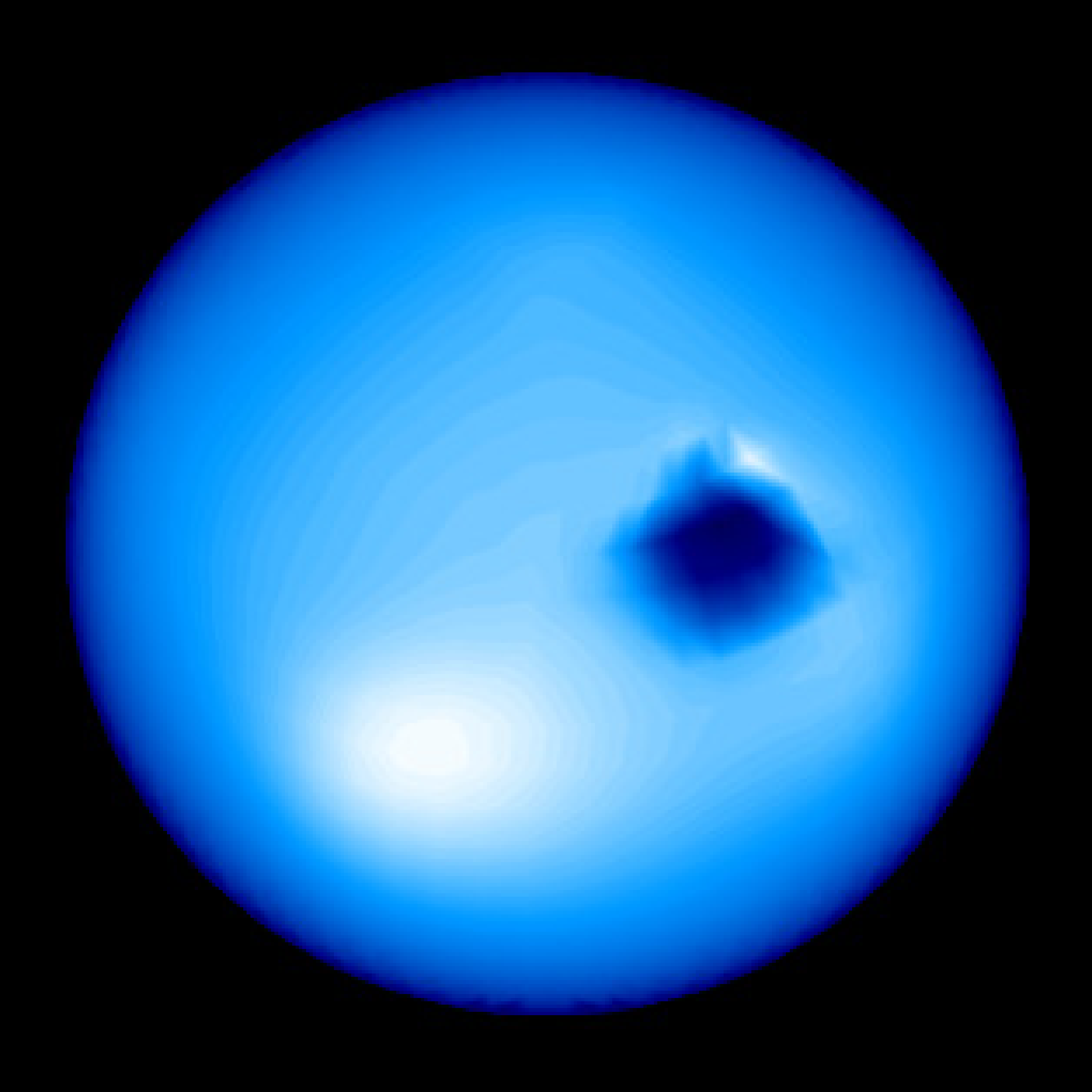}
\includegraphics[width=0.1\textwidth]{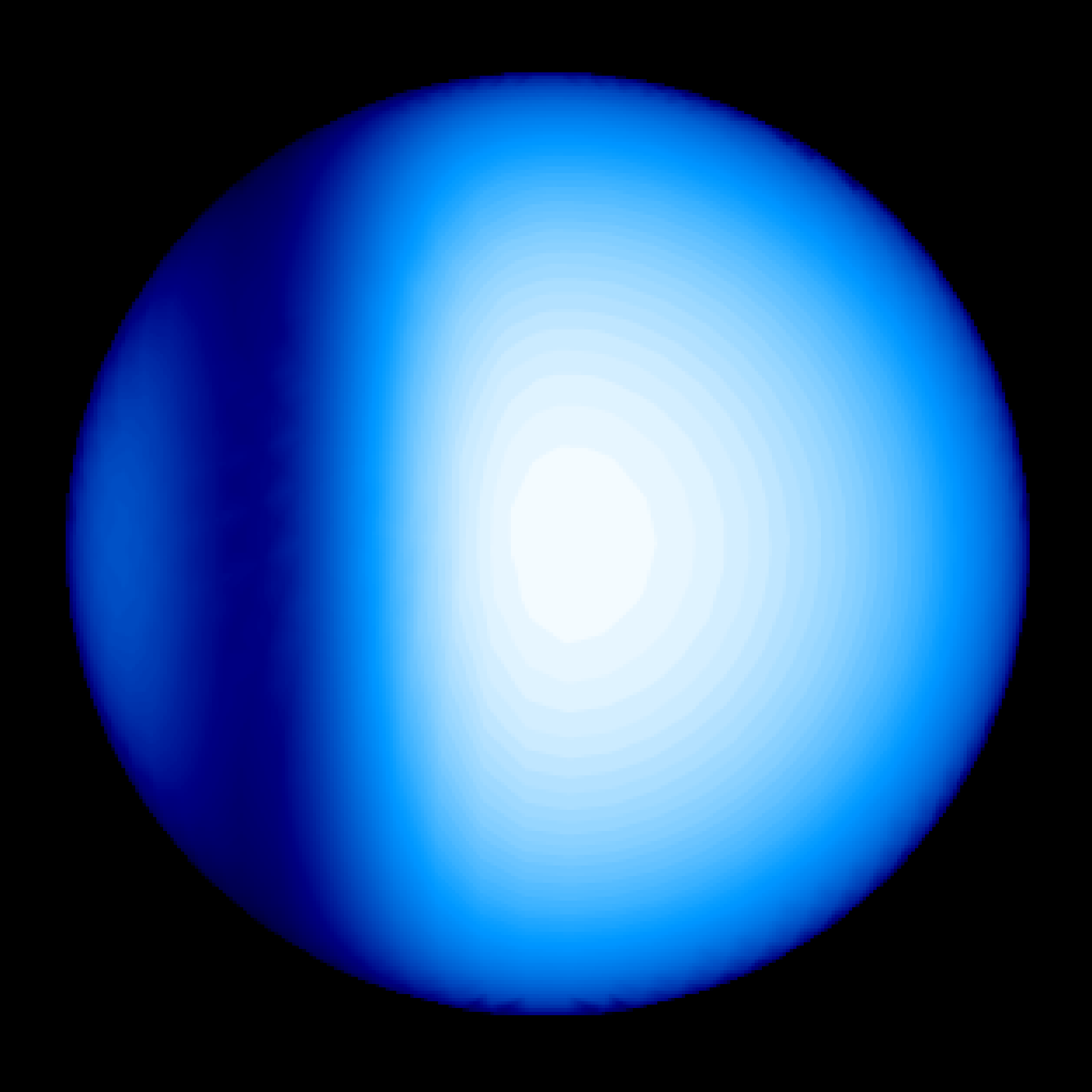}
\includegraphics[width=0.1\textwidth]{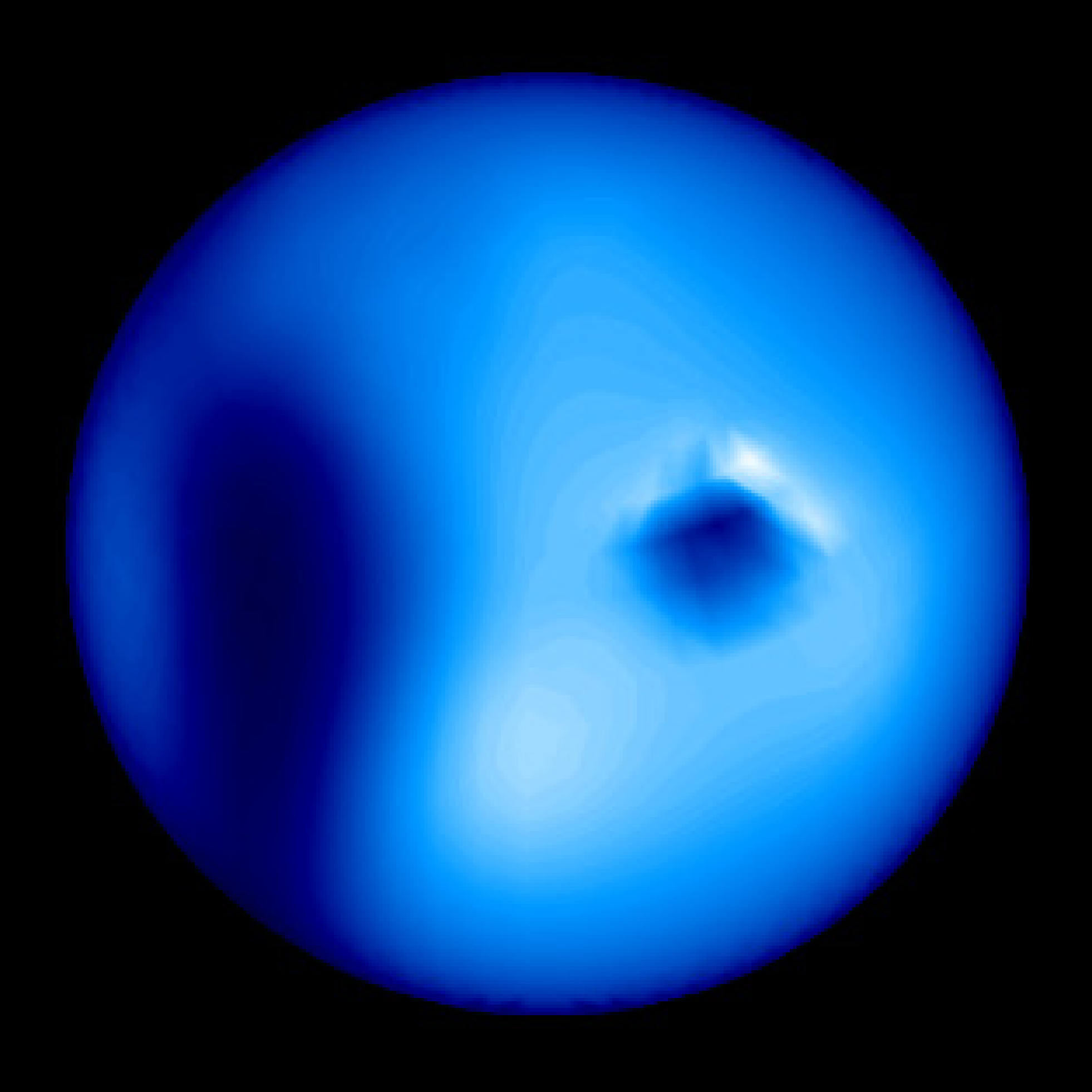}
}
\vspace{0.2cm}
\centerline{
\includegraphics[width=0.33\hsize]{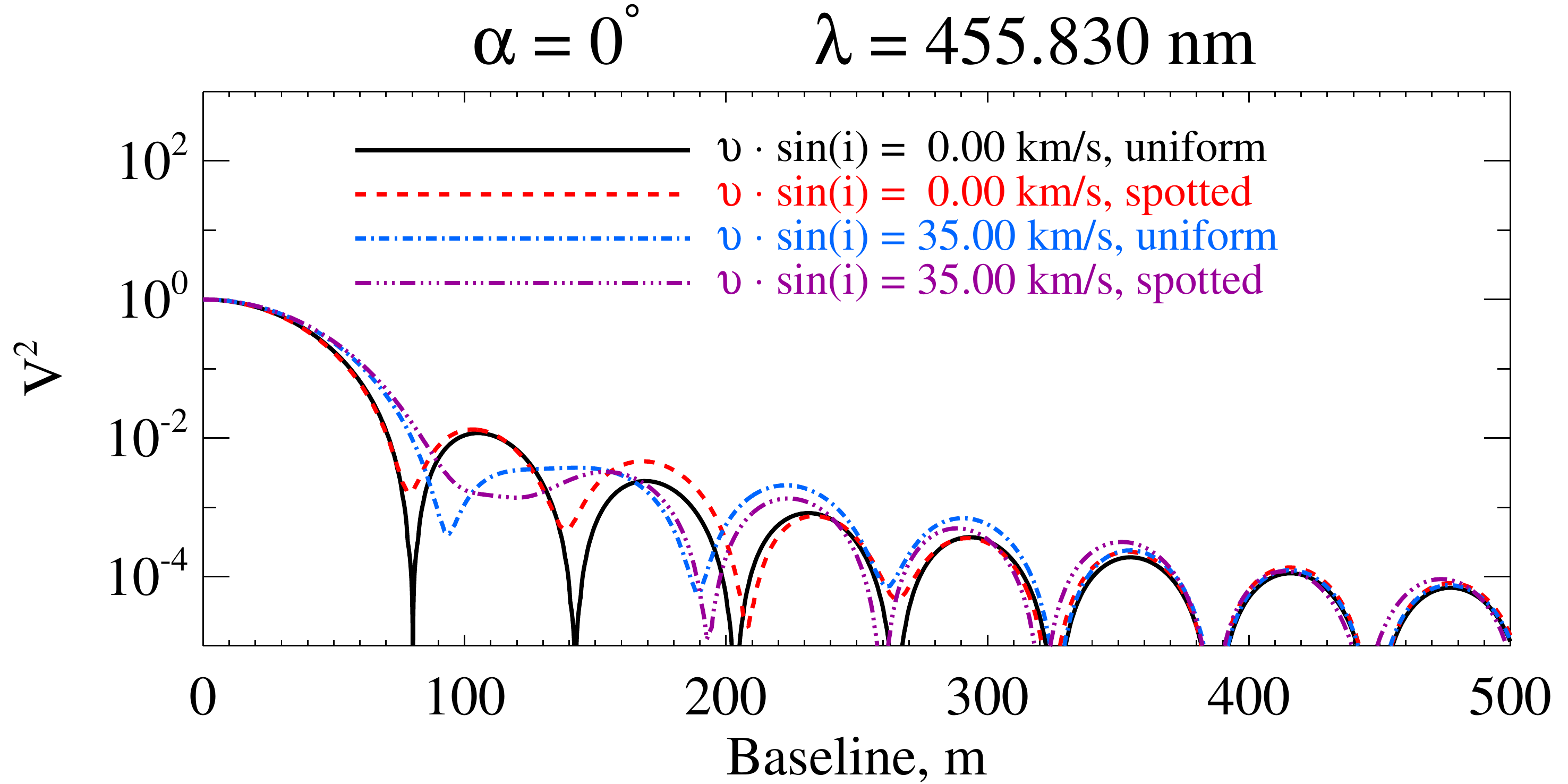}
\includegraphics[width=0.33\hsize]{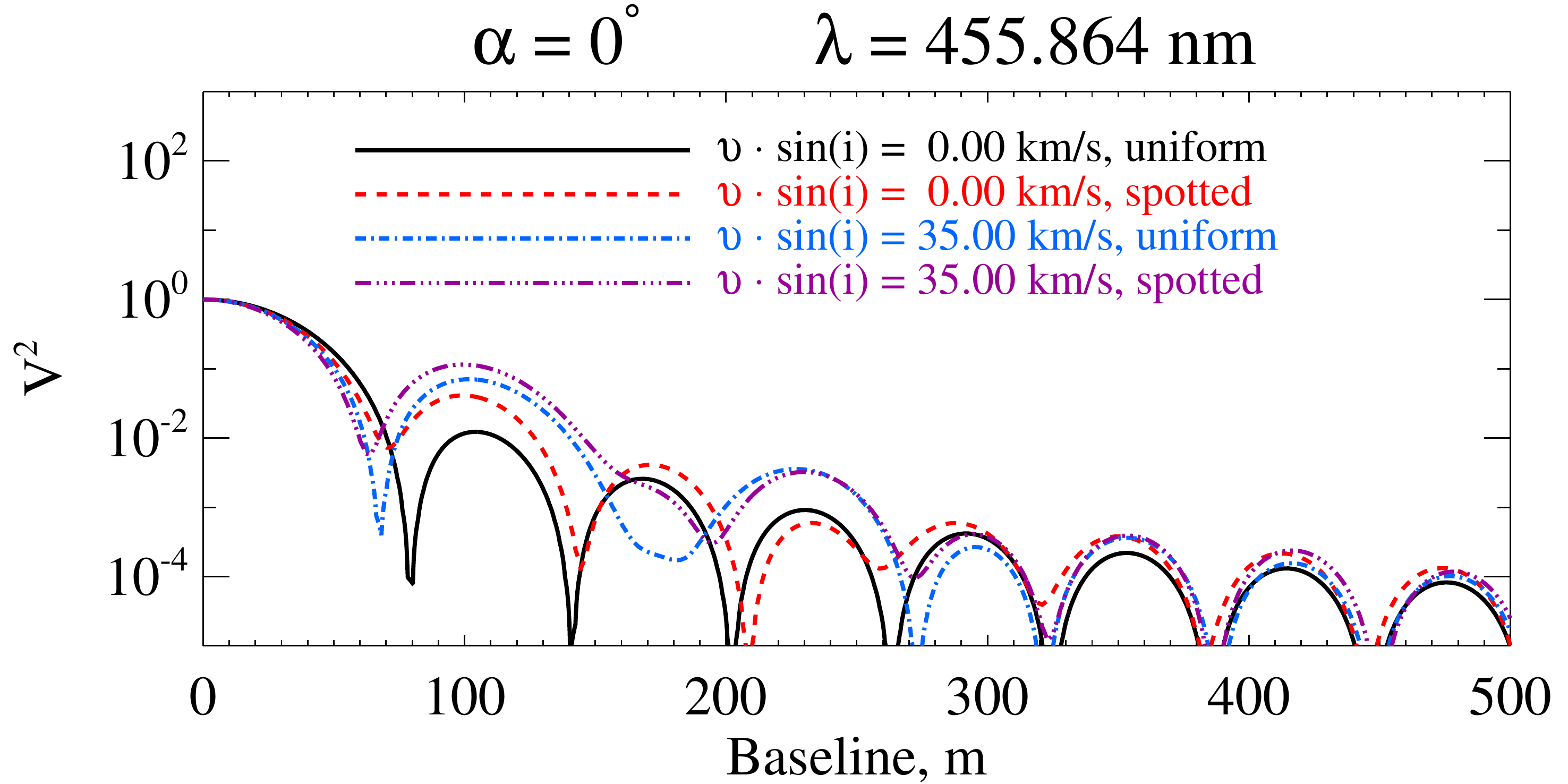}
\includegraphics[width=0.33\hsize]{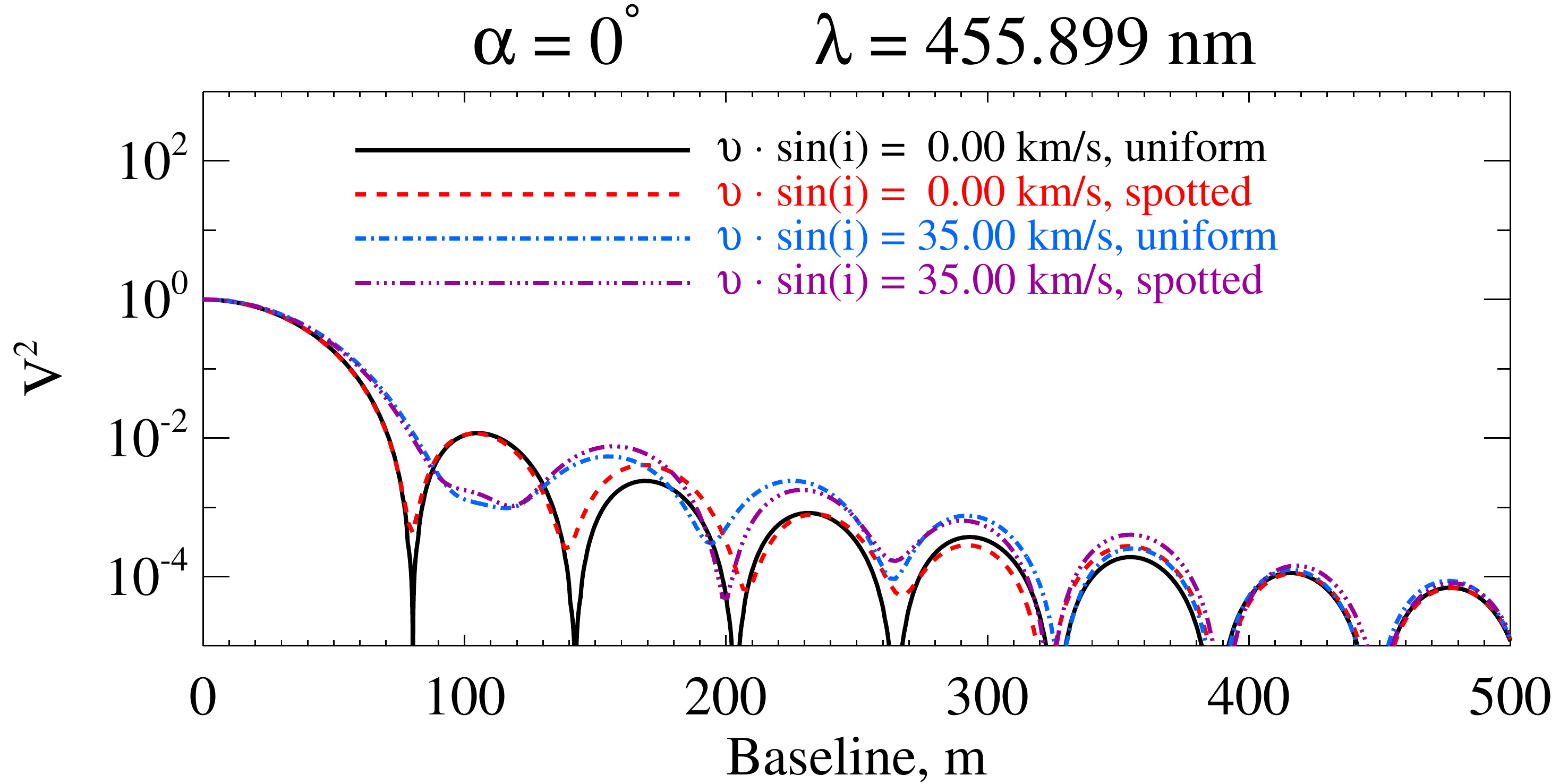}
}
\centerline{
\includegraphics[width=0.33\hsize]{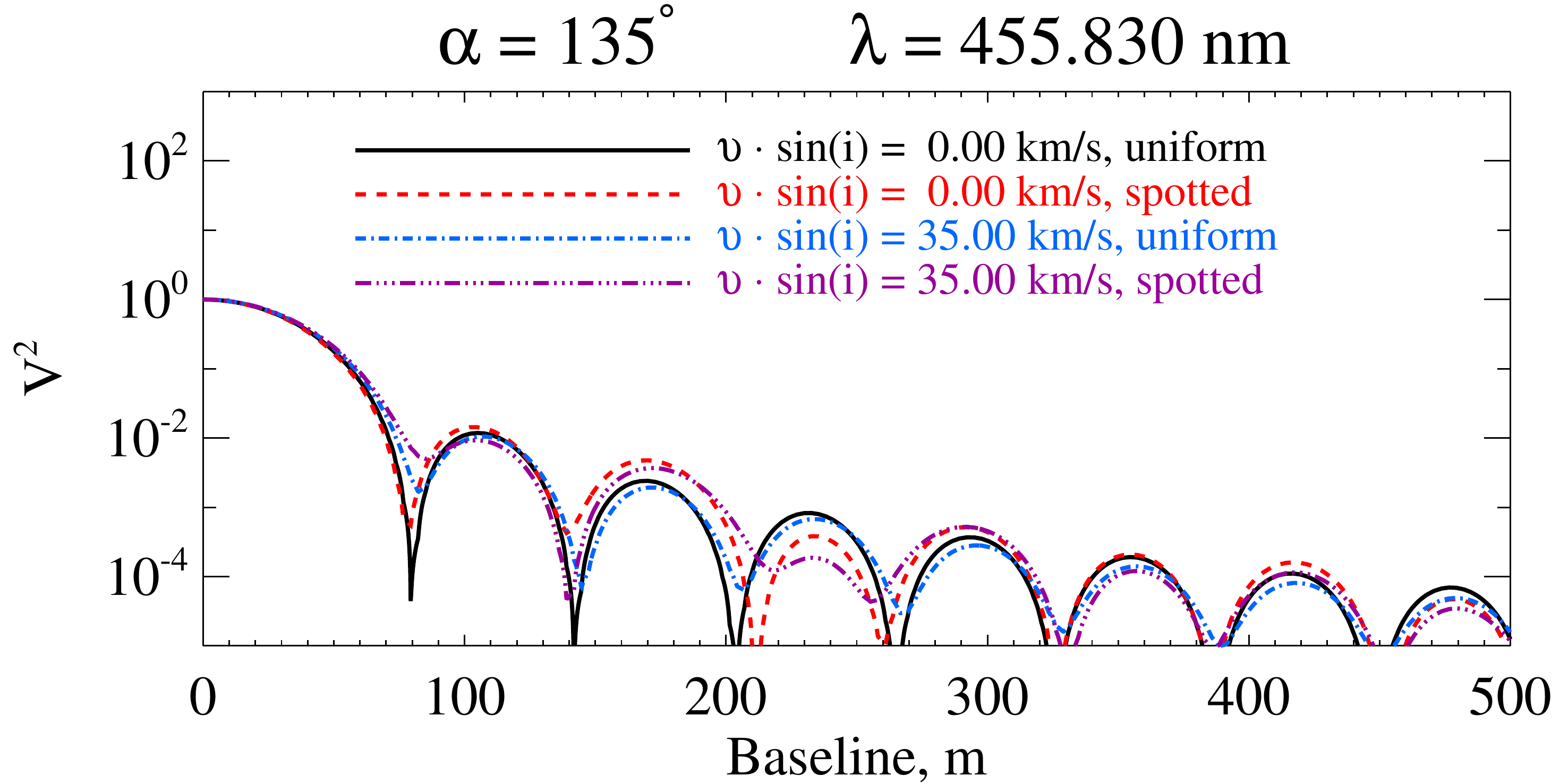}
\includegraphics[width=0.33\hsize]{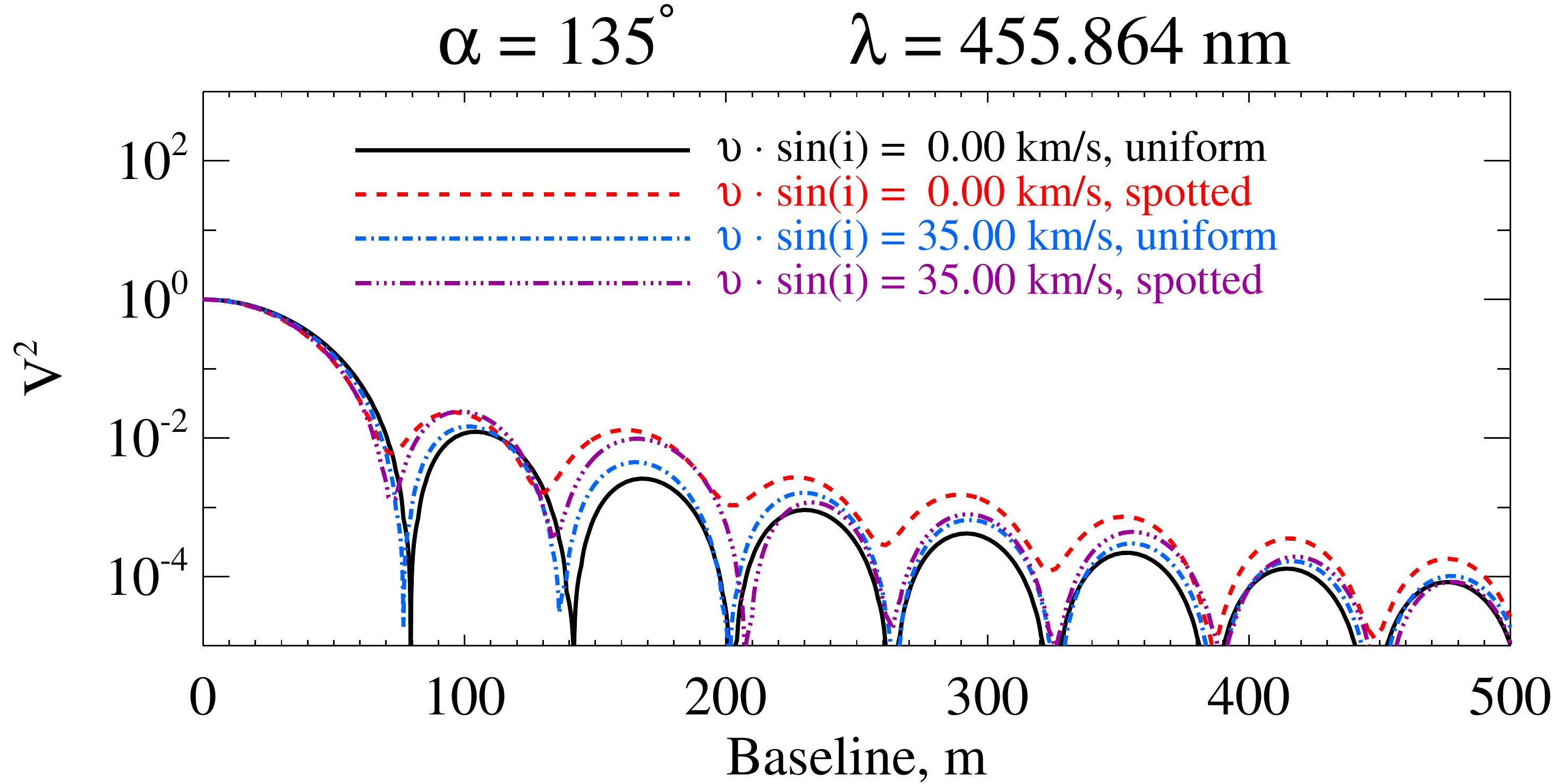}
\includegraphics[width=0.33\hsize]{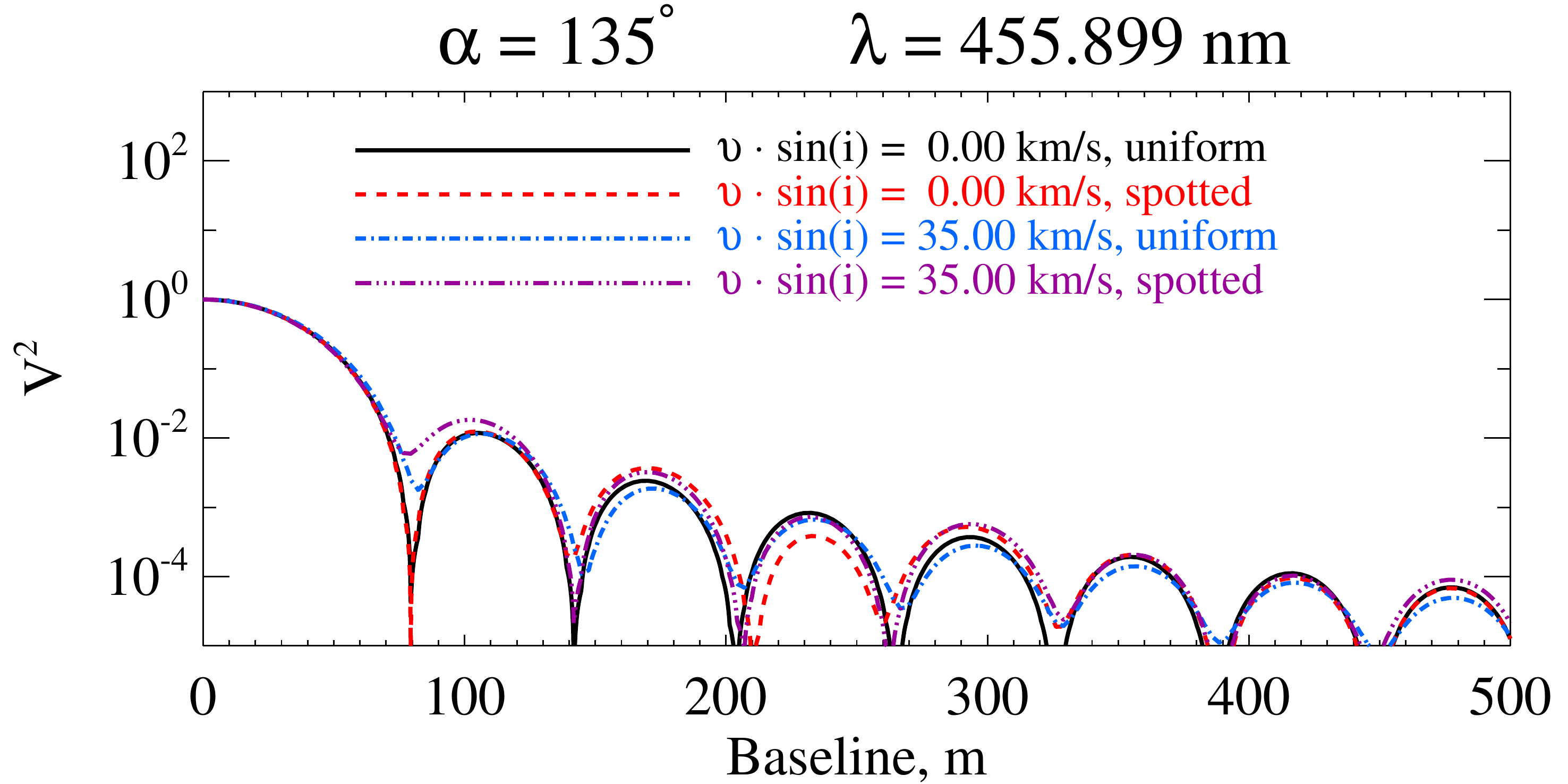}
}
\centerline{
\includegraphics[width=0.33\hsize]{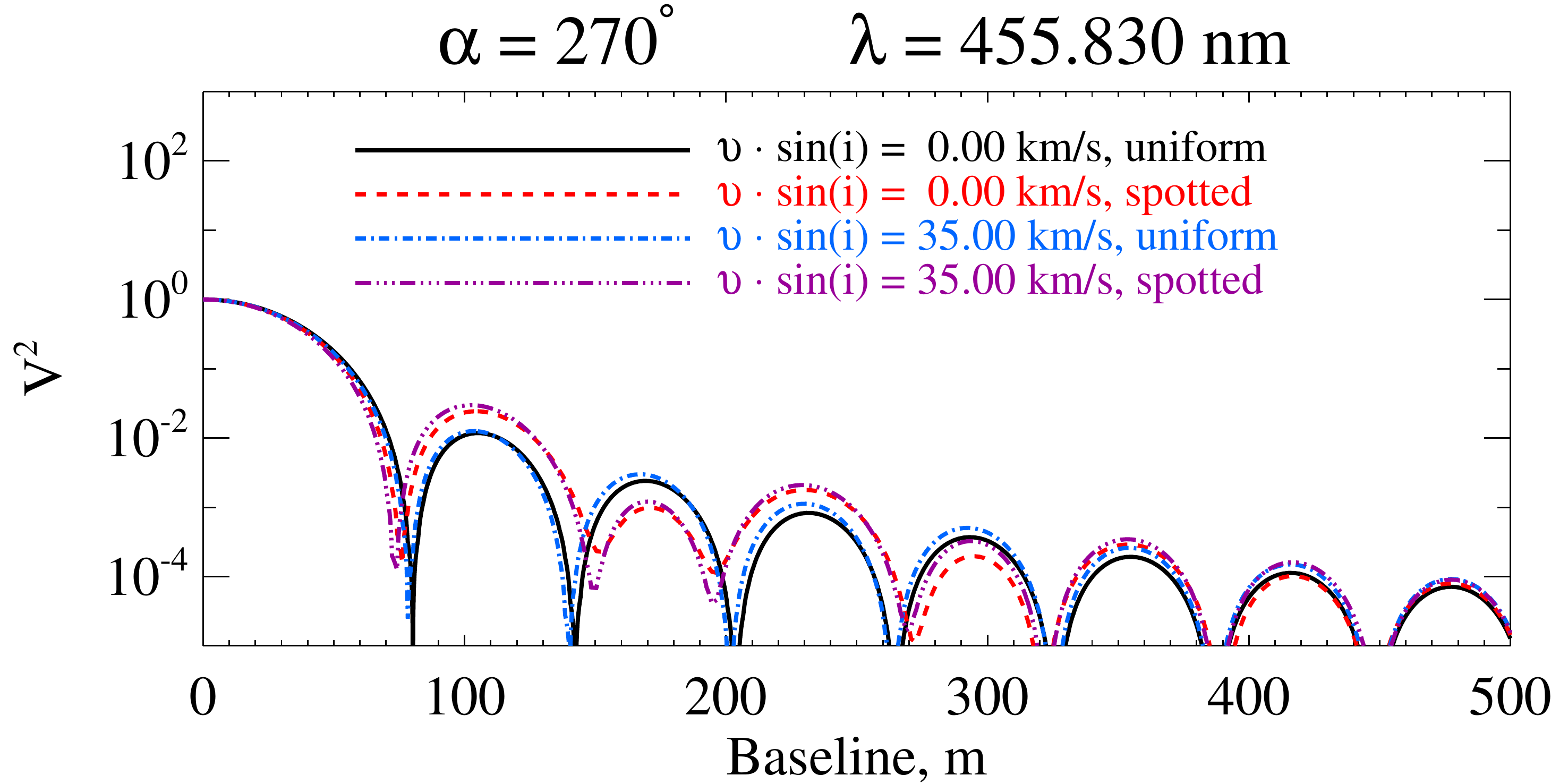}
\includegraphics[width=0.33\hsize]{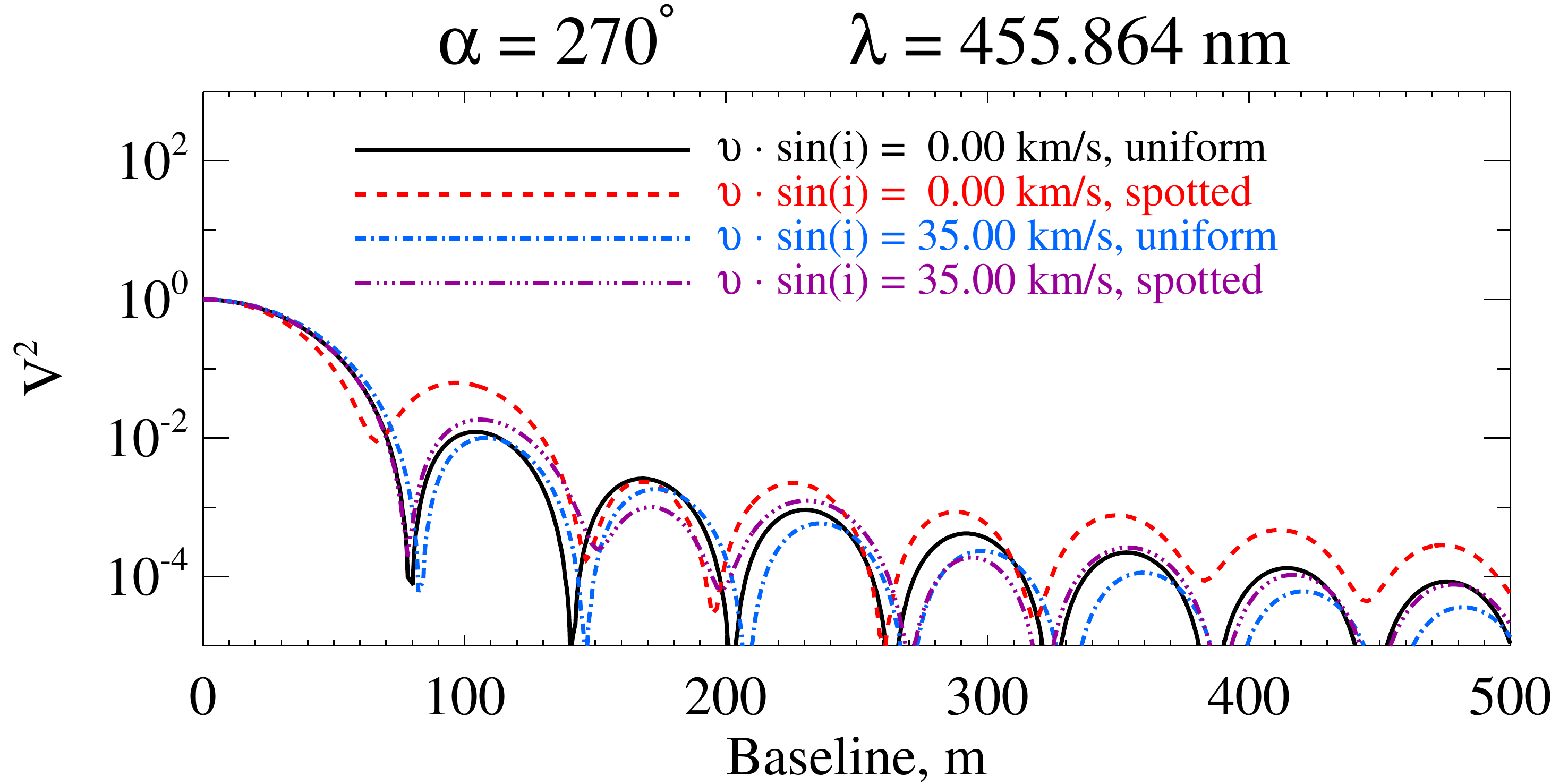}
\includegraphics[width=0.33\hsize]{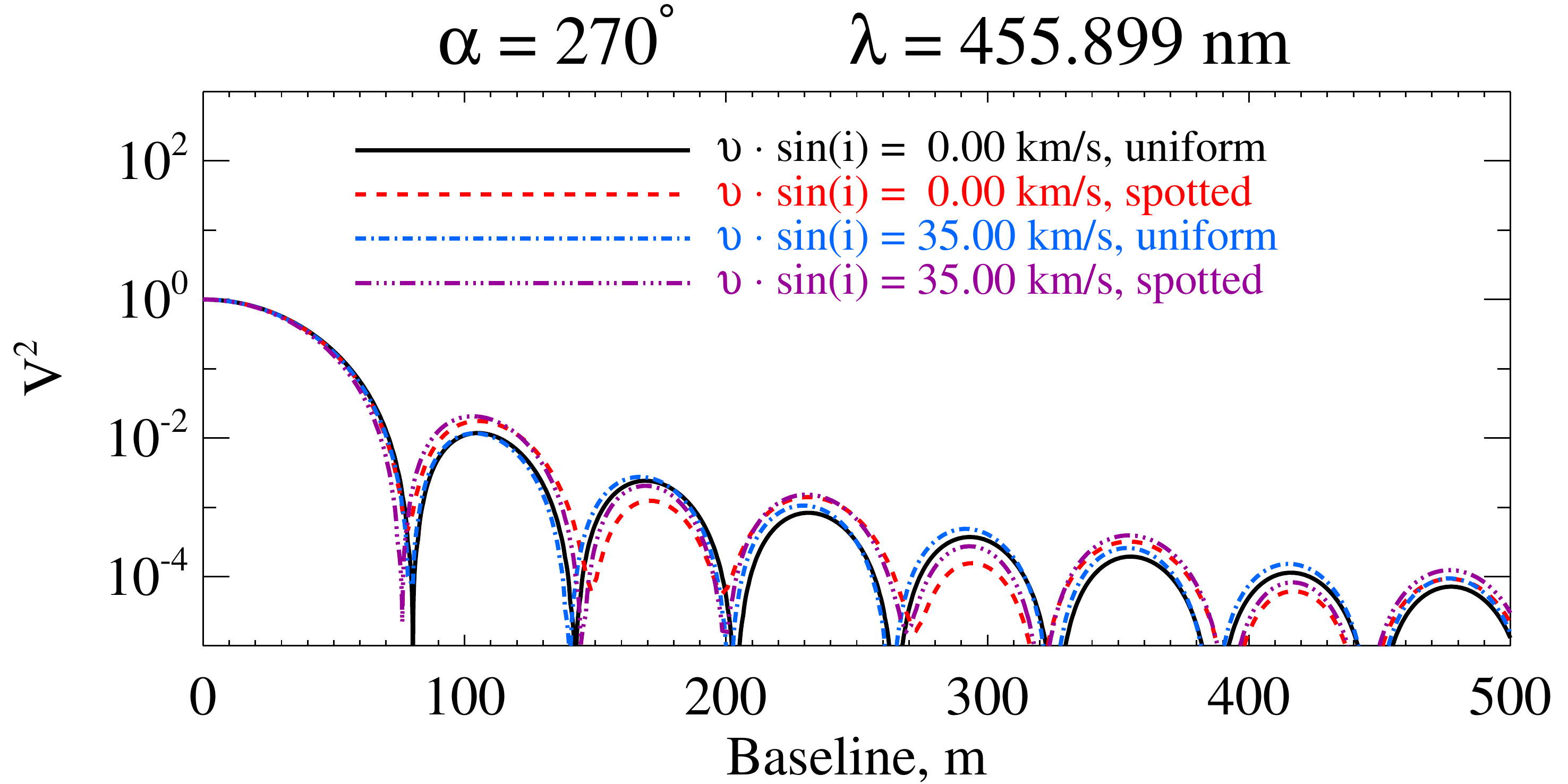}
}
\caption{Stellar intensity images and {squared} visibility  at three wavelengths (three columns from left to right) 
around \ion{Cr}{ii}~$455.86$~nm line and spectral resolution $R=30\,000$. 
The three images on top of each column show normalized intensity distributions of the star
computed assuming (from left to right) $\vsini=0$~\kms\ with spots, $\vsini=35$~\kms\ without spots, 
and $\vsini=35$~\kms\ with spots, respectively.
Three rows with {squared} visibility  plots correspond to position angles
of (from top to bottom) $0^\circ$, $135^\circ$, and $270^\circ$ respectively. Calculations are shown 
for the case of $\vsini=0$~\kms\ and $\vsini=35$~\kms\ for homogeneous and spotted stellar surfaces (see plot legends).}
\label{fig:vis-mono-v-1}
\end{minipage}
}
\end{figure*}

Visual part of CP stars' spectrum contain many strong lines of Cr and Fe that are (together with Si) the major
opacity sources \citep{2007A&A...469.1083K}, as was confirmed by recent studies of light variability of CP stars
{\citep{2012A&A...537A..14K,2010A&A...524A..66S,2009A&A...499..567K,2007A&A...470.1089K}}.
Other elements do not affect or have only marginal influence on the continuum flux
and therefore the spots of these elements can be seen only in corresponding spectral lines
at medium or high spectral resolution.
Any of these lines can be subject of interferometric investigation. 
However, at high spectral resolution, in cases like \uma, 
the stellar rotation starts to affect visibility signal in spectral lines.
The reason for this is that the stellar brightness even in the spotless case 
is not homogeneous in monochromatic light any more, and is represented
by a dark stripe of constant $\vsini$.
This stripe is shifted across the stellar surface by Doppler effect when looking
at different wavelengths inside a profile of a spectral line, as shown in the top panel of
Fig.~\ref{fig:vis-mono-v-1} for the center of strong \ion{Cr}{ii}~$455.86$~nm line
and two additional wavelengths blue- and red-ward from the line center respectively.

The remaining plots in Fig.~\ref{fig:vis-mono-v-1} 
show the predicted  {squared} visibility vs. baseline for the different maps.
As expected, the largest difference in {squared}  visibility 
between uniform and spotted surfaces for the case of zero rotation is predicted in the line core
for the $\alpha=0^\circ$ and $\alpha=270^\circ$ because of one large and a few smaller spots seen on the stellar surface
(middle column of Fig.~\ref{fig:vis-mono-v-1}).
In this case, spots can already be detected at $V^2\lesssim0.3$.
The characteristic signature of spots can be recognized with a modulation 
of the $V^2$  that does not go to zero before the third lobe.
The effect of spots when analyzing the first visibility lobe is to make the star look
larger compared to the case without spots (i.e. shift of the fist visibility minimum towards shorter baselines).
This effect is more pronounced for the case of zero rotation, but can also be seen at the line core
when $\vsini=35$~\kms.

The pure rotational effect (black line vs. blue in Fig.~\ref{fig:vis-mono-v-1}) 
is strong and observable clearly at $\alpha=0^\circ$, because this {position angle} is 
perpendicular to the stripe caused by Doppler shift. 
The visibility curve does not go to zero in the first lobe of visibility, 
and the effect is clearly visible even in the first lobe. 
If rotation and spot are summed, the rotation dominates on the spot, 
and the latter is detectable at a shorter wavelength $455.830$~nm and $\alpha=0^\circ$, 
however at substantially lower visibility $V^2\lesssim10^{-2}$. 
This information cannot be generalized as it depends on the
brightness of the spot, and on the rotation rate.

\begin{figure*}
\rotatebox{90}{
\begin{minipage}{\textheight}
\centerline{
\includegraphics[width=0.1\hsize]{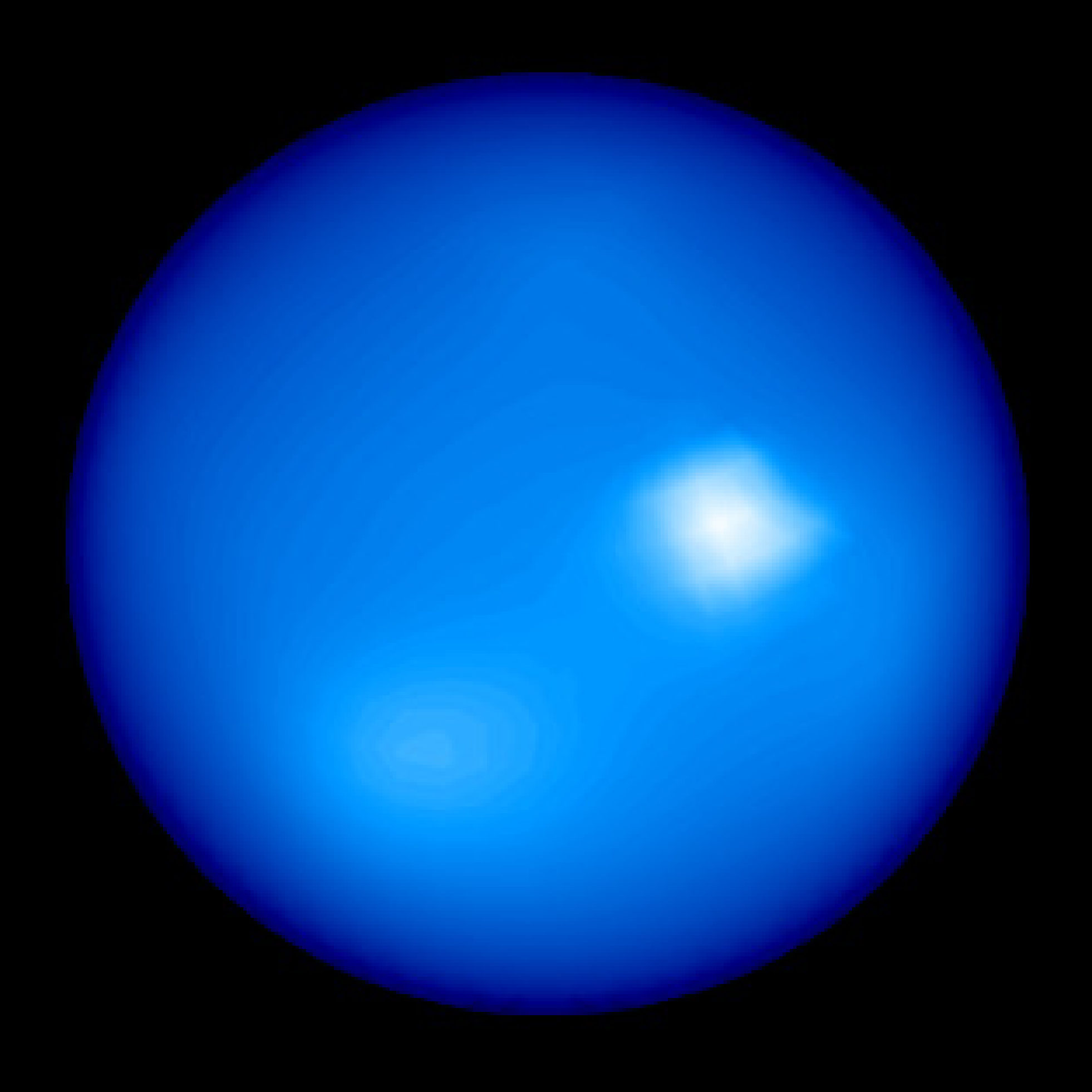}
\includegraphics[width=0.1\hsize]{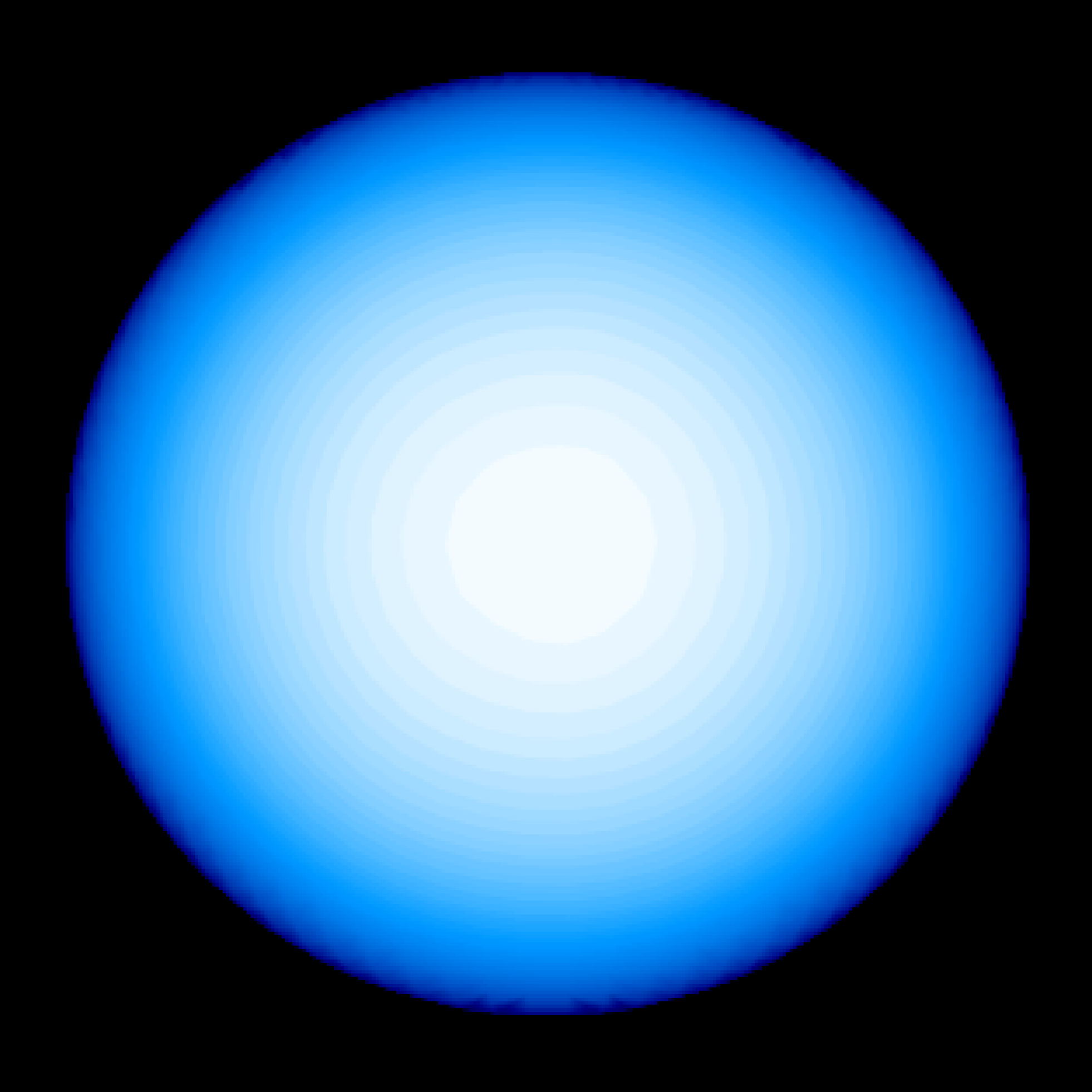}
\includegraphics[width=0.1\hsize]{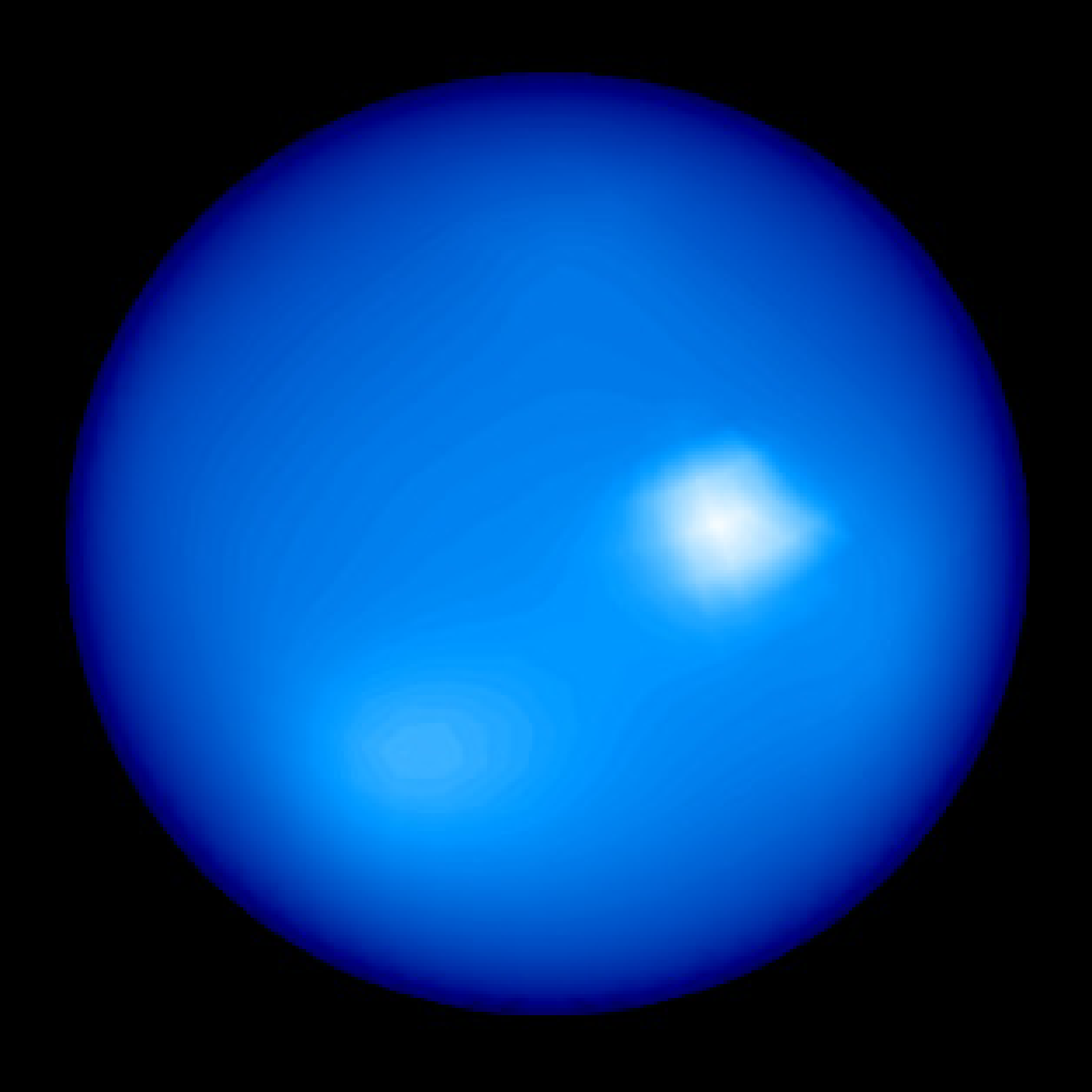}\hfill
\includegraphics[width=0.1\hsize]{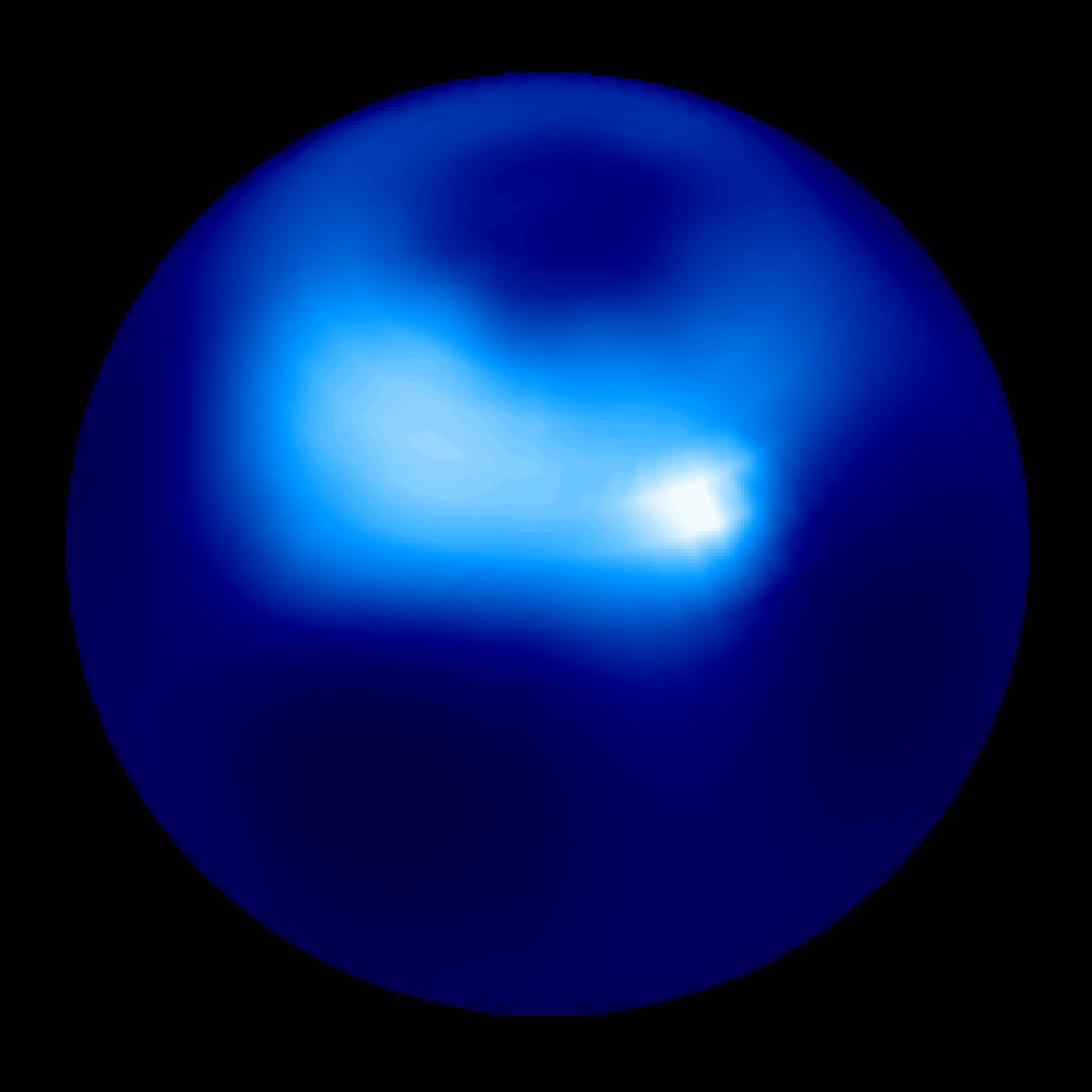}
\includegraphics[width=0.1\hsize]{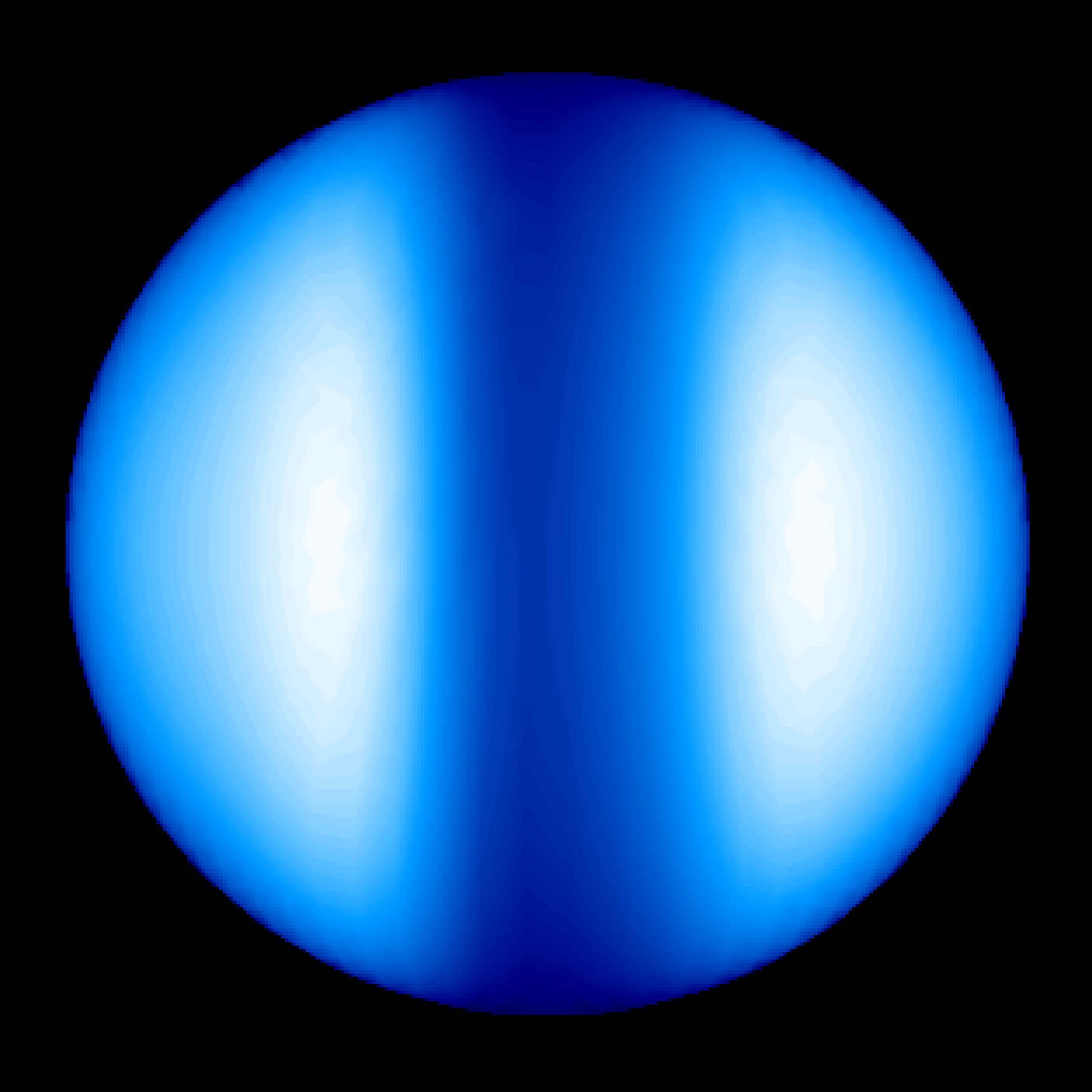}
\includegraphics[width=0.1\hsize]{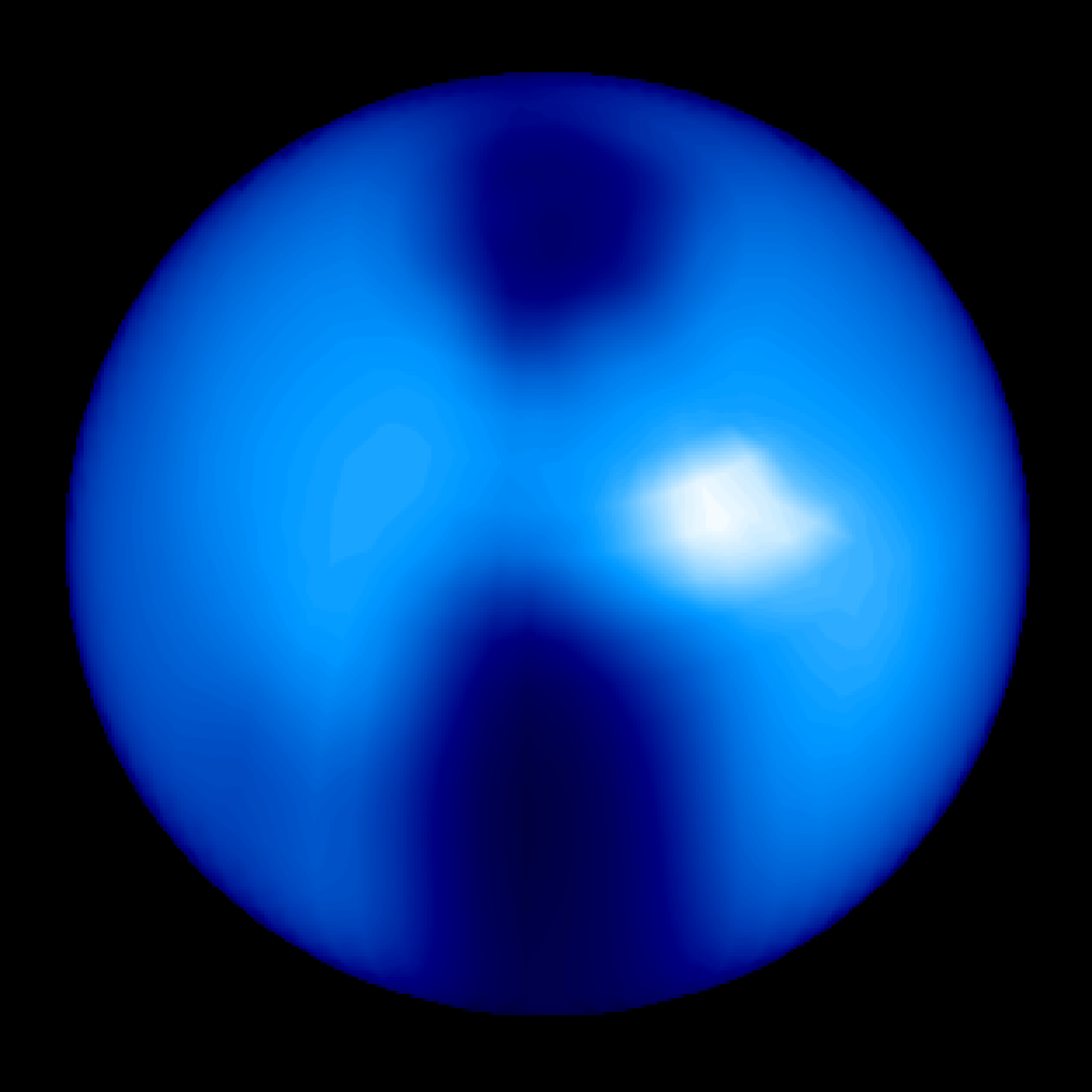}\hfill
\includegraphics[width=0.1\hsize]{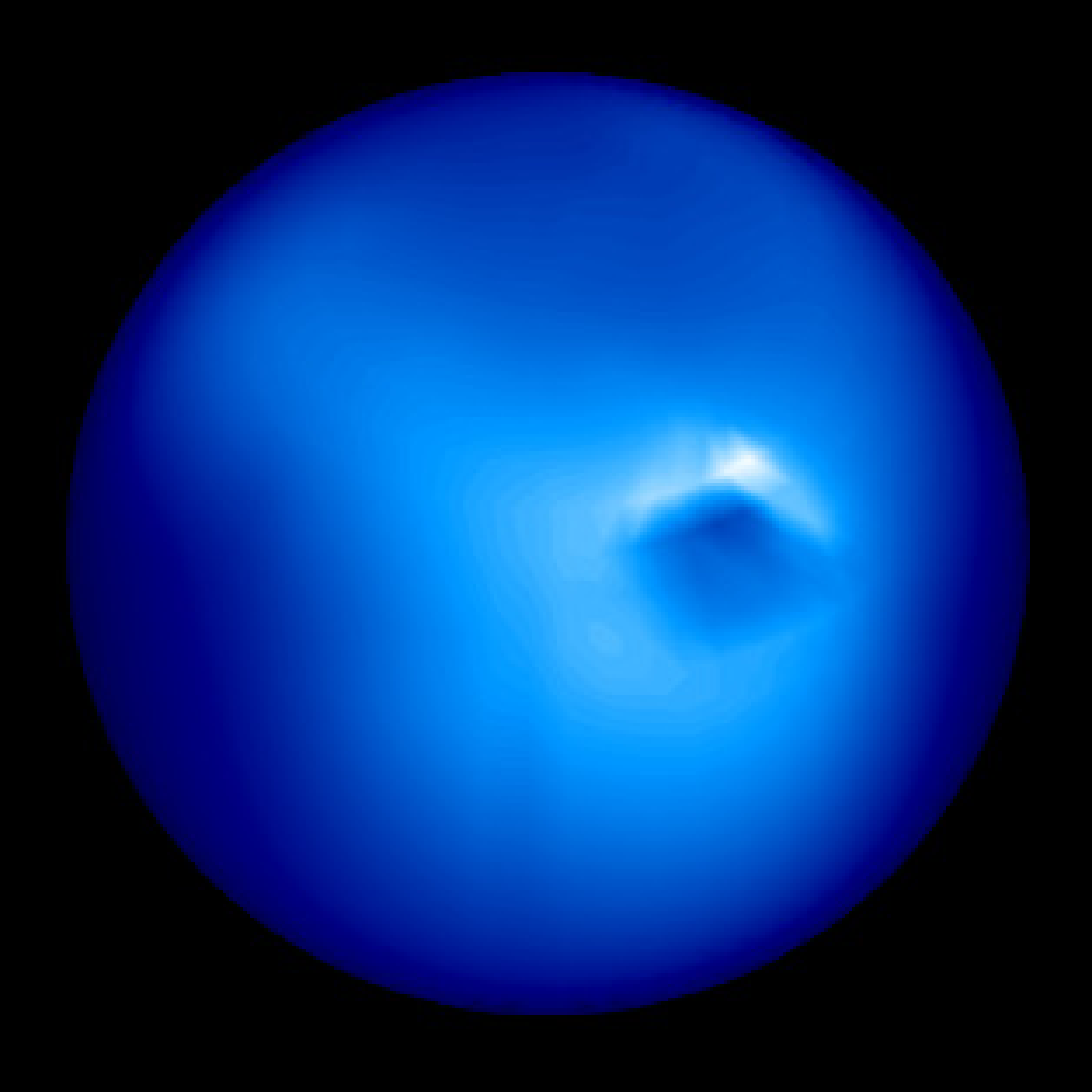}
\includegraphics[width=0.1\hsize]{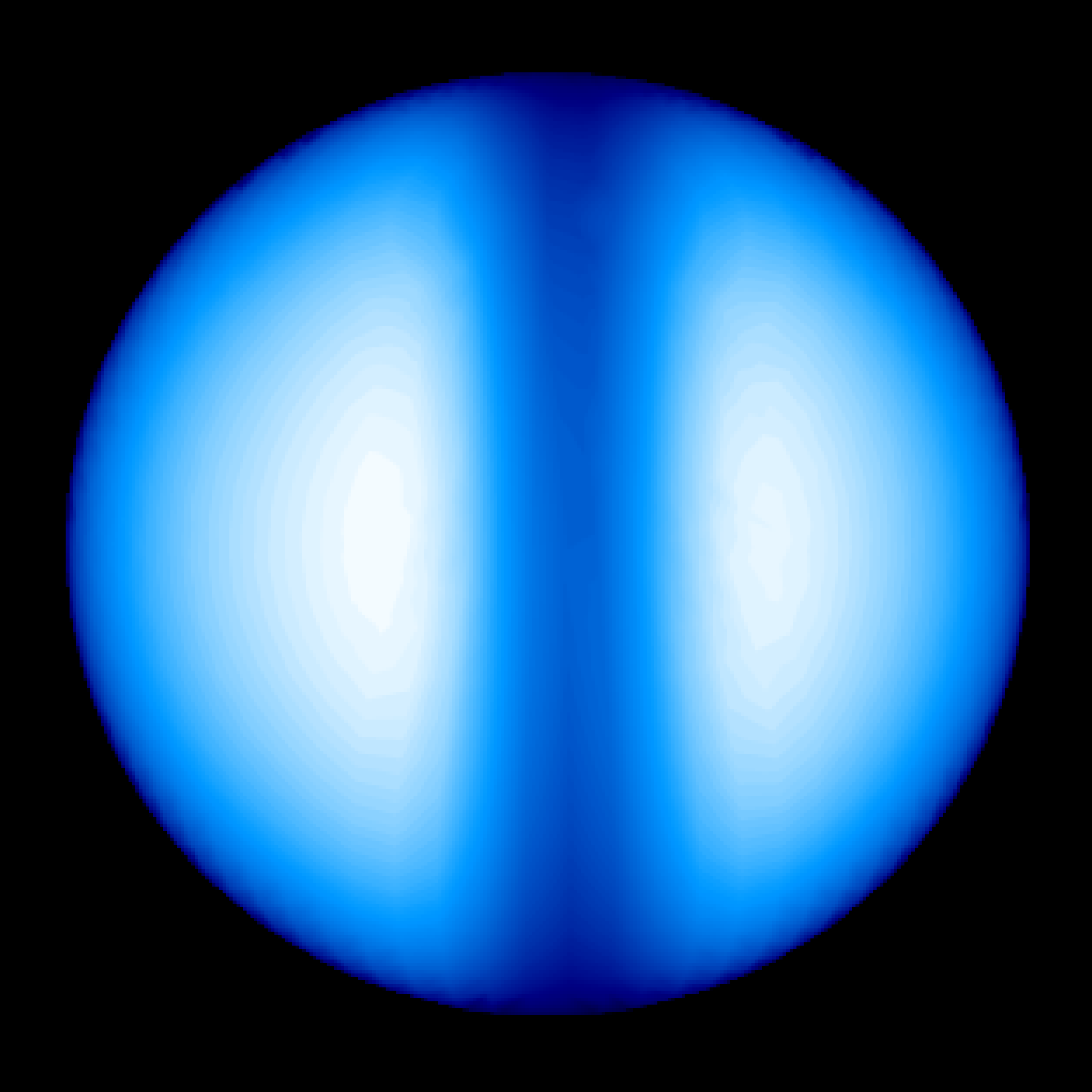}
\includegraphics[width=0.1\hsize]{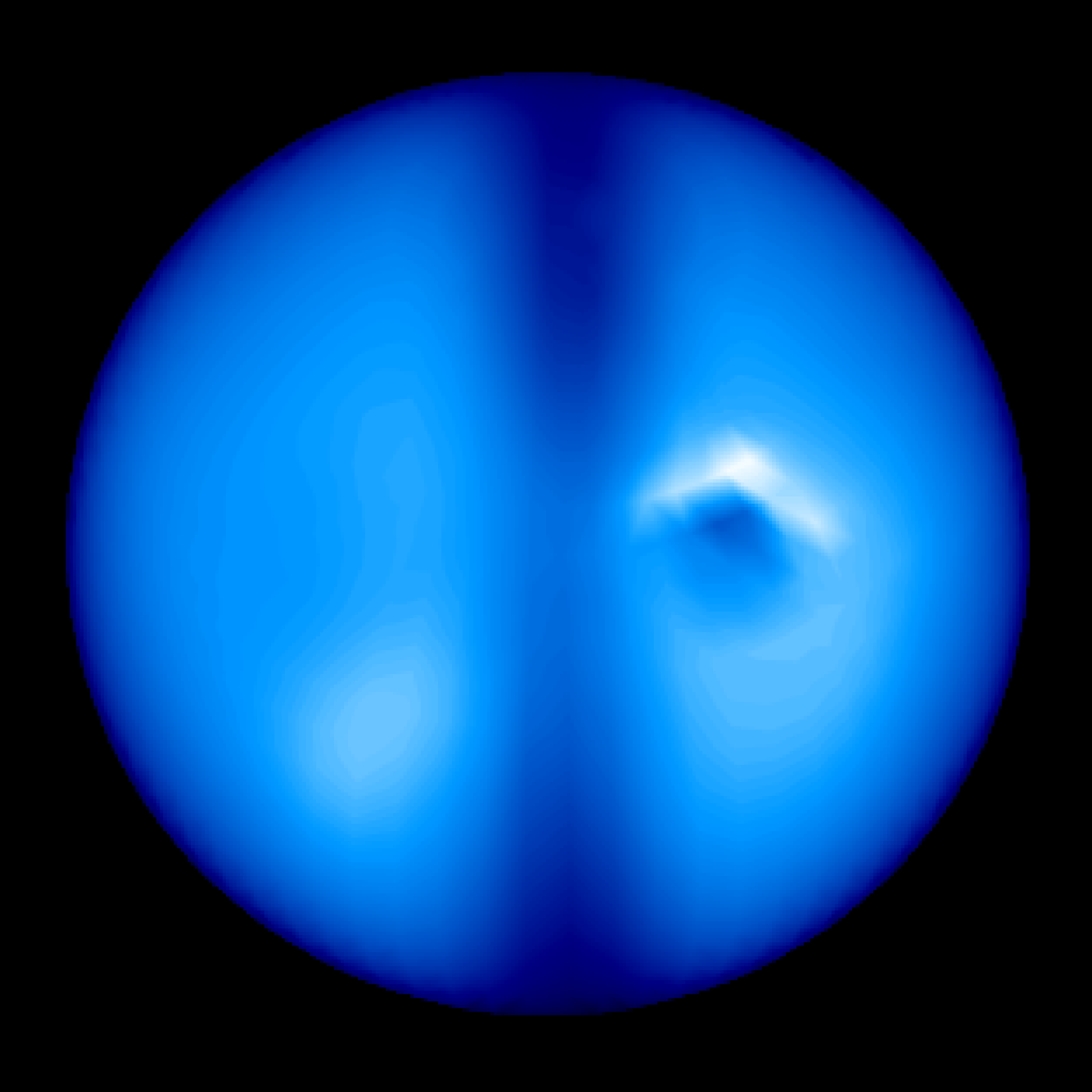}
}
\vspace{0.2cm}
\centerline{
\includegraphics[width=0.33\hsize]{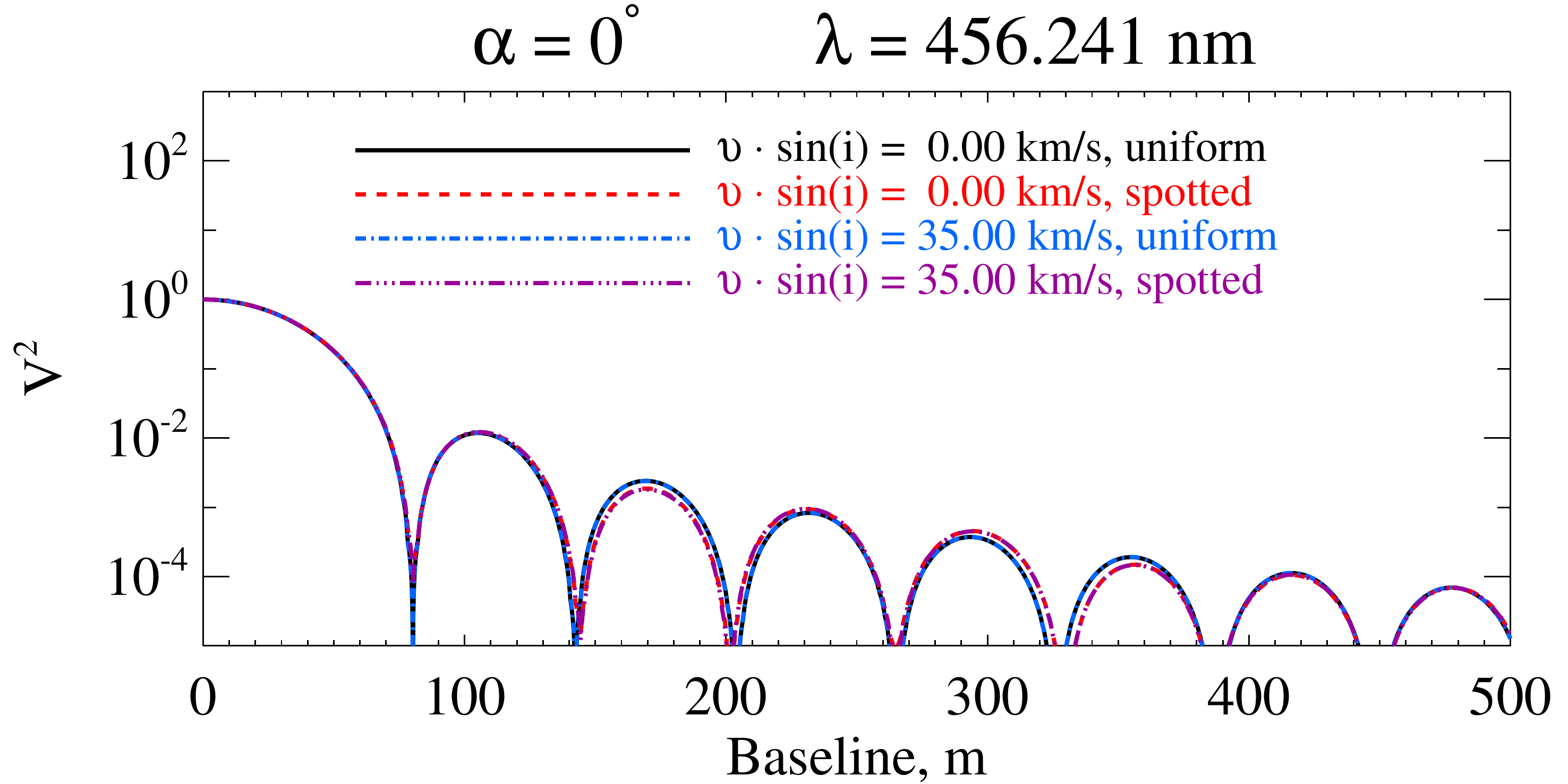}
\includegraphics[width=0.33\hsize]{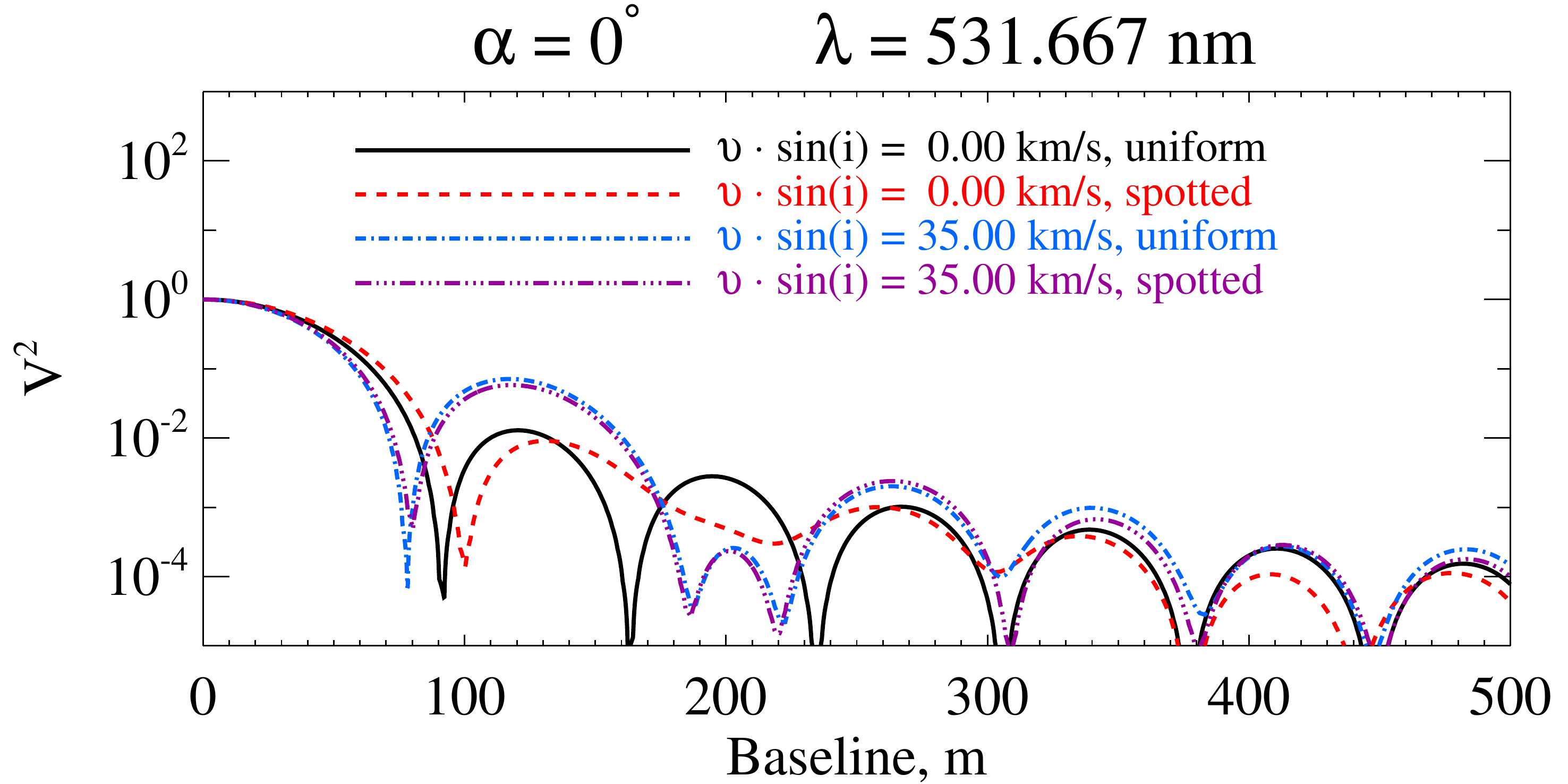}
\includegraphics[width=0.33\hsize]{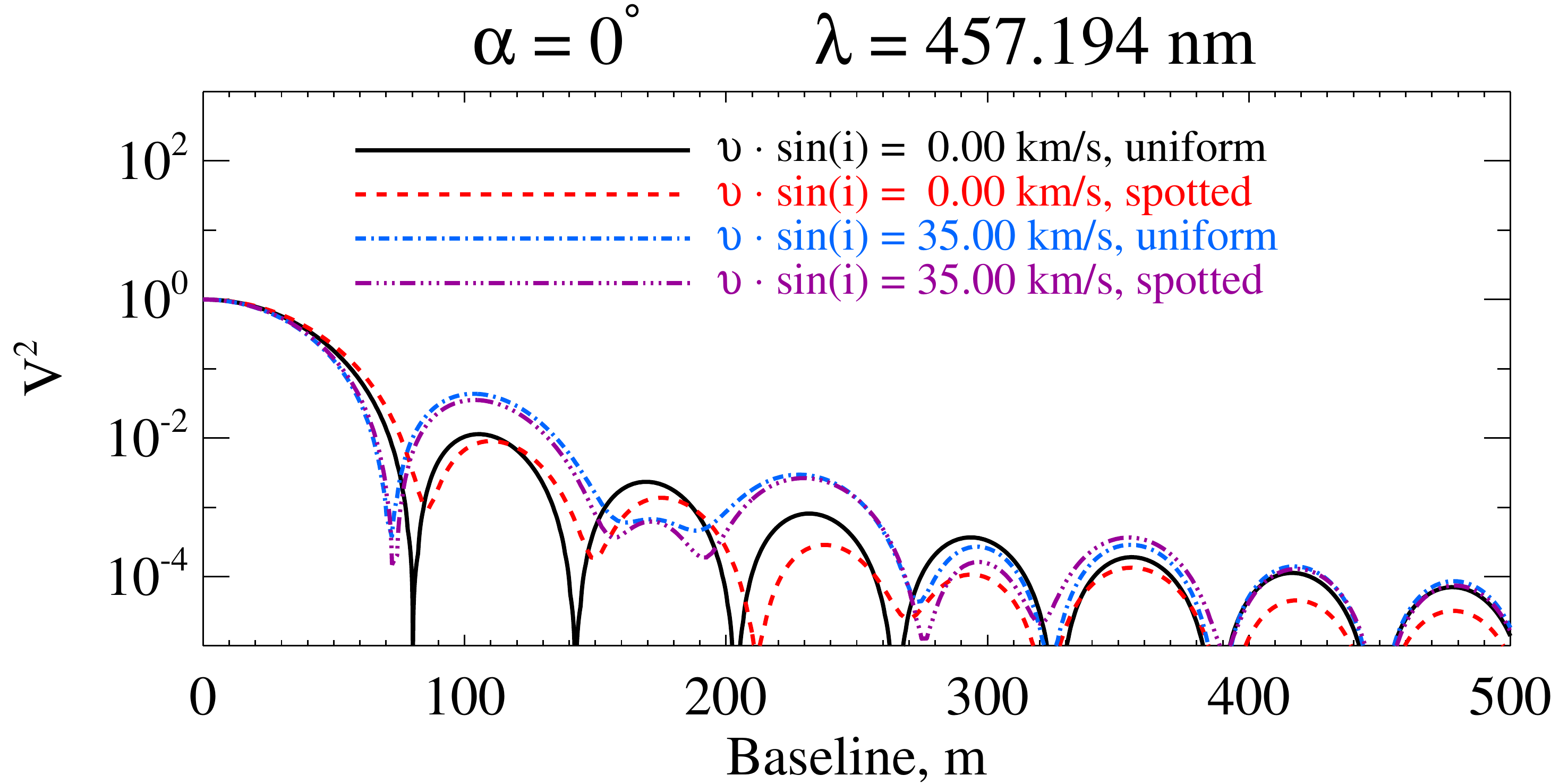}
}
\centerline{
\includegraphics[width=0.33\hsize]{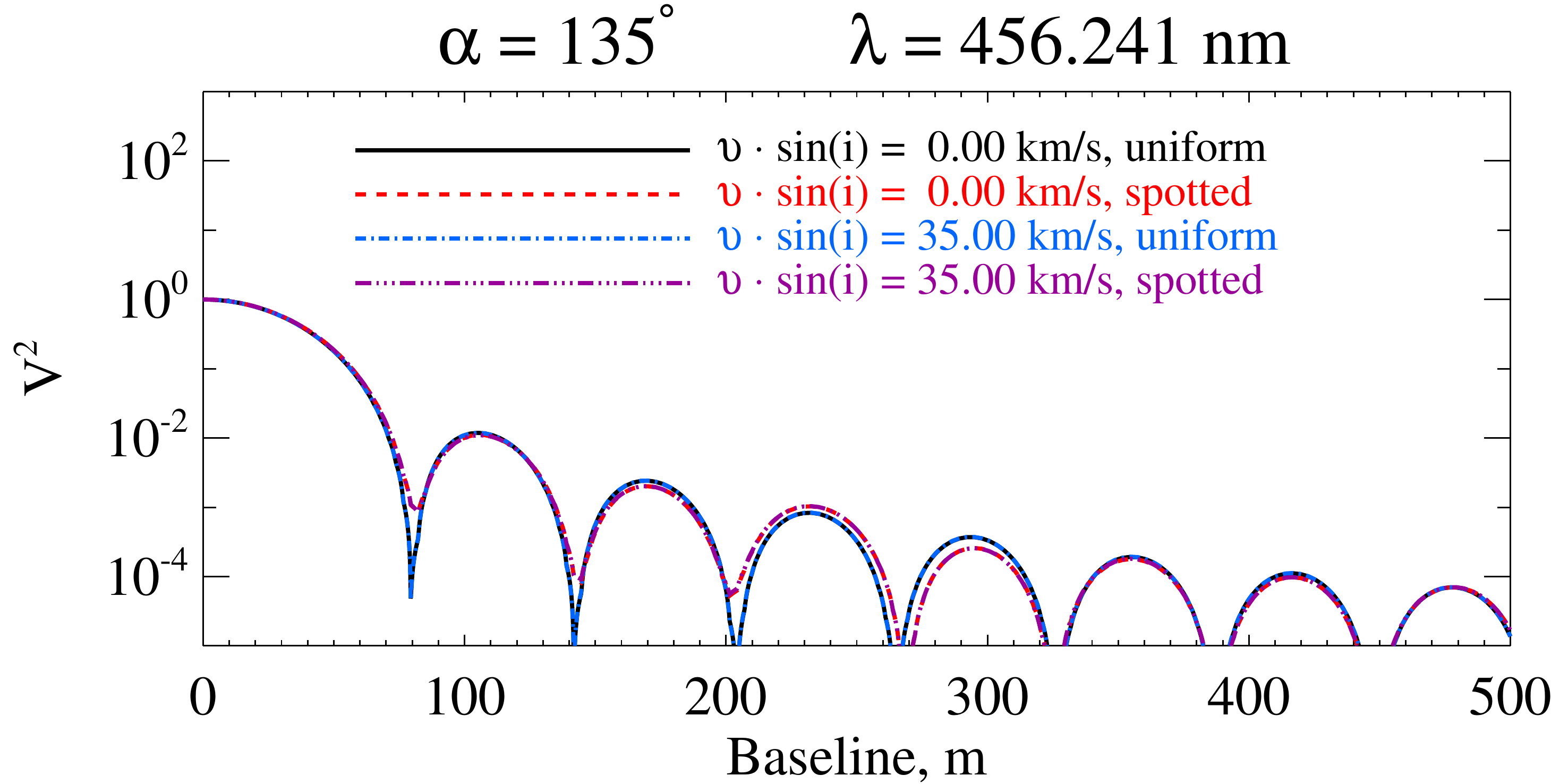}
\includegraphics[width=0.33\hsize]{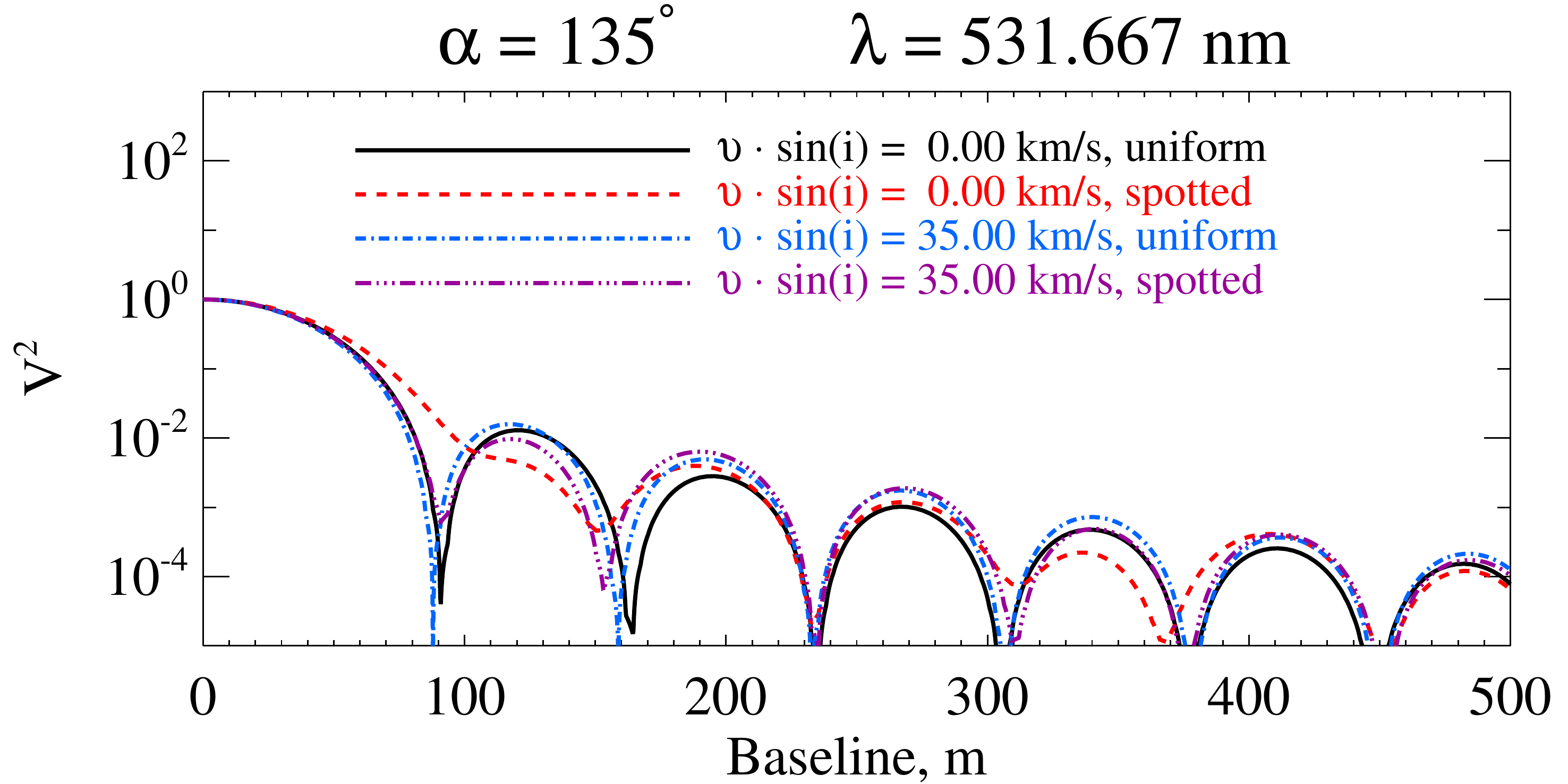}
\includegraphics[width=0.33\hsize]{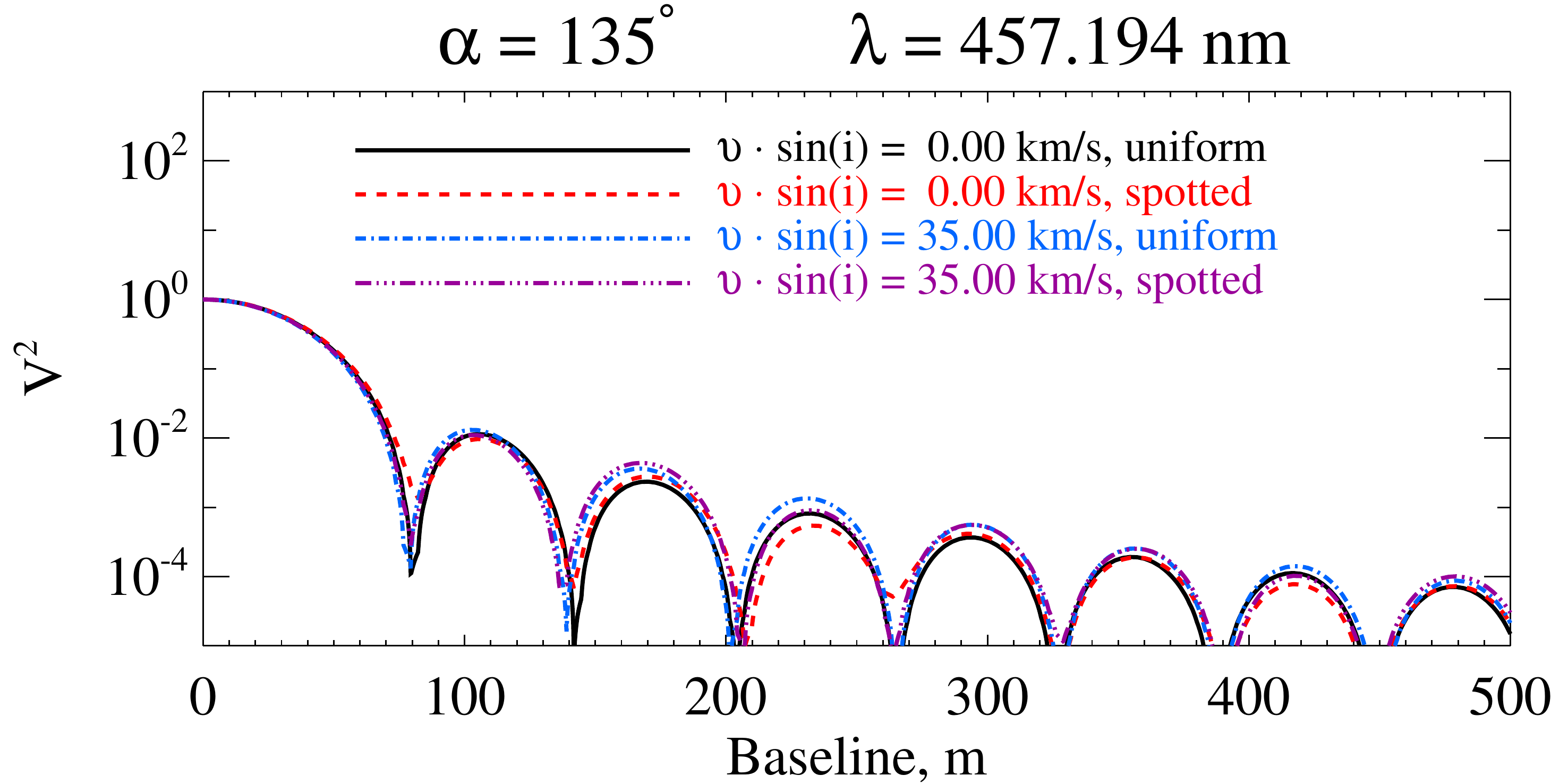}
}
\centerline{
\includegraphics[width=0.33\hsize]{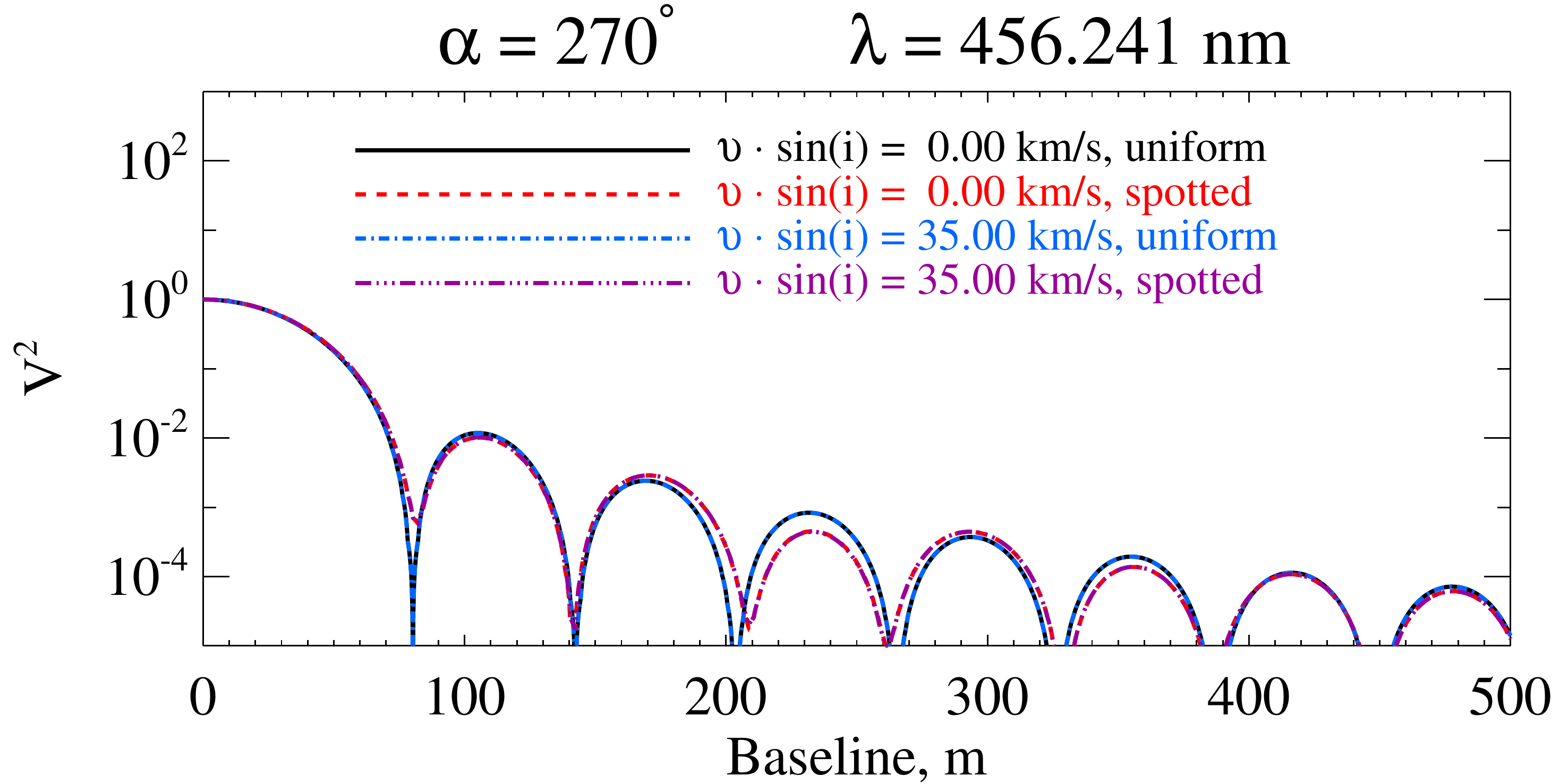}
\includegraphics[width=0.33\hsize]{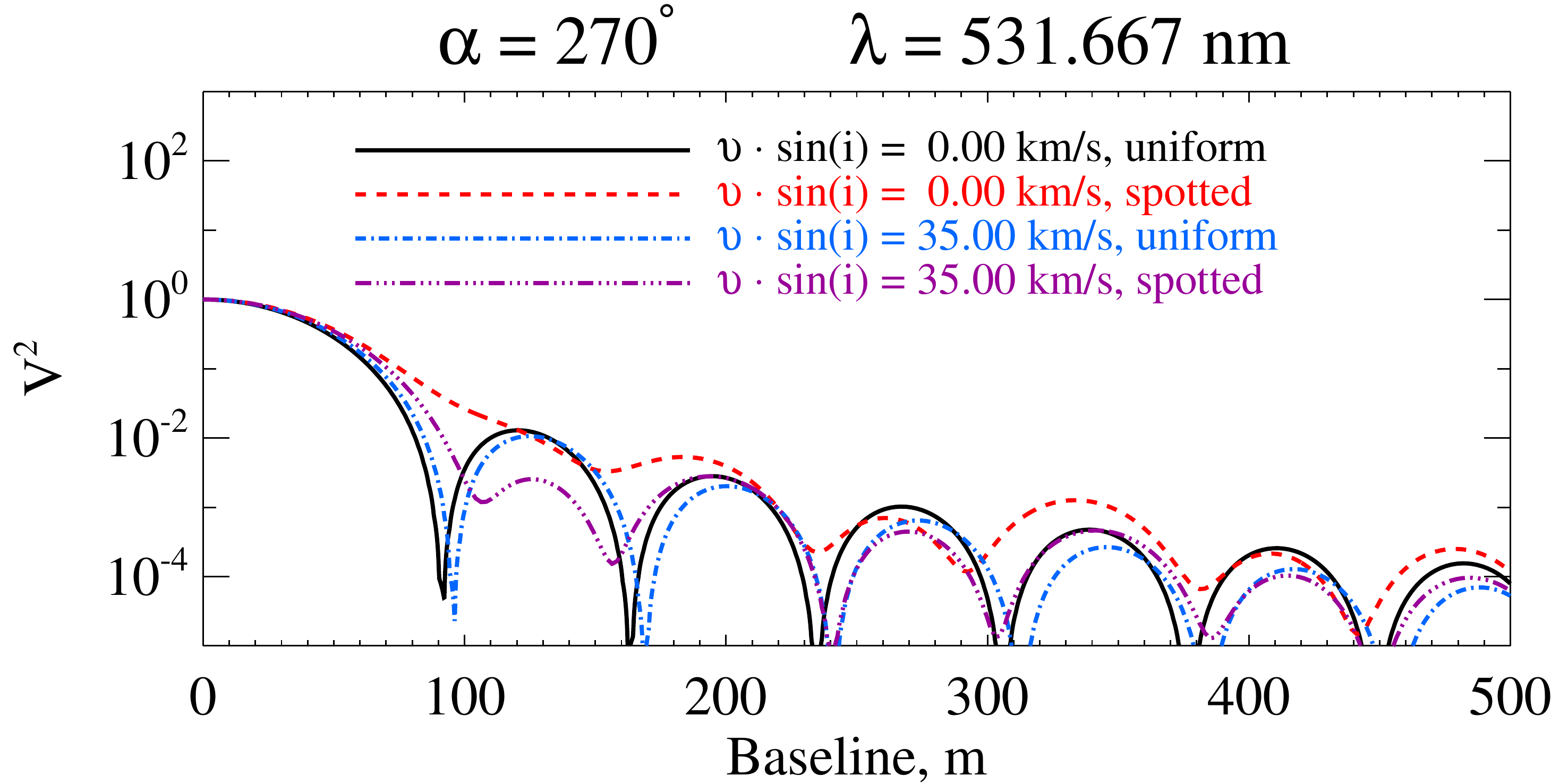}
\includegraphics[width=0.33\hsize]{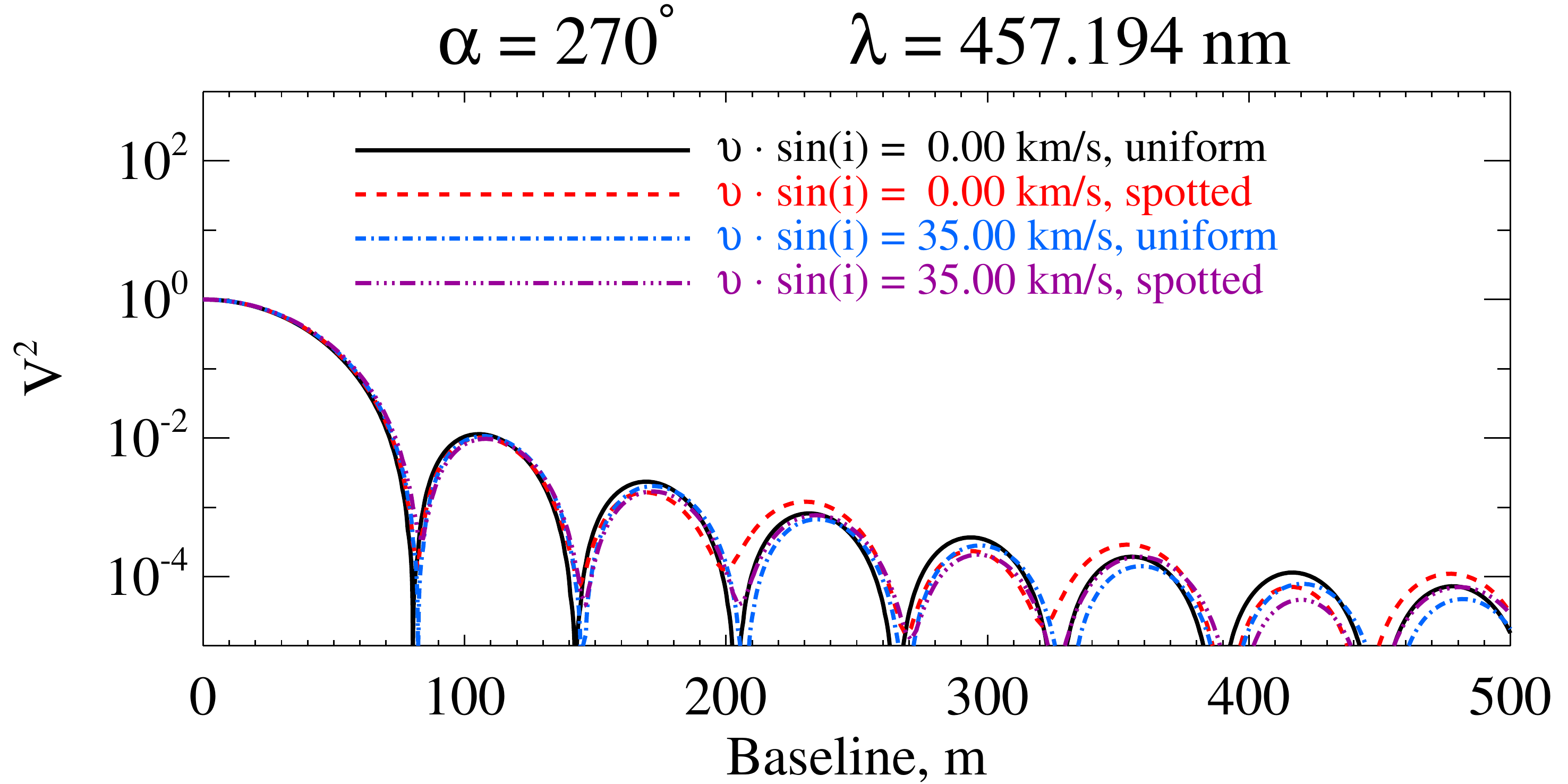}
}
\caption{Same as on Fig.~\ref{fig:vis-mono-v-1}, but at the continuum wavelength $456.241$~nm,
\ion{Fe}{ii}~$531.667$~nm, and a blend of \ion{Cr}{i}$+$\ion{Ti}{2} $457.194$~nm lines.}
\label{fig:vis-mono-v-2}
\end{minipage}
}
\end{figure*}

Naturally, the best way to disentangle rotation and spots (avoiding time-series observations) is to look
at continuum wavelengths where the stellar visibility is not subjected to the Doppler effect. This is illustrated
on the left column of Fig.~\ref{fig:vis-mono-v-2}. Note that spots are still visible even at continuum wavelengths.
This is because, as discussed in \citet{2010A&A...524A..66S}, the flux variability in all wavelength domains happens 
mainly due to the modulation of opacity of elements Fe, Cr, and Si. 
Therefore, the continuum intensities are larger in the spots 
of these elements and in the visual domain, 
as seen from the top-left plot of Fig.~\ref{fig:vis-mono-v-2}. However, the spot signatures are very weak, i.e.
$V^2\lesssim10^{-3}$ at $\alpha=135^\circ$ and $\alpha=270^\circ$.
Because of rich chemistry of CP star's atmospheres, it may be difficult to find a true continuum level, especially
in case of fast rotating stars and when observing in narrow spectral windows. 
This can make the observation at pure continuum wavelength to be a complicated task.

It is known that the spot detection depends on the orientation of the projected baseline. An example of this behaviour is illustrated
using \ion{Fe}{ii}~$531.66$~nm line and is plotted on the second column of Fig.~\ref{fig:vis-mono-v-2}. 
At zero rotational phase, the Fe abundance distribution is characterized by a wide spot at latitudes $0^\circ$ to $40^\circ$
which intersects partially with Cr spot but located slightly below it \citep[for detailed abundance maps, see][]{2003A&A...406.1033L}. 
Therefore, the area occupied by a Cr spot looks brighter in the core of the
\ion{Fe}{ii}~$531.66$~nm line because of increase of the continuum flux, while the rest of the stellar surface is dimmer
due to the wide region of enhanced Fe abundance along with the dark stripe crossing the stellar disk caused by rotation itself.
Its hard to see the spot signal at $\alpha=0^\circ$ and $\alpha=135^\circ$ if star rotates, whereas
at slow rotation (i.e., close to $0$~\kms) the spot could in principle be detected below $0.3$ {squared} visibility. 
In fact the spots act making the star slightly smaller by shifting of the zero of {squared} visibility  towards longer baselines.
At $\alpha=270^\circ$, however,
the spot induces a strong signal in both $\vsini=0$~\kms\ and $\vsini=35$~\kms. Thus the combination of a few
position angles and observations at nearby wavelengths will ideally help to constrain not only the spot position, 
but also the stellar rotation.

Because of the combined effect of stellar rotation and spots, there are certain wavelengths where the spot signal
almost vanishes in {squared} visibility  independent of how strong the brightest contrast is. On the third column
of Fig.~\ref{fig:vis-mono-v-2} we show the stellar images and {squared} visibility  in the strong line of
\ion{Ti}{ii}~$457.19$~nm which is blended with a few less strong \ion{Cr}{i} lines. In this particular case,
the Cr spot coincides exactly with the area of depleted Ti abundance. This leads to a superposition of
bright and dark areas that, incidentally, produce almost no difference in $V^2$  for any position angle considered.
The spot signal can only be seen at $V^2\lesssim10^{-2}$ and $\alpha=0^\circ$ and only if the star is a very slow rotation.
The effect of spots is again making the star appearing smaller compared to the case without spots.
It is essential to have interferometric observations at different spectral windows that
will allow to capture lines of different elements~--~an important step for the definite spot detection.

\begin{figure*}
\rotatebox{90}{
\begin{minipage}{\textheight}
\centerline{
\includegraphics[width=0.1\hsize]{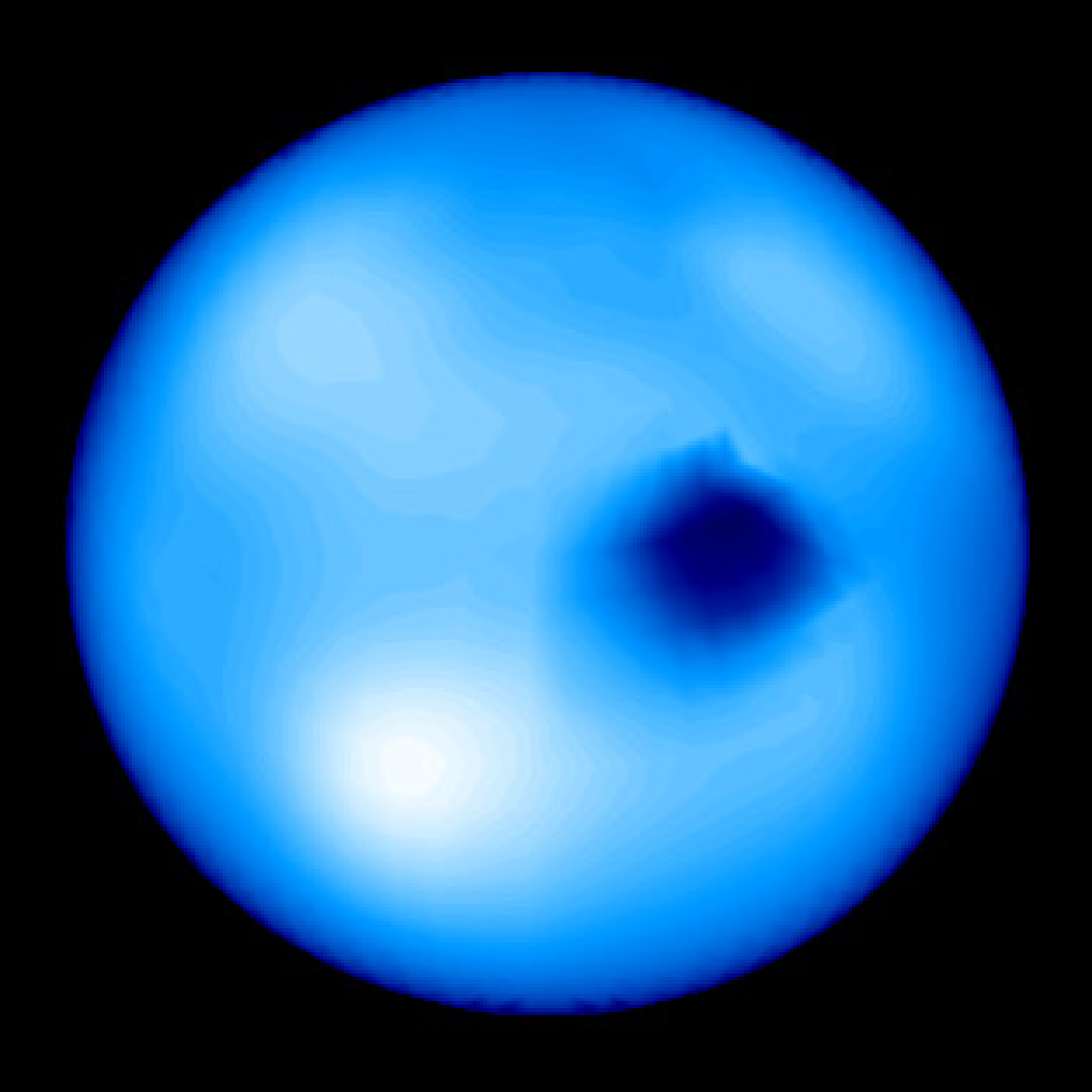}
\includegraphics[width=0.1\hsize]{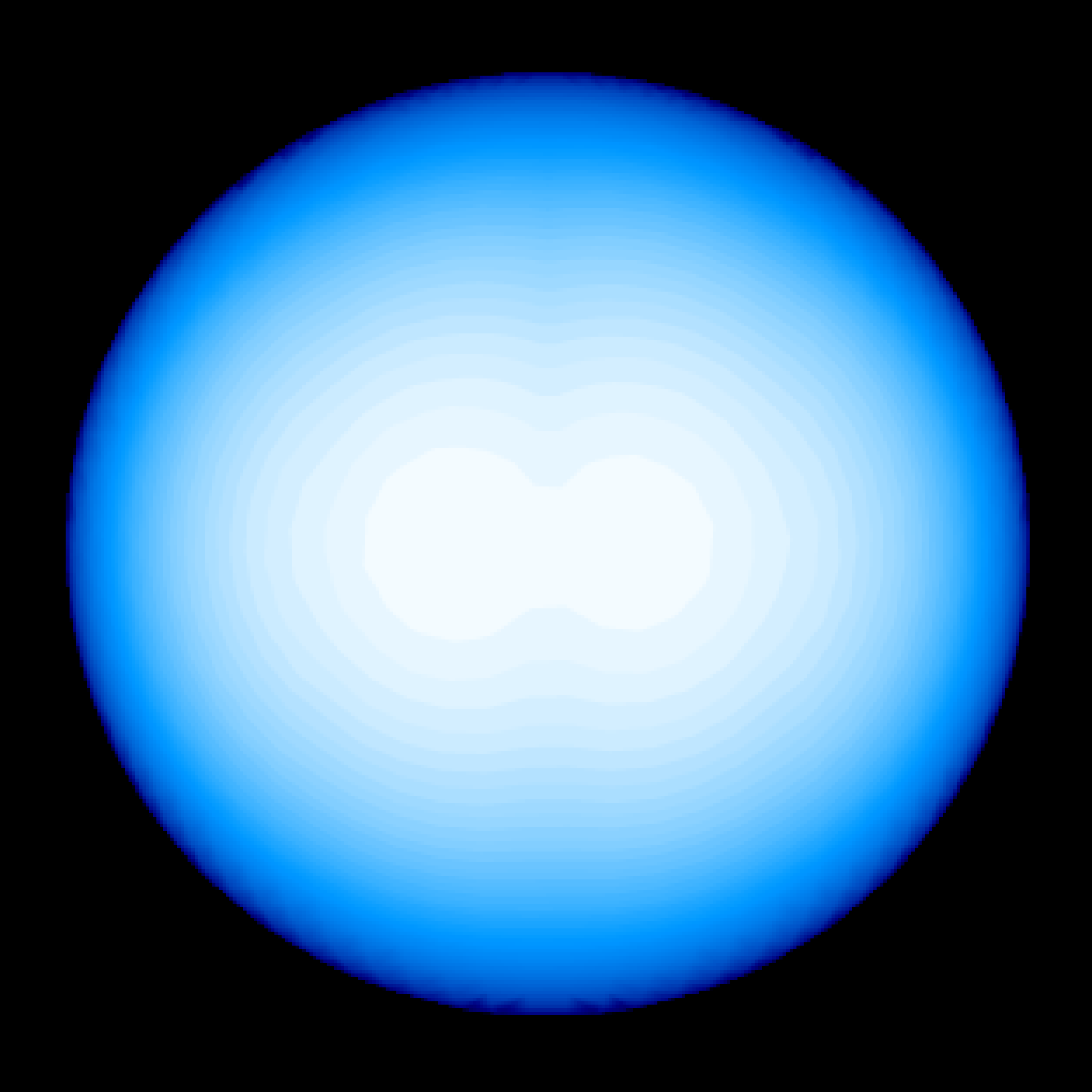}
\includegraphics[width=0.1\hsize]{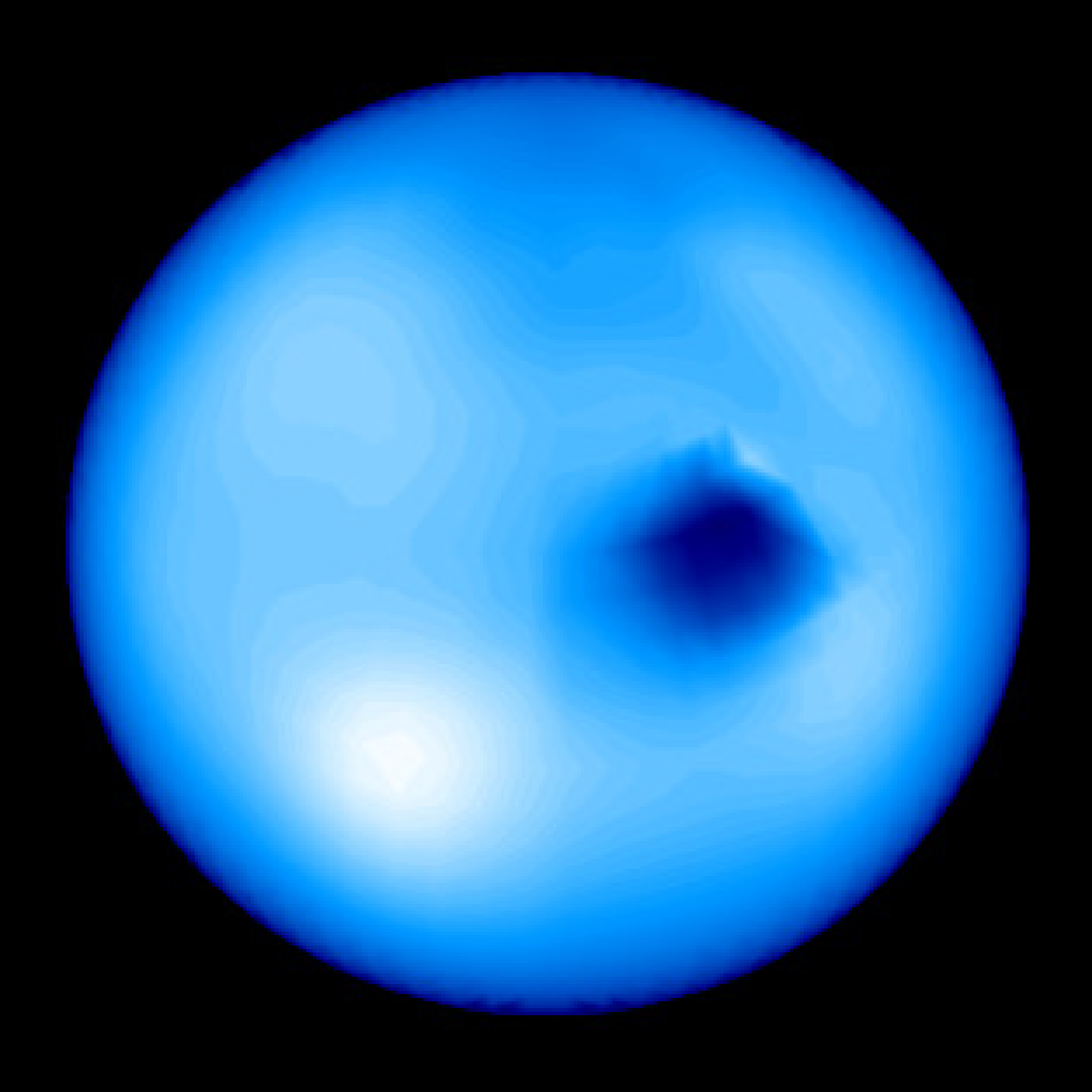}\hfill
\includegraphics[width=0.1\hsize]{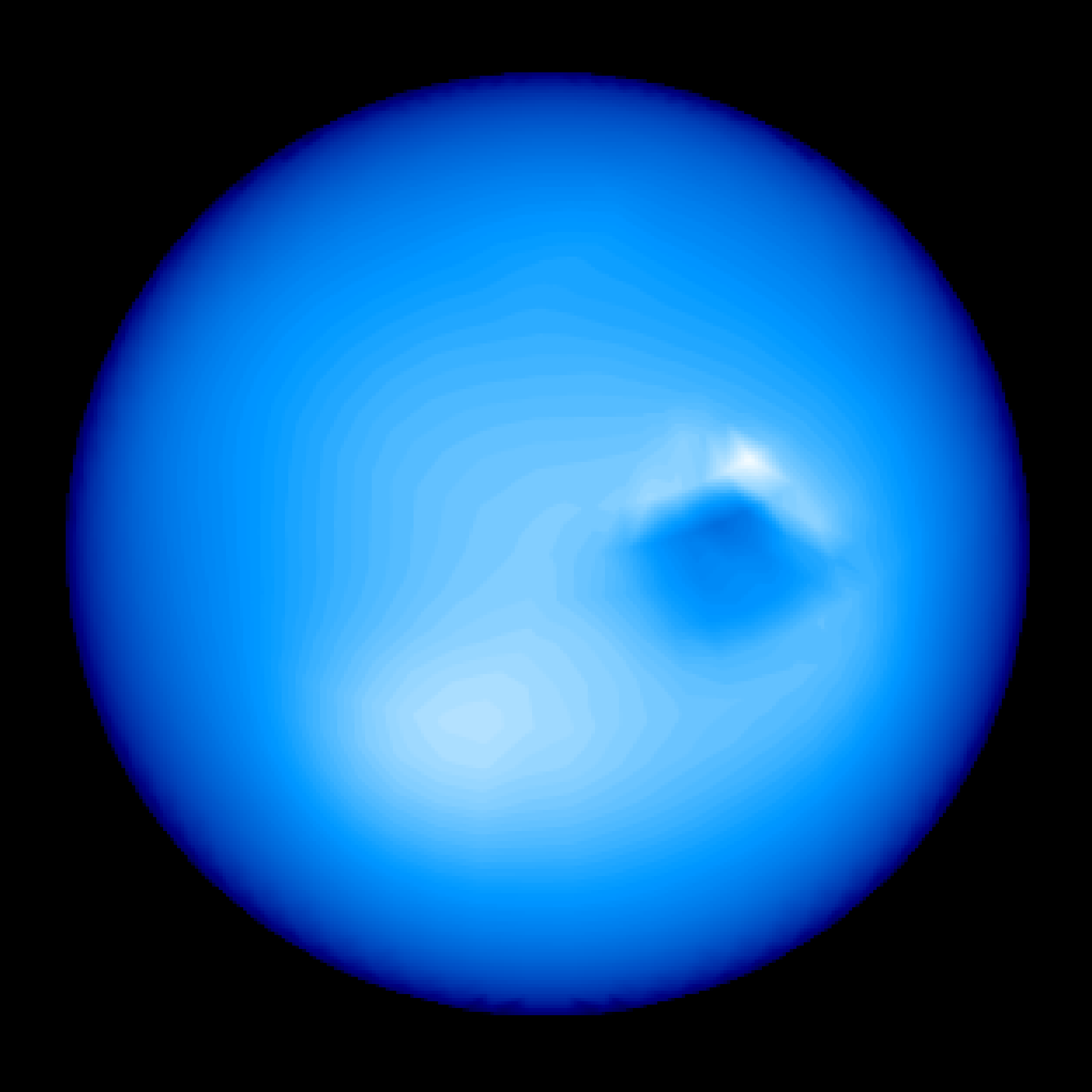}
\includegraphics[width=0.1\hsize]{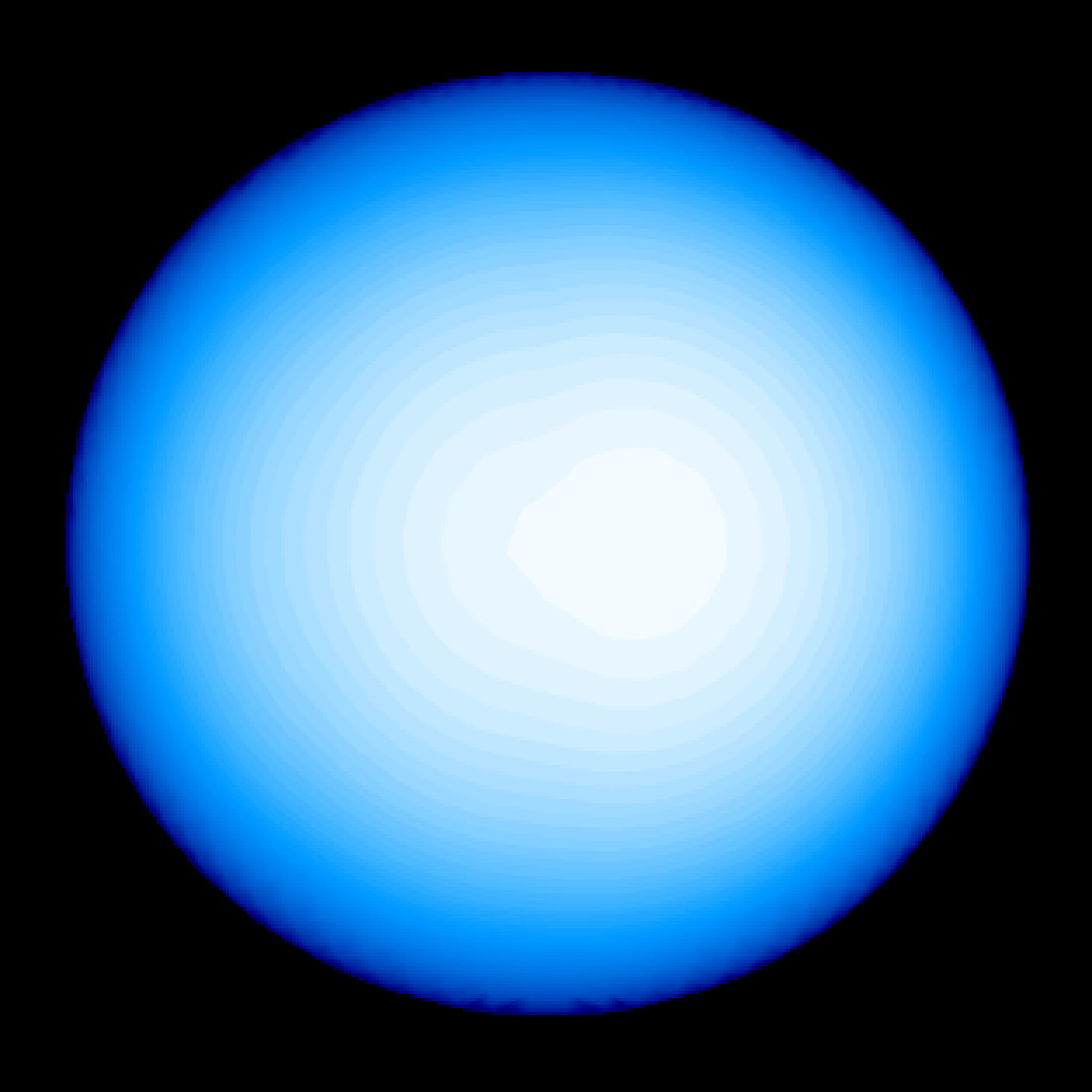}
\includegraphics[width=0.1\hsize]{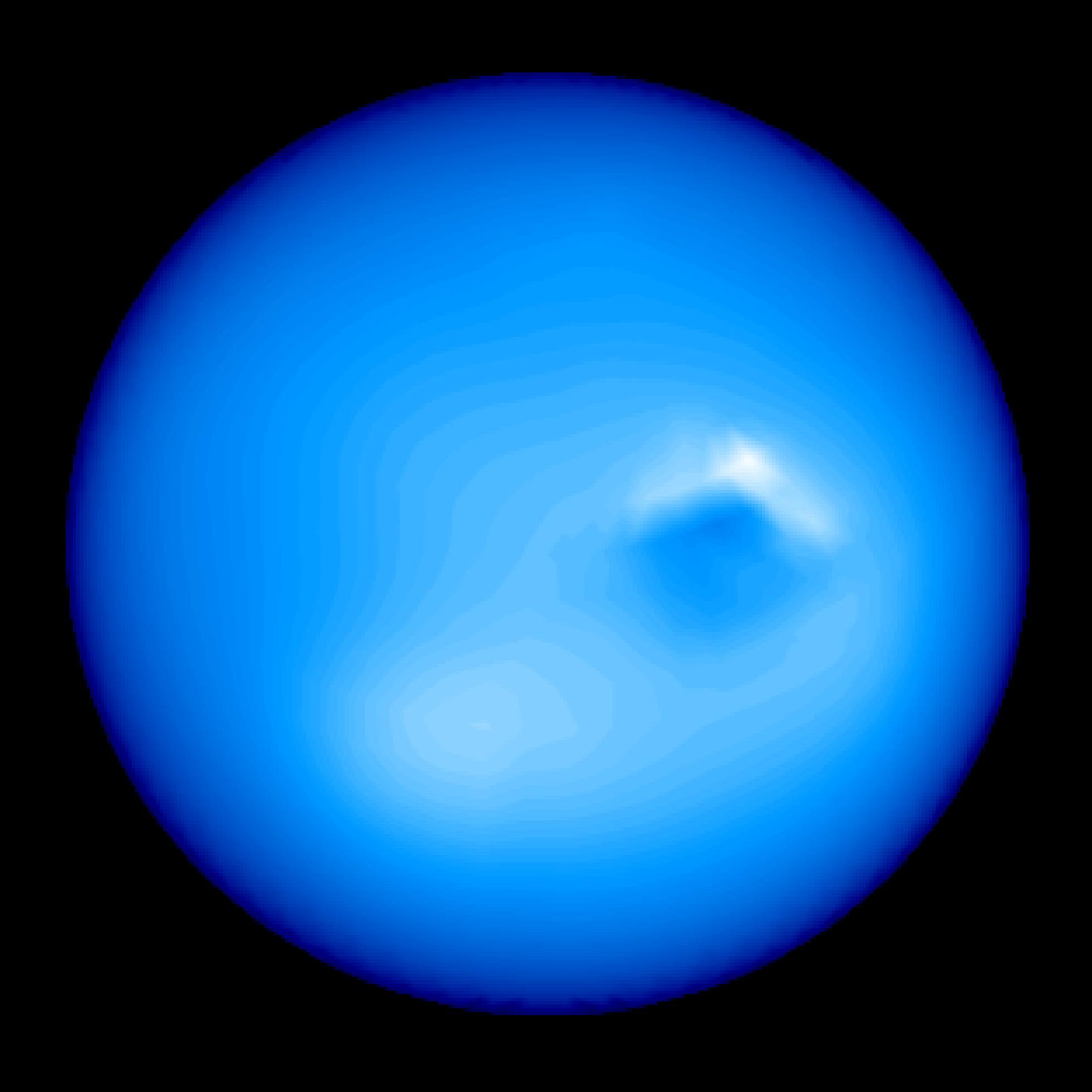}\hfill
\includegraphics[width=0.1\hsize]{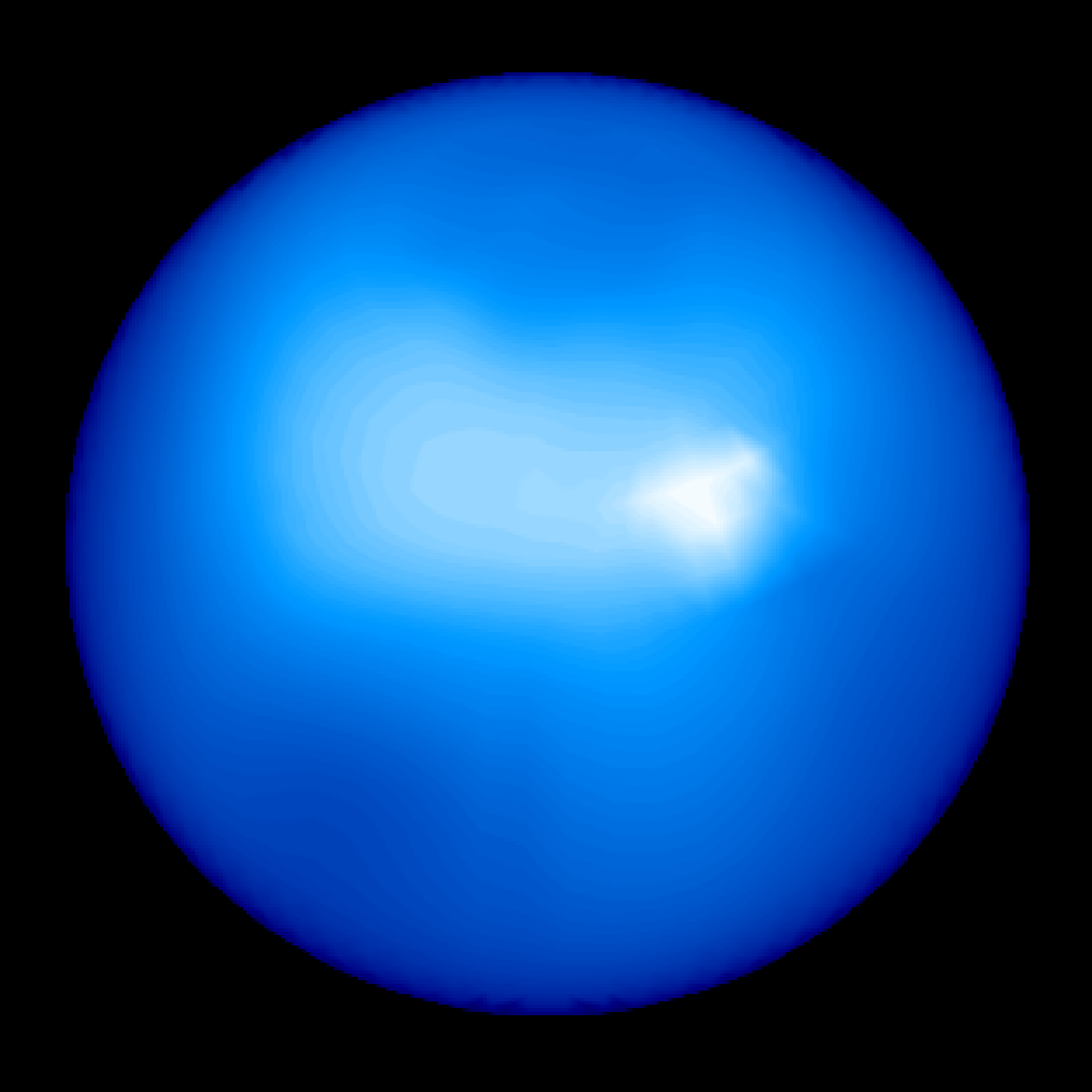}
\includegraphics[width=0.1\hsize]{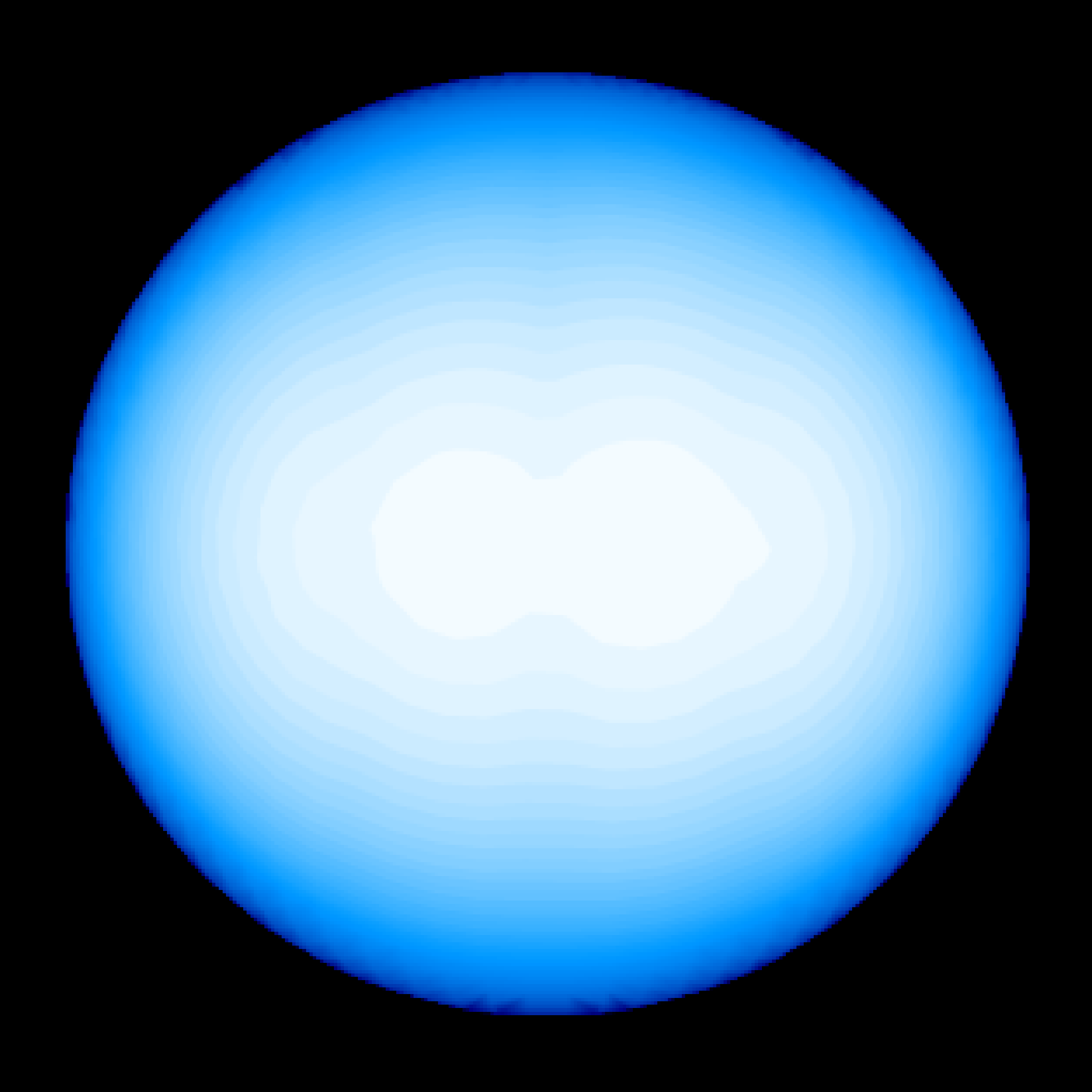}
\includegraphics[width=0.1\hsize]{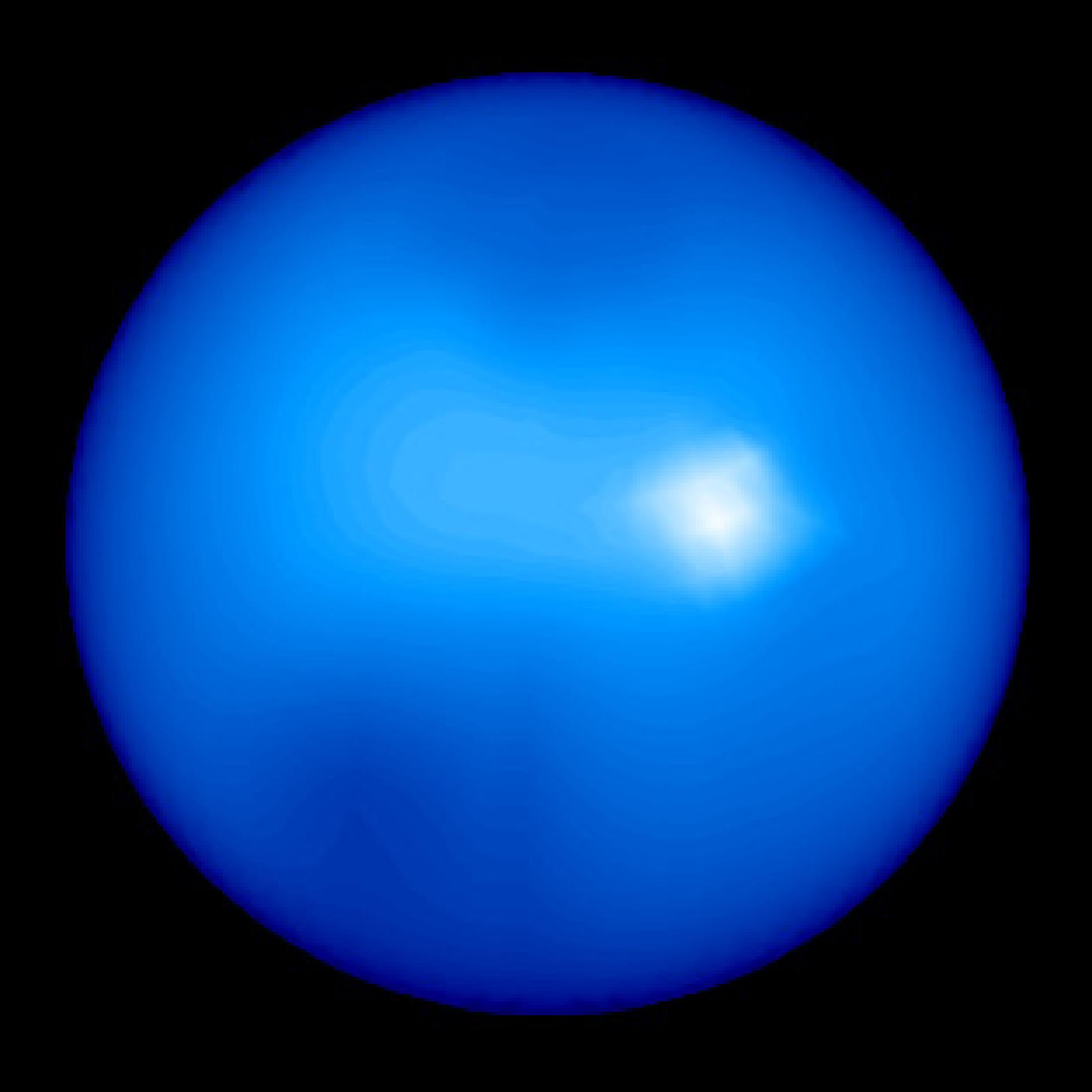}
}
\vspace{0.2cm}
\centerline{
\includegraphics[width=0.33\hsize]{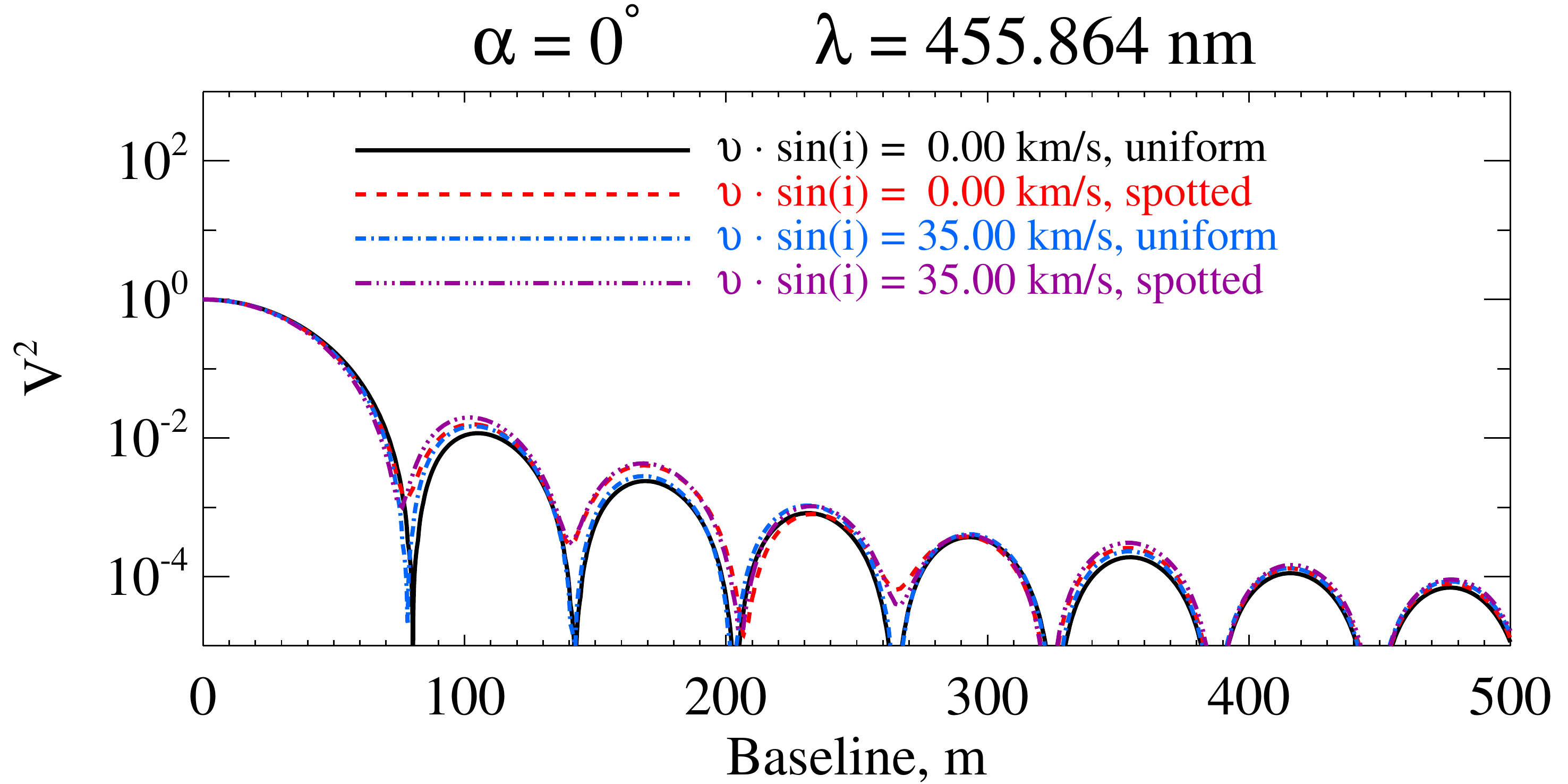}
\includegraphics[width=0.33\hsize]{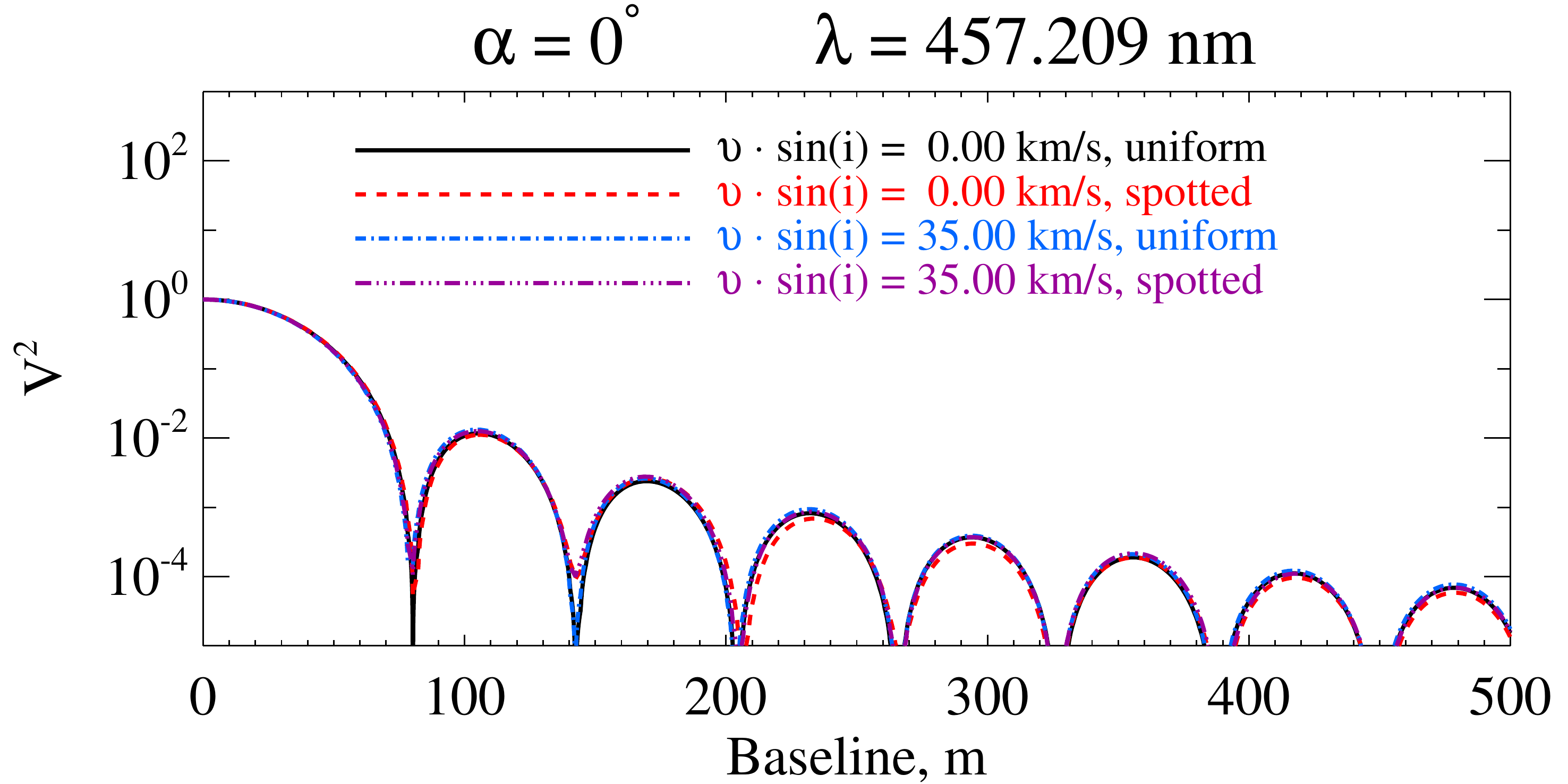}
\includegraphics[width=0.33\hsize]{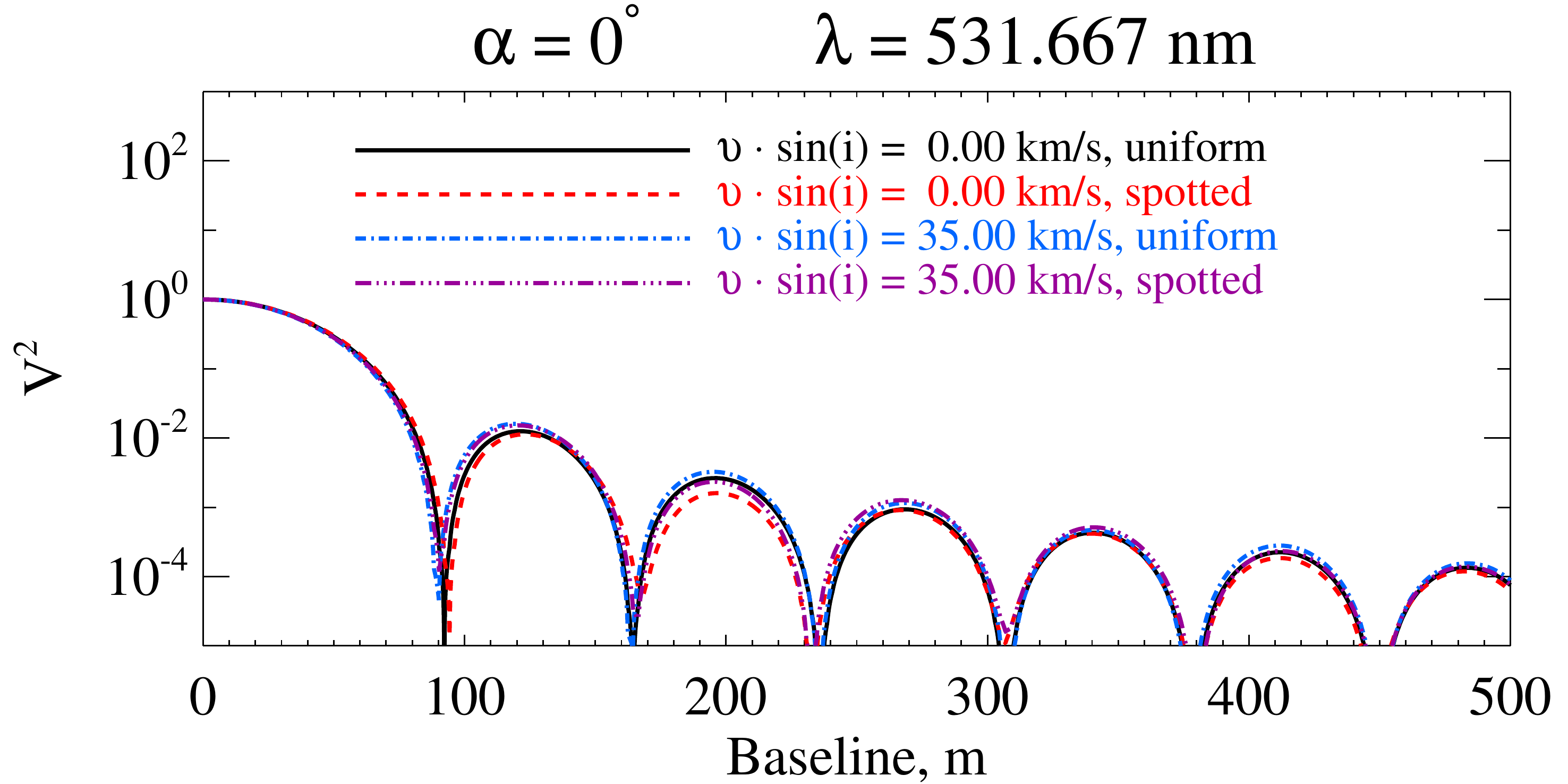}
}
\centerline{
\includegraphics[width=0.33\hsize]{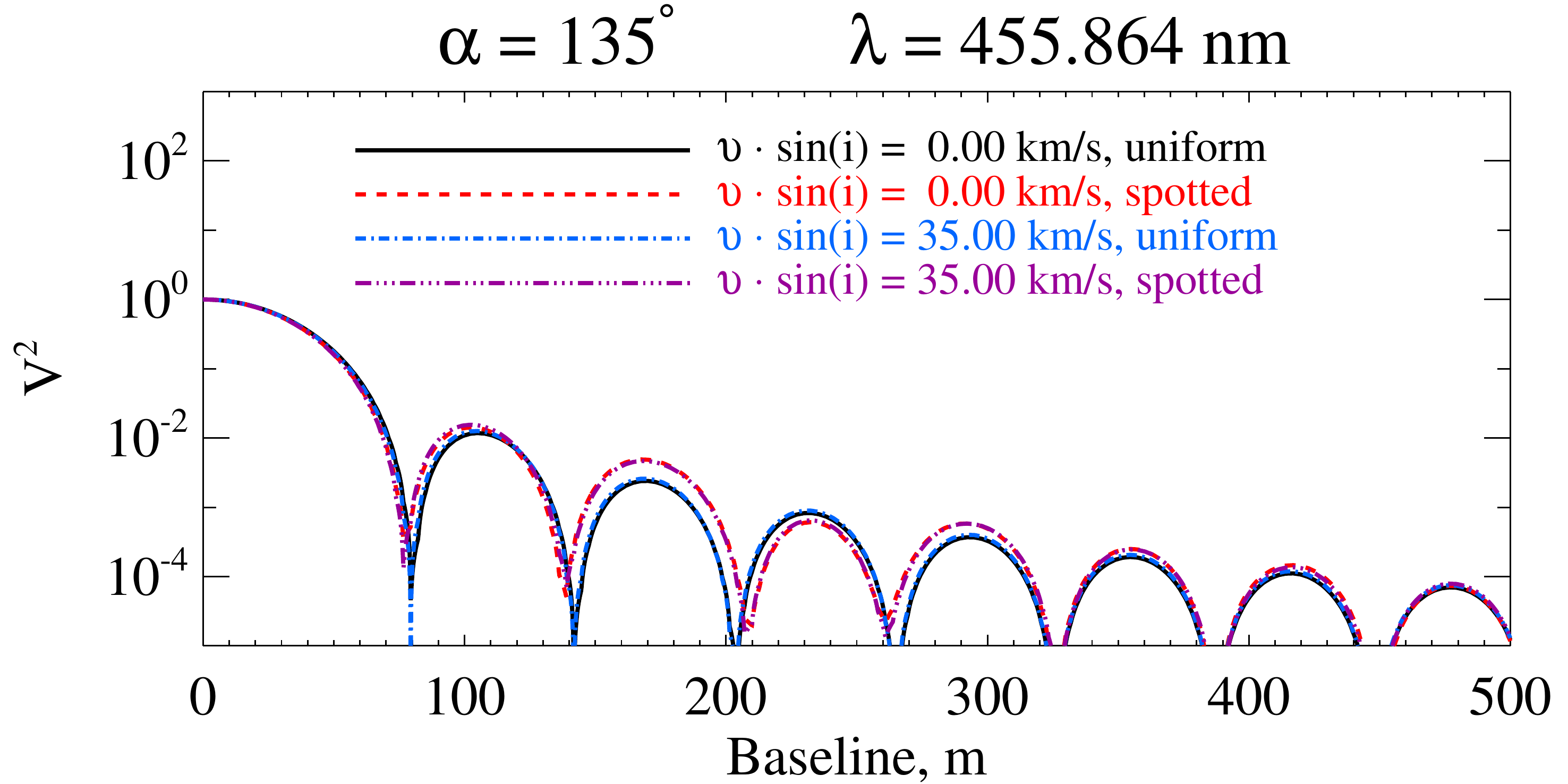}
\includegraphics[width=0.33\hsize]{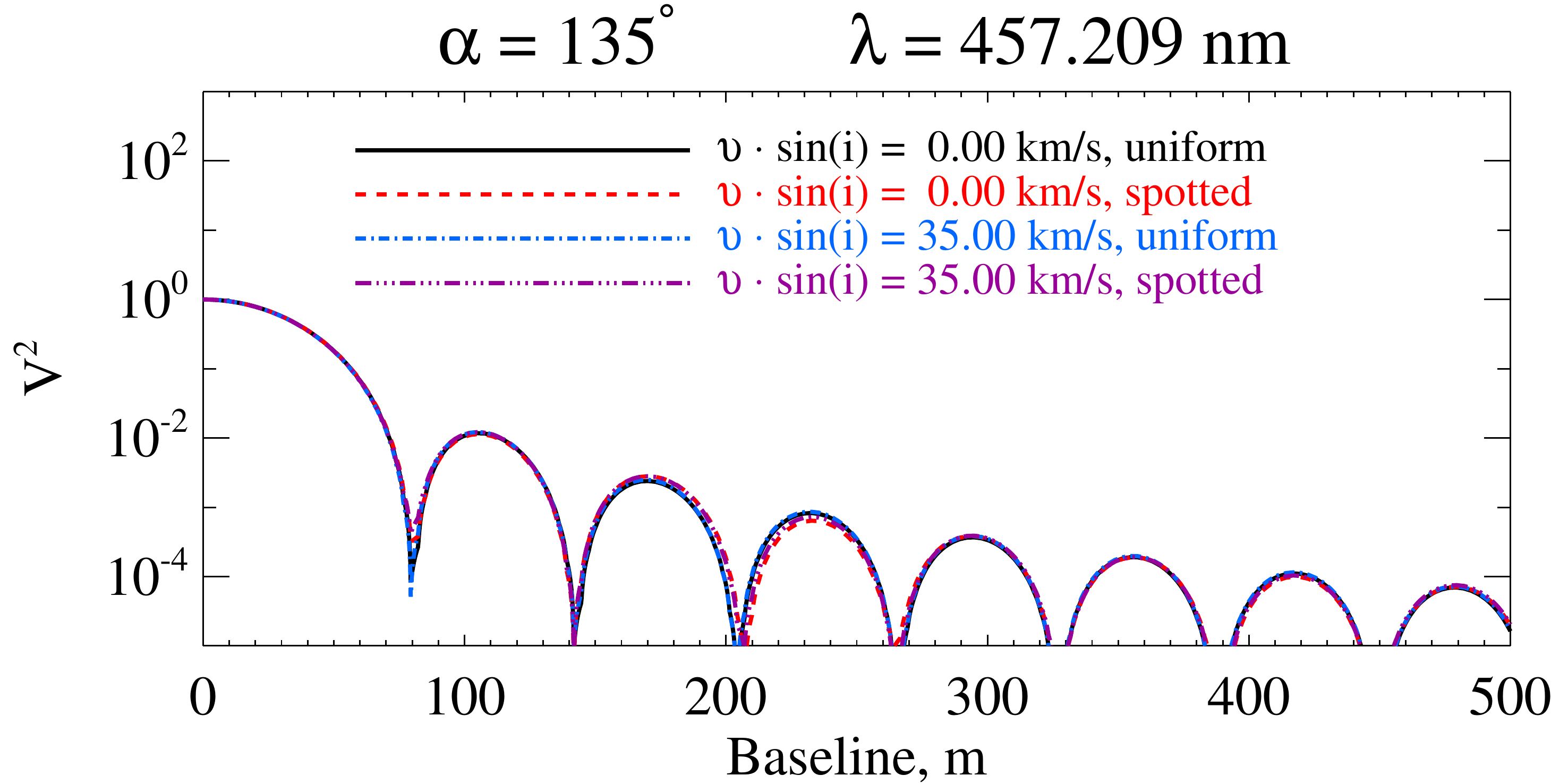}
\includegraphics[width=0.33\hsize]{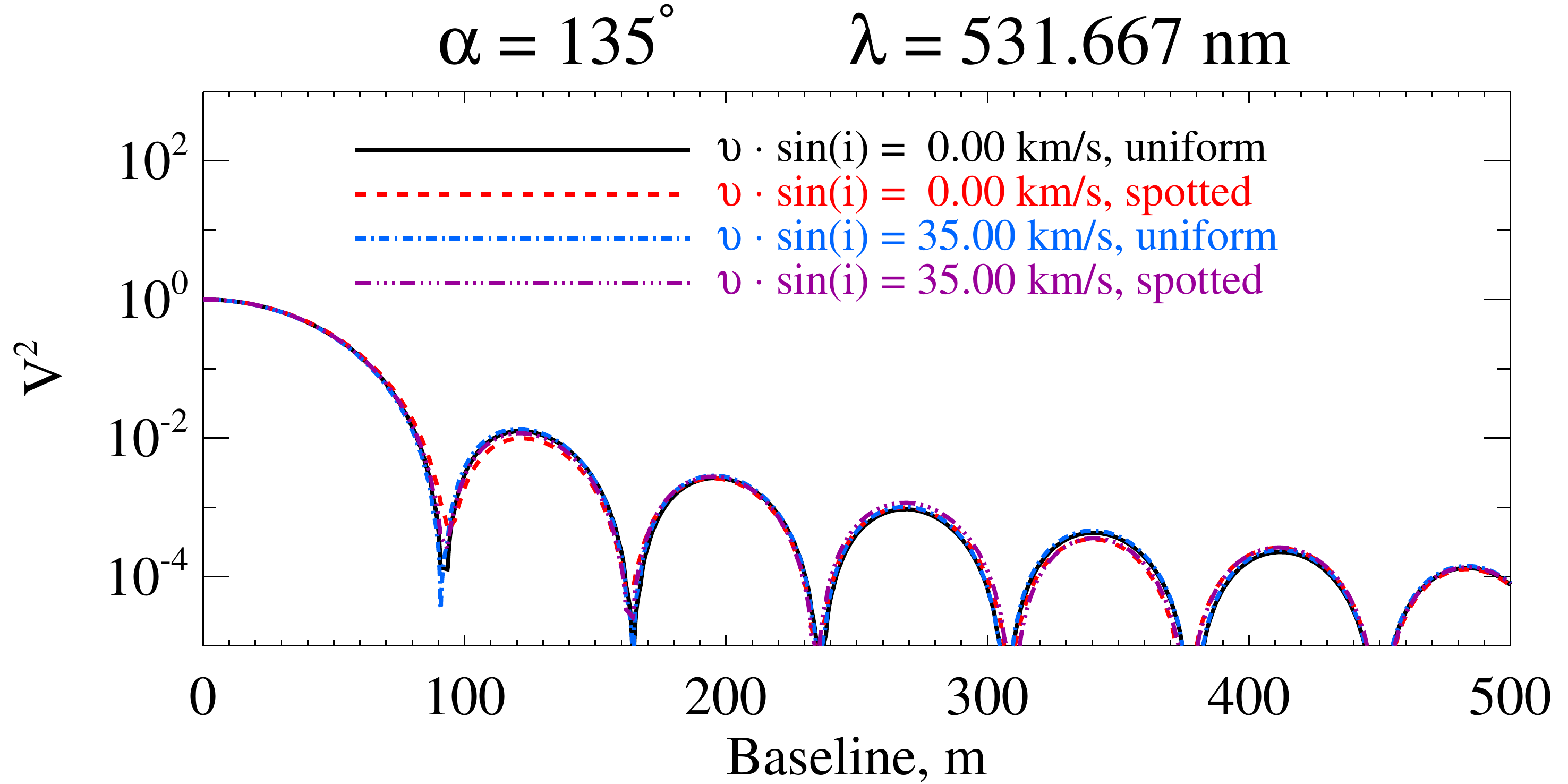}
}
\centerline{
\includegraphics[width=0.33\hsize]{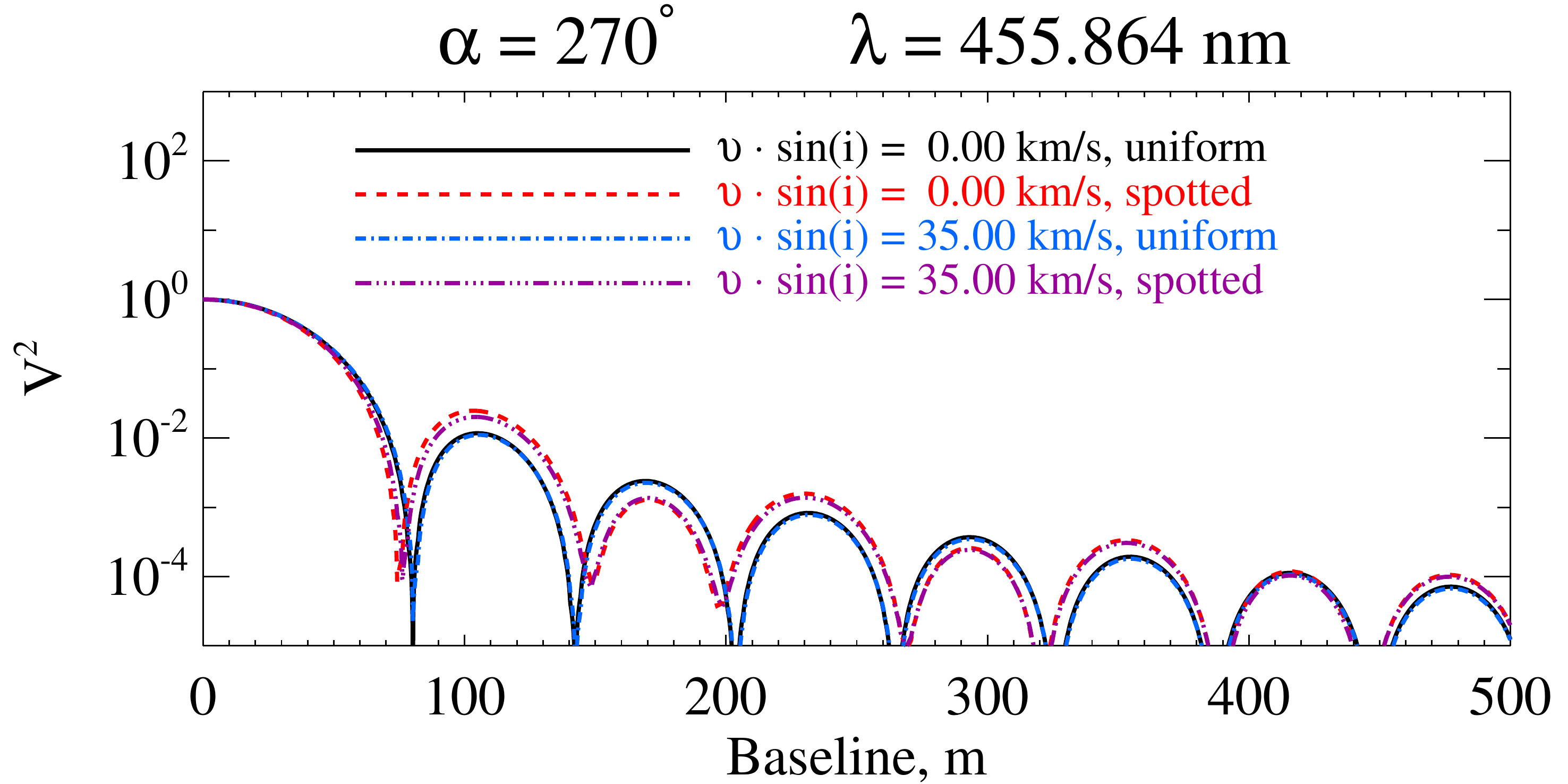}
\includegraphics[width=0.33\hsize]{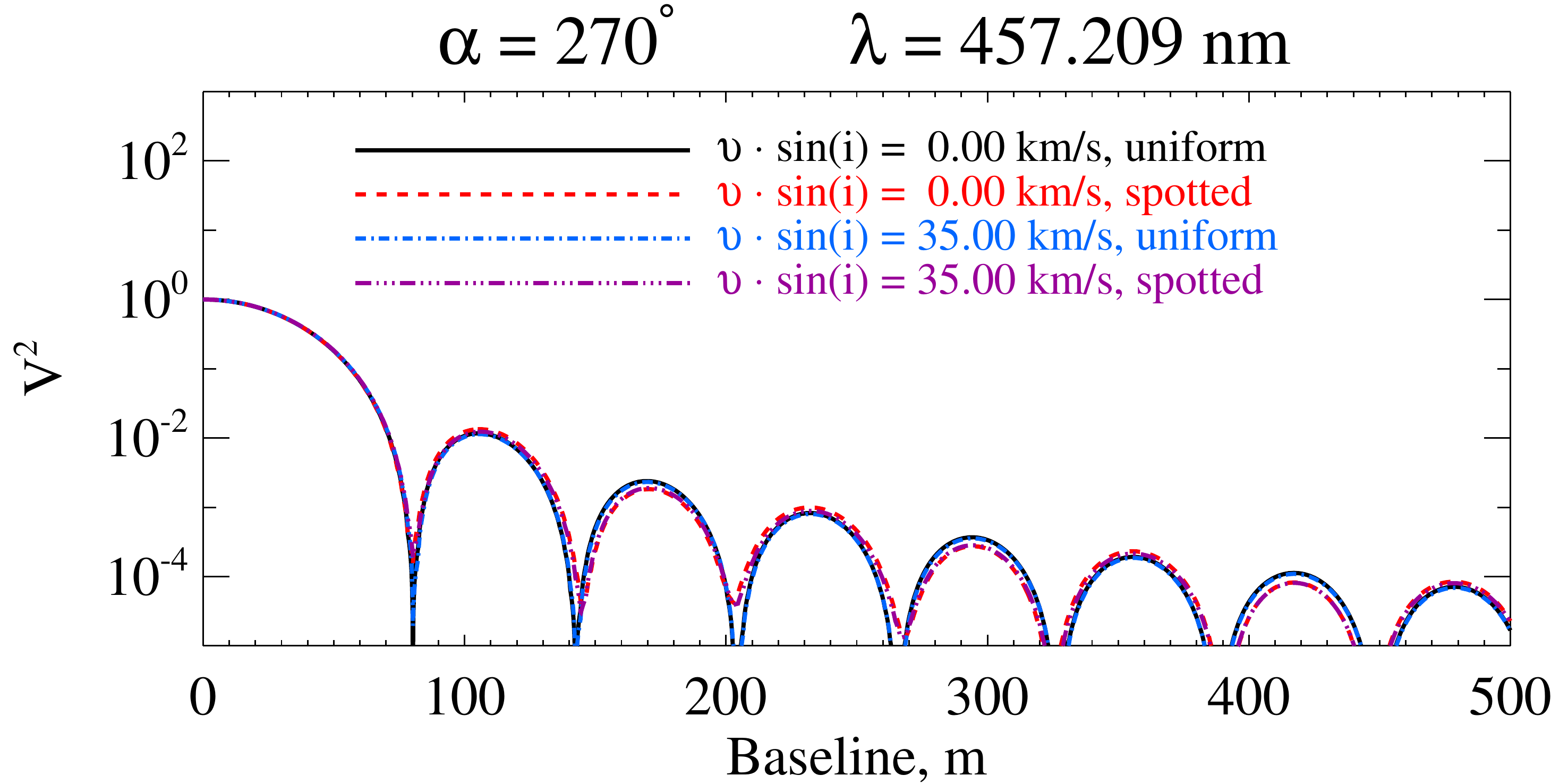}
\includegraphics[width=0.33\hsize]{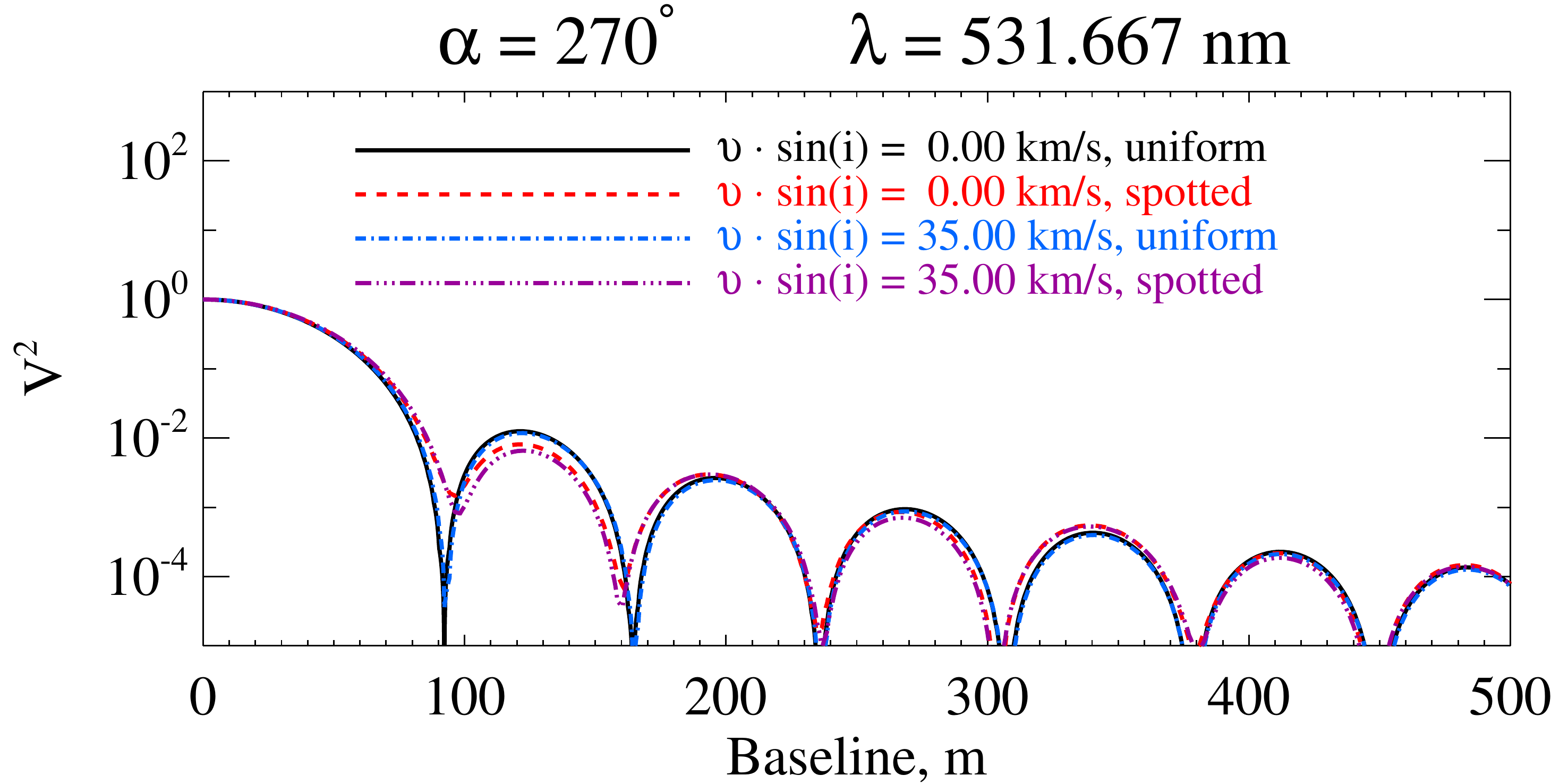}
}
\caption{Same as on Fig.~\ref{fig:vis-mono-v-1}, but for  \ion{Cr}{ii}~$455.86$~nm,
blend of \ion{Cr}{i}$+$\ion{Ti}{2} lines at $457.194$~nm, and \ion{Fe}{ii}~$531.667$~nm line
at spectral resolution $R=6\,000$.}
\label{fig:vis-mono-v-3}
\end{minipage}
}
\end{figure*}

The impact of rotation vanishes when the spectral resolution is degraded. At a lowest resolution 
of $R=6\,000$ which we assumed in our simulations for the visual region, most of the lines become blended and the intensity contrast
between adjacent wavelength points decreases. Therefore, no stripes of constant $\vsini$  are seen 
on the stellar surface no more, and {squared} visibility  curves look similar independent upon the choice of rotation rate.
This is illustrated on Fig.~\ref{fig:vis-mono-v-3}.
Similar to the case of higher resolution, the spot is better detected in the lines of Cr and the reason is that
Cr has a largest abundance gradient over the visible stellar disk. The strong \ion{Fe}{ii}~$531.667$~nm line
also induced noticeable signal at $\alpha=270^\circ$. In either case the spots are detected already at the first
visibility lobe but at small $V^2\lesssim0.02$. The blend of \ion{Cr}{i}$+$\ion{Ti}{ii} lines at $457.194$~nm
gives only a marginal signal.

\subsubsection{Visibility vs. wavelength}

\begin{figure*}
\centerline{
\includegraphics[width=0.33\hsize]{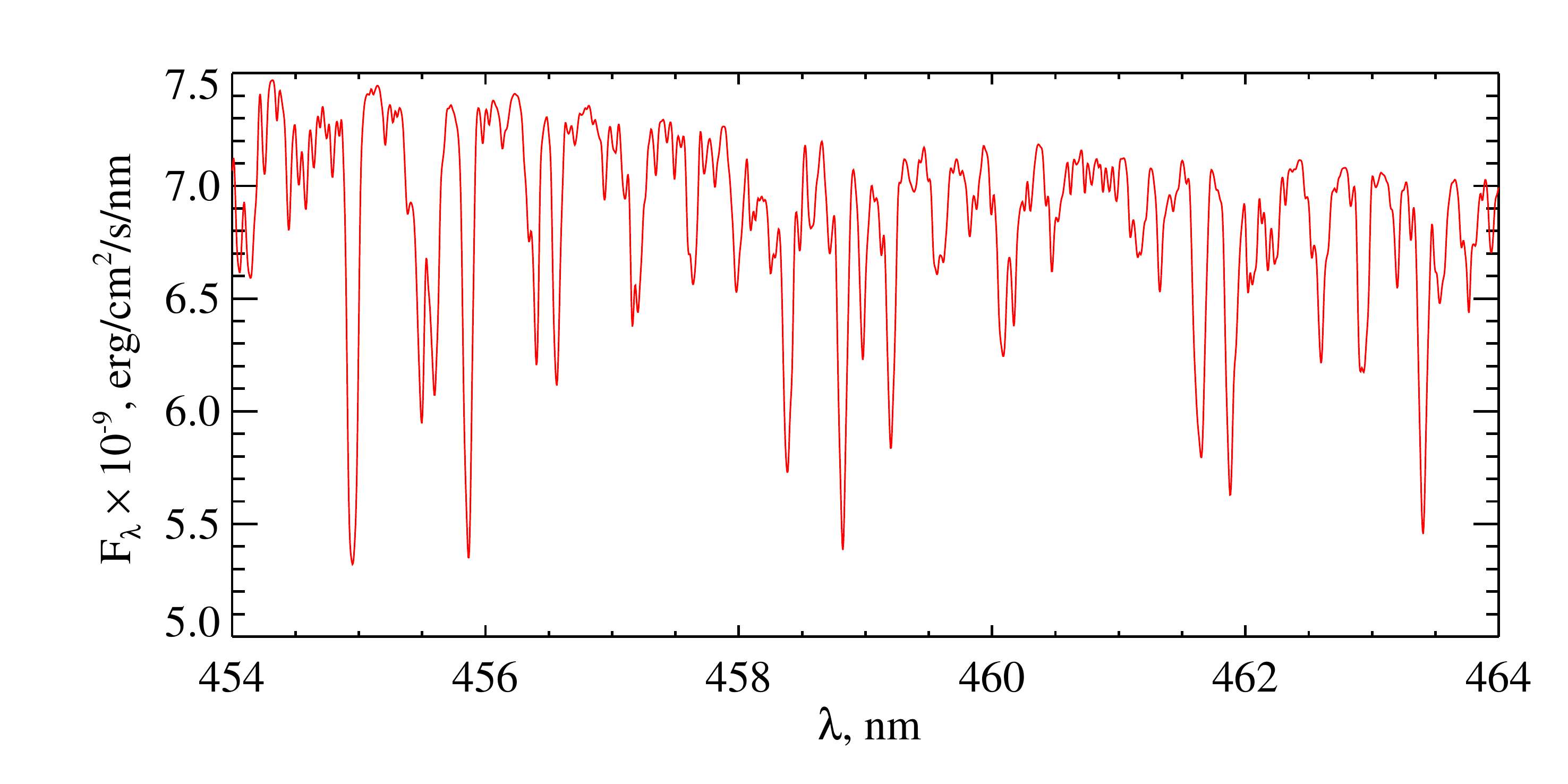}
\includegraphics[width=0.33\hsize]{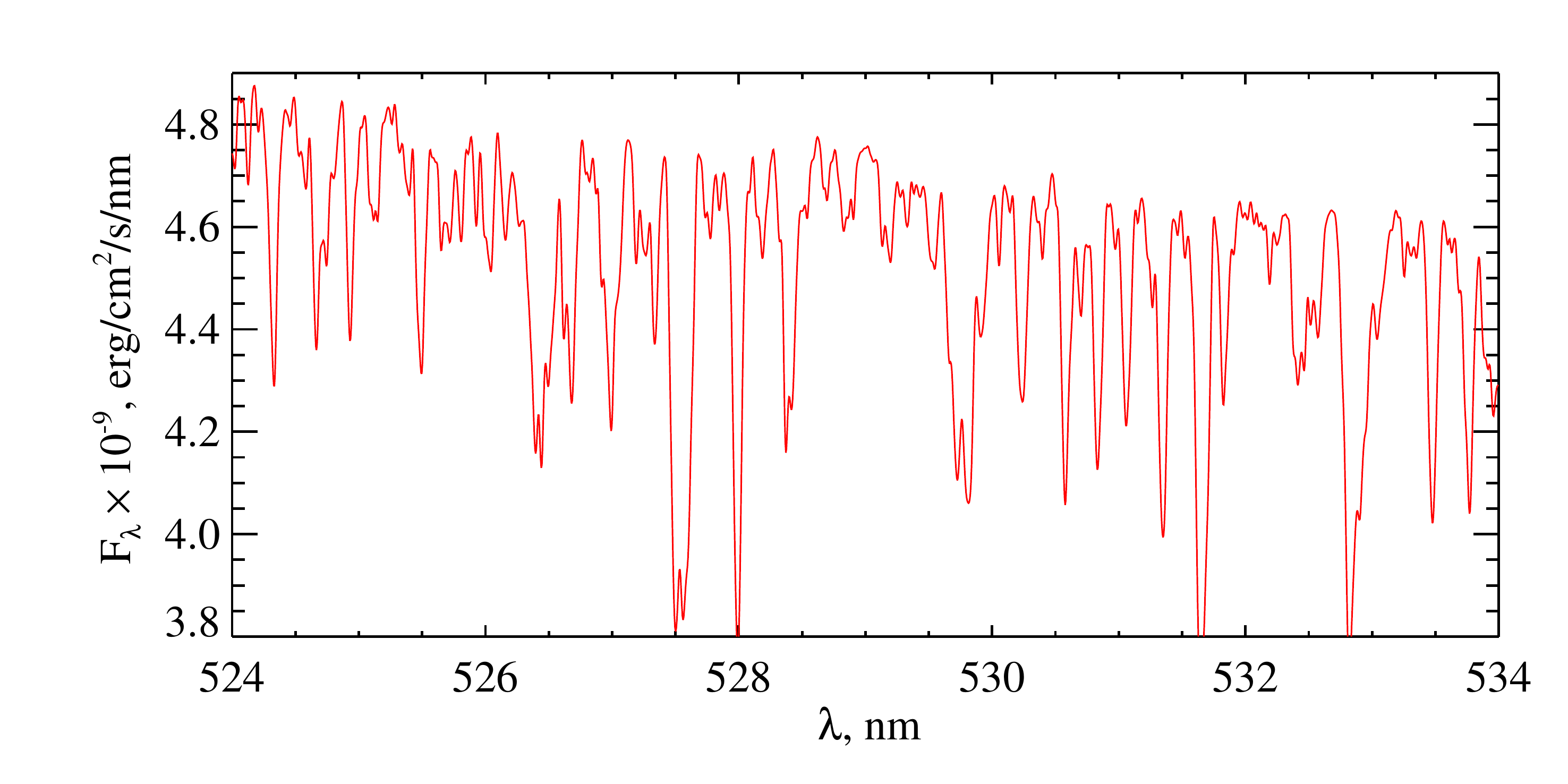}
\includegraphics[width=0.33\hsize]{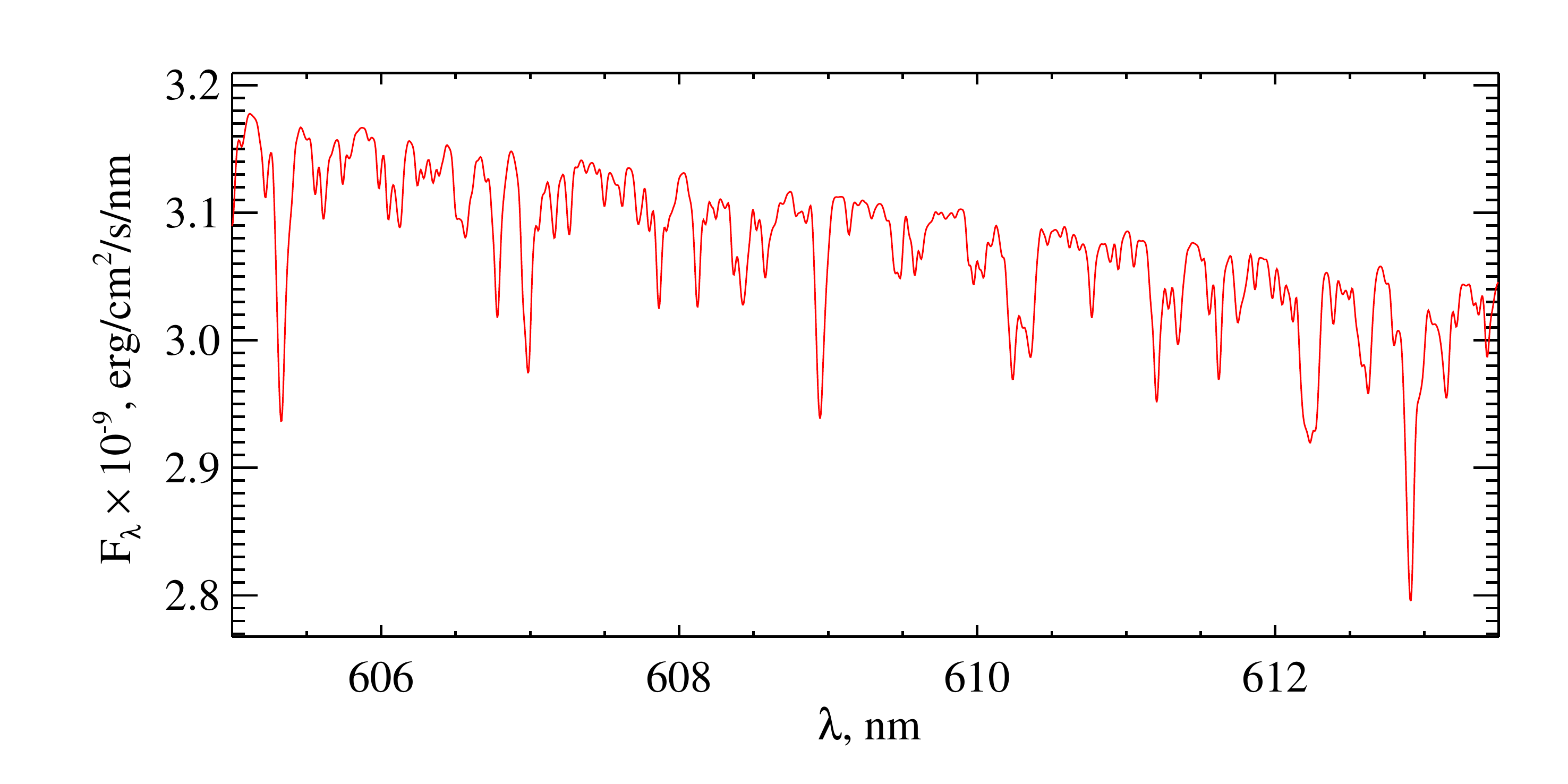}
}
\centerline{
\includegraphics[width=0.33\hsize]{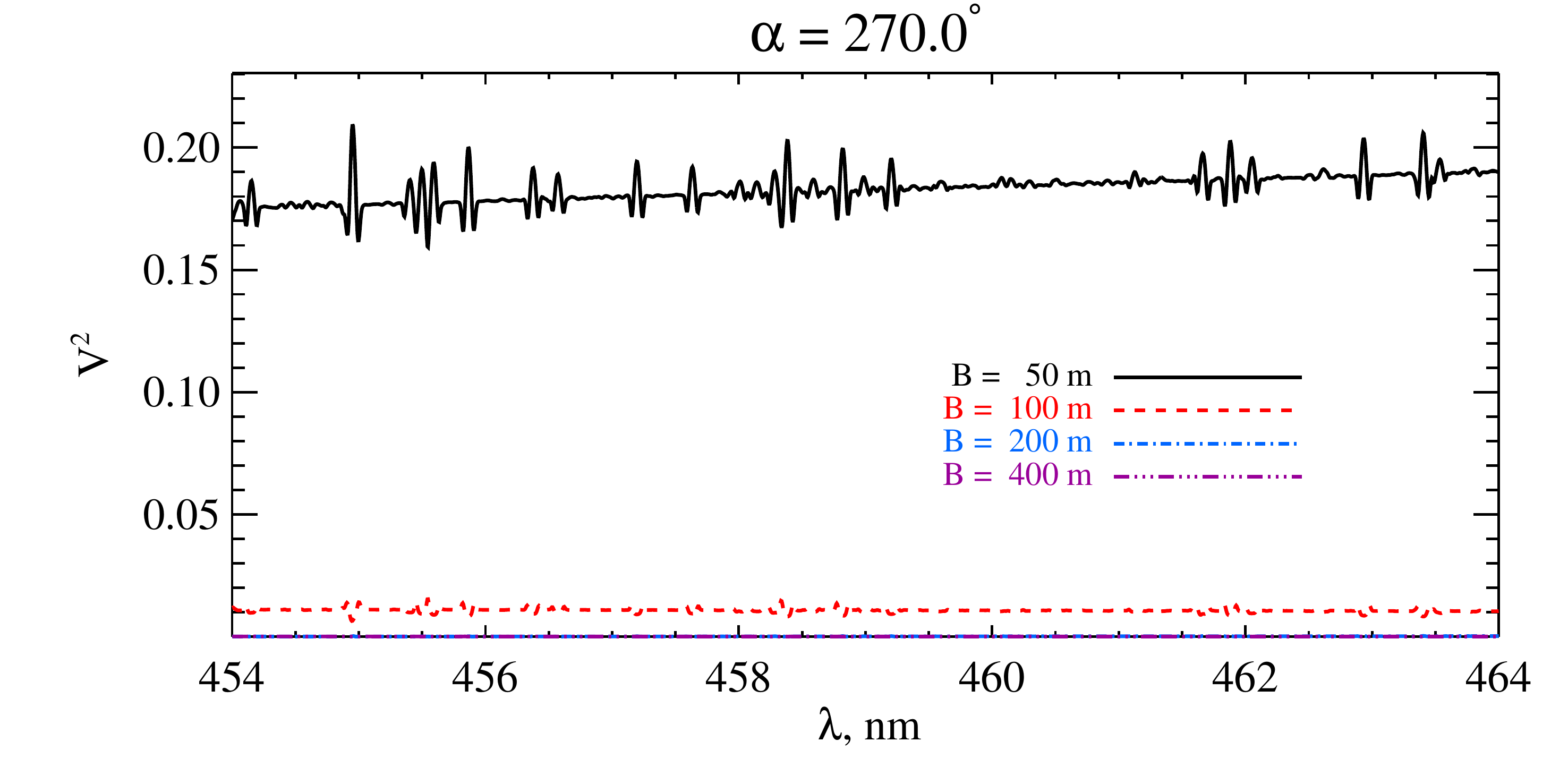}
\includegraphics[width=0.33\hsize]{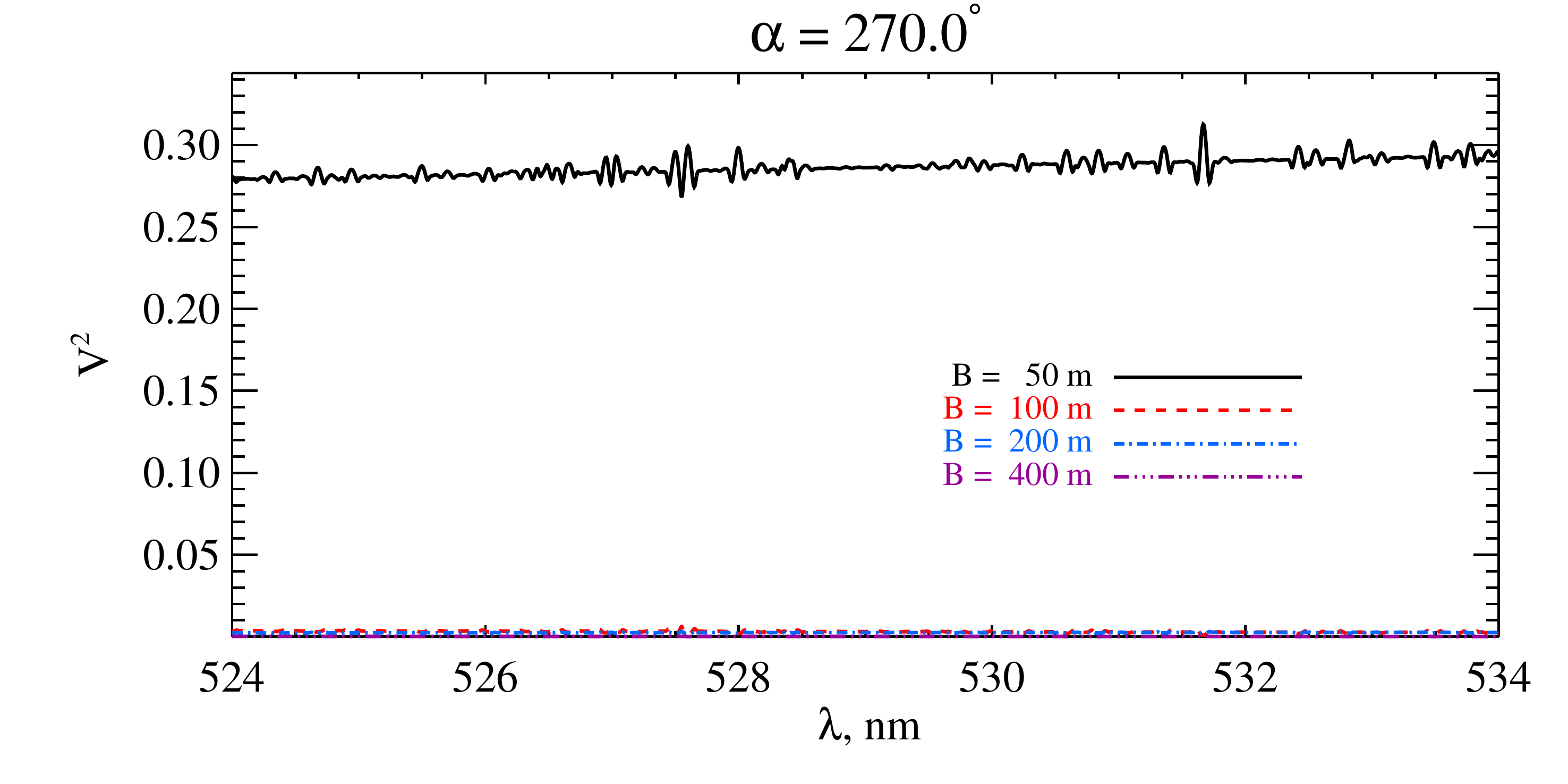}
\includegraphics[width=0.33\hsize]{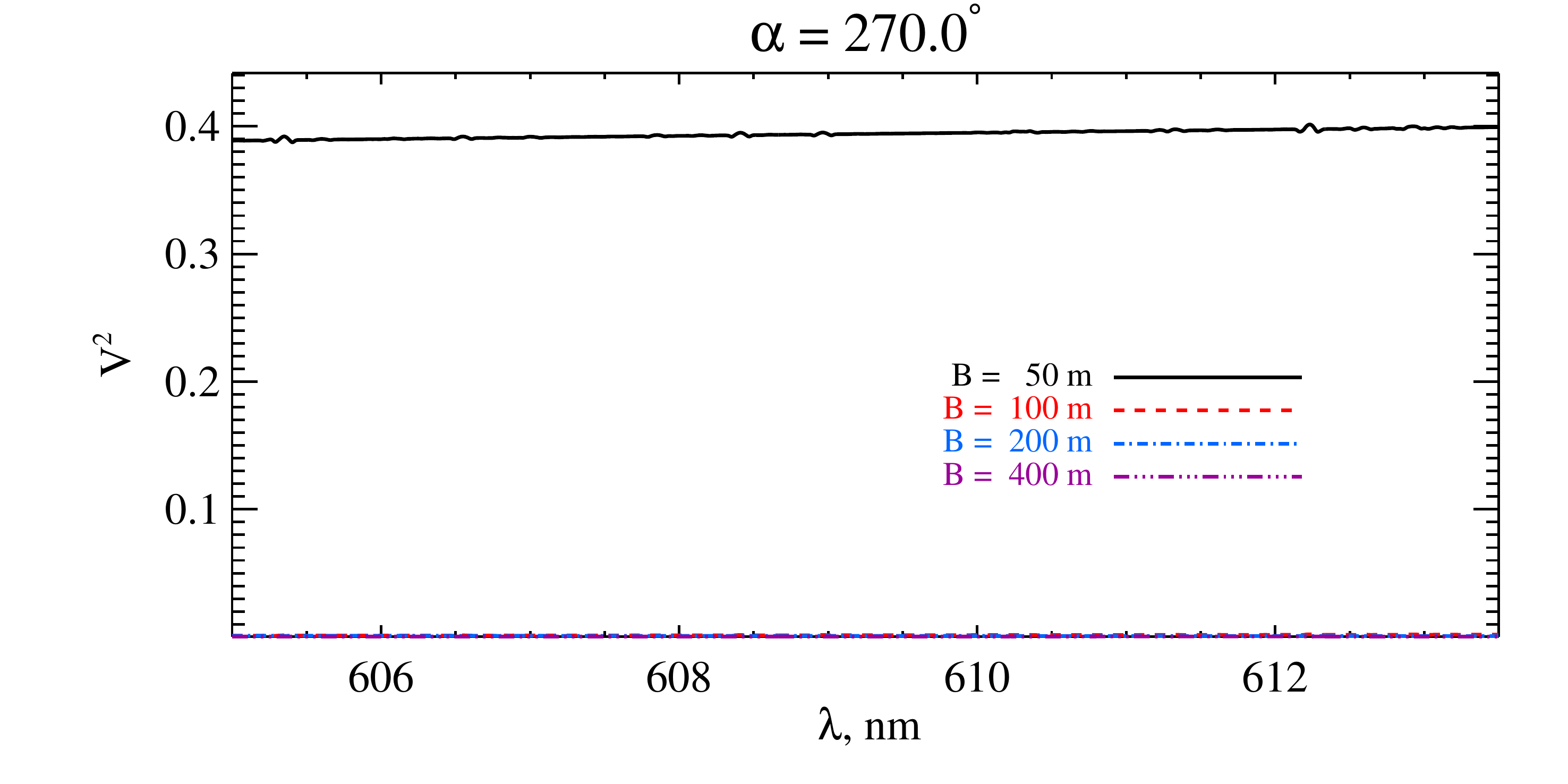}
}
\centerline{
\includegraphics[width=0.33\hsize]{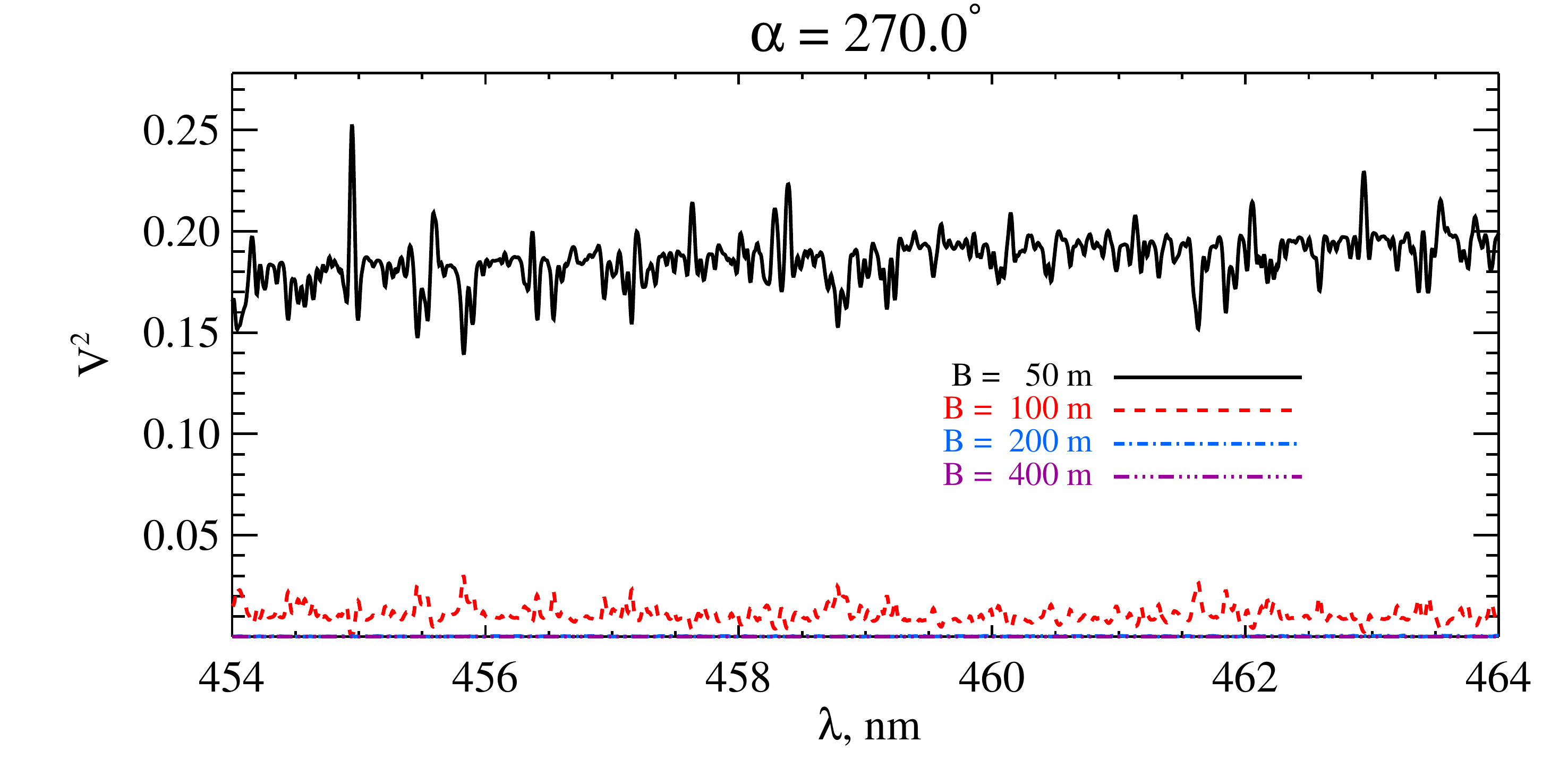}
\includegraphics[width=0.33\hsize]{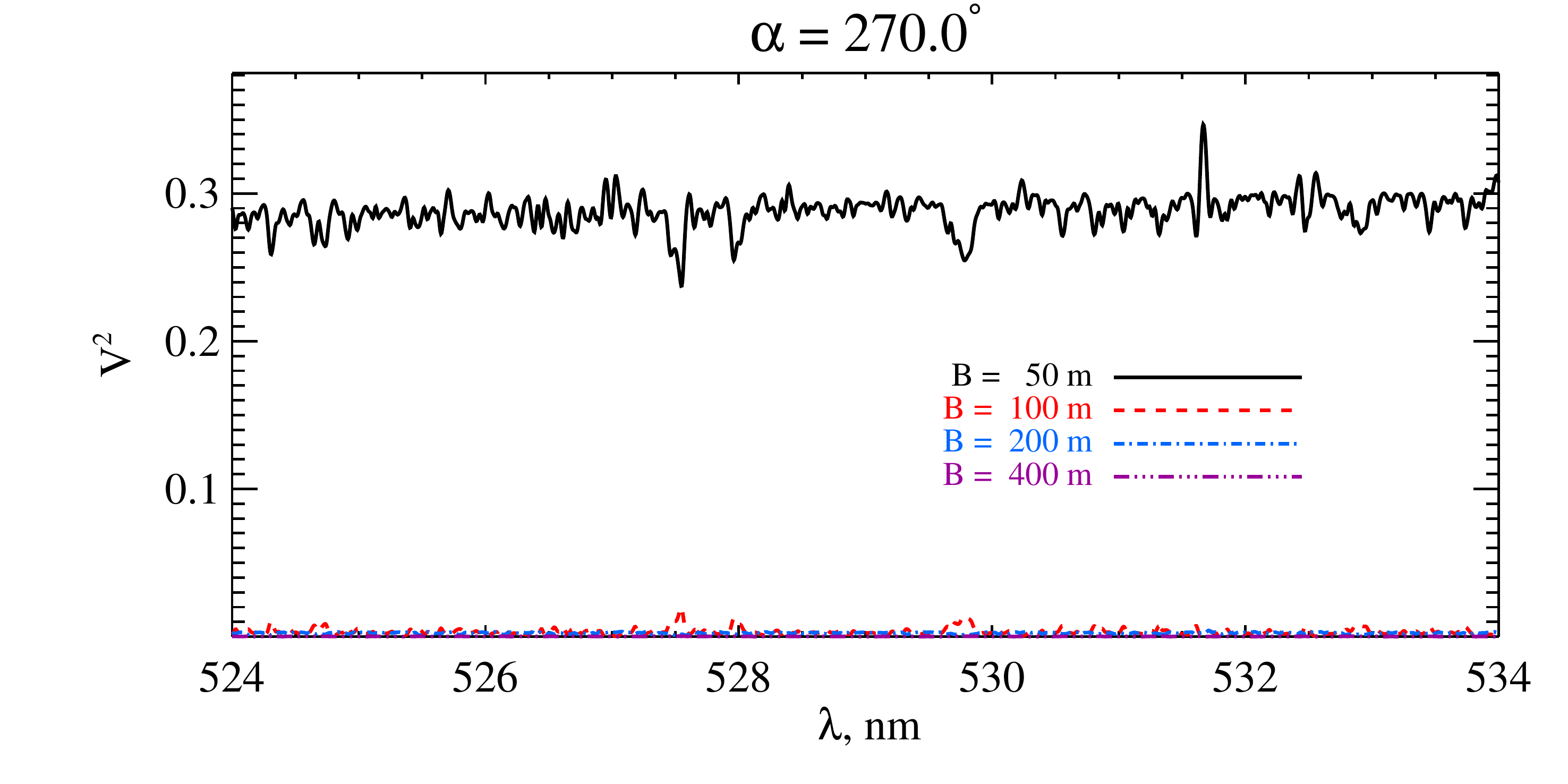}
\includegraphics[width=0.33\hsize]{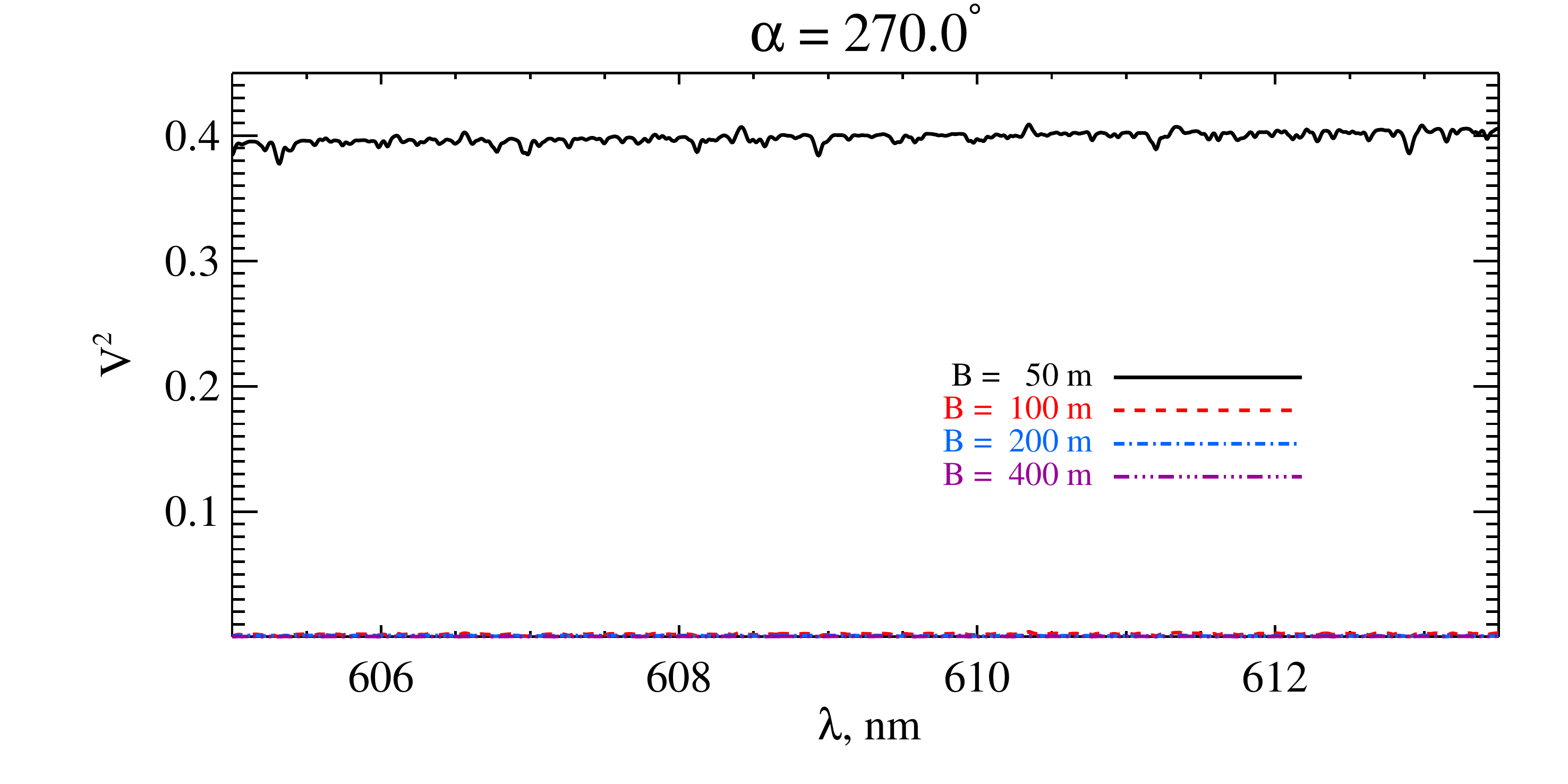}
}
\caption{{Squared} visibility  as a function of wavelength calculated at four selected baselines.
First row~--~spectrum predicted by spotted model; second and third rows~--~{squared} visibility predicted by homogeneous
and spotted models respectively. In all plots $R=30\,000$, $\vsini=35$~\kms. 
{Squared} visibility  plots show predictions for the orientation with position angle of $270^\circ$.}
\label{fig:vis-lambda-v-r30000}
\end{figure*}

\begin{figure*}
\centerline{
\includegraphics[width=0.33\hsize]{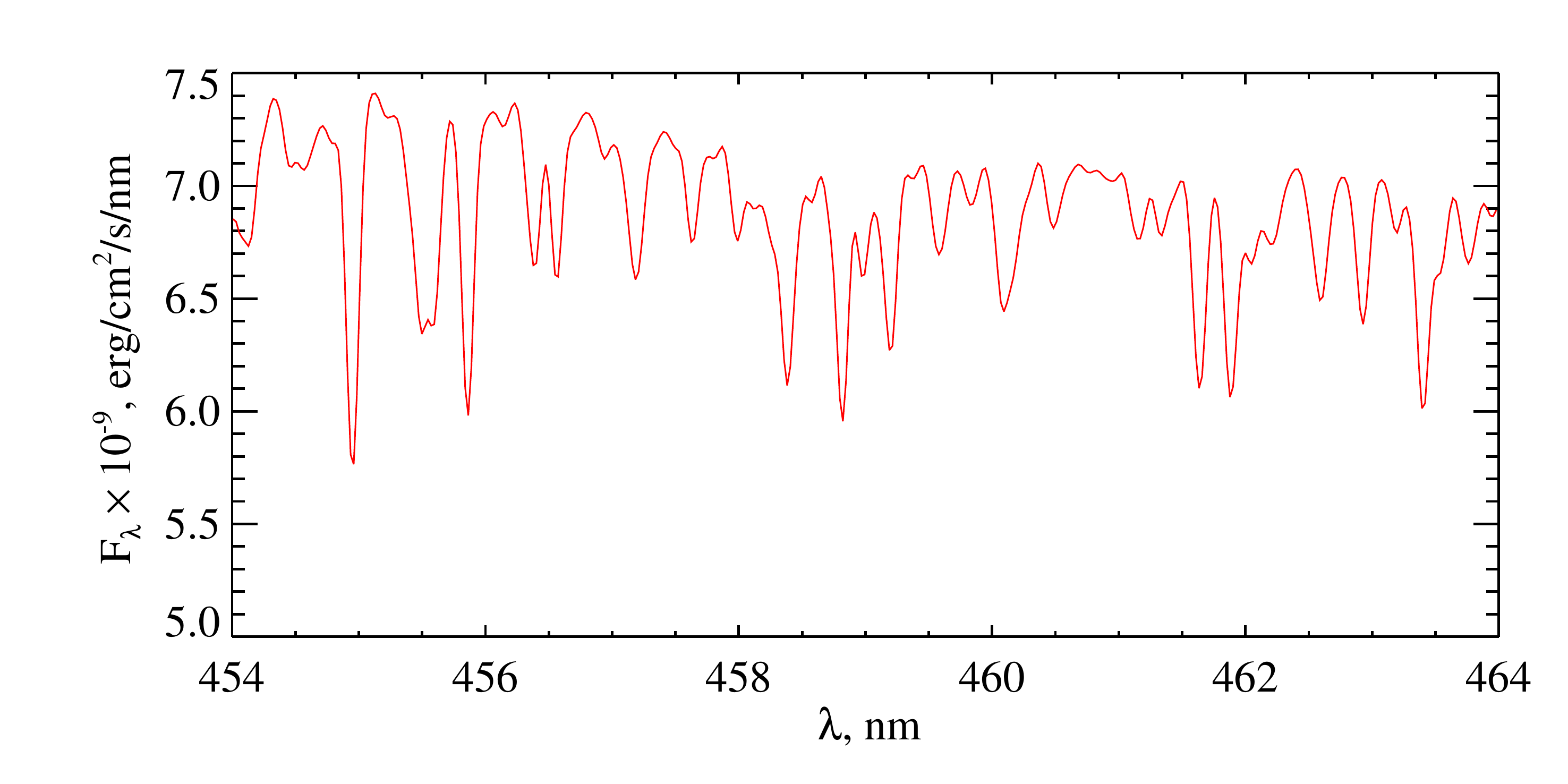}
\includegraphics[width=0.33\hsize]{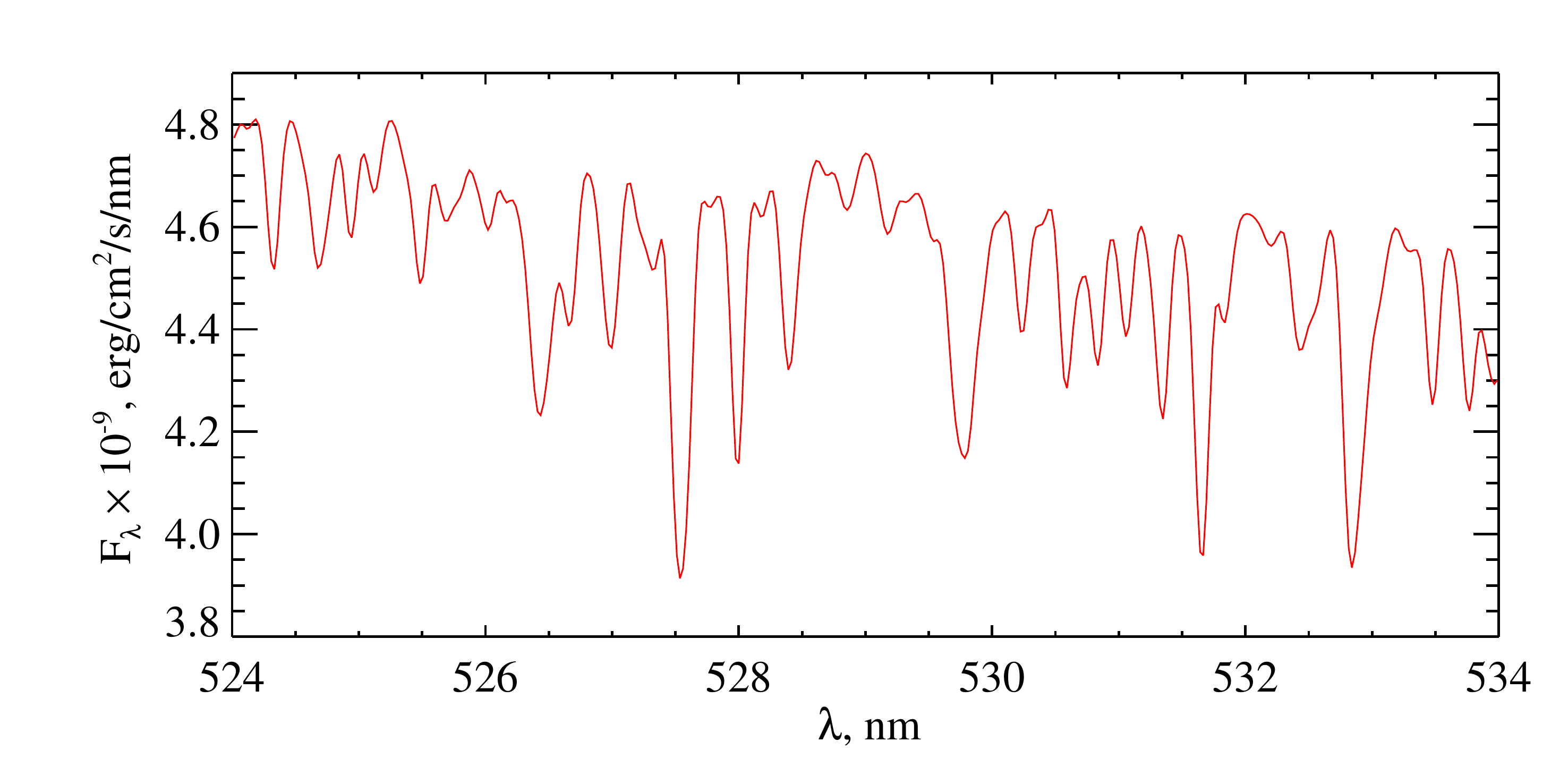}
\includegraphics[width=0.33\hsize]{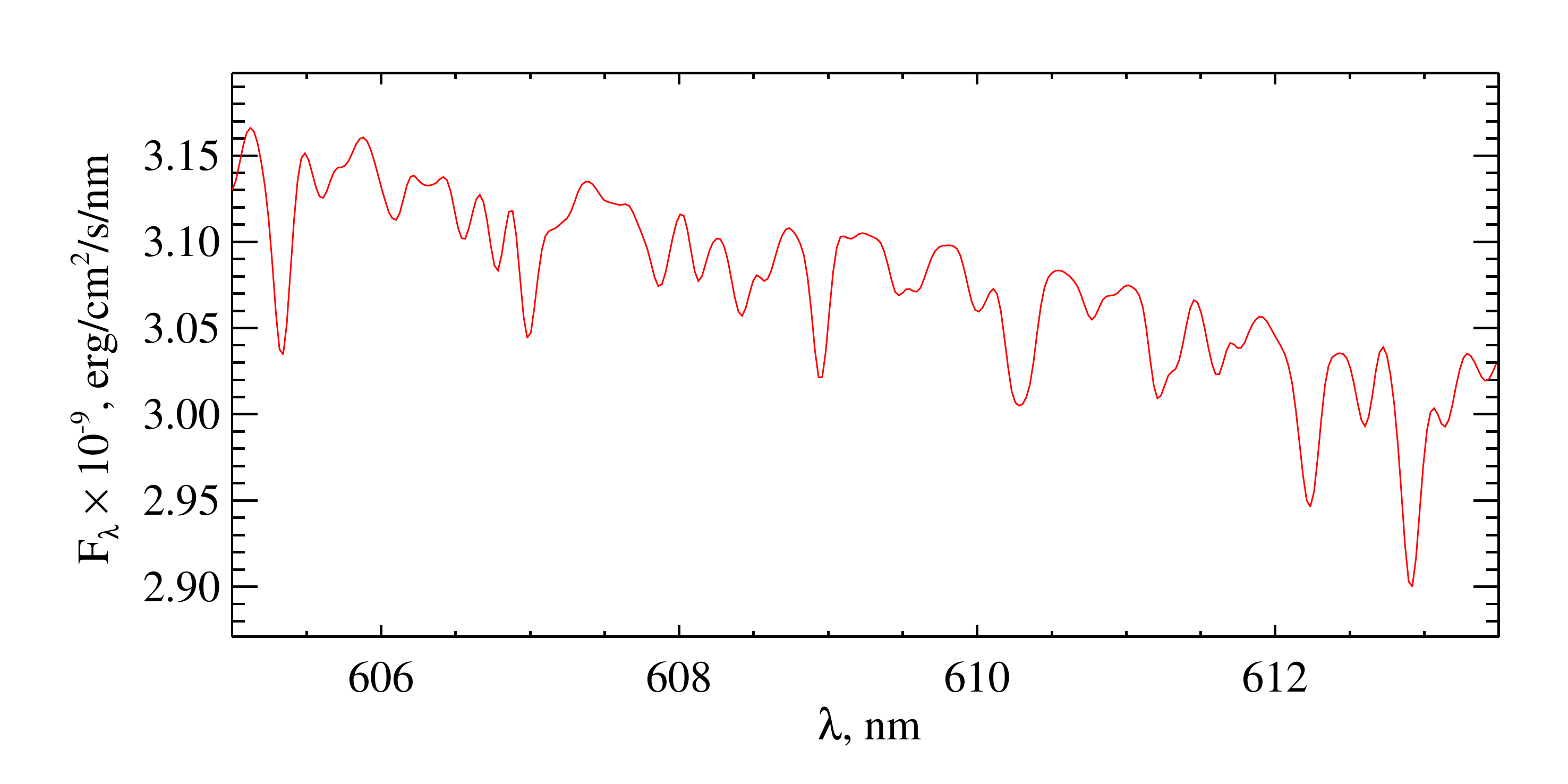}
}
\centerline{
\includegraphics[width=0.33\hsize]{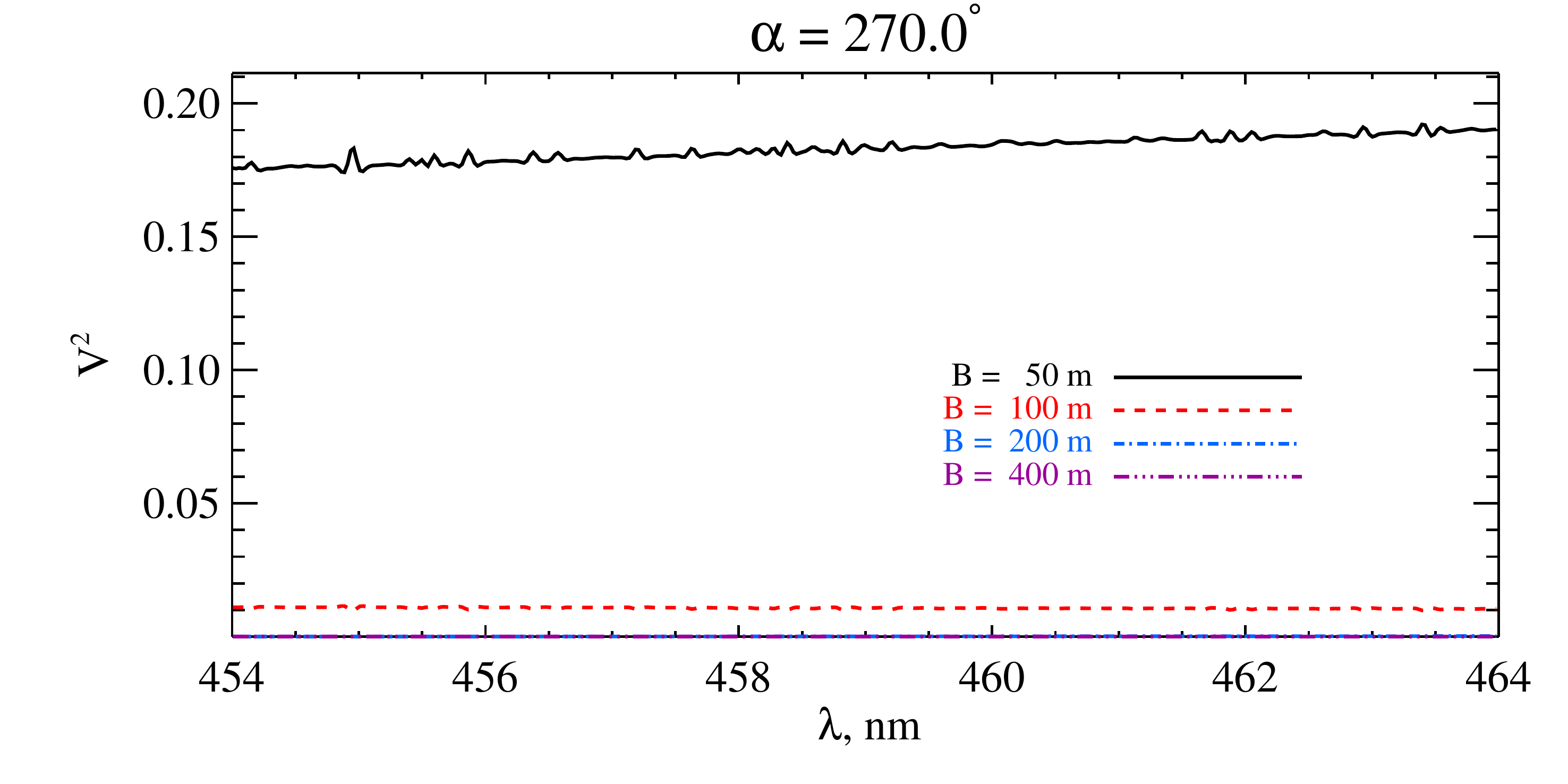}
\includegraphics[width=0.33\hsize]{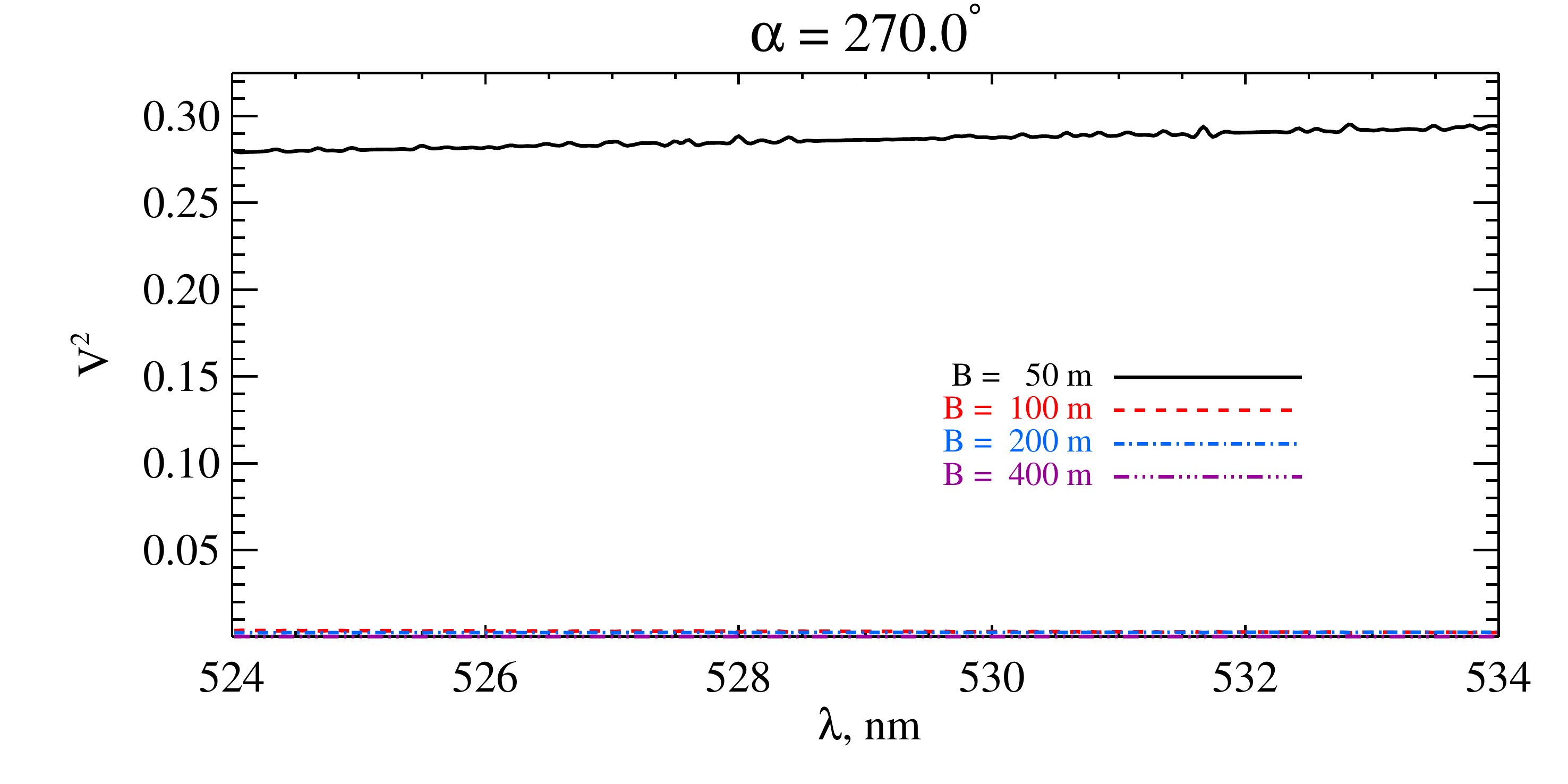}
\includegraphics[width=0.33\hsize]{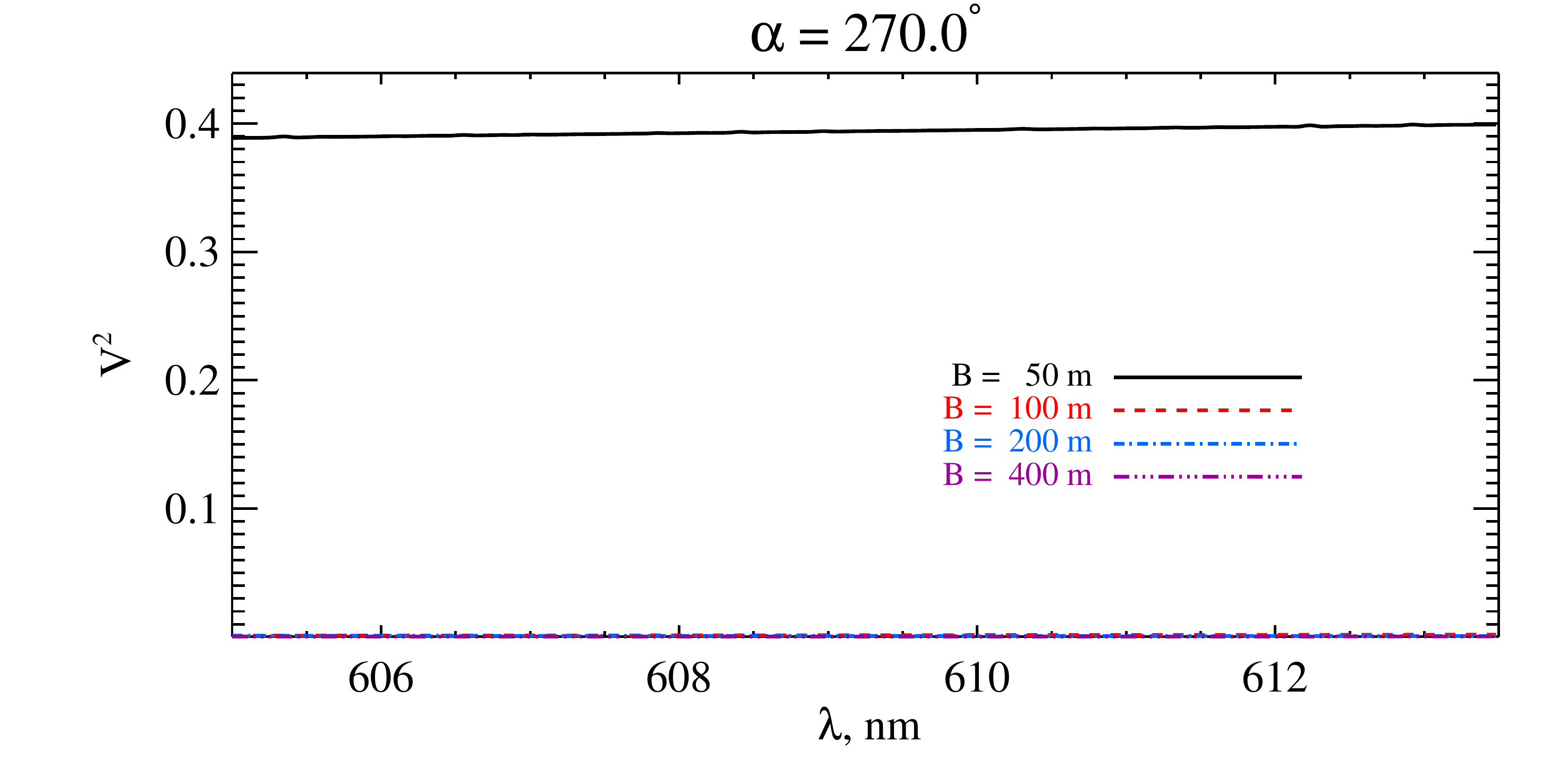}
}
\centerline{
\includegraphics[width=0.33\hsize]{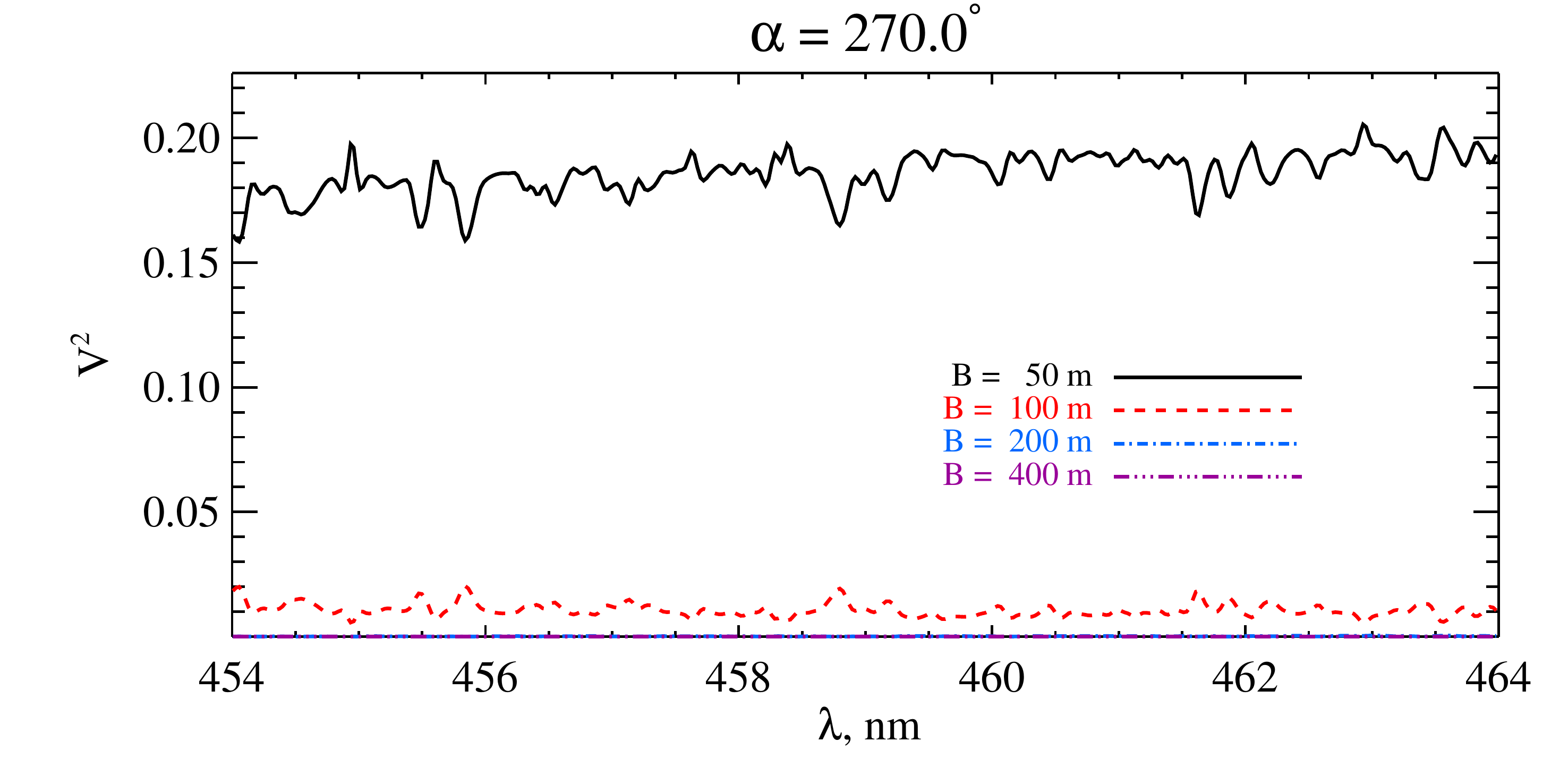}
\includegraphics[width=0.33\hsize]{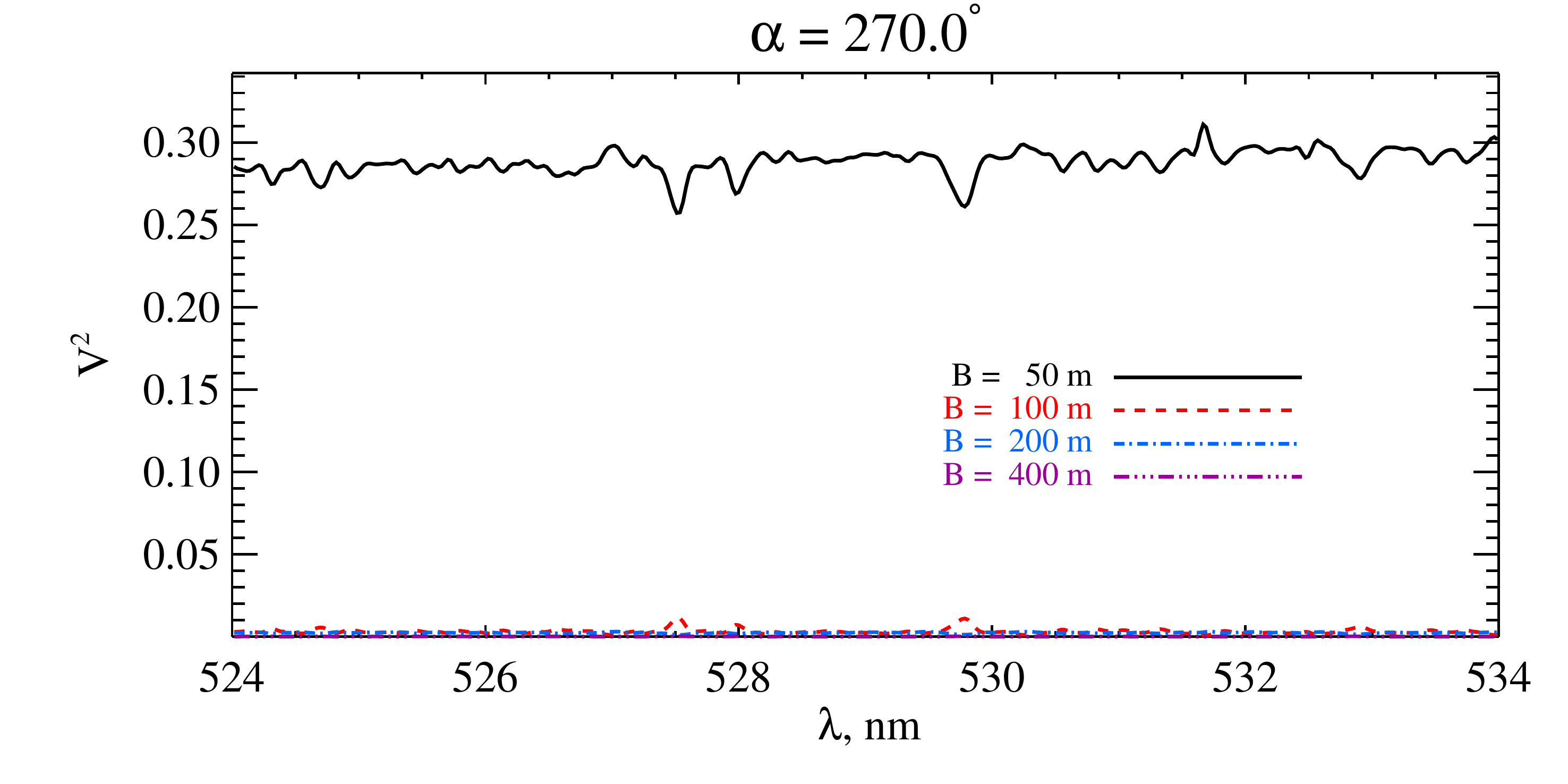}
\includegraphics[width=0.33\hsize]{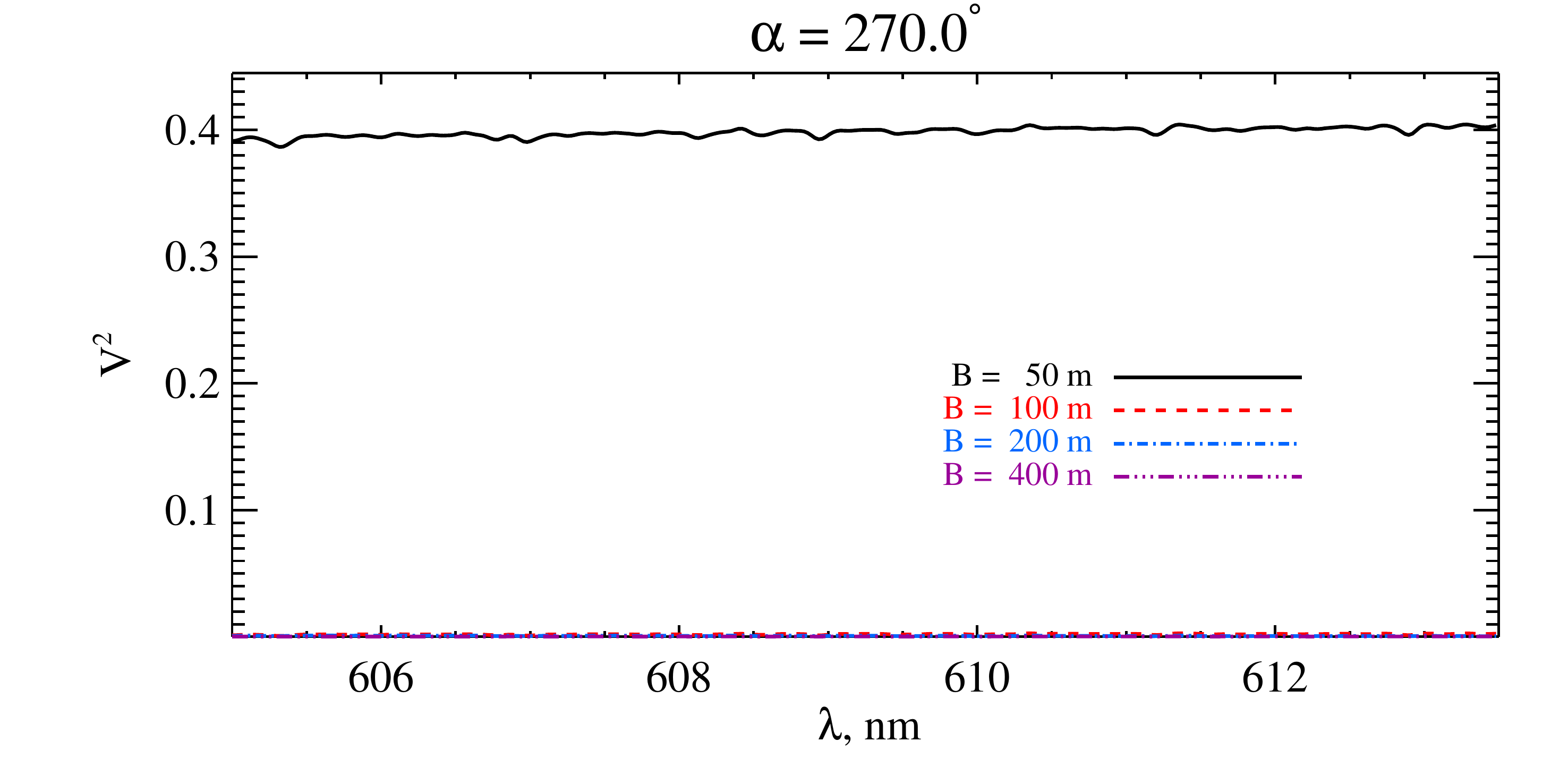}
}
\caption{Same as on Fig.~\ref{fig:vis-lambda-v-r30000}, but with $R=6\,000$.}
\label{fig:vis-lambda-v-r6000}
\end{figure*}

\begin{figure*}
\centerline{
\includegraphics[width=0.33\hsize]{figures/sp-wavelength-images-R30000-vsini35.00-mode2-lambda-4540.0000-4640.0000.pdf}
\includegraphics[width=0.33\hsize]{figures/sp-wavelength-images-R30000-vsini35.00-mode2-lambda-5240.0000-5340.0000.pdf}
\includegraphics[width=0.33\hsize]{figures/sp-wavelength-images-R30000-vsini35.00-mode2-lambda-6050.0000-6135.0000.pdf}
}
\centerline{
\includegraphics[width=0.33\hsize]{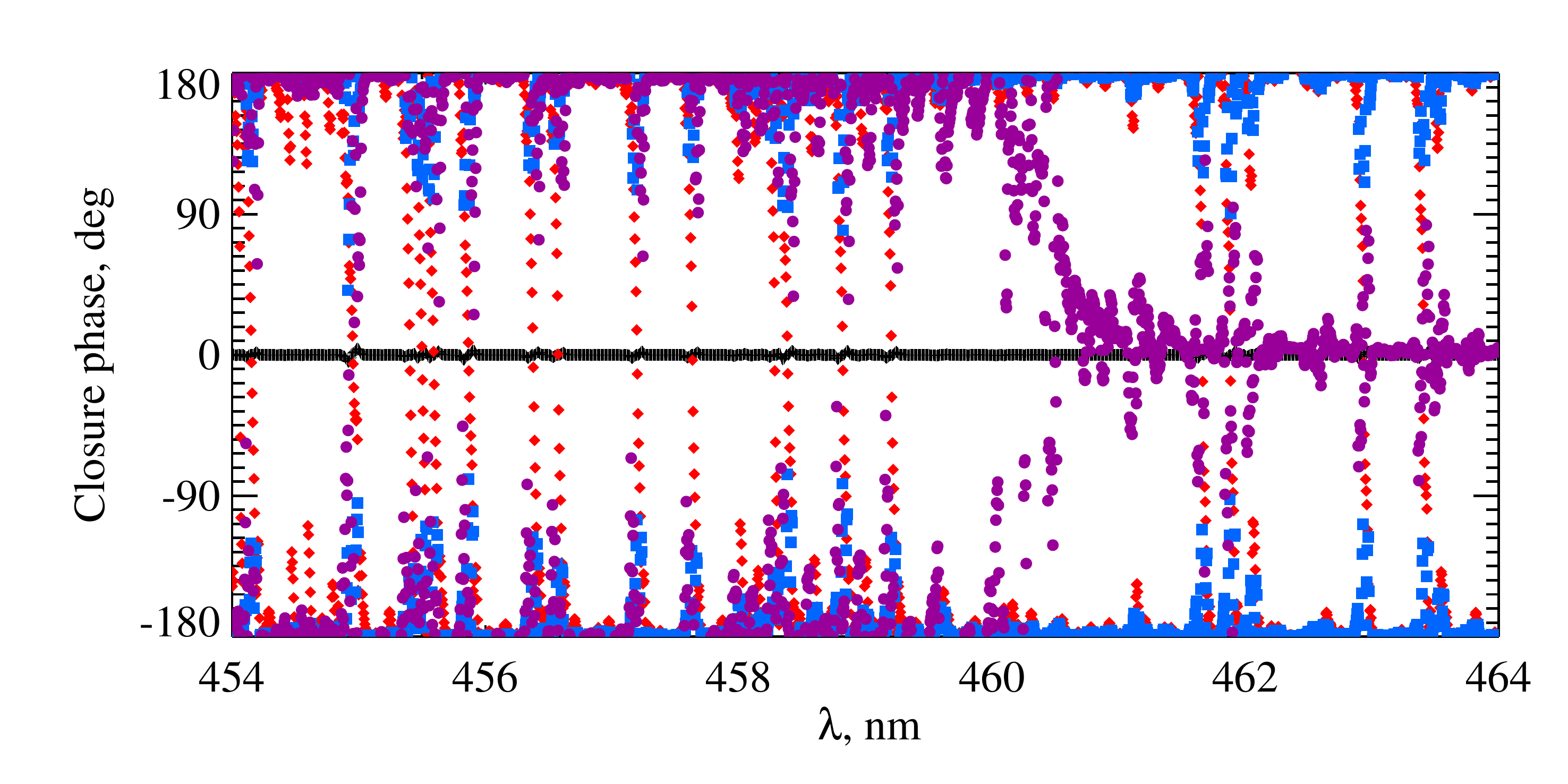}
\includegraphics[width=0.33\hsize]{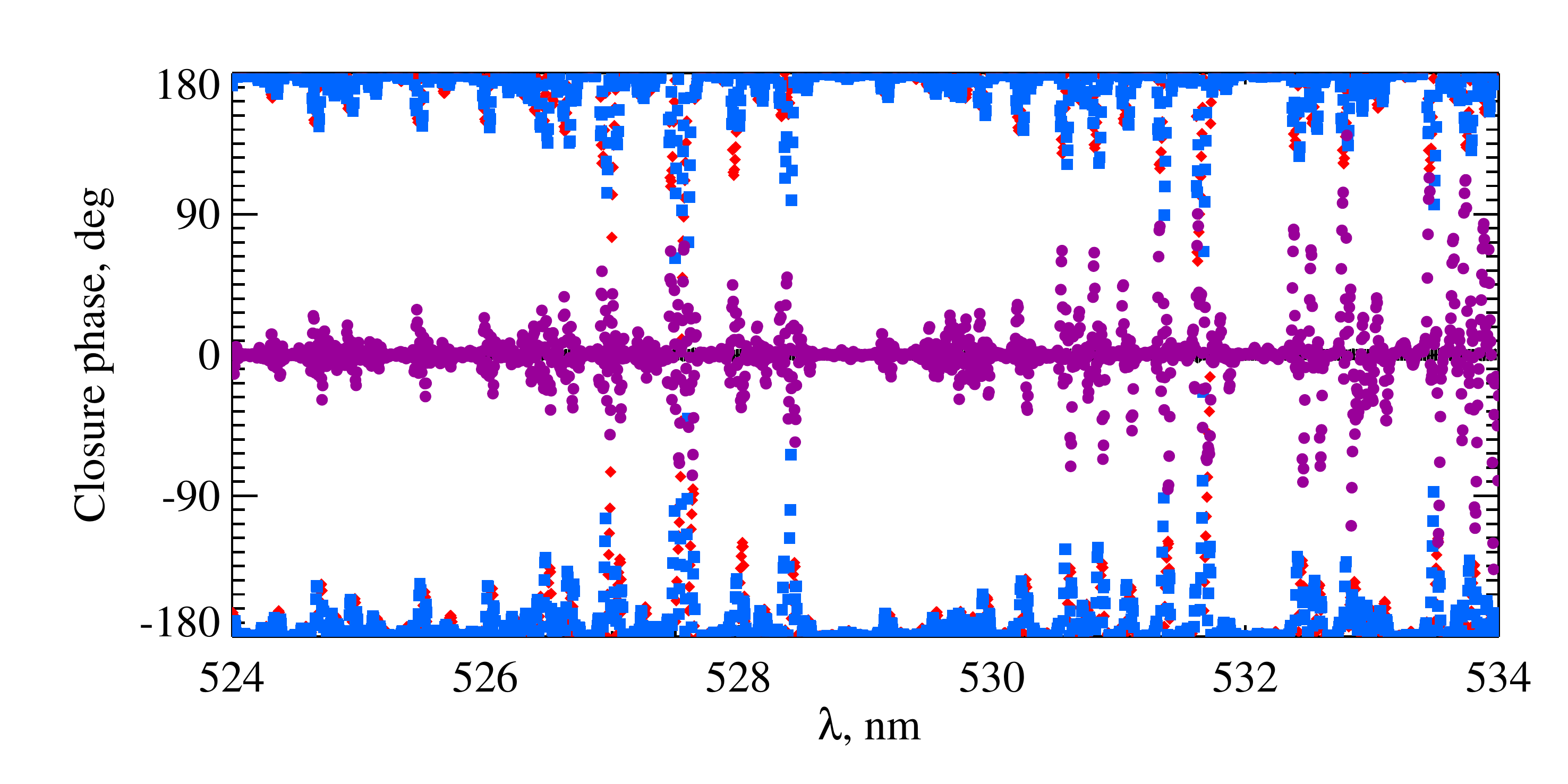}
\includegraphics[width=0.33\hsize]{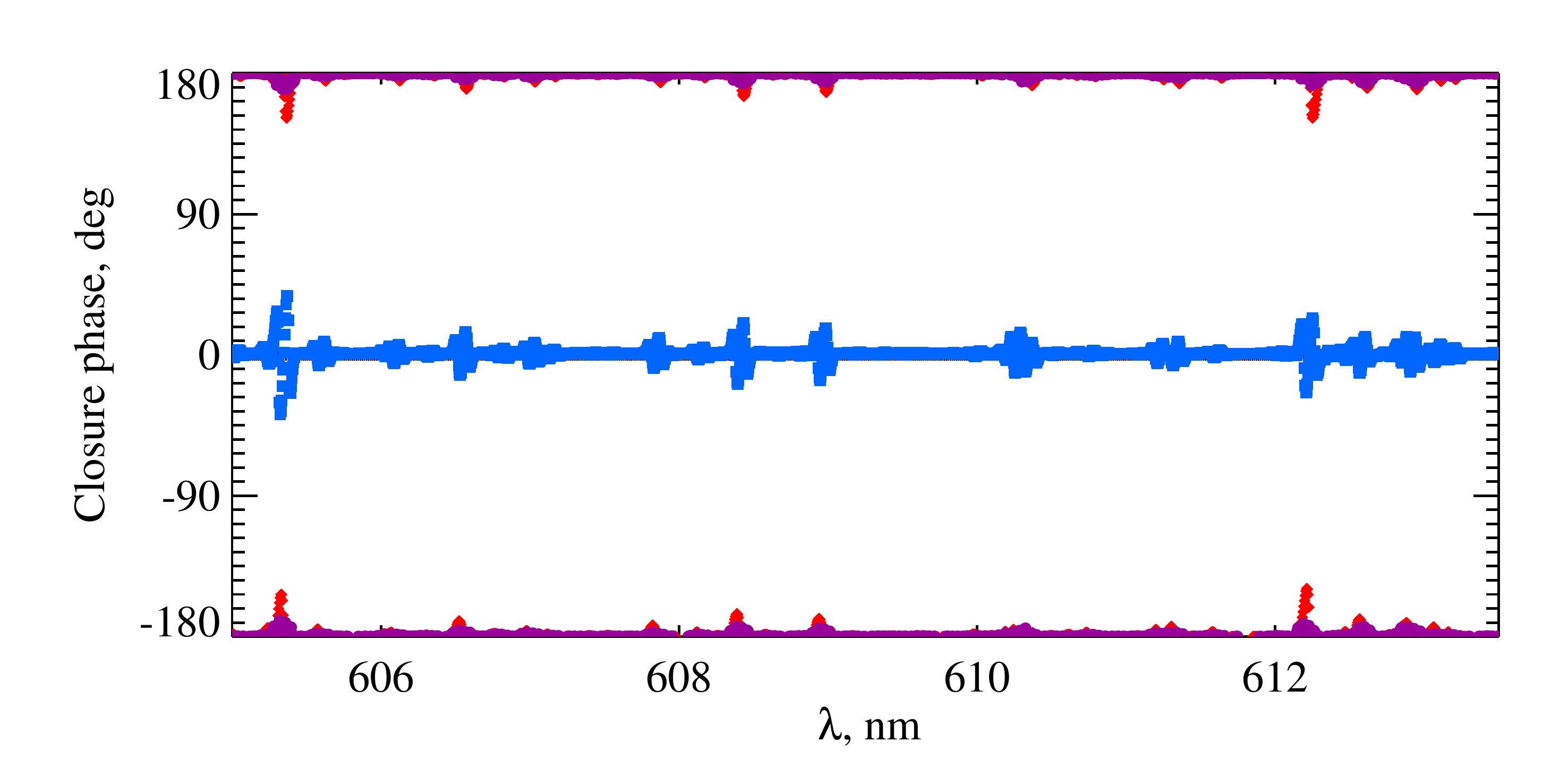}
}
\centerline{
\includegraphics[width=0.33\hsize]{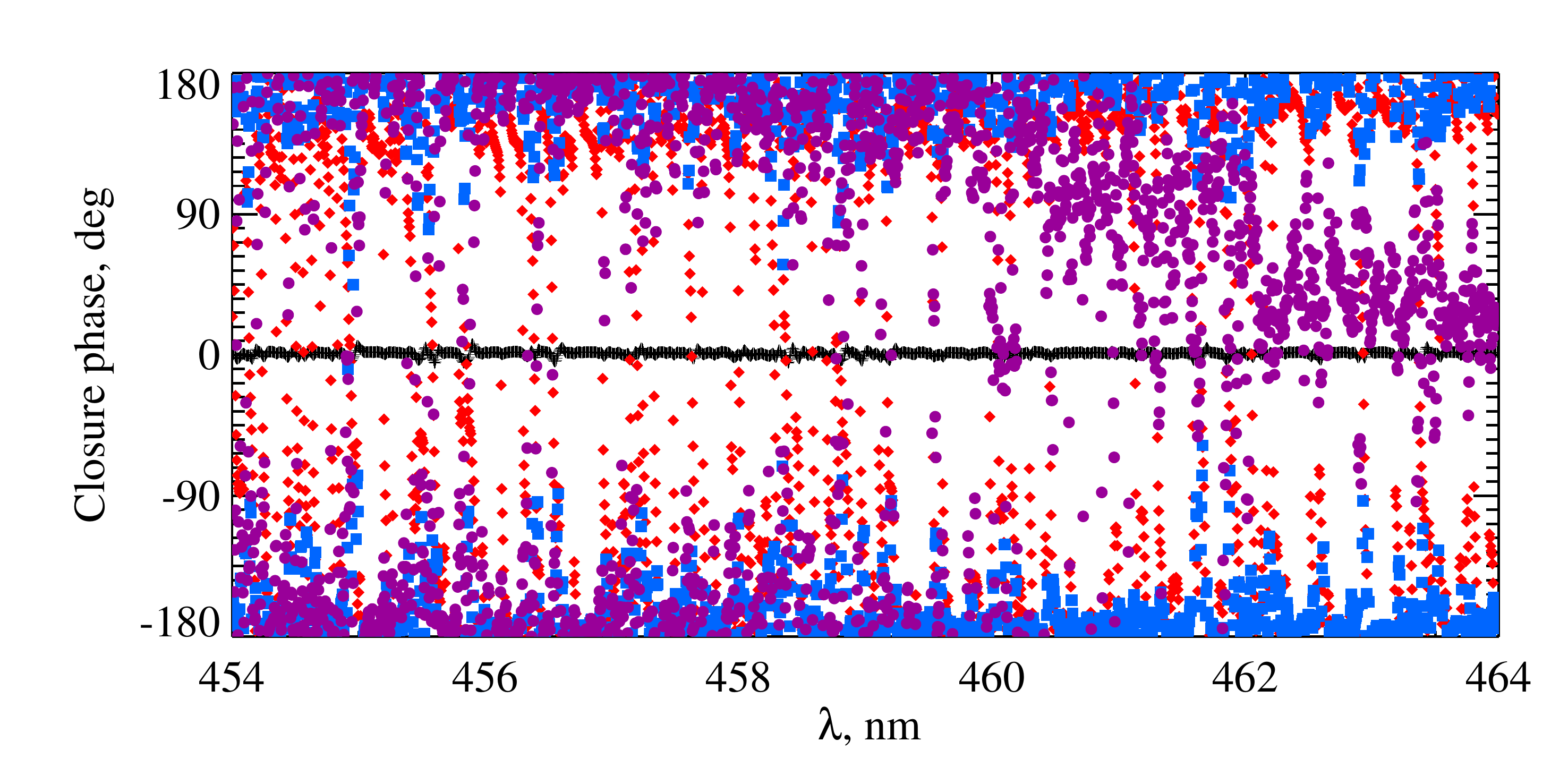}
\includegraphics[width=0.33\hsize]{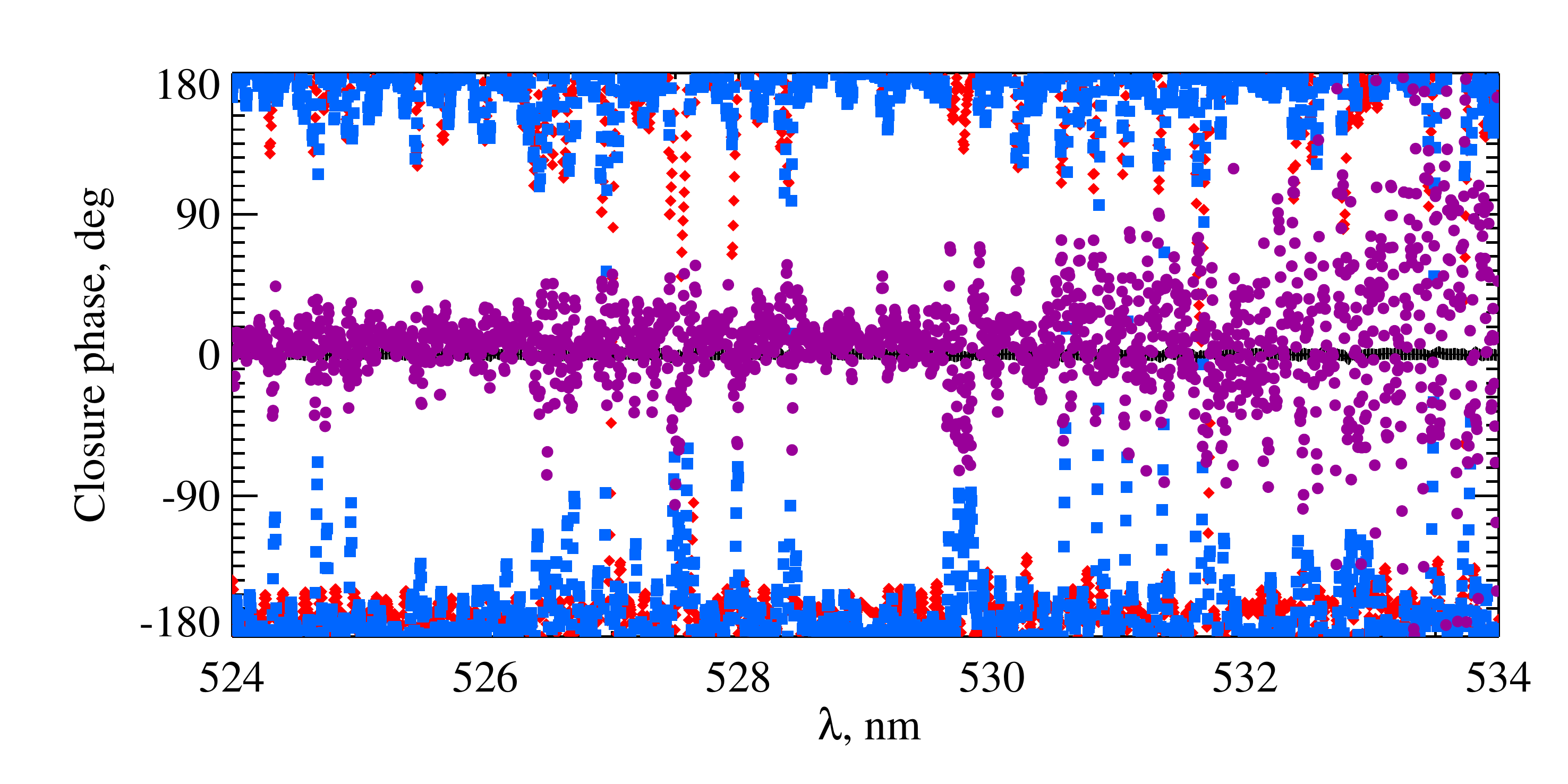}
\includegraphics[width=0.33\hsize]{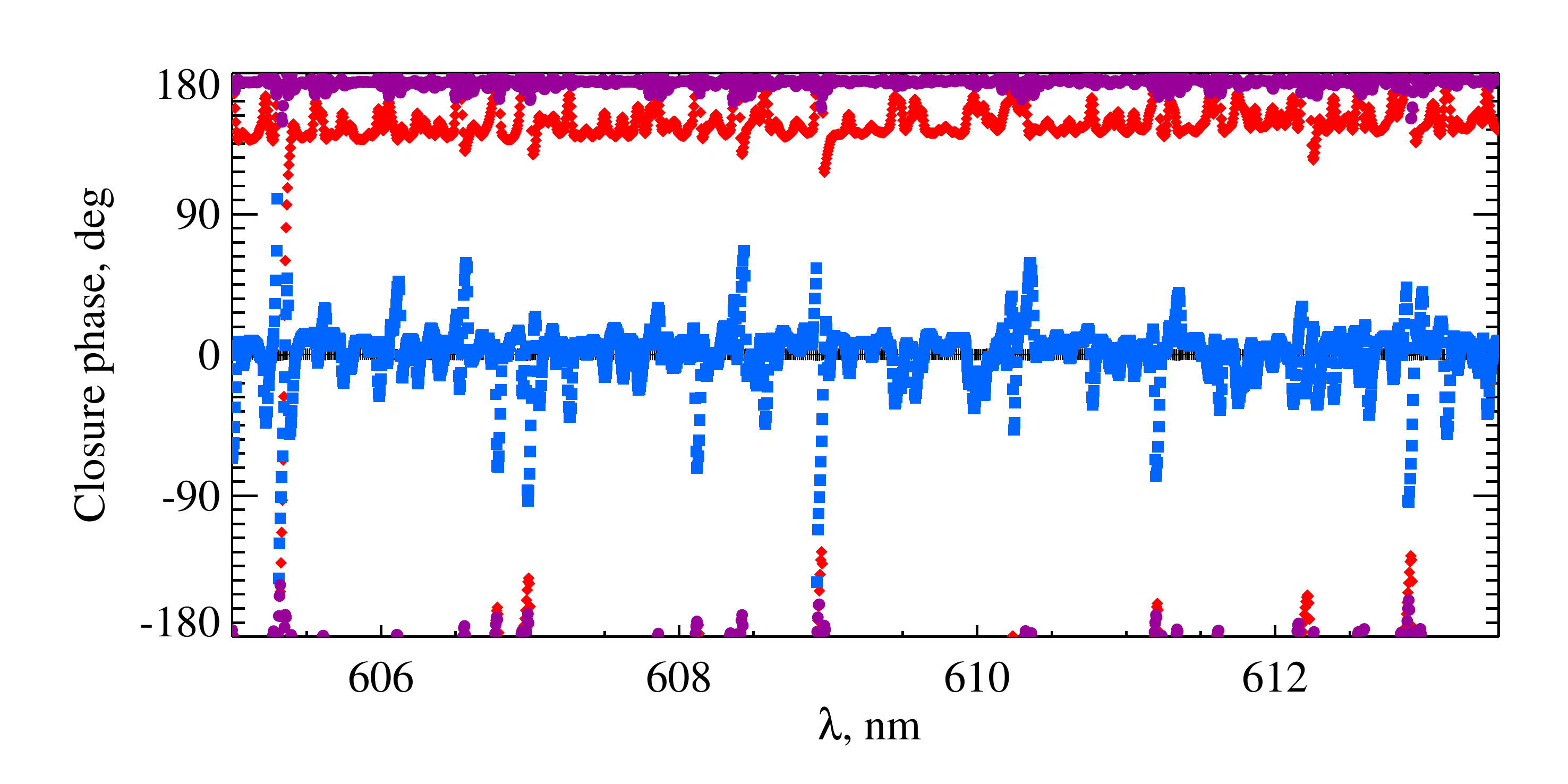}
}
\caption{Closure phases as a function of wavelength.
First row~--~spectrum predicted by spotted model; second and third rows~--~closure phase predicted by homogeneous
and spotted models, respectively. In all plots $R=30\,000$, $\vsini=35$~\kms. 
Closure phases were computed for the following configurations:
($0^\circ,40$m)$+$($270^\circ,40$m)$+$($135^\circ,57$m)~--~black crosses; 
($0^\circ,100$m)$+$($270^\circ,100$m)$+$($135^\circ,141$m)~--~red diamonds;
($0^\circ,180$m)$+$($270^\circ,180$m)$+$($135^\circ,255$m)~--~blue squares;
($0^\circ,320$m)$+$($270^\circ,320$m)$+$($135^\circ,453$m)~--~violet circles. See online version for colored symbols.}
\label{fig:cp-lambda-v-r30000}
\end{figure*}

\begin{figure*}
\centerline{
\includegraphics[width=0.33\hsize]{figures/sp-wavelength-images-R6000-vsini35.00-mode2-lambda-4540.0000-4640.0000.pdf}
\includegraphics[width=0.33\hsize]{figures/sp-wavelength-images-R6000-vsini35.00-mode2-lambda-5240.0000-5340.0000.pdf}
\includegraphics[width=0.33\hsize]{figures/sp-wavelength-images-R6000-vsini35.00-mode2-lambda-6050.0000-6135.0000.pdf}
}
\centerline{
\includegraphics[width=0.33\hsize]{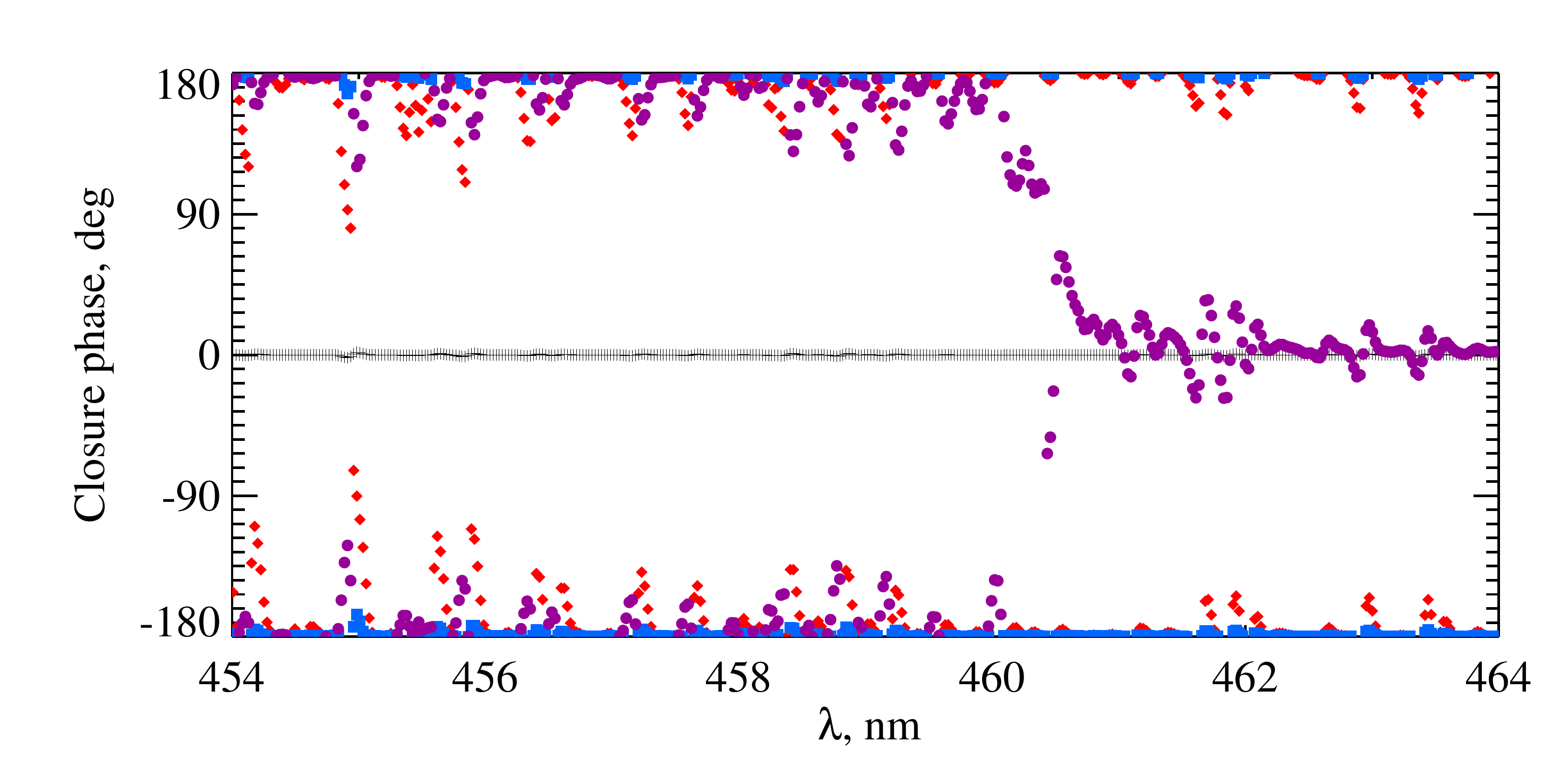}
\includegraphics[width=0.33\hsize]{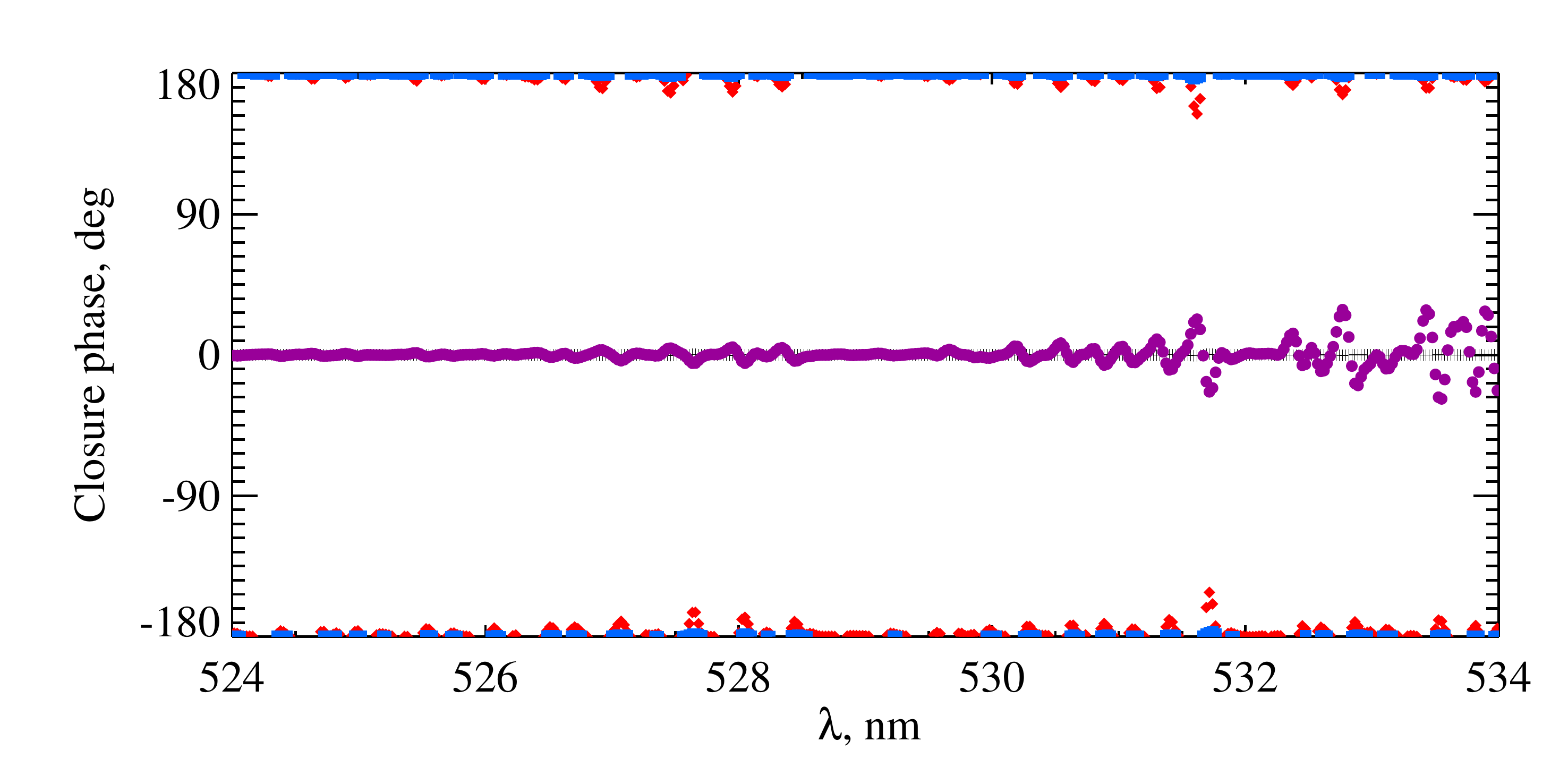}
\includegraphics[width=0.33\hsize]{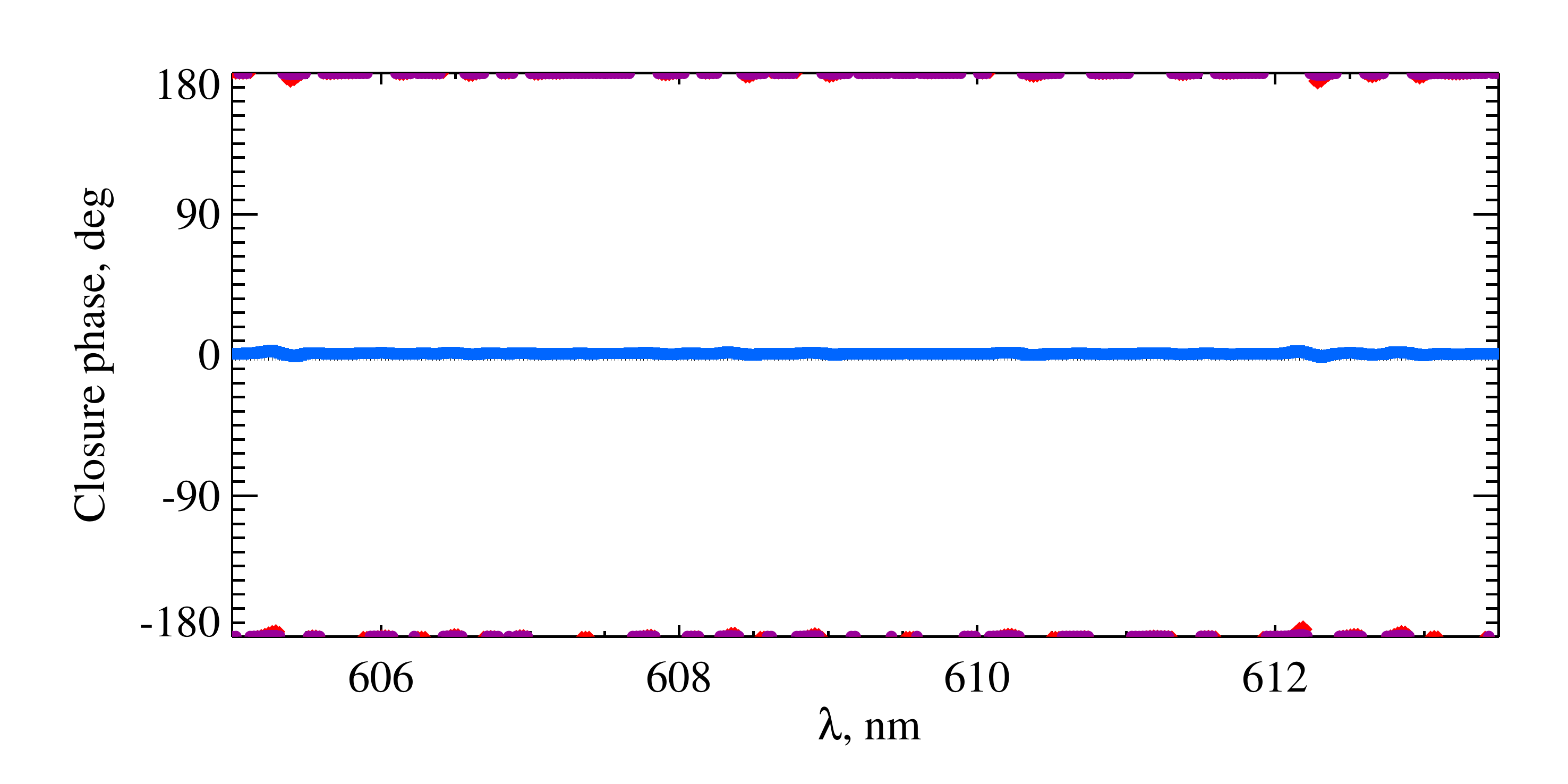}
}
\centerline{
\includegraphics[width=0.33\hsize]{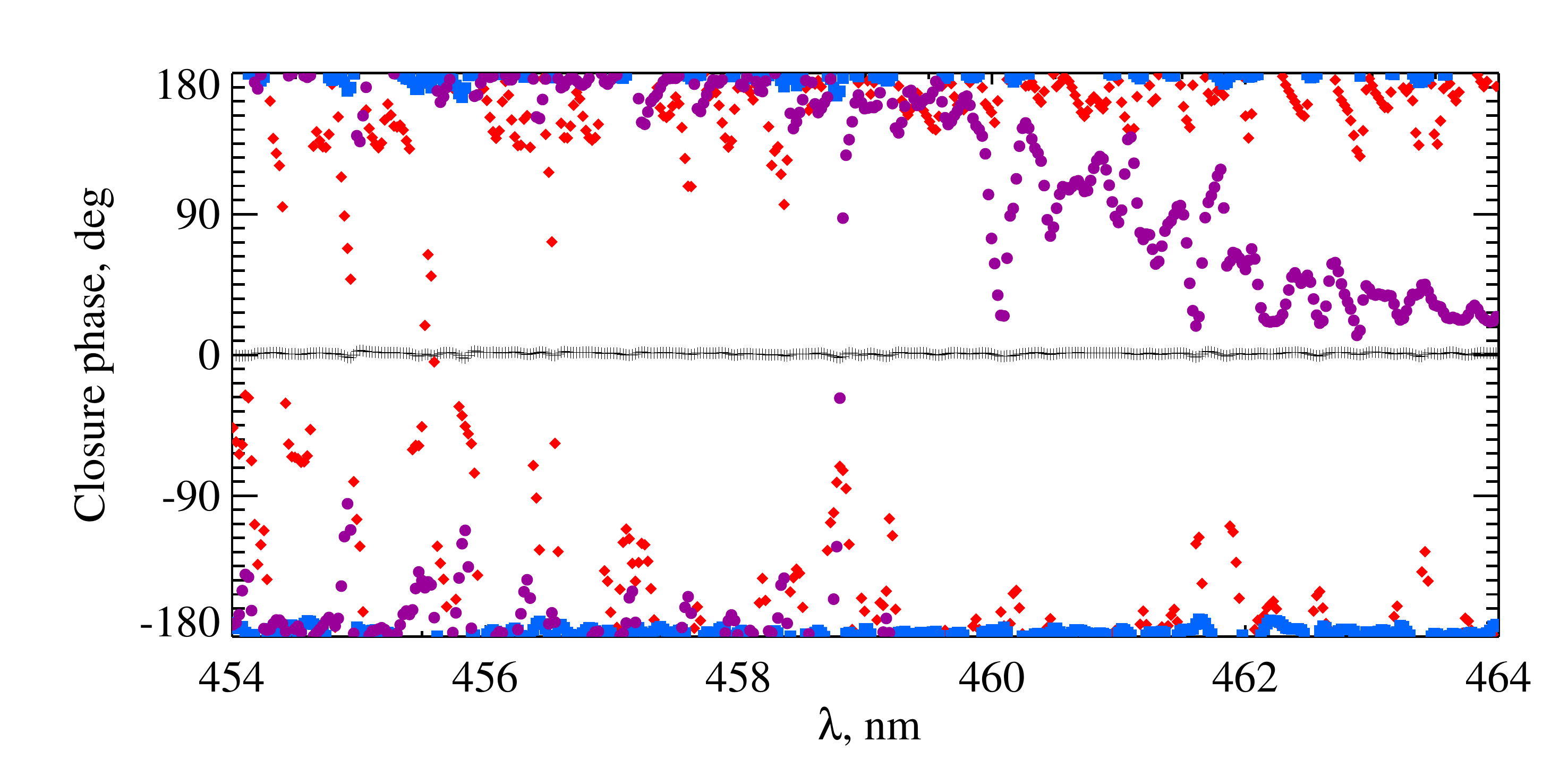}
\includegraphics[width=0.33\hsize]{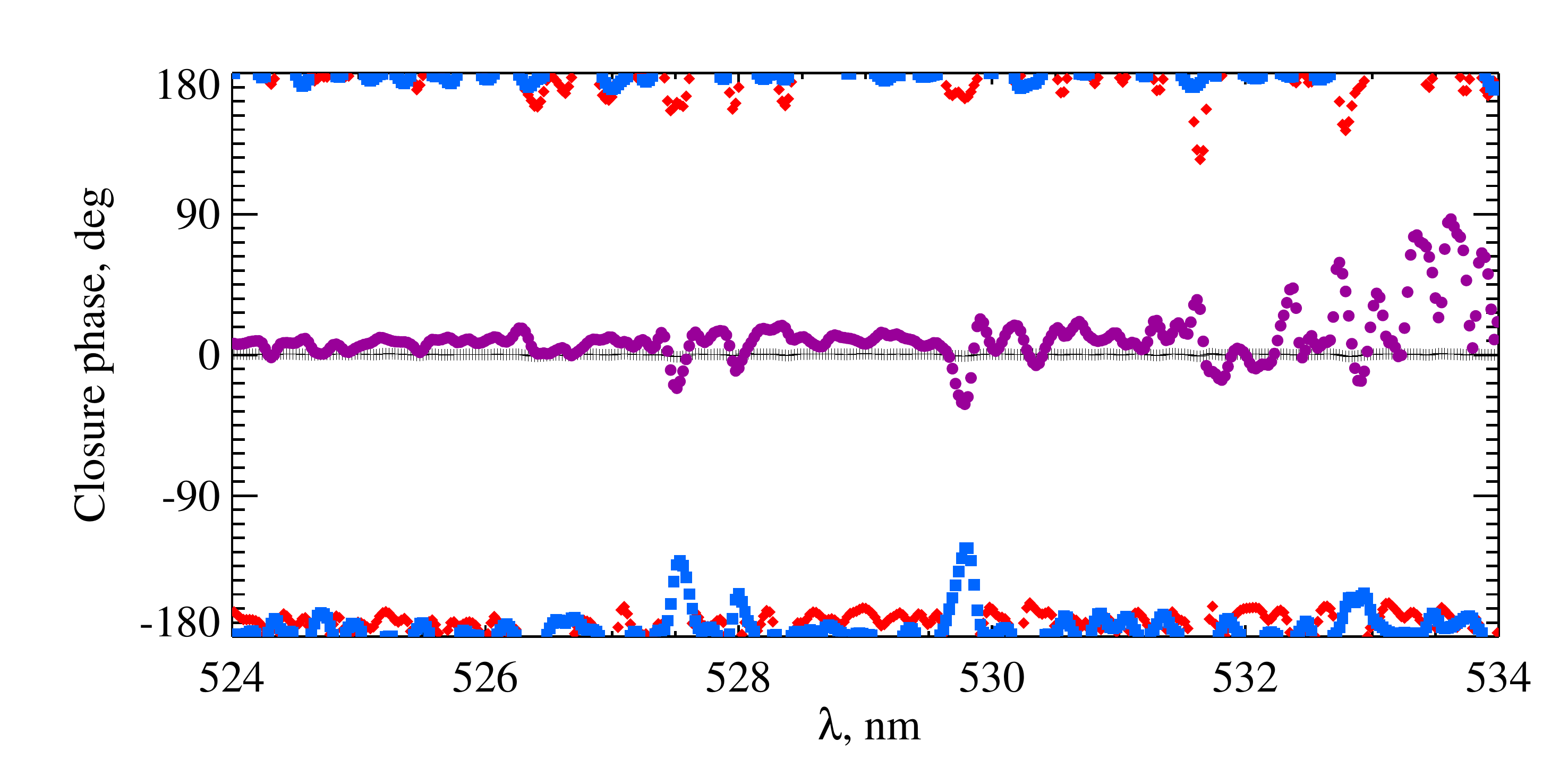}
\includegraphics[width=0.33\hsize]{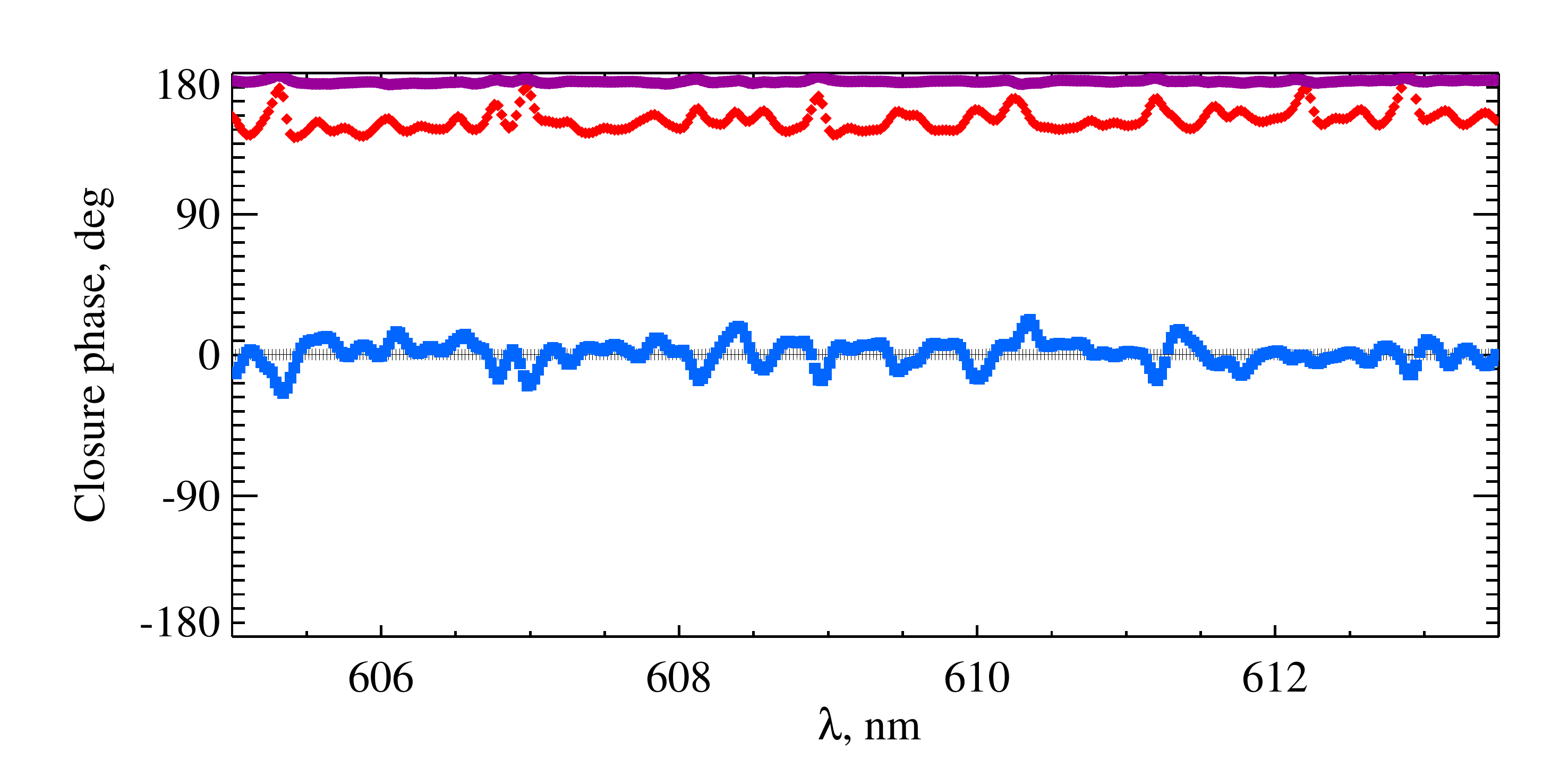}
}
\caption{Same as on Fig.~\ref{fig:cp-lambda-v-r30000}, but with $R=6\,000$. See online version for colored symbols.}
\label{fig:cp-lambda-v-r6000}
\end{figure*}

First row of Fig.~\ref{fig:vis-lambda-v-r30000} illustrates the three 
wavelength intervals in visual domain that were used to compute
wavelength dispersed {squared} visibility  for $R=30\,000$. These visibilities are plotted 
in the second and third rows of Fig.~\ref{fig:vis-lambda-v-r30000} 
for the uniform and spotted surfaces and for $\vsini=35$~\kms.
As example, predictions only for $\alpha=270^\circ$ are shown because this configuration
provides stronger signals compared with the other two.
Features are observed at optical wavelengths while at longer wavelengths ($605.0-613.5$~nm)
the intensity contrast become weaker and the signal amplitude drops significantly.
{By comparing visibility corresponding to star with and without spots, it is obvious that the presence
of spots can be located already with short baselines ($\approx50$~m, see the legend on Fig.~\ref{fig:vis-lambda-v-r30000}).}

At a lower spectral resolution of $R=6\,000$ rotation {becomes} less important and signals in $V^2$
are weak and rare, as shown in the Fig.~\ref{fig:vis-lambda-v-r6000}.
There are several deeps seen only for the spotted case at $454.0-464.0$~nm and $524.0-534.0$~nm regions,
while only marginal features can be seen in the $605.0-613.5$~nm region.

\subsubsection{Closure phase}

To illustrate the closure phase signatures we assumed a simple  isosceles triangle configuration of imaginary telescopes
with two position angles of $0^\circ$ and $270^\circ$.
Such a configuration implies that the third baseline must be oriented at $135^\circ$
and its length ($B_{\rm max}$ hereafter) is estimated from the $(u,v)$ coordinates of the first two projected baselines.

Figure~\ref{fig:cp-lambda-v-r30000} illustrates predictions for the same three spectral regions considered
in the previous section, for uniform and spotted models, $R=30\,000$, and $\vsini=35$~\kms.
We find that at this resolution there is a clear signal in closure phase detected at maximum baseline
of a triangle $B_{\rm max}>100$~m. However, at certain configurations it is indeed hard to see the difference  
between uniform and spotted stellar surfaces. Fortunately, there are many configurations for which the spotted star looks different
compared to a homogeneous one. In general, a rotating star with
uniform surface produces closure phases that are symmetric
relative to the core of spectral lines, whereas spots induce more rich and complex closure phase patterns.
Moreover, for a configuration with the sides of isosceles triangle of $100$~m and $200$~m 
(i.e. $B_{\rm max}=141$~m and $B_{\rm max}=255$~m, respectively) 
a clear spot signal is detected in $605.0-613.5$~nm spectral window where the rotation does not seem to affect
the closure phase very much compared to the other two spectral regions of shorter wavelengths.

As already mentioned, at a lower spectral resolution of $R=6\,000$ rotation becomes less important and signals in closure phase
are grossly due to spots, as shown on Fig.~\ref{fig:cp-lambda-v-r6000}. Still, rotation is capable of inducing
pretty strong features at several baselines and short wavelengths, but much more features are detected if star has spots.
Similar to the previous case with $R=30\,000$, we find many configurations where
the spot signals can be unambiguously detected.

In general, the spots are detected in all three wavelength windows,
however at short wavelengths the spectral line density is higher and many signals are recovered.
In case of very slow rotations, the short wavelengths will provide a rich and strong spot signals compared to
longer wavelengths.

\subsection{Infrared wavelength domain}

\subsubsection{Visibility vs. baseline}

\begin{figure*}
\rotatebox{90}{
\begin{minipage}{\textheight}
\centerline{
\includegraphics[width=0.1\hsize]{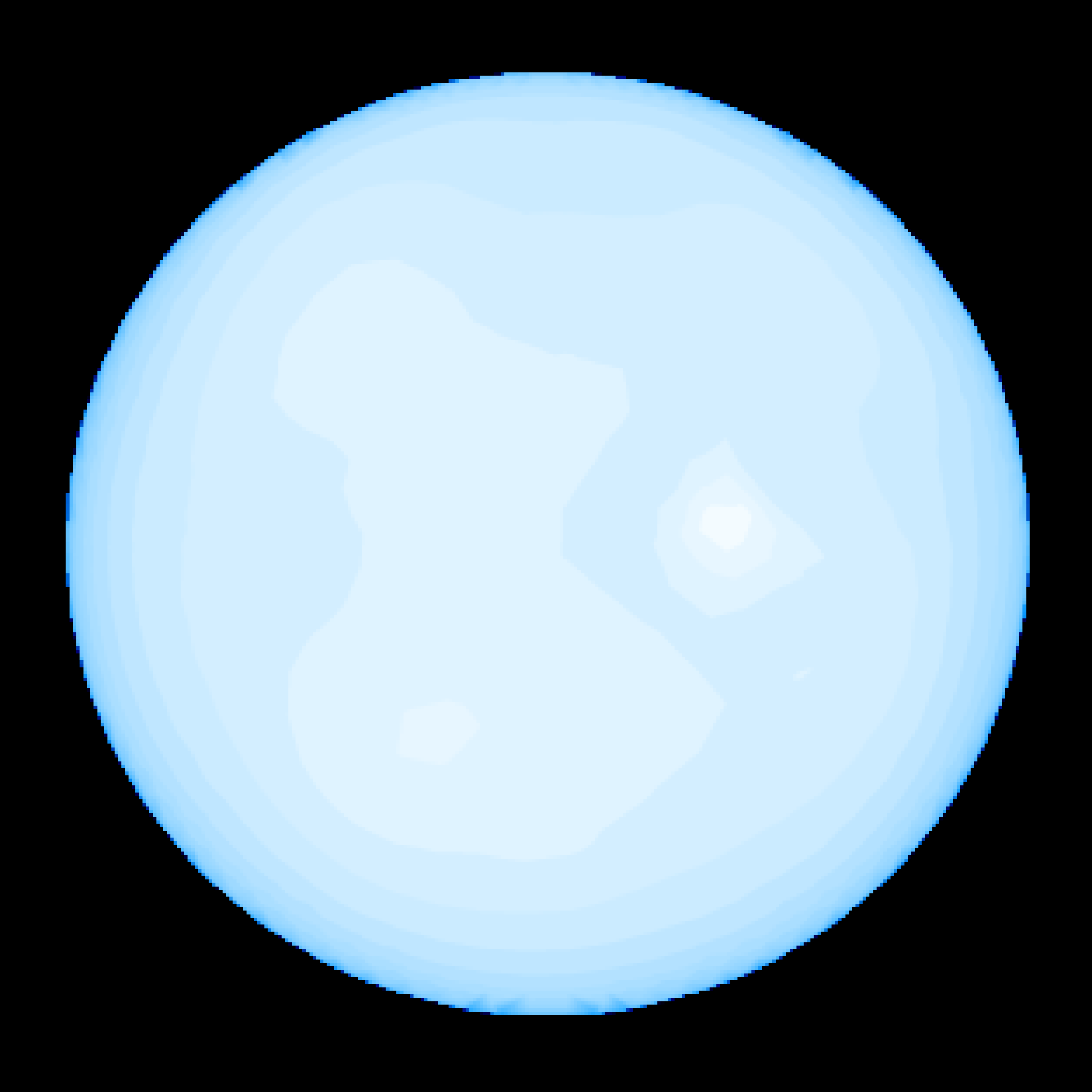}
\includegraphics[width=0.1\hsize]{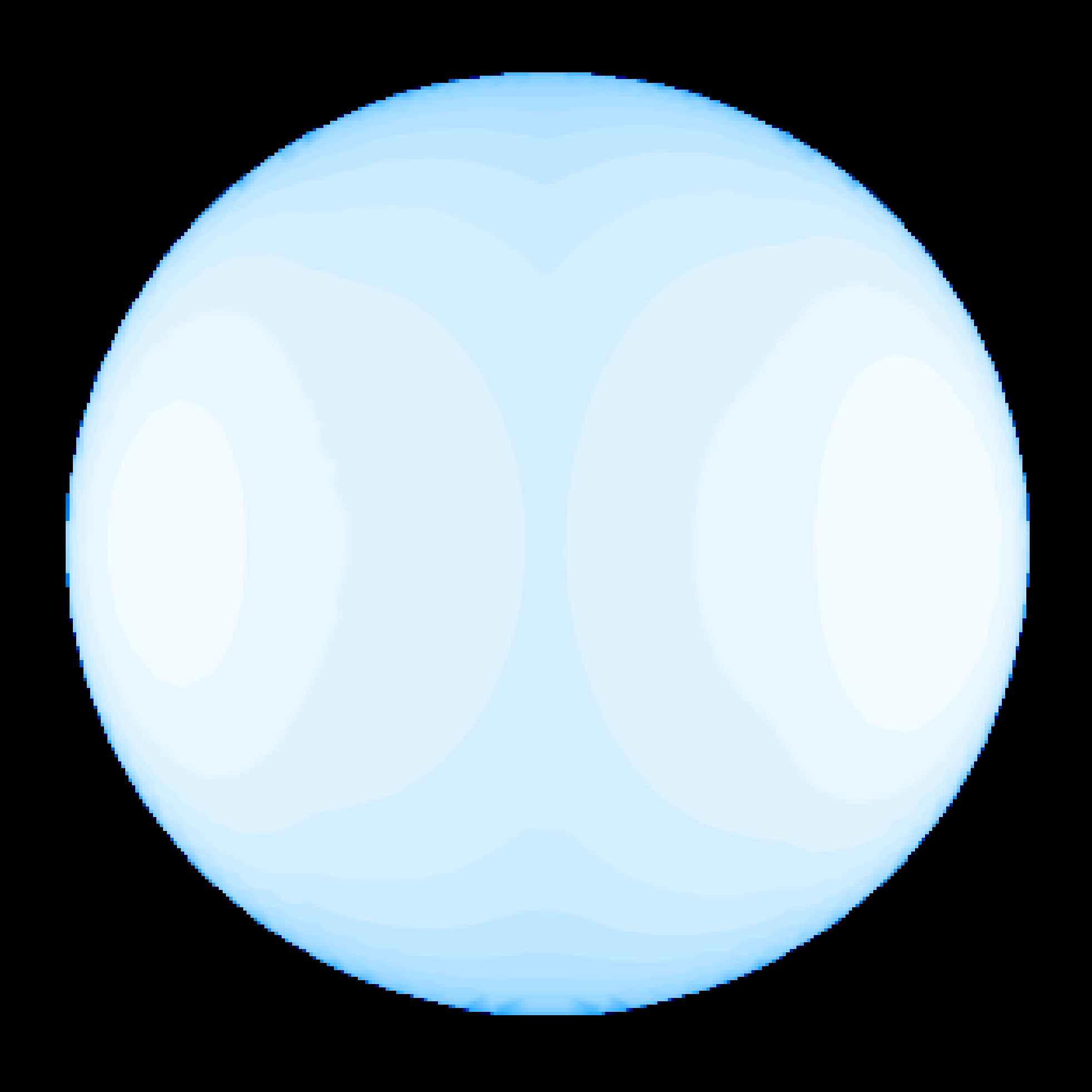}
\includegraphics[width=0.1\hsize]{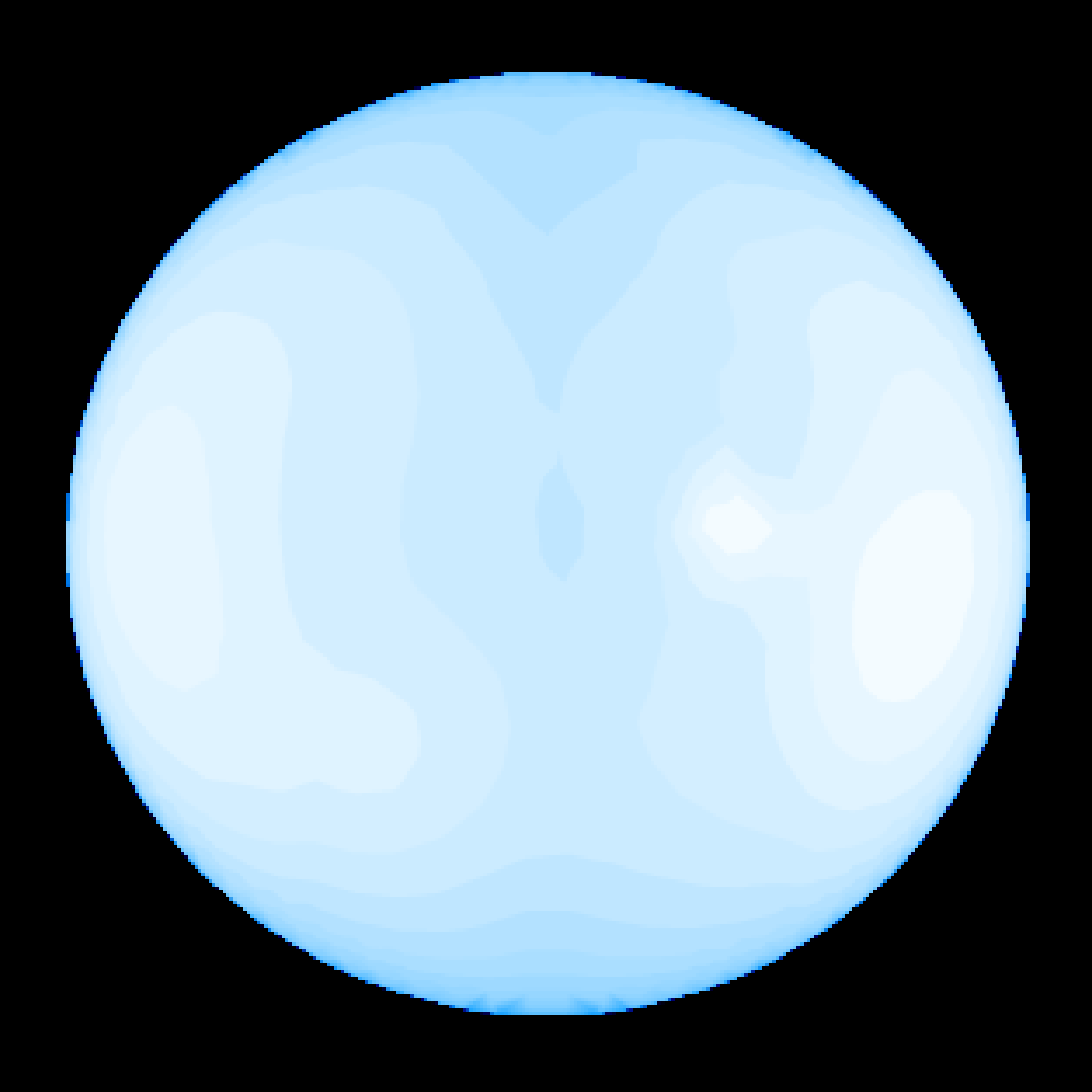}\hfill
\includegraphics[width=0.1\hsize]{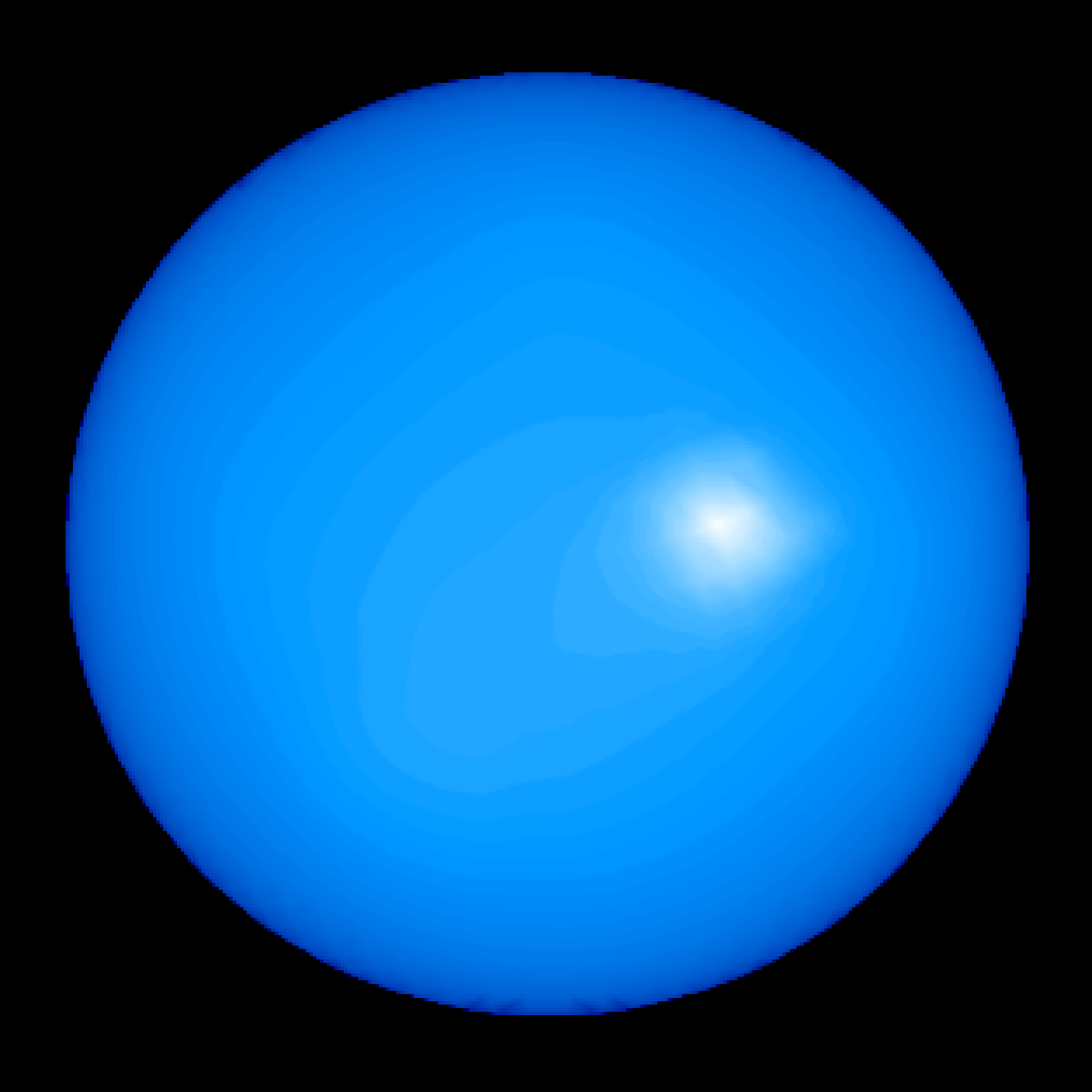}
\includegraphics[width=0.1\hsize]{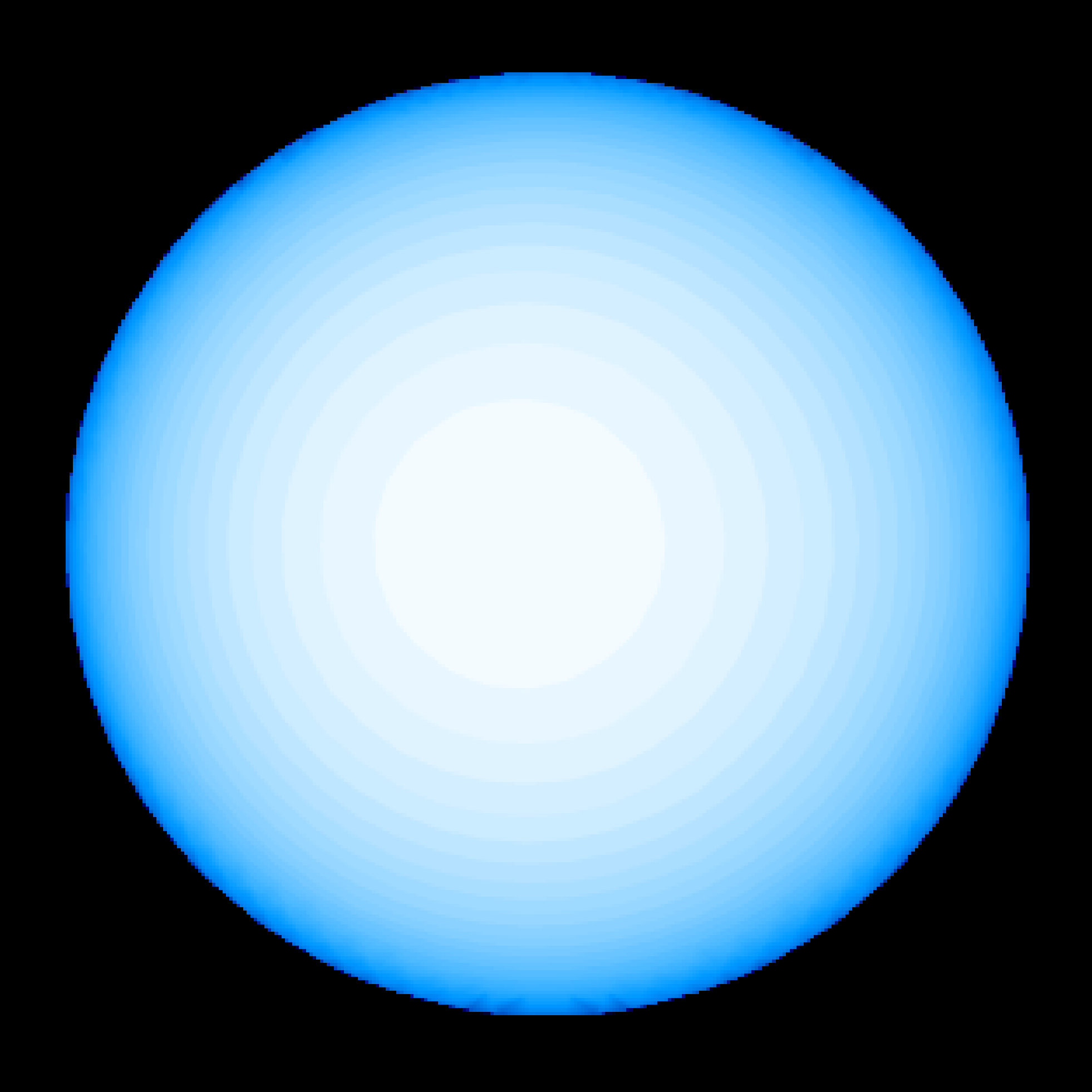}
\includegraphics[width=0.1\hsize]{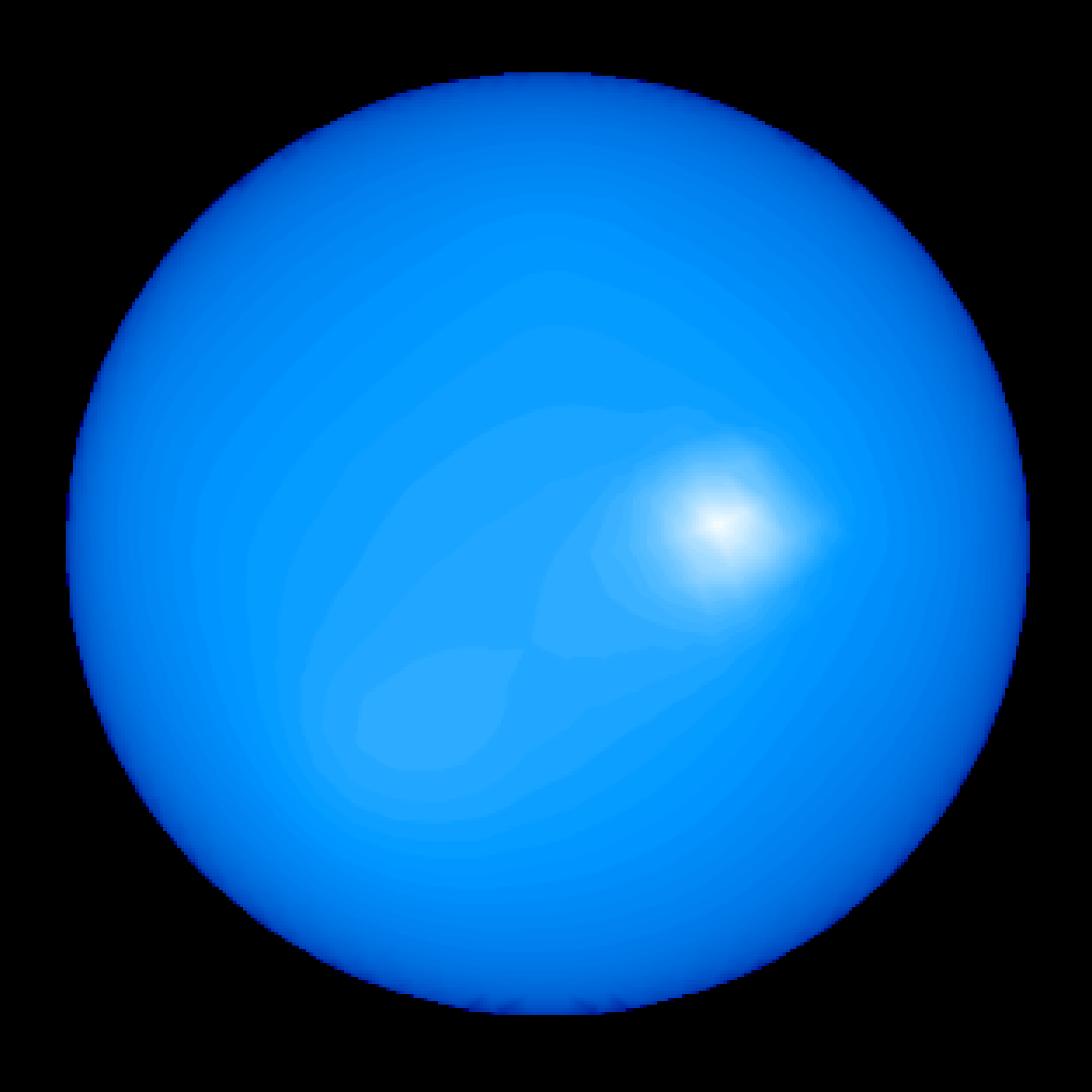}\hfill
\includegraphics[width=0.1\hsize]{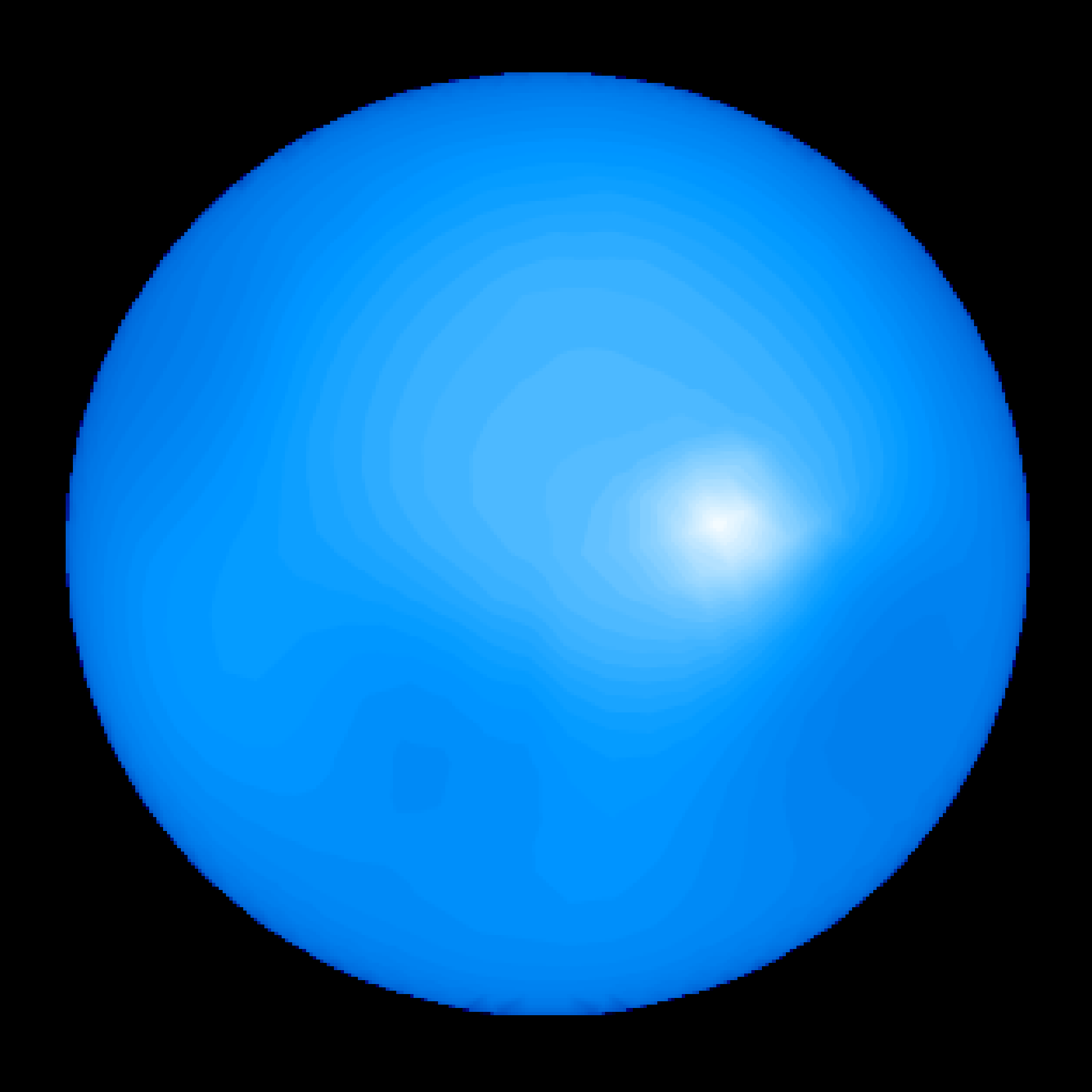}
\includegraphics[width=0.1\hsize]{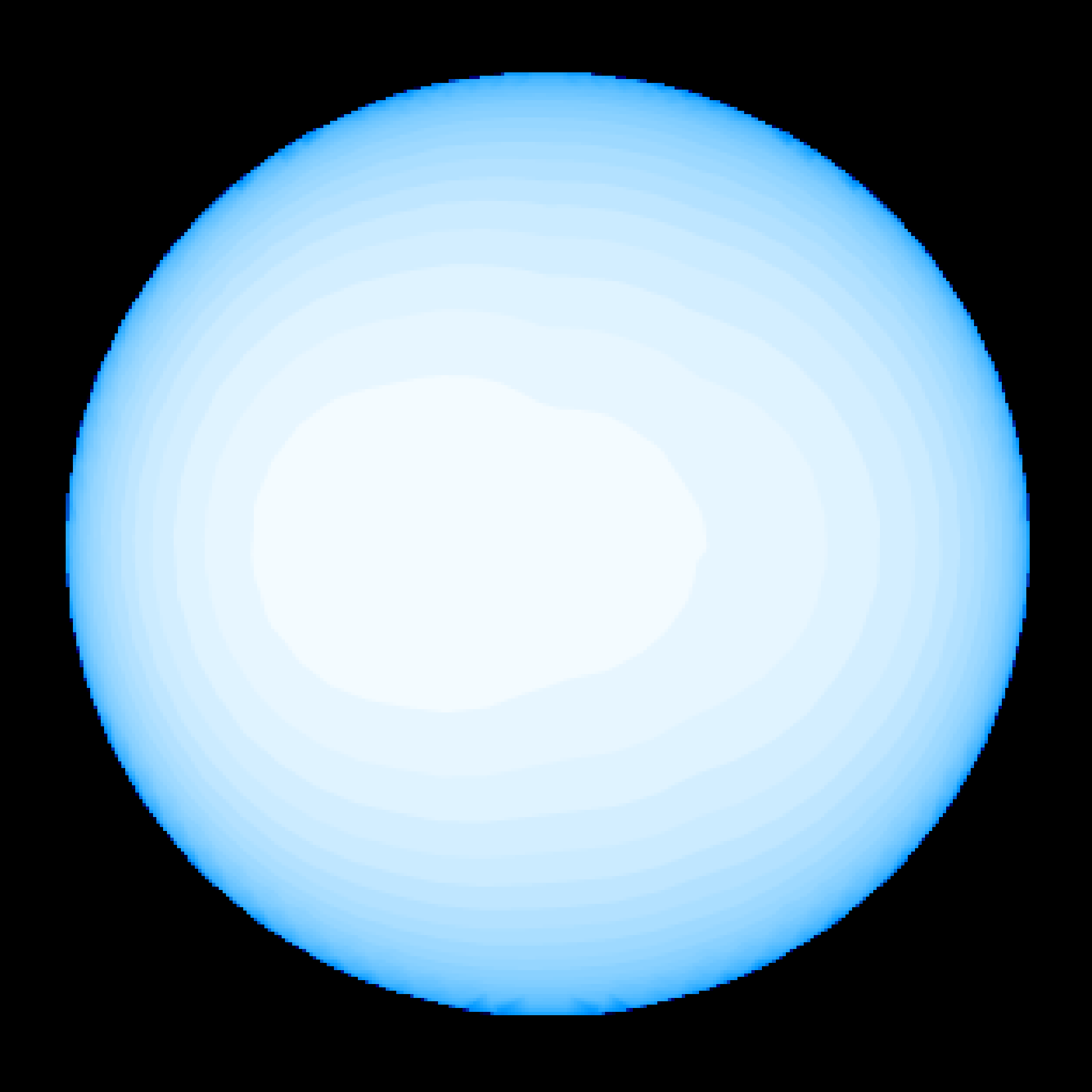}
\includegraphics[width=0.1\hsize]{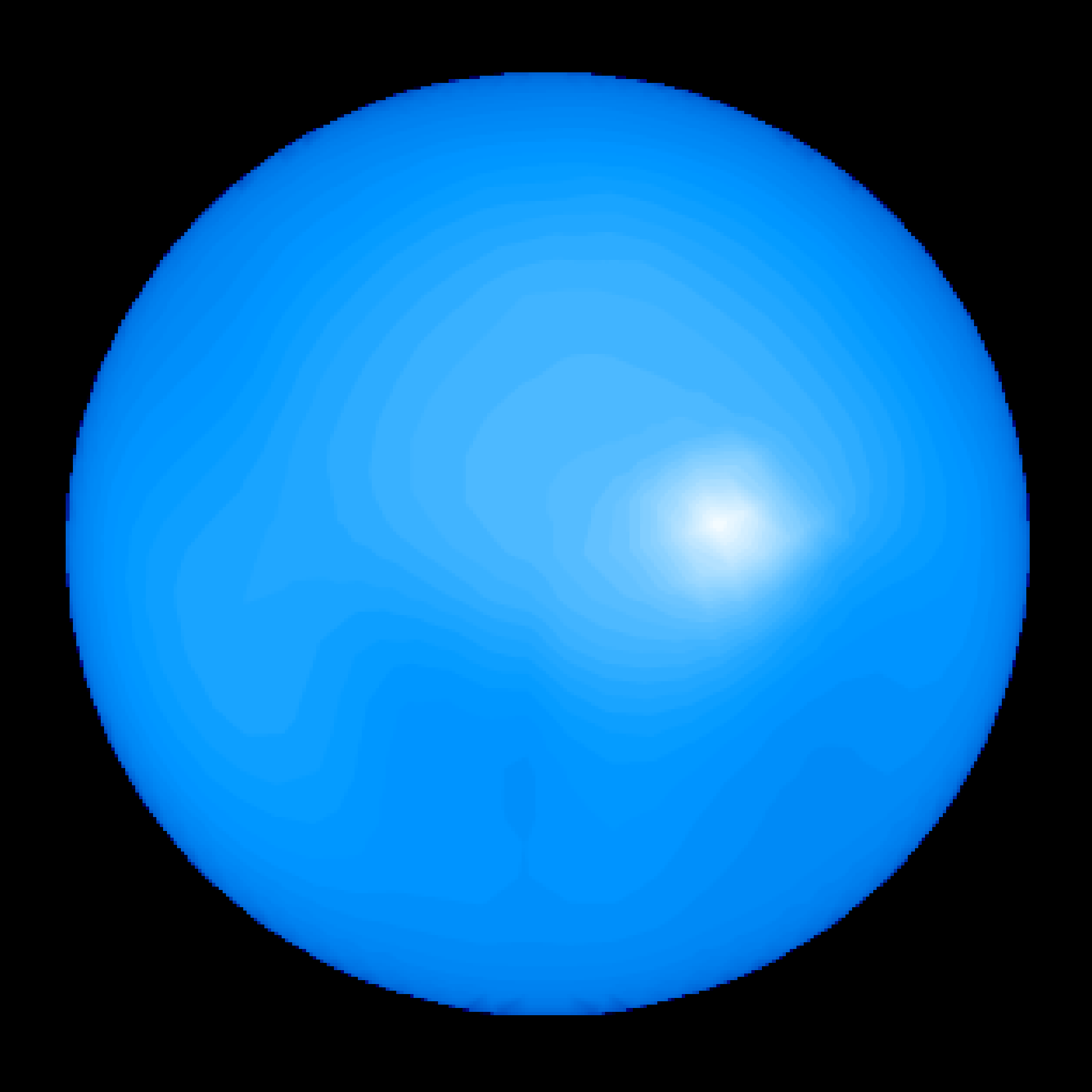}
}
\vspace{0.2cm}
\centerline{
\includegraphics[width=0.33\hsize]{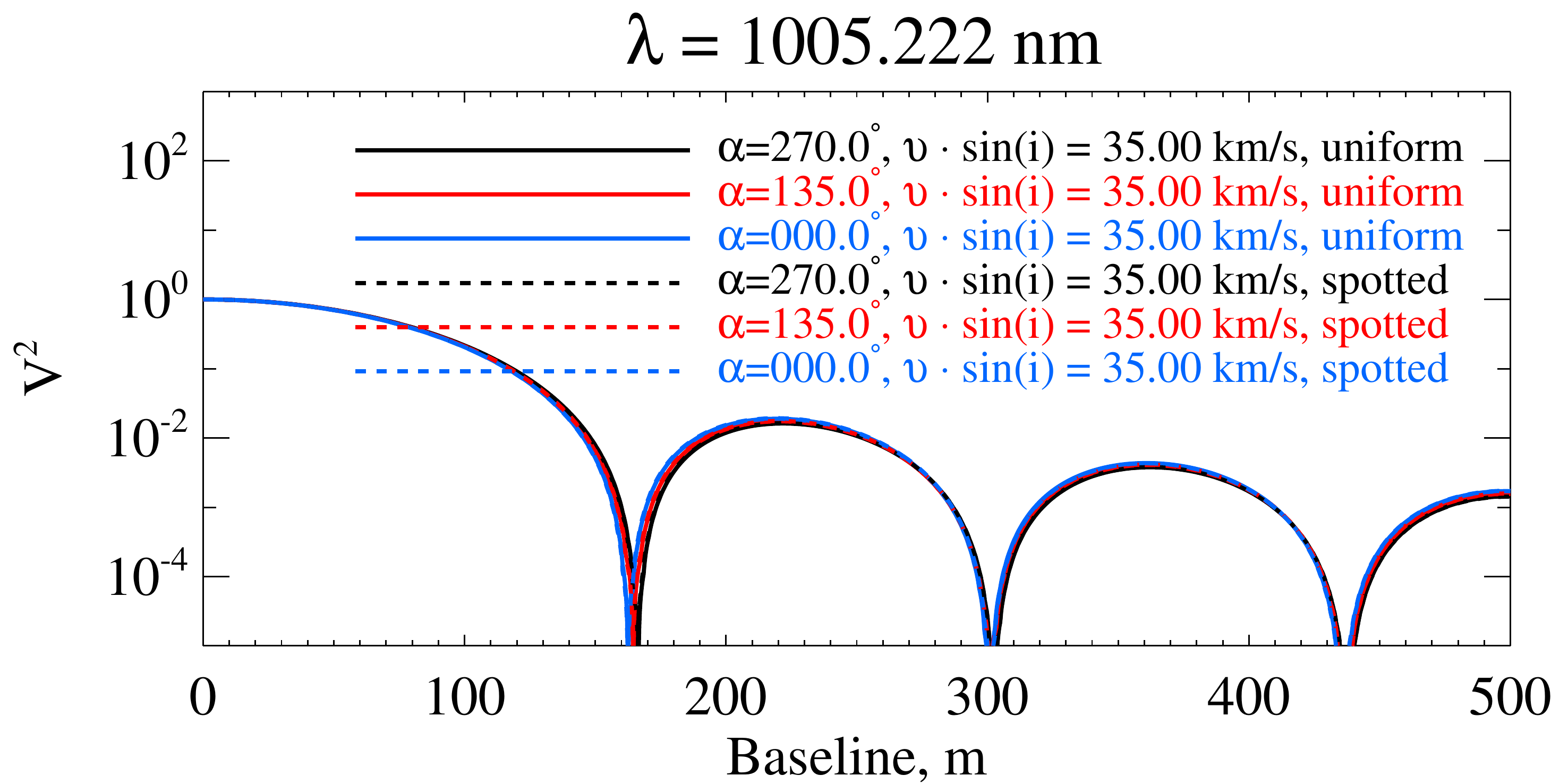}
\includegraphics[width=0.33\hsize]{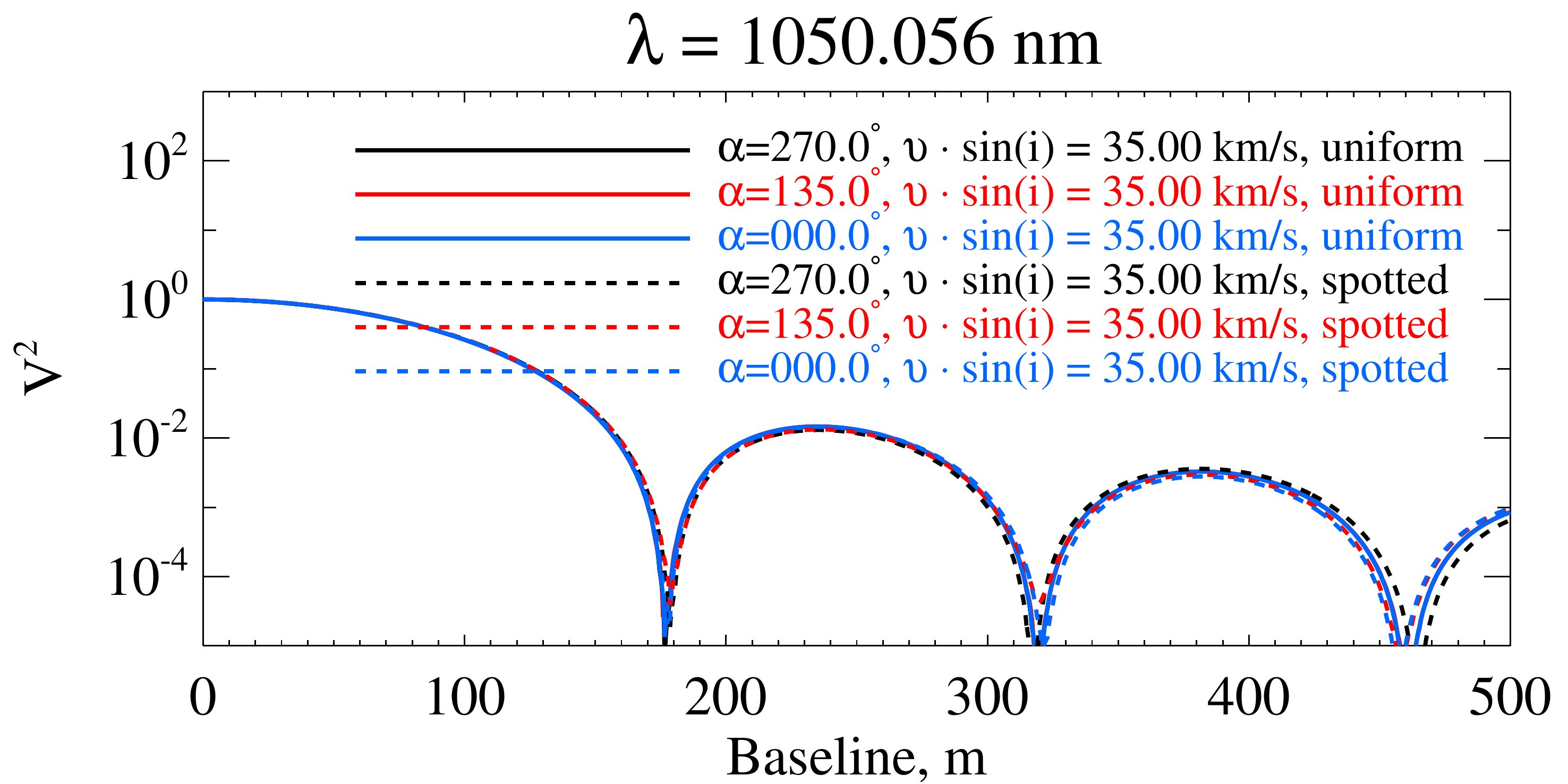}
\includegraphics[width=0.33\hsize]{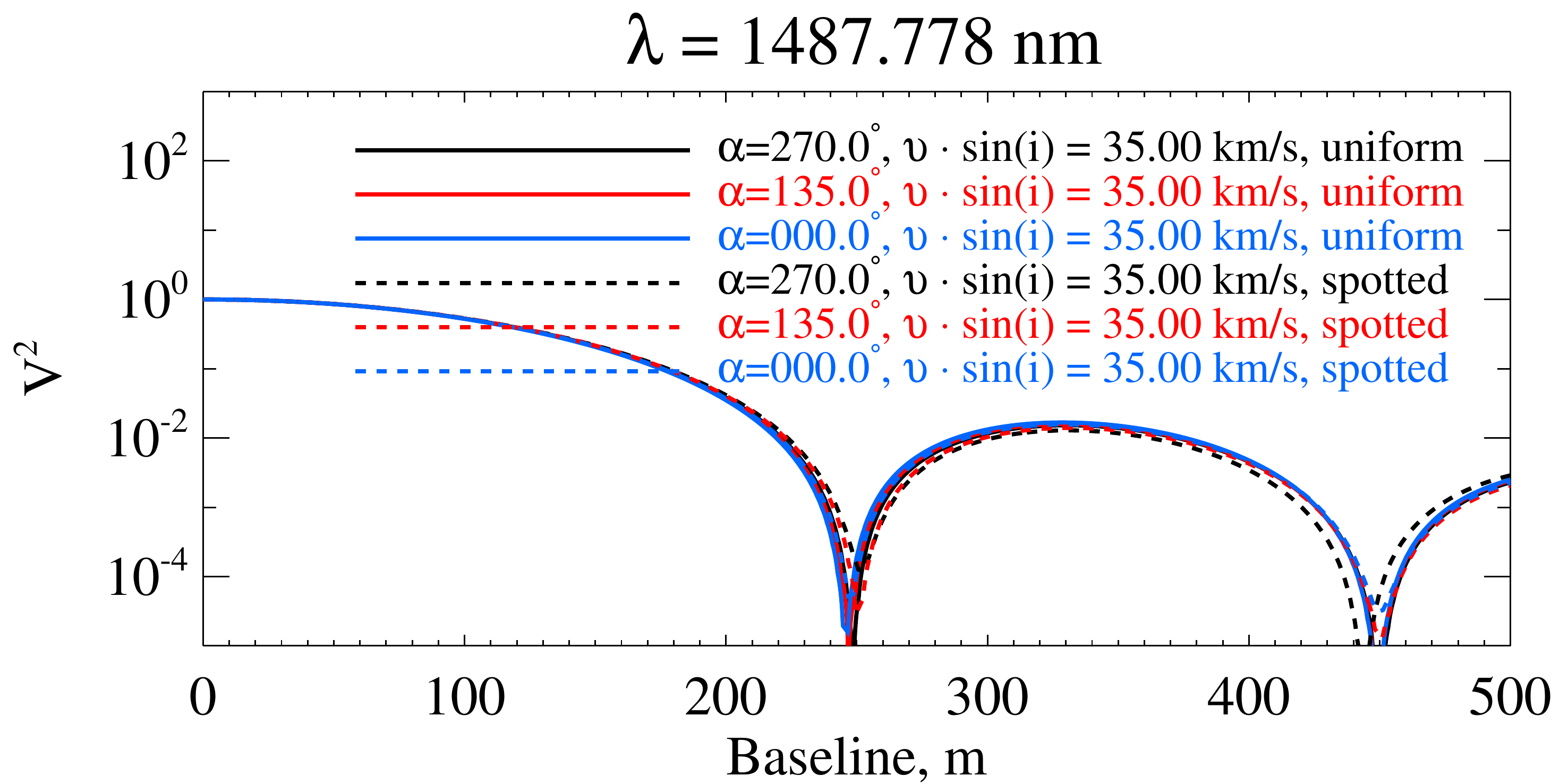}
}
\caption{{Squared} visibility  predicted at the cores of the \ion{H}{i}~$1005.2$~nm,
\ion{Fe}{ii}~$1050.0$~nm, and \ion{Mg}{i}~$1487.7$~nm lines and spectral resolution $R=6\,000$.
The rotational velocity is $\vsini=35$~\kms. The {squared} visibility  from homogeneous and spotted surfaces
are shown by full and dashed lines respectively.
Top panel shows intensity images with homogeneous and spotted abundance distributions.}
\label{fig:vis-mono-ir-1}
\end{minipage}
}
\end{figure*}

Hydrogen lines are the only strong spectroscopic features seen in infrared $J$, $H$, and $K$ bands. At $R=6\,000$ there are
some lines of metals but their number decreases from towards longer wavelengths.
As an example, on Fig.~\ref{fig:vis-mono-ir-1} we show model predictions at the cores of H, Fe, and Mg lines.
Even in line cores the intensity contrast is weak.
There is a clear yet small difference in {squared} visibility 
obtained at different position angles seen at baselines longer than a few hundred meters where {squared} visibility  drops below
$V^2\lesssim10^{-2}$ in case of Fe and Mg lines, and nothing can be seen at the core of H line, at least above $V^2\geqslant10^{-4}$.

\subsubsection{Visibility vs. wavelength}

\begin{figure*}
\centerline{
\includegraphics[width=0.33\hsize]{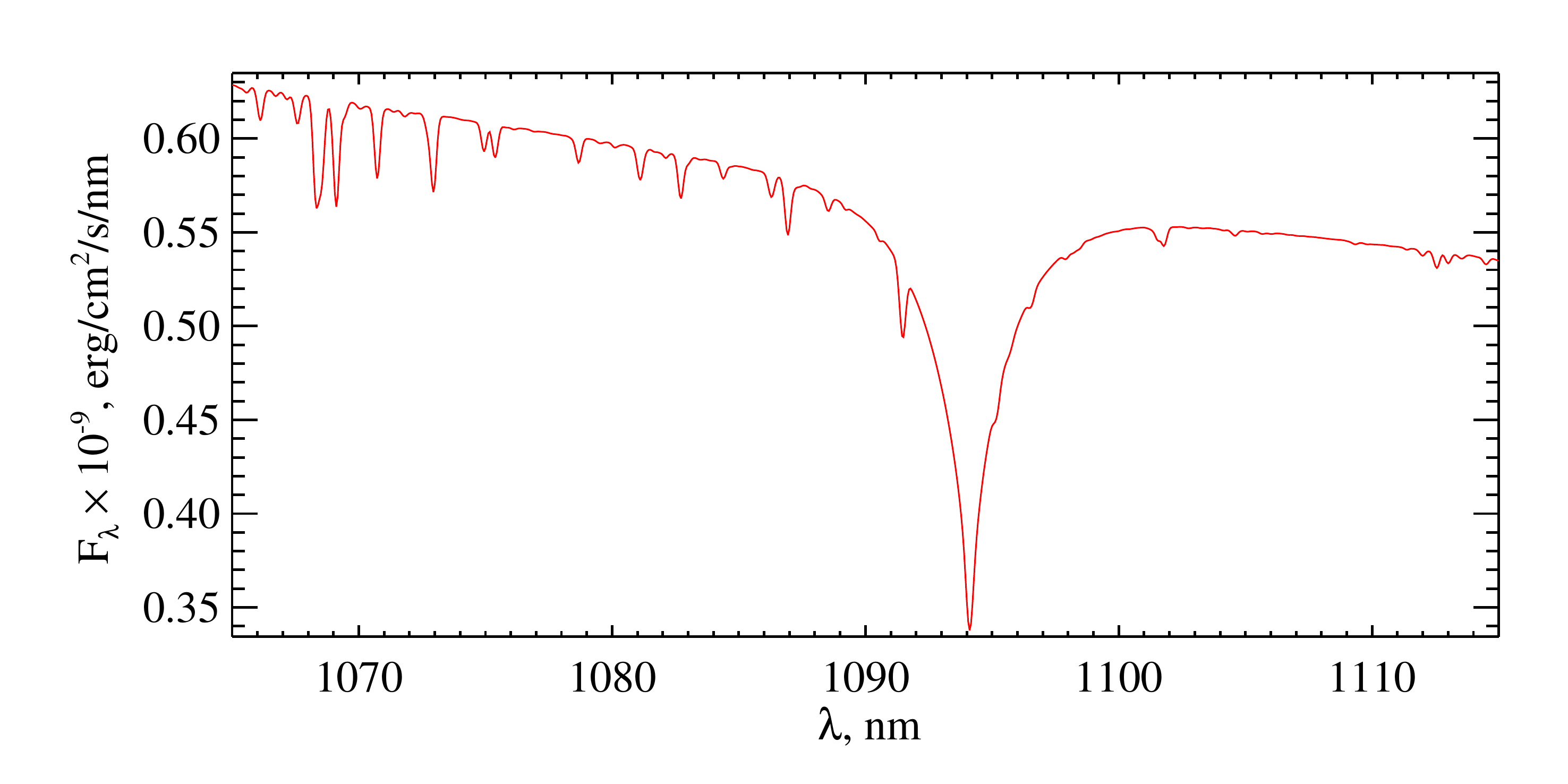}
\includegraphics[width=0.33\hsize]{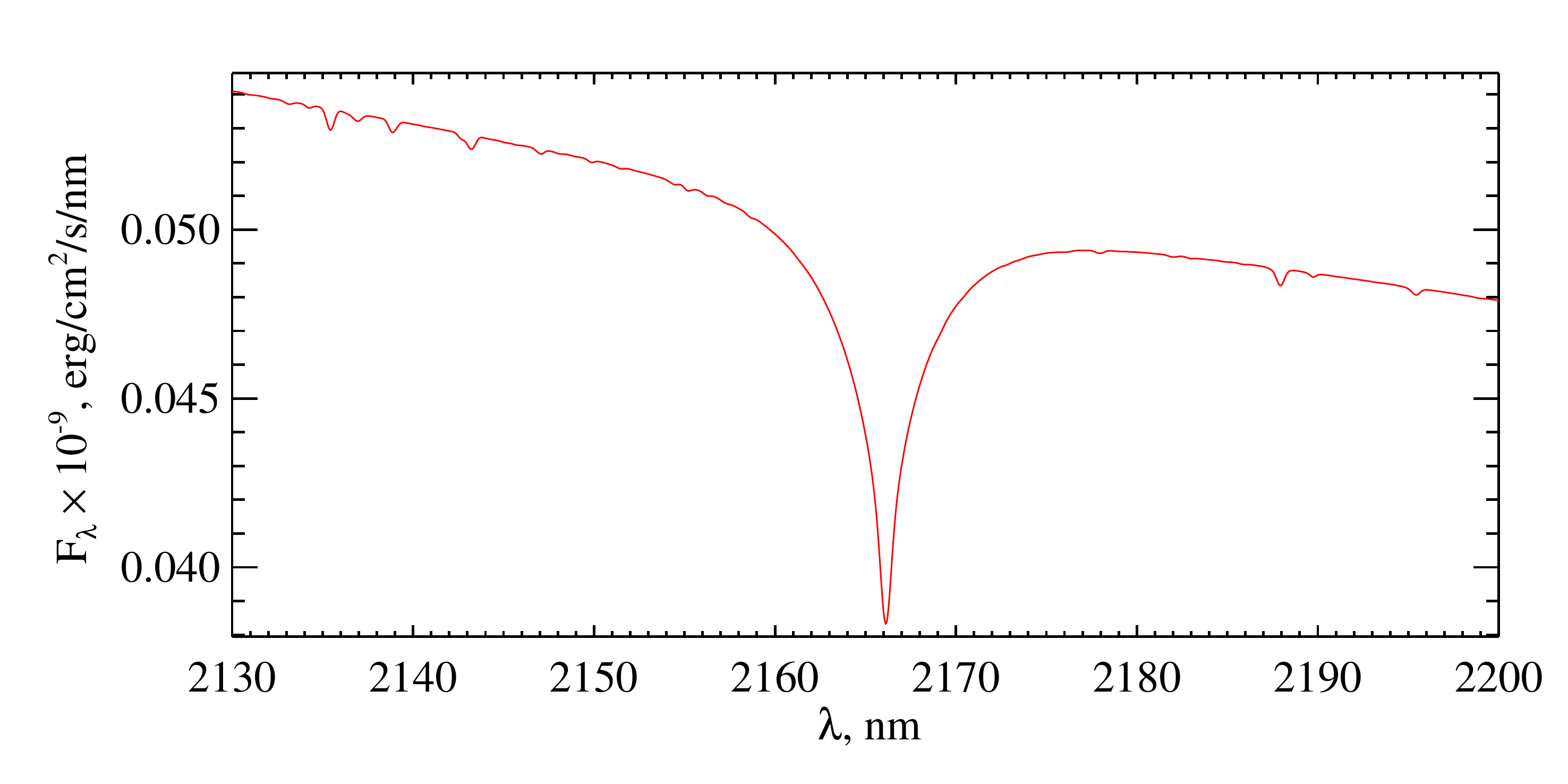}
\includegraphics[width=0.33\hsize]{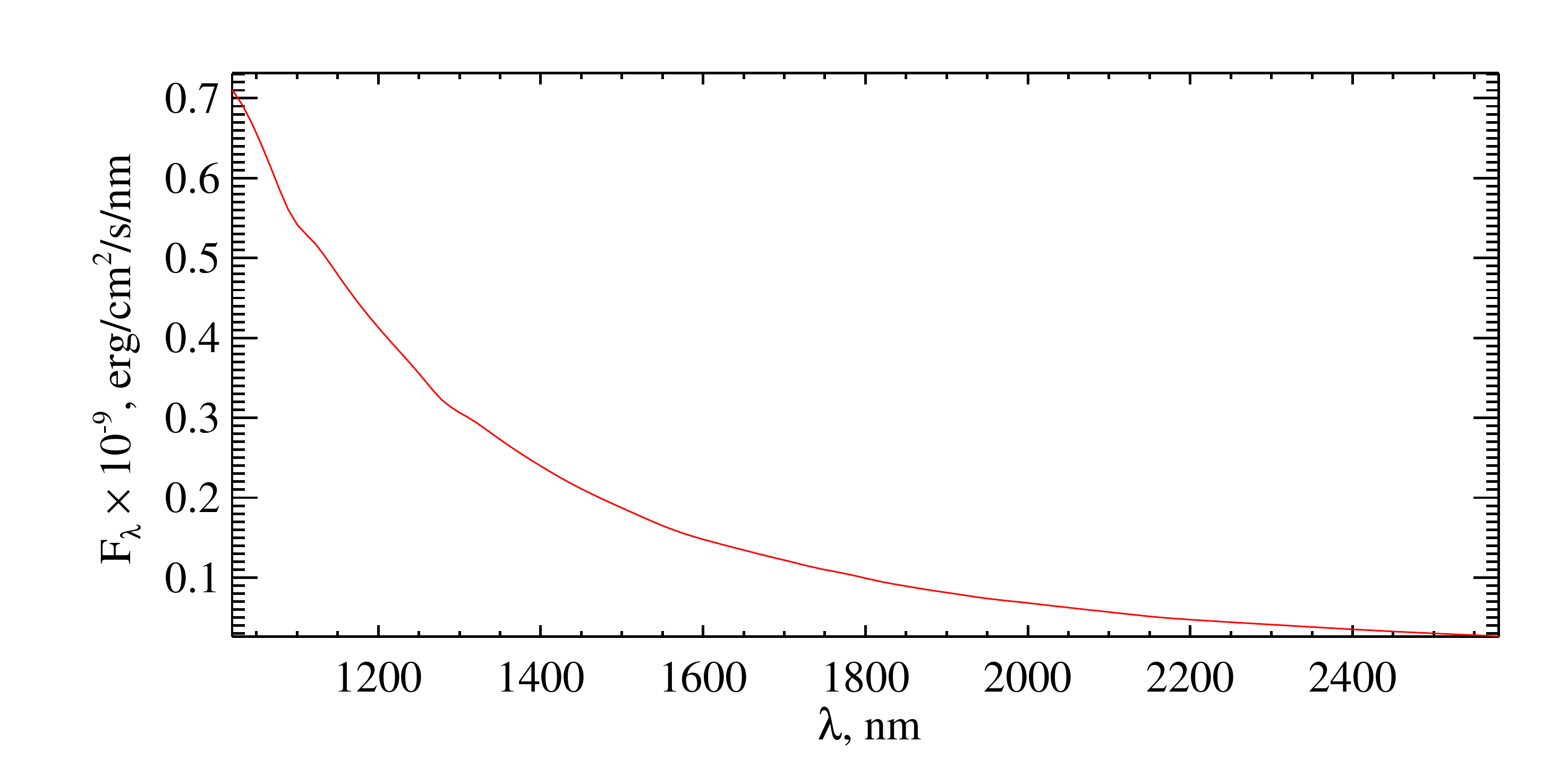}
}
\centerline{
\includegraphics[width=0.33\hsize]{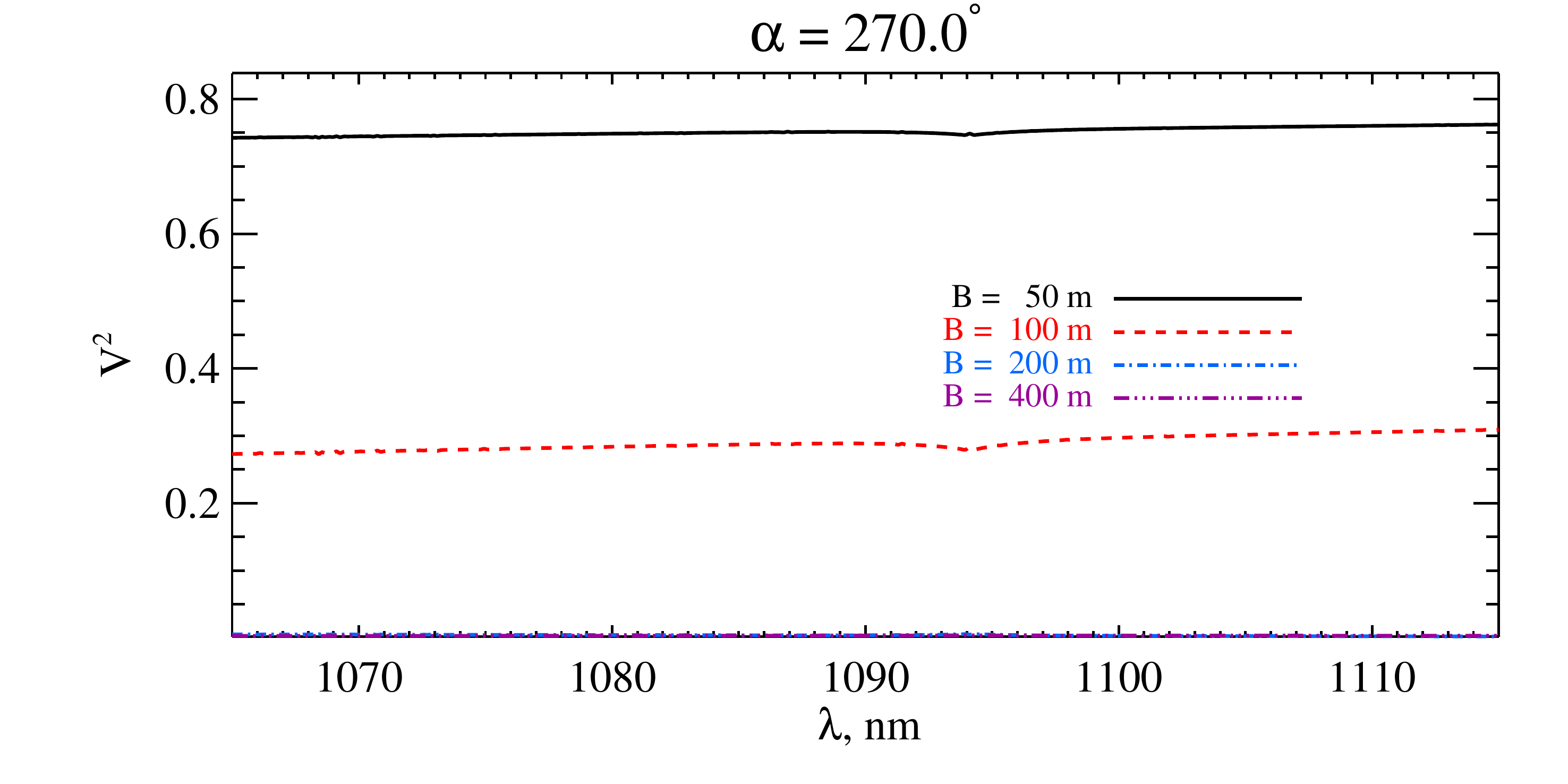}
\includegraphics[width=0.33\hsize]{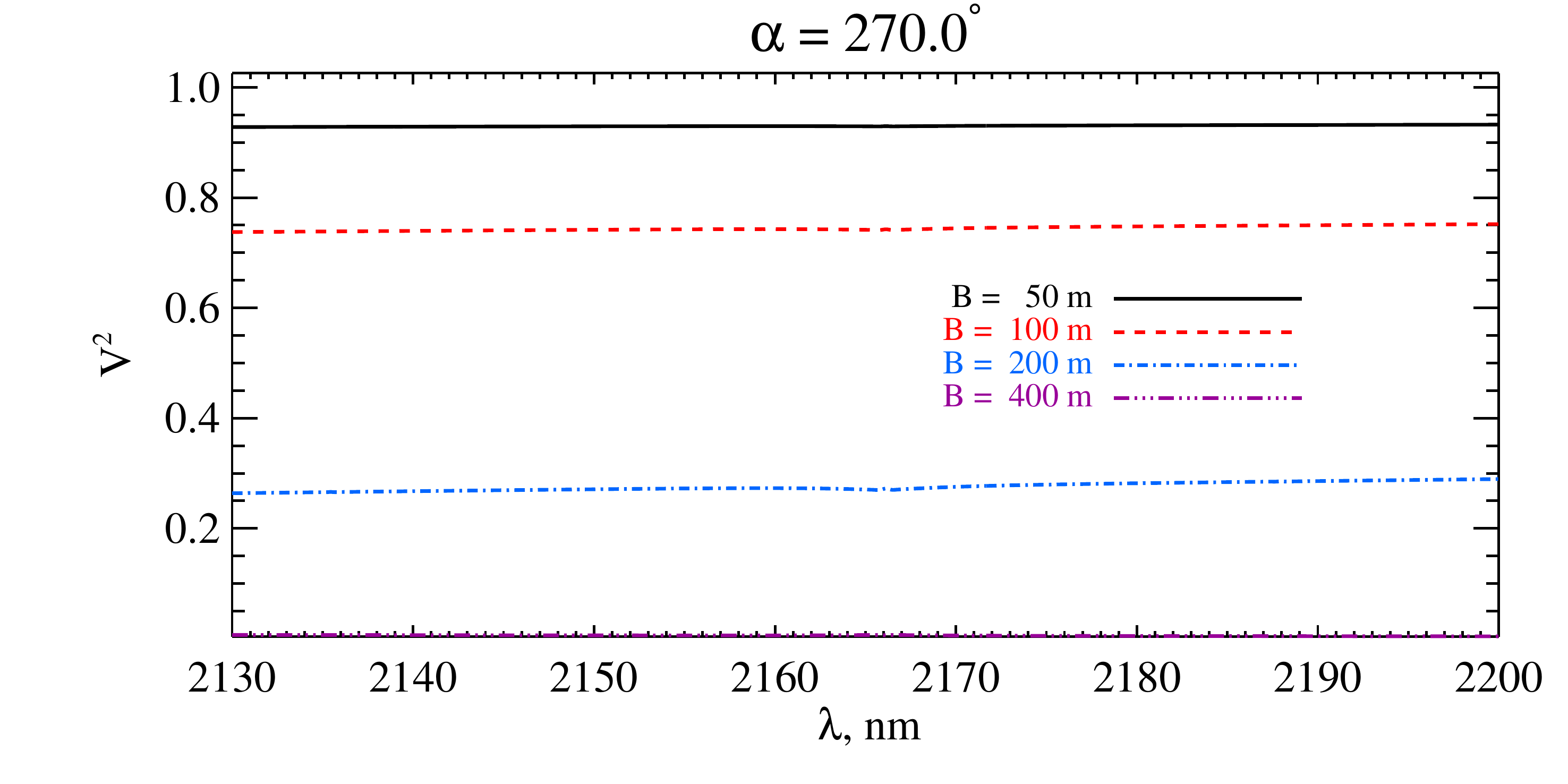}
\includegraphics[width=0.33\hsize]{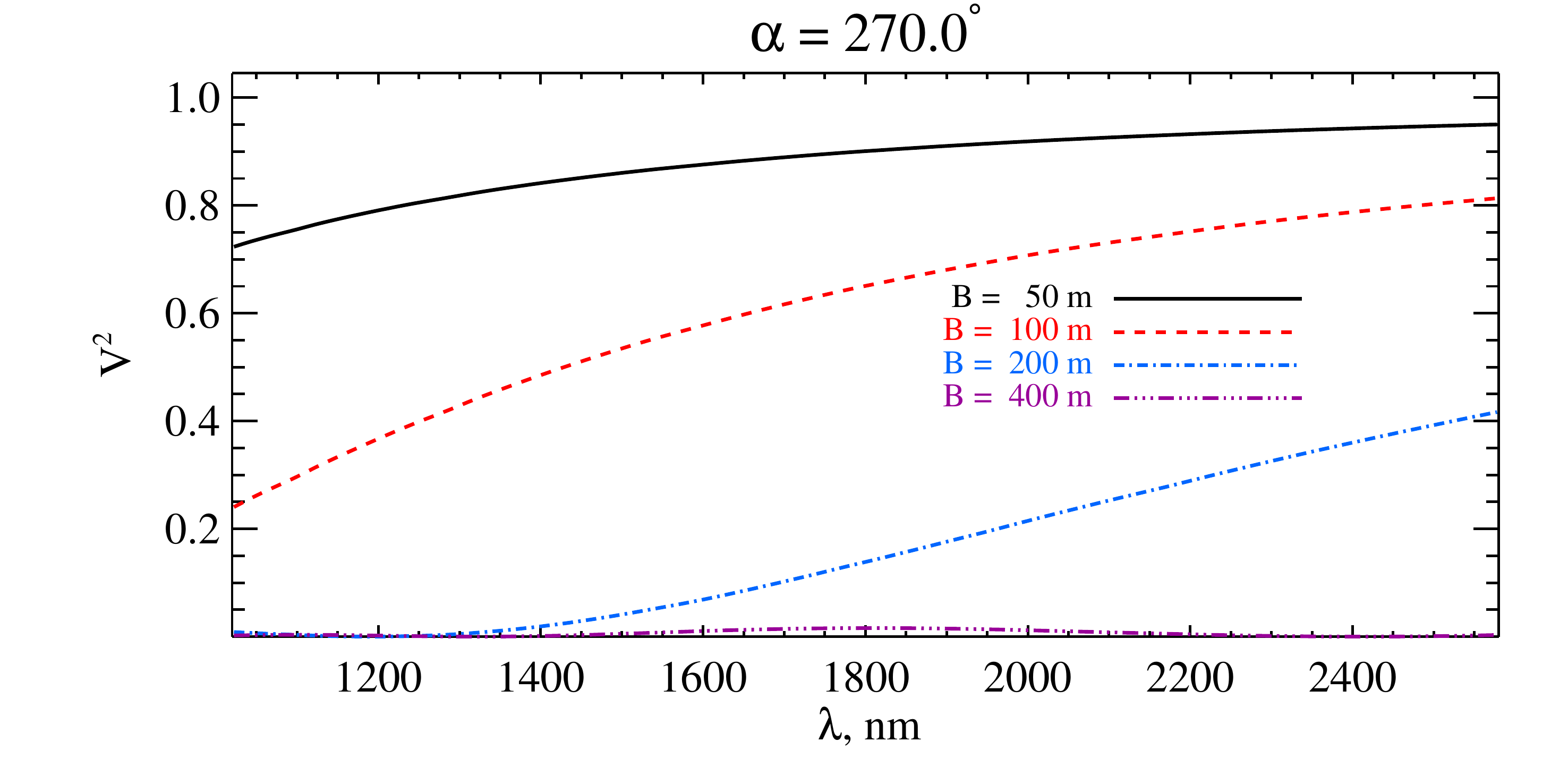}
}
\centerline{
\includegraphics[width=0.33\hsize]{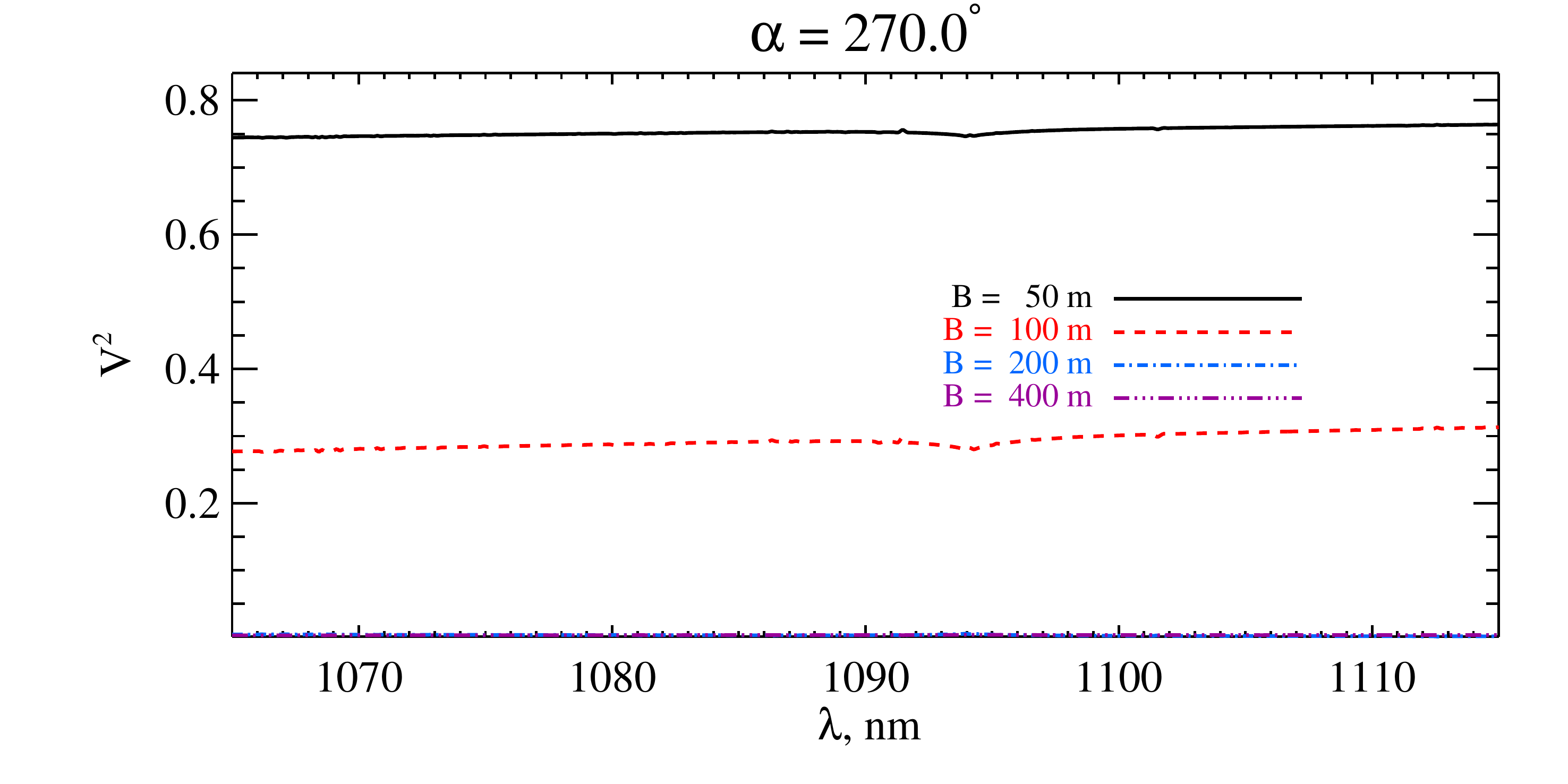}
\includegraphics[width=0.33\hsize]{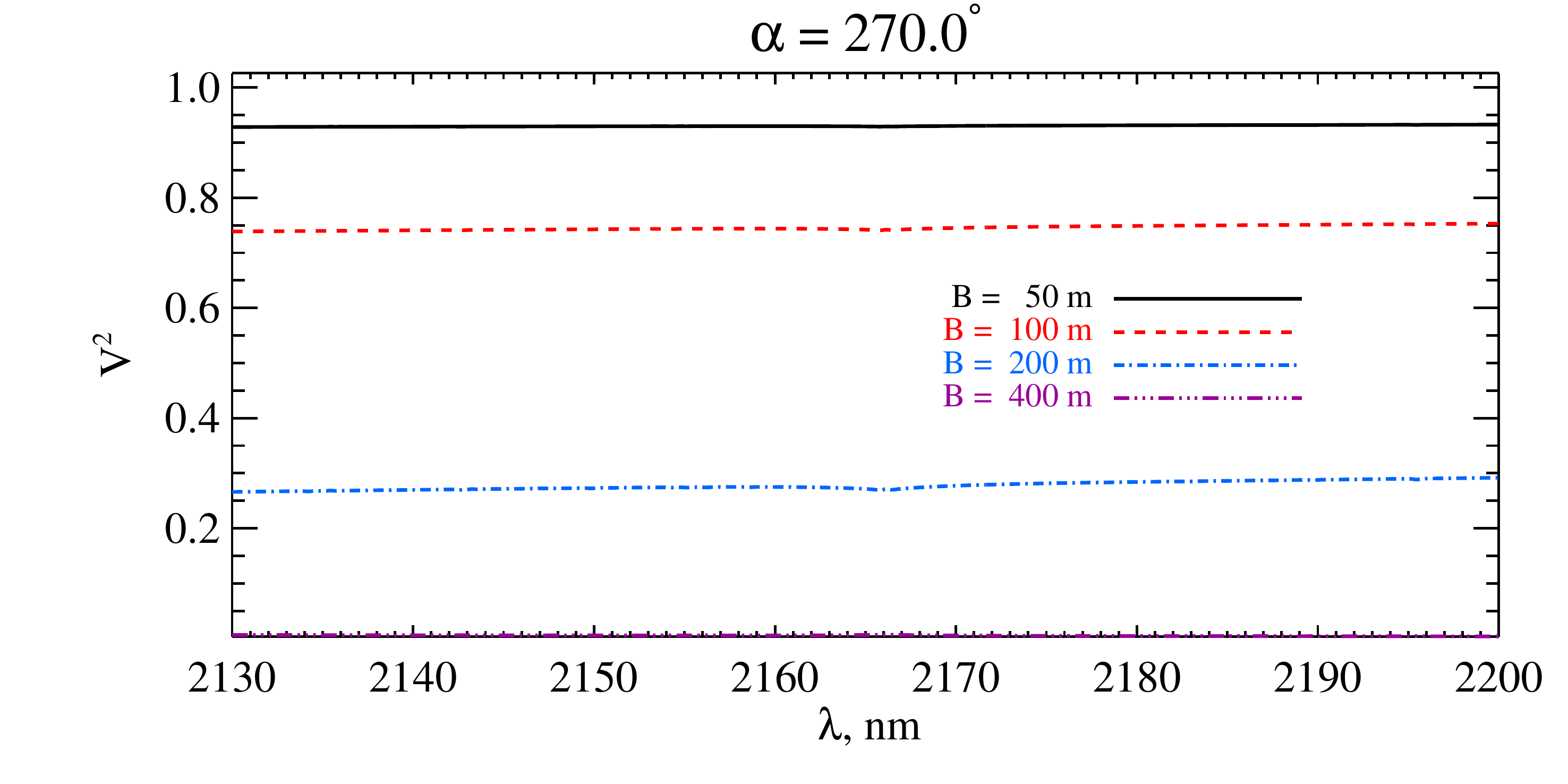}
\includegraphics[width=0.33\hsize]{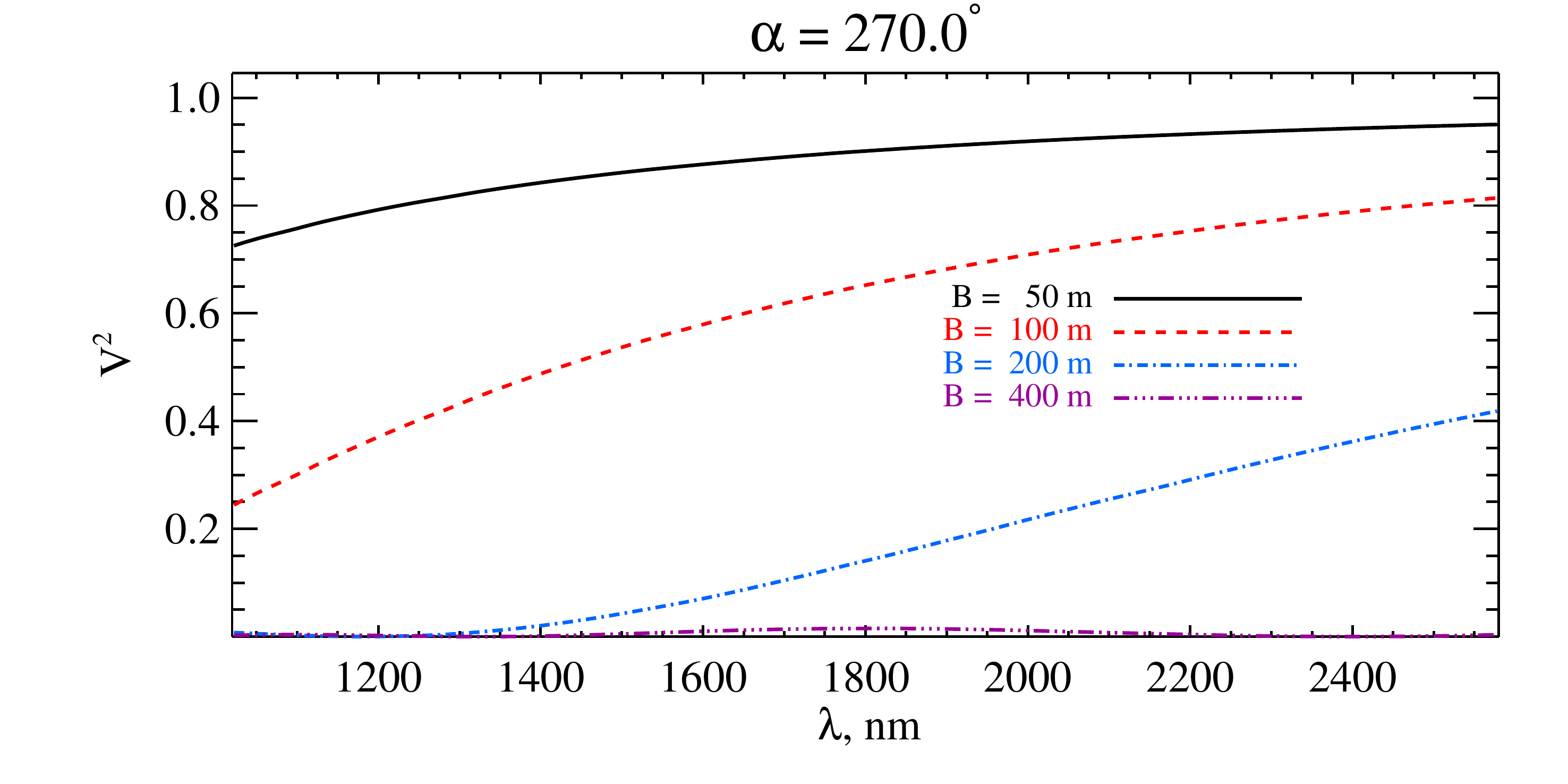}
}
\caption{{Squared} visibility  as a function of wavelength calculated at four selected baselines.
First row~--~spectrum predicted by spotted model; second and third rows~--~{squared} visibility  predicted by homogeneous
and spotted models respectively.
Three spectral windows are shown: $1065.0-1115.0$~nm, $R=6\,000$ (left column),
$2130.0-2200.0$~nm, $R=6\,000$ (middle column), and $1020.0-2580.0$~nm, $R=30$ (right column).
In all plots $\vsini=35$~\kms.}
\label{fig:vis-lambda-ir}
\end{figure*}

{Squared} visibility  computed at different spectral channels and resolutions $R=6\,000$
and $R=30$ are shown on Fig.~\ref{fig:vis-lambda-ir}. At both resolutions, {squared} visibility 
plots reveal no characteristic spectral line features except hydrogen lines and only with $R=6\,000$.
At a lowest resolution of $R=30$ the spectral lines are not resolved, as shown
in the third column of Fig.~\ref{fig:vis-lambda-ir}, and the uniform and spotted surface show very
similar visibility curves.

\subsubsection{Closure phase}

\begin{figure*}
\centerline{
\includegraphics[width=0.33\hsize]{figures/sp-wavelength-images-R6000-vsini35.00-mode2-lambda-10650.000-11150.000.pdf}
\includegraphics[width=0.33\hsize]{figures/sp-wavelength-images-R6000-vsini35.00-mode2-lambda-21300.000-22000.000.pdf}
\includegraphics[width=0.33\hsize]{figures/sp-wavelength-images-R30-vsini35.00-mode2-lambda-10200.000-25800.000.pdf}
}
\centerline{
\includegraphics[width=0.33\hsize]{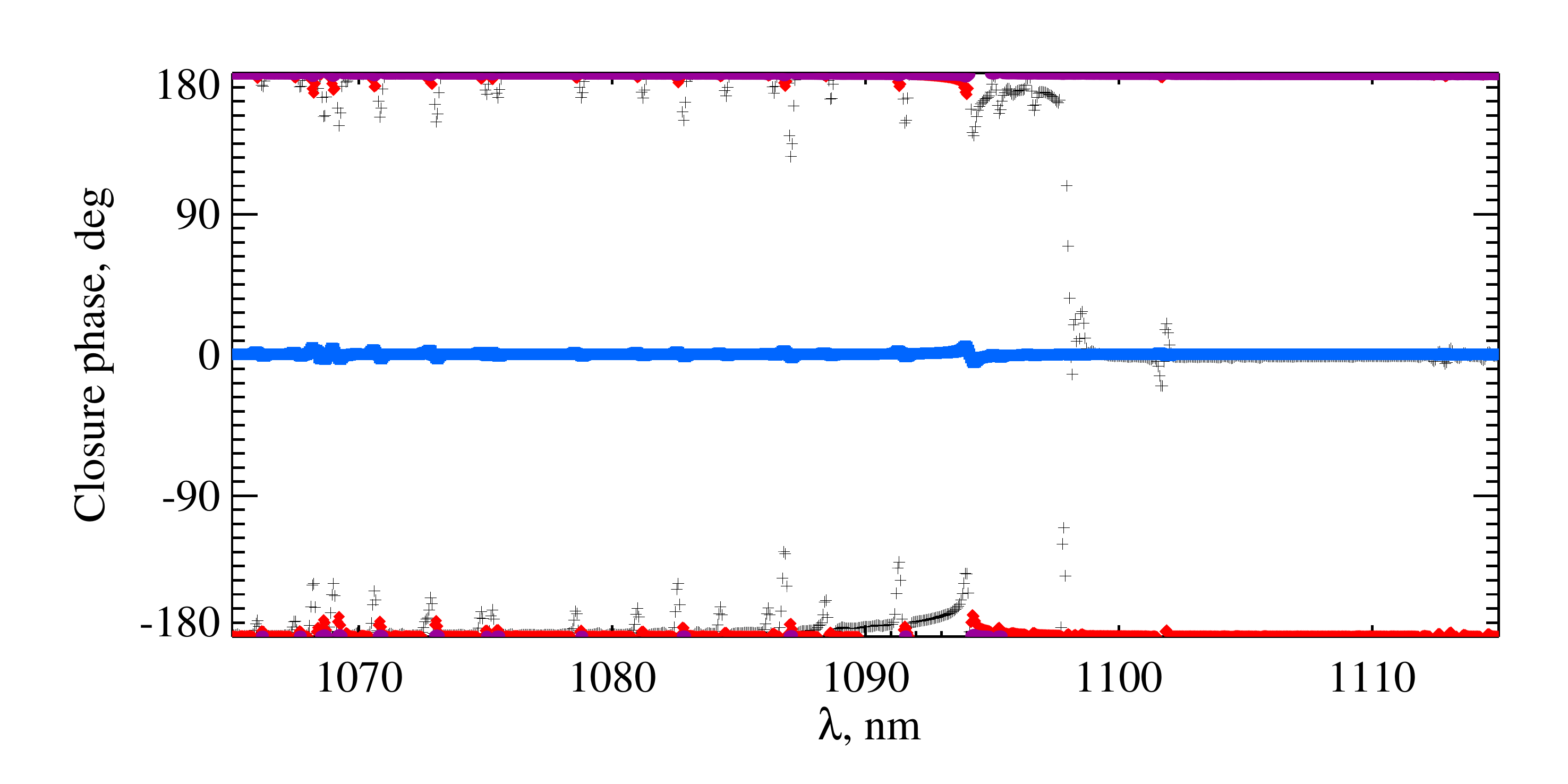}
\includegraphics[width=0.33\hsize]{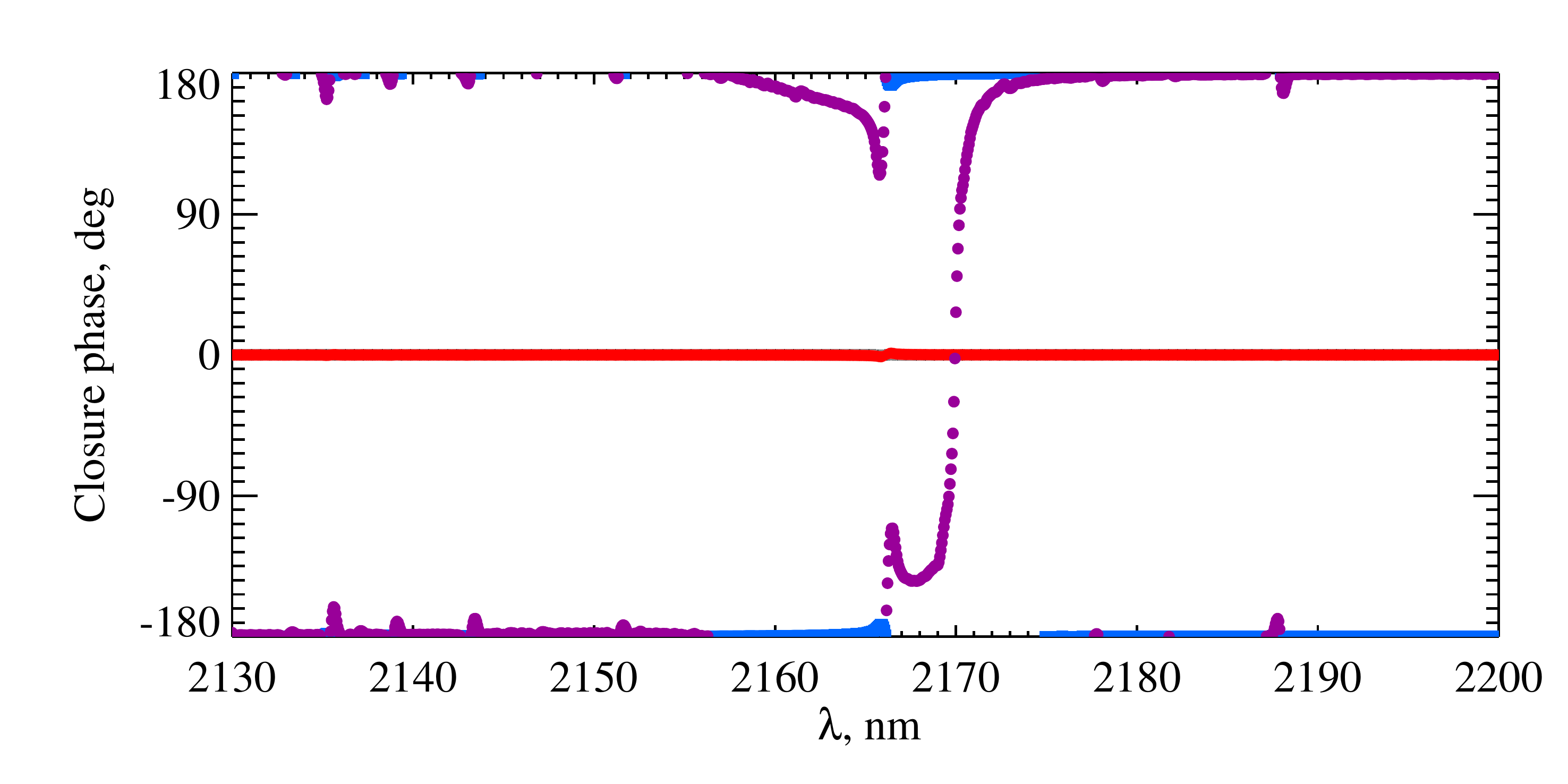}
\includegraphics[width=0.33\hsize]{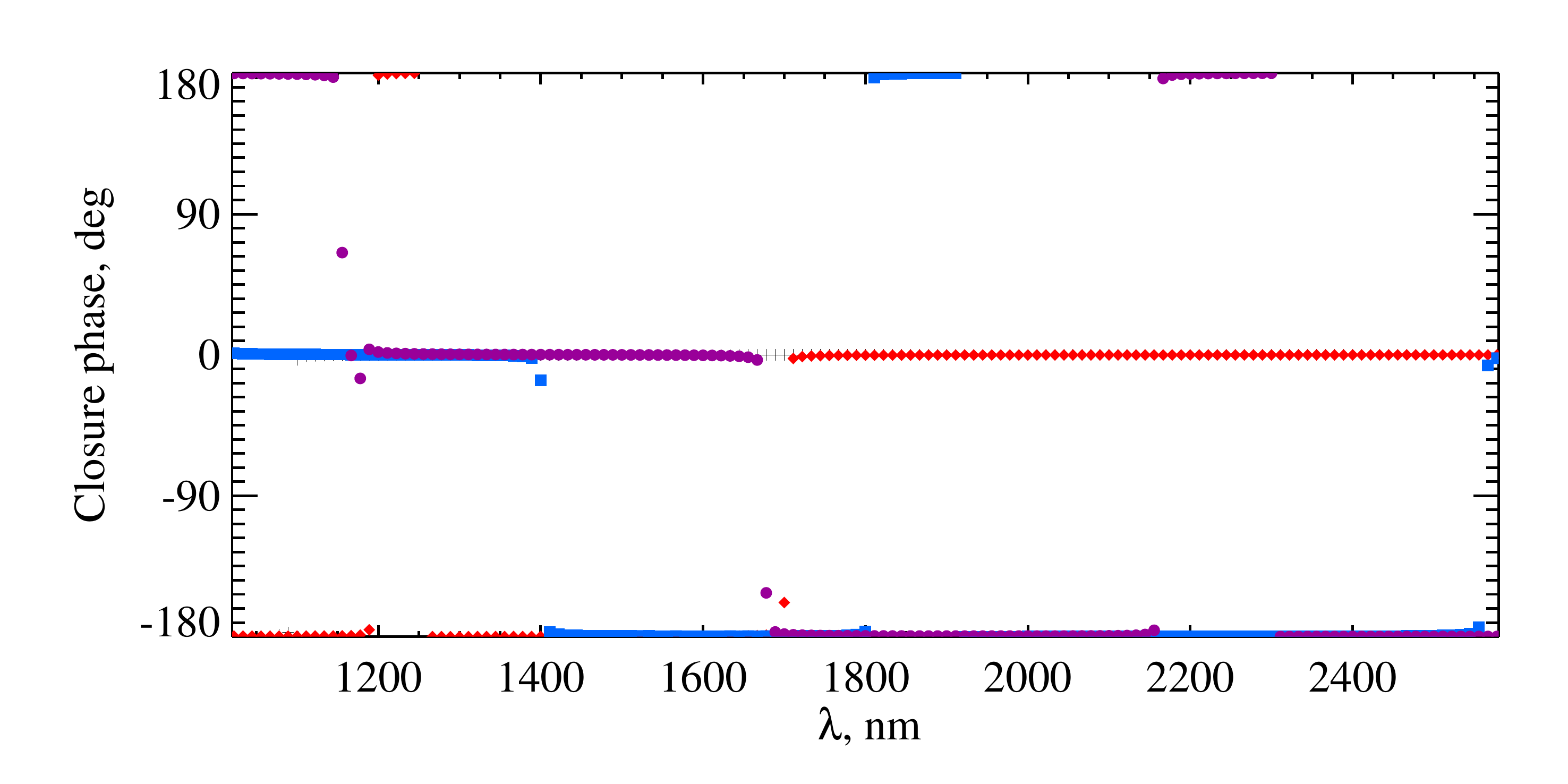}
}
\centerline{
\includegraphics[width=0.33\hsize]{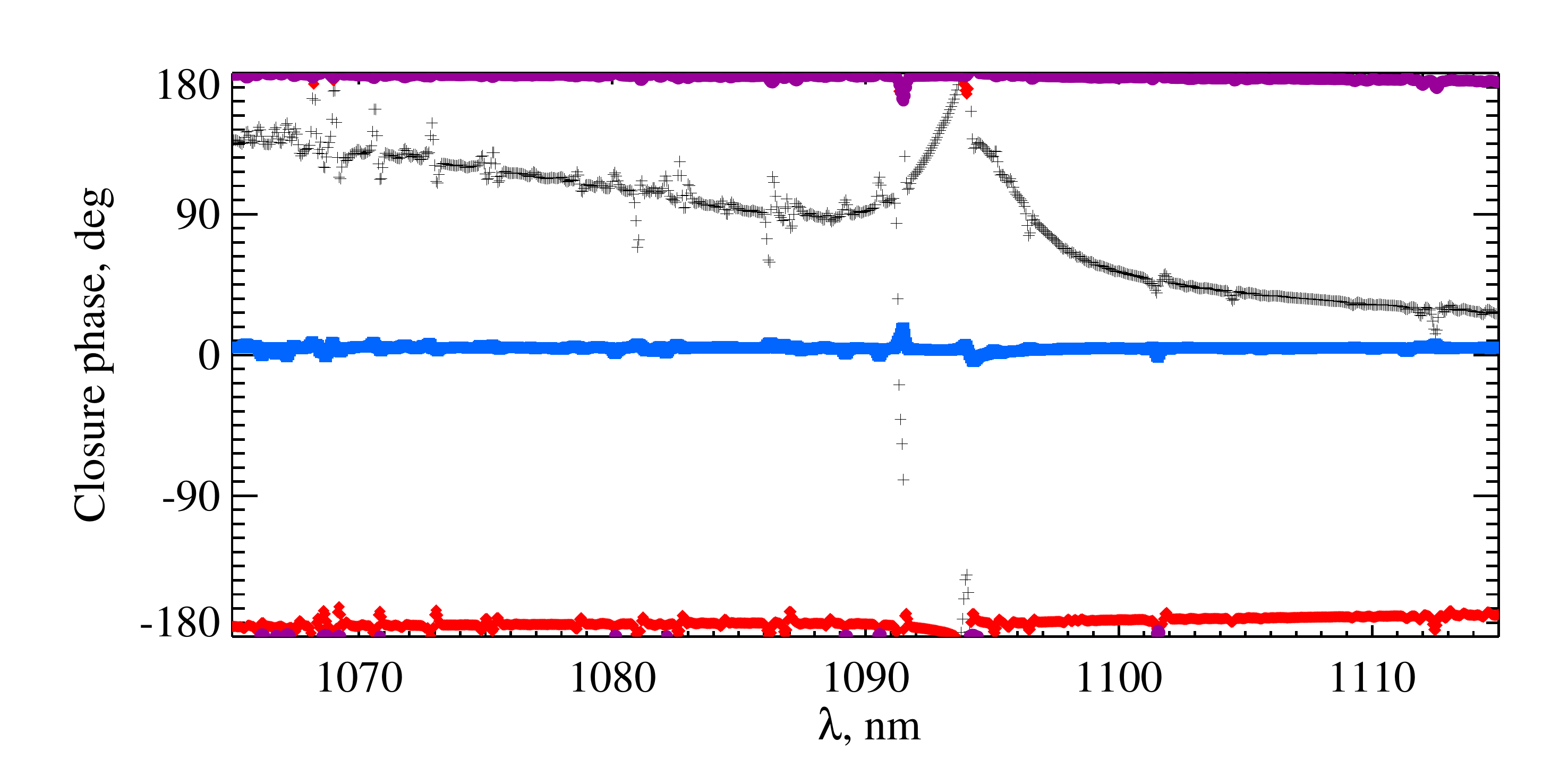}
\includegraphics[width=0.33\hsize]{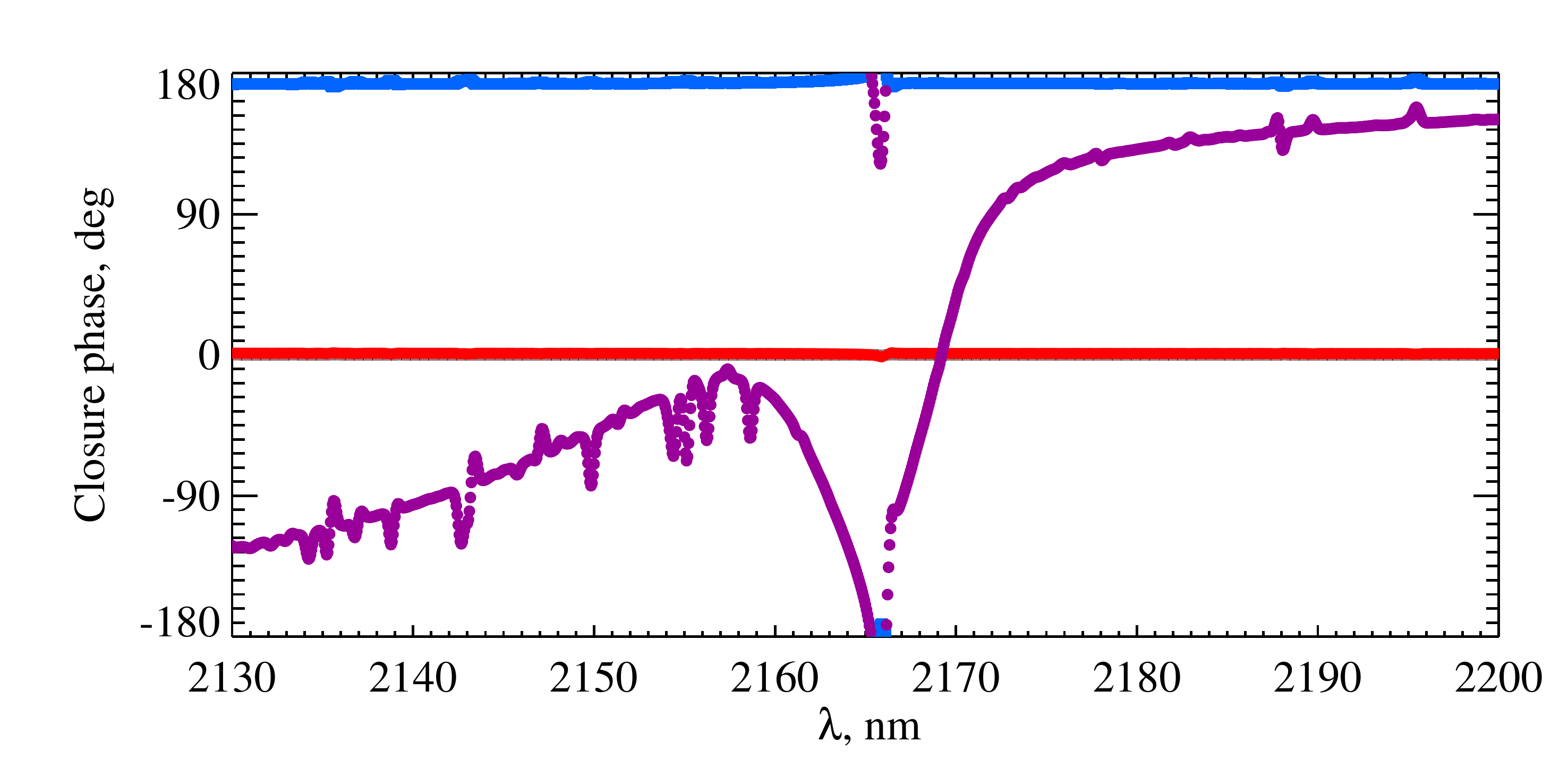}
\includegraphics[width=0.33\hsize]{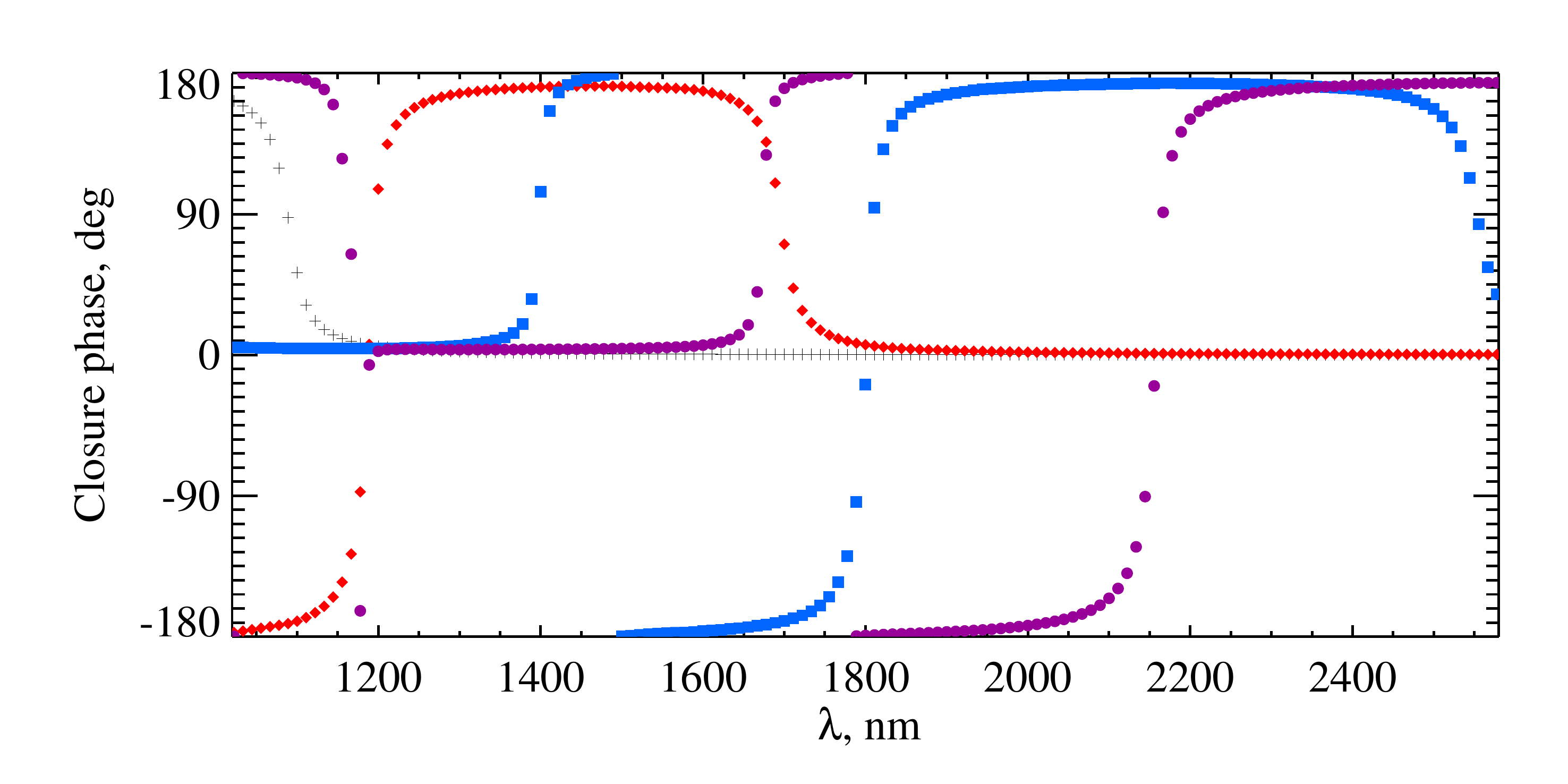}
}
\caption{
Closure phases as a function of wavelength.
First row~--~spectrum predicted by spotted model; second and third rows~--~closure phase predicted by homogeneous
and spotted models, respectively. 
Three spectral windows are shown: $1065.0-1115.0$~nm, $R=6\,000$ (left column),
$2130.0-2200.0$~nm, $R=6\,000$ (middle column), and $1020.0-2580.0$~nm, $R=30$ (right column).
In all plots $\vsini=35$~\kms. 
Closure phases were computed for the following configurations:
($0^\circ,130$m)$+$($270^\circ,130$m)$+$($135^\circ,184$m)~--~black crosses; 
($0^\circ,200$m)$+$($270^\circ,200$m)$+$($135^\circ,283$m)~--~red diamonds;
($0^\circ,300$m)$+$($270^\circ,300$m)$+$($135^\circ,424$m)~--~blue squares;
($0^\circ,358$m)$+$($270^\circ,358$m)$+$($135^\circ,506$m)~--~violet circles.
See online version for colored symbols.}
\label{fig:cp-lambda-ir}
\end{figure*}

The analysis of the closure phase signals demonstrate that the surface inhomogeneities can already be
detected in the $J$ band between $1000$~nm and $1100$~nm, $R=6\,000$ and $B_{\rm max}>180$~m.
An example of the strongest signal is shown in the left-hand column of Fig.~\ref{fig:cp-lambda-ir} for
$B_{\rm max}=184$~m. Spots are also detected at longer wavelengths. For instance,
middle row of Fig.~\ref{fig:cp-lambda-ir} illustrates predictions for a narrow spectral channel in $K$ band with
a hydrogen line from Brackett series.
Rotation signatures are visible at
particular configurations, e.g. with $B_{\rm max}=184$~m in the $J$ band and $B_{\rm max}=506$~m in the $K$ band,
but the shape of these signatures is strongly modified when spots are present.

There is no signal in closure phase seen for the
homogeneous star at lowest resolution $R=30$. 
The spotted star also illustrates phase changes by $\pm180^\circ$, 
but the closure phase pattern differs substantially from the homogeneous case: it does not show
any sharp jumps but rather smooth transitions between $0^\circ$ and $180^\circ$.
(see bottom left hand plot in Fig.~\ref{fig:cp-lambda-ir}). We therefore conclude
that spots can be detected in infrared even with very low spectral resolution, but
with baselines longer than $B_{\rm max}>180$~m.


\section{Discussion and conclusions}

In this paper we examined the possibility to detect abundance spots in atmospheres of CP stars
by running numerical simulations of such interferometric observable like visibility and closure phase.
These quantities were computed at different wavelength domains and spectral resolutions.
As a case study, we used abundance maps 
of the well-known CP star \uma\ which has one of the largest angular diameter among all CP stars known by date.

We confirm that the best spectral regions to search for abundance spots and rotation are in the visual domain close to
the Balmer jump, i.e. where the star radiates maximum of its flux. In that region, the intensity contrast
is higher in spectral lines of spotted elements thus providing most easy way to detect and characterize them
both from the analysis of {squared} visibility  and closure phase.
The spots and rotation signatures {can also be detected} in NIR. However, in our simulations this happens 
only at low {squared} visibility  $V^2\leqslant10^{-2}$ which corresponds to maximum baselines of triangle telescope configuration
of $B_{\rm max}\geqslant180$~m as seen in Fig.~\ref{fig:cp-lambda-ir}. Important that spots can be detected with
very low spectral resolution of $R=30$ (see third column of Fig.~\ref{fig:cp-lambda-ir}).
The stellar rotation is observed only if spectral lines, especially hydrogen lines (which are strongest spectral features in NIR), 
are resolved, i.e. with  spectral resolution on the order of $R\sim6\,000$ (see first and second columns of Fig.~\ref{fig:cp-lambda-ir}).

Because of star's large angular diameter, its surface can already be resolved with relatively
short baselines $B>50$~m in visual. Baselines longer than
$B\gtrsim200$~m are needed to resolve the star in infrared wavelengths.

Observing with high spectral resolution in visual domain allows one to constrain positions
of abundance spots and stellar rotation velocity. To achieve this goal from {squared} visibility  curves
observations with different position angles are needed. The wavelength dispersed closure phases
can also be used to derive stellar rotation and spots. This is possible because rotation induces
a characteristic symmetric phase jump at both sides of cores of spectral lines 
(see, e.g., Fig.~\ref{fig:cp-lambda-v-r30000}).
When spots are presented, these signals are modified, as well as more features appear at other wavelengths.

The analysis of {squared} visibility  shows that spots are clearly detected already at a first visibility
lobe, at least in strong spectral features of such elements as Cr and Fe. Individual details
depend on the position angle and, more critically, on spectral resolution (i.g. contrast). In most optimistic
cases, the difference between spots and rotation becomes clearly noticeable at {squared} visibility  $V^2\lesssim0.3$ 
as shown using examples of \ion{Cr}{ii}~$455.86$~nm (Fig.~\ref{fig:vis-mono-v-1}) and
\ion{Fe}{ii}~$531.67$~nm (Fig.~\ref{fig:vis-mono-v-2}) lines. 
From the behaviour of the position of the first visibility lobe we find that the effect of spots
is to make the star look larger (compared to the spotless case) if spots are dark 
and smaller if spots are bright, respectively.

One of our goals was to verify whether the abundance inhomogeneities on the surface of \uma\ 
can be detected with modern interferometric facilities. According to our simulations
an instrument like VEGA or its successor based on the principle of the demonstrator FRIEND
(Fibered and spectrally Resolved Interferometric Equipment New design, \citet{berio2014})
should be able to detect the effect of spots and spots+rotation, provided
that the instrument is able to measure {squared visibility  down to $\approx10^{-3}$}, and/or closure phase in visual. 
An instrument with the spectral resolution around $6\,000$ like 
AMBER or GRAVITY but baselines longer than $180$~m  would be able to measure rotation, and also rotation+spots.


In Table~\ref{tab:stars} we summarize the application of modern and planned interferometric facilities to a sample of CP stars.
Majority of stars in this table are magnetic CP stars listed in \citet{2006A&A...450..763K} 
with four additional stars HD~37776, HD~72106, 
HD~103498, and HD~177410,
and five presumably non-magnetic HgMn stars (shown in the end of the table). 
We used information from Table~\ref{tab:inters} to estimate the observability of each star as applied to a particular instrument. 
As seen from Table~\ref{tab:stars}, for most of CP stars their estimated angular diameters are below $1$~mas. 
Therefore, both long baselines of hundreds of meters
and detectors sensitive to values of $V^2<10^{-2}$ are required. 
One can see that there are many stars that cannot be observed with neither modern nor planned facilities.
Such benchmark stars as, say, HD~101065 (Przybylski's star) and HD~37776 (Landstreet's star) are among them.
On the other hand, we predict that it is possible to observe many stars with NPOI and/or SUSI
if the baseline of the latter will be increased.
This is the case for another well-known stars as, say, HD~137949 ($33$~Lib) and HD~65339 ($53$~Cam).

Finally, we find that a considerable fraction of CP stars in our sample can already be subject for spot detection
using existing interferometric facilities. Magnetic stars HD~24712, HD~40312, HD~128898, HD~137909, HD~201601, as well as
HgMn stars HD~358, HD~33904 are all well-known objects and for some of them DI maps are available in the literature.
But there are many more in the table. Note that none of the stars from our sample can be observed with VLTI
and the reason for this is a short baseline range provided by VLTI compared to other existing interferometers 
(see Table~\ref{tab:inters}). In fact, among 203 stars listed in Table~\ref{tab:stars}, 157 are visible at VLTI location,
and several of them could be observed if VLTI had maximum baselines longer than $300$~m. 
Alternatively, there are two objects that could be observed already with available maximum baseline of $140$~m
but detectors operating in visual were needed.


One should not forget that the brightness contrast in spectral lines that result from our simulations of 
\uma\ may differ from the real one, at least at certain wavelengths. This may happen because of internal inaccuracies
in atomic line parameters that affect the depth of respective spectroscopic features. This does not matter
for the medium and low resolutions tested in our investigation, but may be an issue for the highest resolution
of $R=30\,000$. We stress that observing with high resolution is important for studying
the stratification of chemical elements over stellar surfaces and to constrain theoretical models because
many features in visibility and closure phase can be studied.

A word of caution should be said regarding DI abundance maps themselves. 
In the DI code, the rotational modulation of spectral lines is interpreted as caused by abundance inhomogeneities only.
In reality, however, other effects such as, say, magnetic fields (if not included in DI analysis), 
may also contribute to the absolute values of the surface abundances recovered (however, the relative abundance changes
should not be affected much!). In case of \uma\ the magnetic field
is very weak, on the order of a few hundred Gauss \citep{2000MNRAS.313..851W}, and thus cannot seriously affect the results of DI. 
Also, all modern DI codes map only horizontal distribution of chemical elements and ignore their vertical
variations. This means that resulting abundance maps represent some vertically averaged abundance value.
Nevertheless, modelling the  light variability of \uma\ \citet{2010A&A...524A..66S} obtained a good agreement between model predictions
and observations based on the same DI maps that we used in this study.
The authors, however, predicted the variability in narrow
and broad-band photometric filters where the possible inaccuracies in atomic data and contrast in certain spectroscopic
features, even if present, do not play critical role.

Finally, the above consideration also suggests that comparing the observed visibility with synthetic ones will allow us
to constrain atmospheric models and to provide an independent validation of the DI results.


\section*{Acknowledgments}
DS acknowledges financial support from CRC~963~--~Astrophysical Flow Instabilities and Turbulence 
(project A16 and A17) to DS. CP acknowledges the support by the Belgian Federal Science
Policy Office via the PRODEX Programme of ESA.
OK is a Royal Swedish Academy of Sciences Research Fellow,
supported by the grants from the Knut and Alice Wallenberg Foundation, Swedish Research
Council, and the G\"oran Gustafsson Foundation. This research made use of the
computer corporate facility of the Georg-August University of G\"ottingen 
and the Max-Planck-Gesellschaft (GWDG), as well as web services: SIMBAD, NASA ADS, VALD.



\onecolumn
\begin{center}
\begin{longtable}{ccccccccccccc}

\caption{Observability of CP stars with modern and future interferometers.}\label{tab:stars}\\

\hline
HD     & $V$  &  $J$  & $H$  & $K$  &  $d$  & $R$     & $\theta$  & \multicolumn{4}{c}{$B_{\rm max}^{\dagger}$}             & ref. \\ 
number & mag  &  mag  & mag  & mag  &  pc   & $\Rsun$ & mas       & \multicolumn{4}{c}{m}                                   & \\ 
       &      &       &      &      &       &         &           & $0.55$~\mum  & $1.25$~\mum & $1.65$~\mum  & $2.15$~\mum & \\ 
\hline                                       
\endfirsthead                                
\caption{continued.}\\
\hline
HD     & $V$  &  $J$  & $H$  & $K$  &  $d$  & $R$     & $\theta$  & \multicolumn{4}{c}{$B_{\rm max}^{\dagger}$}             & ref. \\ 
number & mag  &  mag  & mag  & mag  &  pc   & $\Rsun$ & mas       & \multicolumn{4}{c}{m}                                   & \\ 
       &      &       &      &      &       &         &           & $0.55$~\mum  & $1.25$~\mum & $1.65$~\mum  & $2.15$~\mum & \\ 
\hline
\endhead
\hline
\endfoot
\hline
\multicolumn{13}{c}{Magnetic CP stars}\\
\hline
1048     & 6.25  & 6.14  & 6.24  & 6.22  & 140.8  & 2.37  & 0.16  &   723 &  1644 &  2170 &  2828 & 1  \\ 
2453     & 6.91  & 6.71  & 6.76  & 6.74  & 221.7  & 2.57  & 0.11  &  1050 &  2388 &  3152 &  4108 & 1  \\ 
3980     & 5.70  & 5.59  & 5.58  & 5.57  &  67.0  & 5.76  & 0.80  &   402 &   915 &  1208 &  1574 & 1  \\ 
4778     & 6.15  & 6.09  & 6.15  & 6.14  & 100.3  & 1.91  & 0.18  &   641 &  1457 &  1923 &  2507 & 1  \\ 
5737     & 4.27  & 4.63  & 4.81  & 4.67  & 238.1  & 7.52  & 0.29  &   385 &   877 &  1157 &  1508 & 1  \\ 
8441     & 6.68  & 6.46  & 6.53  & 6.50  & 204.9  & 3.48  & 0.16  &   718 &  1632 &  2155 &  2808 & 1  \\ 
9996     & 6.39  & 6.34  & 6.40  & 6.43  & 160.5  & 2.24  & 0.13  &   875 &  1989 &  2626 &  3422 & 1  \\ 
10221    & 5.59  & 5.62  & 5.70  & 5.69  & 119.2  & 3.45  & 0.27  &   420 &   956 &  1262 &  1644 & 1  \\ 
10783    & 6.43  & 6.54  & 6.58  & 6.56  & 176.7  & 3.17  & 0.17  &   679 &  1543 &  2037 &  2654 & 1  \\ 
11187    & 7.94  & 7.13  & 7.22  & 7.23  & 321.5  & 2.92  & 0.08  &  1342 &  3051 &  4028 &  5249 & 1  \\ 
11503    & 4.52  &  ...  &  ...  & 4.72  &  50.3  & 2.47  & 0.46  &   247 &   563 &   743 &   969 & 1  \\ 
12288    & 7.74  & 7.42  & 7.40  & 7.43  & 308.6  & 2.55  & 0.08  &  1476 &  3355 &  4429 &  5771 & 1  \\ 
12447    & 4.11  &  ...  &  ...  &  ...  &  46.2  & 2.07  & 0.42  &   272 &   618 &   817 &  1064 & 1  \\ 
12767    & 4.69  & 5.00  & 5.05  & 5.01  & 113.8  & 3.44  & 0.28  &   403 &   916 &  1210 &  1577 & 1  \\ 
14437    & 7.27  & 7.26  & 7.29  & 7.31  & 159.7  & 2.46  & 0.14  &   792 &  1801 &  2378 &  3099 & 1  \\ 
15089    & 4.53  &  ...  &  ...  &  ...  &  40.7  & 2.13  & 0.49  &   233 &   529 &   699 &   911 & 1  \\ 
15144    & 5.83  & 5.70  & 5.68  & 5.63  &  77.0  & 1.72  & 0.21  &   545 &  1238 &  1635 &  2130 & 1  \\ 
17775    & 8.03  & 7.76  & 7.78  & 7.77  & 173.3  & 1.69  & 0.09  &  1249 &  2838 &  3747 &  4882 & 1  \\ 
18296    & 5.10  & 5.11  & 5.24  & 5.23  &  98.0  & 3.77  & 0.36  &   317 &   720 &   951 &  1239 & 1  \\ 
18610    & 8.14  & 7.89  & 7.95  & 7.88  & 213.2  & 2.53  & 0.11  &  1029 &  2339 &  3088 &  4024 & 1  \\ 
19712    & 7.35  & 7.37  & 7.47  & 7.41  & 165.3  & 1.74  & 0.10  &  1156 &  2627 &  3468 &  4519 & 1  \\ 
19805    & 7.96  & 7.69  & 7.70  & 7.66  & 222.2  & 1.87  & 0.08  &  1447 &  3289 &  4341 &  5657 & 1  \\ 
19832    & 5.76  & 5.93  & 5.99  & 6.01  & 154.1  & 5.28  & 0.32  &   684 &  1556 &  2054 &  2676 & 1  \\ 
21699    & 5.46  & 5.57  & 5.67  & 5.67  & 185.5  & 3.94  & 0.20  &   574 &  1305 &  1723 &  2245 & 1  \\ 
22316    & 6.28  & 6.38  & 6.51  & 6.51  & 157.2  & 2.80  & 0.17  &   684 &  1555 &  2053 &  2675 & 1  \\ 
22374    & 6.73  & 6.44  & 6.49  & 6.42  & 130.7  & 2.74  & 0.19  &   582 &  1323 &  1746 &  2276 & 1  \\ 
22470    & 5.23  & 5.46  & 5.56  & 5.53  & 149.3  & 3.15  & 0.20  &   577 &  1312 &  1733 &  2258 & 1  \\ 
22920    & 5.53  & 5.76  & 5.79  & 5.87  & 152.2  & 4.62  & 0.28  &   401 &   913 &  1205 &  1571 & 1  \\ 
23207    & 7.54  & 7.27  & 7.29  & 7.25  & 207.0  & 2.64  & 0.12  &   956 &  2174 &  2870 &  3740 & 1  \\ 
23408    & 3.87  & 3.93  & 4.12  & 3.99  & 117.5  & 5.76  & 0.46  &   248 &   565 &   746 &   972 & 1  \\ 
24155    & 6.38  & 6.32  & 6.36  & 6.39  & 123.2  & 2.21  & 0.17  &   679 &  1544 &  2038 &  2656 & 1  \\ 
24188    & 6.25  & 6.52  & 6.57  & 6.58  & 143.3  & 2.25  & 0.15  &   777 &  1767 &  2333 &  3040 & 1  \\ 
24712    & 6.00  & 5.43  & 5.31  & 5.26  &  49.2  & 1.46  & 0.28  &   296 &   674 &   890 &  1160 & 1  \\ 
25267    & 4.66  & 4.83  & 4.89  & 4.80  & 100.4  & 3.26  & 0.30  &   375 &   853 &  1126 &  1467 & 1  \\ 
25354    & 7.84  & 7.78  & 7.87  & 7.82  & 155.3  & 1.49  & 0.09  &  1273 &  2893 &  3819 &  4976 & 1  \\ 
25823    & 5.17  & 5.36  & 5.41  & 5.43  & 128.9  & 3.54  & 0.26  &   443 &  1008 &  1330 &  1733 & 1  \\ 
27309    & 5.34  & 5.52  & 5.61  & 5.64  & 100.0  & 2.35  & 0.22  &   519 &  1181 &  1559 &  2031 & 1  \\ 
28843    & 5.81  & 5.99  & 6.09  & 6.07  & 145.8  & 2.44  & 0.16  &   728 &  1655 &  2185 &  2847 & 1  \\ 
30466    & 7.28  & 6.86  & 6.88  & 6.81  & 149.5  & 2.22  & 0.14  &   822 &  1870 &  2468 &  3216 & 1  \\ 
32633    & 7.07  & 7.00  & 7.03  & 7.06  & 179.5  & 1.92  & 0.10  &  1142 &  2596 &  3428 &  4466 & 1  \\ 
34452    & 5.37  & 5.63  & 5.75  & 5.76  & 126.7  & 2.81  & 0.21  &   550 &  1250 &  1651 &  2151 & 1  \\ 
34797    & 6.54  & 6.70  & 6.80  & 6.77  & 244.5  & 3.35  & 0.13  &   891 &  2025 &  2674 &  3484 & 1  \\ 
37776    & 6.96  & 7.29  & 7.42  & 7.42  & 330.0  & 2.75  & 0.08  &  1215 &  2762 &  3646 &  4751 & 2  \\ 
38823    & 7.32  & 6.94  & 6.95  & 6.89  &  97.5  & 2.46  & 0.23  &   483 &  1099 &  1451 &  1891 & 1  \\ 
39317    & 5.59  & 5.58  & 5.61  & 5.58  & 130.0  & 4.05  & 0.29  &   391 &   890 &  1174 &  1530 & 1  \\ 
40312    & 2.62  & 2.69  & 2.70  & 2.75  &  50.8  & 4.58  & 0.84  &   148 &   336 &   444 &   578 & 1  \\ 
42616    & 7.17  & 6.82  & 6.86  & 6.84  & 183.8  & 2.66  & 0.13  &   841 &  1913 &  2525 &  3290 & 1  \\ 
42659    & 6.75  & 6.44  & 6.41  & 6.36  & 131.6  & 2.91  & 0.21  &   552 &  1254 &  1656 &  2157 & 1  \\ 
49333    & 6.08  & 6.42  & 6.47  & 6.53  & 241.5  & 2.73  & 0.11  &  1078 &  2450 &  3235 &  4215 & 1  \\ 
49976    & 6.29  & 6.25  & 6.27  & 6.23  &  99.9  & 2.00  & 0.19  &   610 &  1386 &  1830 &  2384 & 1  \\ 
54118    & 5.17  & 5.26  & 5.32  & 5.29  &  92.3  & 2.66  & 0.27  &   423 &   962 &  1270 &  1655 & 1  \\ 
55522    & 5.89  & 6.23  & 6.39  & 6.36  & 257.1  & 3.25  & 0.12  &   965 &  2194 &  2897 &  3774 & 1  \\ 
55719    & 5.31  & 5.19  & 5.20  & 5.14  & 126.1  & 3.15  & 0.23  &   488 &  1109 &  1464 &  1908 & 1  \\ 
56350    & 6.69  & 6.78  & 6.88  & 6.84  & 151.3  & 2.37  & 0.15  &   779 &  1770 &  2337 &  3045 & 1  \\ 
60435    & 8.89  & 8.43  & 8.39  & 8.37  & 226.8  & 1.88  & 0.08  &  1473 &  3348 &  4420 &  5759 & 1  \\ 
62140    & 6.47  & 6.16  & 6.18  & 6.17  &  96.7  & 2.09  & 0.20  &   563 &  1281 &  1691 &  2204 & 1  \\ 
63401    & 6.25  & 6.56  & 6.66  & 6.68  & 200.4  & 3.09  & 0.14  &   790 &  1795 &  2370 &  3088 & 1  \\ 
64486    & 5.39  &  ...  &  ...  &  ...  &  99.0  & 2.89  & 0.27  &   418 &   950 &  1254 &  1635 & 1  \\ 
64740    & 4.63  & 5.15  & 5.35  & 5.27  & 232.6  & 4.29  & 0.17  &   661 &  1502 &  1983 &  2584 & 1  \\ 
65339    & 6.02  &  ...  &  ...  &  ...  &  98.7  & 4.55  & 0.43  &   509 &  1157 &  1527 &  1990 & 1  \\ 
71866    & 6.72  & 6.62  & 6.71  & 6.67  & 133.2  & 2.66  & 0.19  &   609 &  1385 &  1829 &  2383 & 1  \\ 
72106    & 8.61  & 8.43  & 8.28  & 7.92  & 278.6  & 0.61  & 0.02  &  2560 &  5818 &  7680 & 10007 & 3  \\ 
72968    & 5.72  & 5.71  & 5.75  & 5.69  &  92.9  & 2.04  & 0.20  &   555 &  1262 &  1666 &  2170 & 1  \\ 
73340    & 5.78  & 6.02  & 6.03  & 6.04  & 136.8  & 2.62  & 0.18  &   636 &  1446 &  1909 &  2488 & 1  \\ 
74521    & 5.66  & 5.72  & 5.79  & 5.82  & 129.7  & 3.00  & 0.22  &   526 &  1197 &  1580 &  2059 & 1  \\ 
75445    & 7.12  & 6.64  & 6.52  & 6.53  & 108.3  & 2.18  & 0.19  &   606 &  1377 &  1818 &  2369 & 1  \\ 
79158    & 5.29  & 5.41  & 5.53  & 5.53  & 178.3  & 4.21  & 0.22  &   516 &  1173 &  1548 &  2017 & 1  \\ 
81009    & 6.53  & 6.22  & 6.24  & 6.15  & 144.5  & 2.71  & 0.17  &   649 &  1476 &  1948 &  2539 & 1  \\ 
83368    & 6.23  &  ...  &  ...  &  ...  &  70.6  & 2.35  & 0.31  &   429 &   977 &  1289 &  1680 & 1  \\ 
83625    & 6.88  & 6.97  & 7.06  & 7.06  & 181.2  & 2.40  & 0.12  &   922 &  2095 &  2766 &  3605 & 1  \\ 
86199    & 6.74  & 6.89  & 7.02  & 6.97  & 232.0  & 3.02  & 0.12  &   938 &  2132 &  2814 &  3667 & 1  \\ 
88158    & 6.44  & 6.56  & 6.63  & 6.62  & 210.5  & 3.69  & 0.16  &   695 &  1579 &  2085 &  2717 & 1  \\ 
88385    & 8.09  & 7.93  & 8.00  & 7.95  & 206.2  & 2.31  & 0.10  &  1089 &  2475 &  3267 &  4257 & 1  \\ 
89103    & 7.78  & 7.97  & 8.08  & 8.06  & 188.7  & 1.66  & 0.08  &  1385 &  3148 &  4155 &  5414 & 1  \\ 
90044    & 5.97  & 5.98  & 6.10  & 6.07  & 104.2  & 2.37  & 0.21  &   536 &  1219 &  1609 &  2096 & 1  \\ 
90569    & 6.04  & 6.02  & 6.08  & 6.04  & 129.2  & 2.47  & 0.18  &   638 &  1450 &  1914 &  2495 & 1  \\ 
92385    & 6.71  & 6.81  & 6.89  & 6.86  & 169.2  & 2.10  & 0.12  &   982 &  2232 &  2946 &  3839 & 1  \\ 
92499    & 8.88  & 8.52  & 8.54  & 8.54  & 282.5  & 2.15  & 0.07  &  1602 &  3641 &  4807 &  6263 & 1  \\ 
92664    & 5.52  & 5.79  & 5.87  & 5.88  & 160.5  & 2.84  & 0.16  &   689 &  1566 &  2067 &  2693 & 1  \\ 
94427    & 7.36  & 6.72  & 6.62  & 6.55  & 143.5  & 2.19  & 0.14  &   797 &  1811 &  2391 &  3116 & 1  \\ 
94660    & 6.11  & 6.21  & 6.17  & 6.20  & 149.9  & 2.81  & 0.17  &   649 &  1476 &  1948 &  2539 & 1  \\ 
96707    & 6.08  & 5.57  & 5.53  & 5.50  & 111.2  & 3.22  & 0.27  &   421 &   958 &  1265 &  1648 & 1  \\ 
98088    & 6.14  & 5.78  & 5.73  & 5.69  & 129.5  & 3.24  & 0.23  &   487 &  1108 &  1463 &  1906 & 1  \\ 
98340    & 7.13  & 7.17  & 7.25  & 7.23  & 221.7  & 2.10  & 0.09  &  1287 &  2925 &  3861 &  5031 & 1  \\ 
101065   & 8.03  & 7.11  & 6.94  & 6.92  & 112.0  & 1.98  & 0.16  &   689 &  1567 &  2069 &  2696 & 1  \\ 
103192   & 4.28  & 4.36  & 4.56  & 4.43  &  95.0  & 3.94  & 0.39  &   294 &   668 &   882 &  1149 & 1  \\ 
103498   & 6.99  & 6.81  & 6.87  & 6.86  & 296.7  & 4.50  & 0.14  &   804 &  1827 &  2412 &  3143 & 4  \\ 
105382   & 4.47  & 4.82  & 4.95  & 4.87  & 134.4  & 3.35  & 0.23  &   490 &  1113 &  1470 &  1915 & 1  \\ 
105770   & 7.37  & 7.02  & 7.03  & 6.96  & 191.2  & 2.53  & 0.12  &   921 &  2093 &  2763 &  3600 & 1  \\ 
108662   & 5.24  & 5.22  & 5.30  & 5.29  &  72.9  & 2.43  & 0.31  &   365 &   829 &  1095 &  1427 & 1  \\ 
108945   & 5.44  & 5.29  & 5.31  & 5.27  &  82.7  & 2.98  & 0.34  &   338 &   768 &  1015 &  1322 & 1  \\ 
109026   & 3.88  & 4.20  & 4.25  & 4.25  &  99.6  & 4.24  & 0.40  &   286 &   651 &   859 &  1119 & 1  \\ 
110066   & 6.40  & 6.29  & 6.33  & 6.33  & 134.6  & 3.19  & 0.22  &   513 &  1167 &  1541 &  2008 & 1  \\ 
111133   & 6.34  & 6.28  & 6.34  & 6.33  & 266.0  & 3.09  & 0.11  &  1050 &  2388 &  3152 &  4108 & 1  \\ 
112185   & 1.77  & 1.72  & 1.73  & 1.76  &  25.3  & 4.14  & 1.52  &    62 &   140 &   186 &   242 & 1  \\ 
112381   & 6.49  &  ...  &  ...  &  ...  & 118.5  & 1.66  & 0.13  &   869 &  1976 &  2609 &  3400 & 1  \\ 
112413   & 2.88  & 3.05  & 3.13  & 3.16  &  35.2  & 3.42  & 0.91  &   186 &   424 &   560 &   730 & 1  \\ 
115226   & 8.50  & 7.94  & 7.90  & 7.85  & 147.1  & 1.54  & 0.10  &  1164 &  2647 &  3494 &  4553 & 1  \\ 
115440   & 8.24  & 8.04  & 8.04  & 8.03  & 199.2  & 1.91  & 0.09  &  1270 &  2888 &  3812 &  4967 & 1  \\ 
115708   & 7.83  & 7.23  & 7.19  & 7.15  & 115.6  & 1.86  & 0.15  &   759 &  1727 &  2279 &  2970 & 1  \\ 
116114   & 7.02  & 6.48  & 6.42  & 6.35  & 129.7  & 2.78  & 0.20  &   569 &  1295 &  1709 &  2227 & 1  \\ 
116458   & 5.66  & 5.67  & 5.70  & 5.67  & 136.1  & 3.46  & 0.24  &   479 &  1089 &  1437 &  1873 & 1  \\ 
116890   & 6.19  & 6.03  & 6.04  & 5.98  & 187.6  & 4.12  & 0.20  &   555 &  1263 &  1667 &  2173 & 1  \\ 
117025   & 6.07  & 5.95  & 5.92  & 5.87  &  86.7  & 2.20  & 0.24  &   479 &  1089 &  1437 &  1873 & 1  \\ 
118022   & 4.94  & 4.90  & 4.90  & 4.88  &  56.7  & 2.18  & 0.36  &   316 &   718 &   948 &  1236 & 1  \\ 
118913   & 7.69  & 7.41  & 7.47  & 7.36  & 256.4  & 2.49  & 0.09  &  1255 &  2852 &  3765 &  4906 & 1  \\ 
119213   & 6.29  & 6.15  & 6.21  & 6.20  &  91.1  & 1.96  & 0.20  &   567 &  1290 &  1703 &  2219 & 1  \\ 
119308   & 7.86  & 7.75  & 7.79  & 7.76  & 245.1  & 1.76  & 0.07  &  1698 &  3860 &  5096 &  6640 & 1  \\ 
119419   & 6.44  & 6.61  & 6.70  & 6.71  & 119.0  & 1.77  & 0.14  &   821 &  1866 &  2463 &  3210 & 1  \\ 
120198   & 5.68  & 5.71  & 5.78  & 5.76  &  89.0  & 2.01  & 0.21  &   541 &  1230 &  1623 &  2115 & 1  \\ 
122532   & 6.08  & 6.26  & 6.37  & 6.33  & 219.8  & 3.10  & 0.13  &   864 &  1964 &  2593 &  3379 & 1  \\ 
122970   & 8.29  & 7.52  & 7.38  & 7.30  & 115.3  & 1.73  & 0.14  &   812 &  1846 &  2437 &  3175 & 1  \\ 
124224   & 5.02  & 5.25  & 5.29  & 5.29  &  79.2  & 1.97  & 0.23  &   488 &  1111 &  1466 &  1911 & 1  \\ 
125248   & 5.90  & 5.83  & 5.87  & 5.84  & 102.0  & 1.95  & 0.18  &   639 &  1452 &  1917 &  2498 & 1  \\ 
125630   & 6.76  & 6.69  & 6.73  & 6.73  & 165.6  & 2.83  & 0.16  &   714 &  1622 &  2142 &  2791 & 1  \\ 
125823   & 4.42  & 4.93  & 4.99  & 4.92  & 140.3  & 3.45  & 0.23  &   496 &  1127 &  1488 &  1939 & 1  \\ 
126515   & 7.07  & 7.10  & 7.15  & 7.15  & 106.4  & 1.84  & 0.16  &   705 &  1603 &  2117 &  2758 & 1  \\ 
127453   & 7.36  & 7.16  & 7.16  & 7.15  & 225.2  & 3.04  & 0.13  &   904 &  2055 &  2713 &  3535 & 1  \\ 
127575   & 7.75  & 7.59  & 7.59  & 7.58  & 142.9  & 1.85  & 0.12  &   943 &  2144 &  2830 &  3687 & 1  \\ 
128775   & 6.62  & 6.80  & 6.87  & 6.87  & 208.8  & 2.27  & 0.10  &  1120 &  2546 &  3361 &  4380 & 1  \\ 
128898   & 3.19  & 2.84  & 2.73  & 2.74  &  16.6  & 1.94  & 1.09  &   104 &   236 &   312 &   407 & 1  \\ 
129899   & 6.44  & 6.43  & 6.49  & 6.46  & 255.1  & 4.95  & 0.18  &   628 &  1429 &  1886 &  2457 & 1  \\ 
130559   & 5.31  & 5.21  & 5.20  & 5.18  &  72.9  & 2.04  & 0.26  &   436 &   991 &  1308 &  1705 & 1  \\ 
132322   & 7.37  & 6.97  & 7.01  & 6.90  & 196.1  & 2.75  & 0.13  &   869 &  1975 &  2608 &  3398 & 1  \\ 
133029   & 6.35  & 6.49  & 6.58  & 6.57  & 169.8  & 2.51  & 0.14  &   825 &  1875 &  2476 &  3226 & 1  \\ 
133652   & 5.97  & 6.04  & 6.07  & 6.05  & 121.4  & 1.83  & 0.14  &   808 &  1838 &  2426 &  3161 & 1  \\ 
133792   & 6.25  & 6.02  & 6.07  & 6.00  & 181.5  & 3.85  & 0.20  &   574 &  1306 &  1724 &  2247 & 1  \\ 
133880   & 5.79  & 5.99  & 5.91  & 5.99  & 110.7  & 2.51  & 0.21  &   537 &  1220 &  1611 &  2099 & 1  \\ 
134214   & 7.46  & 6.82  & 6.68  & 6.67  & 102.7  & 1.71  & 0.15  &   733 &  1666 &  2199 &  2866 & 1  \\ 
134305   & 7.25  & 6.81  & 6.79  & 6.75  & 150.6  & 2.93  & 0.18  &   626 &  1422 &  1878 &  2447 & 1  \\ 
137509   & 6.87  & 7.02  & 7.13  & 7.18  & 195.7  & 2.81  & 0.13  &   847 &  1927 &  2543 &  3314 & 1  \\ 
137909   & 3.68  & 3.38  & 3.28  & 3.28  &  34.3  & 2.47  & 0.67  &   169 &   384 &   507 &   661 & 1  \\ 
137949   & 6.69  & 6.31  & 6.28  & 6.25  &  88.7  & 2.13  & 0.22  &   507 &  1153 &  1522 &  1984 & 1  \\ 
138758   & 7.89  & 7.91  & 7.98  & 7.98  & 208.8  & 2.08  & 0.09  &  1225 &  2786 &  3677 &  4792 & 1  \\ 
140160   & 5.33  & 5.18  & 5.23  & 5.20  &  67.4  & 2.25  & 0.31  &   365 &   829 &  1095 &  1426 & 1  \\ 
140728   & 5.49  & 5.47  & 5.58  & 5.56  &  91.6  & 2.52  & 0.26  &   443 &  1007 &  1329 &  1732 & 1  \\ 
142301   & 5.87  & 5.92  & 6.00  & 5.99  & 158.0  & 2.53  & 0.15  &   762 &  1733 &  2288 &  2981 & 1  \\ 
142990   & 5.43  & 5.58  & 5.67  & 5.65  & 170.4  & 3.02  & 0.16  &   688 &  1565 &  2066 &  2692 & 1  \\ 
143473   & 7.41  & 7.01  & 6.98  & 6.96  & 116.1  & 1.64  & 0.13  &   862 &  1960 &  2587 &  3371 & 1  \\ 
144334   & 5.92  & 6.03  & 5.97  & 5.92  & 161.3  & 2.74  & 0.16  &   718 &  1632 &  2155 &  2808 & 1  \\ 
145501   & 6.30  & 5.78  & 5.69  & 5.61  & 145.3  & 2.49  & 0.16  &   711 &  1617 &  2134 &  2781 & 1  \\ 
147010   & 7.40  & 6.75  & 6.75  & 6.68  & 163.4  & 2.04  & 0.12  &   977 &  2221 &  2931 &  3820 & 1  \\ 
148112   & 4.58  & 4.56  & 4.59  & 4.52  &  76.7  & 3.30  & 0.40  &   283 &   644 &   850 &  1108 & 1  \\ 
148199   & 7.01  & 6.71  & 6.75  & 6.72  & 158.2  & 2.41  & 0.14  &   801 &  1822 &  2405 &  3134 & 1  \\ 
148330   & 5.75  & 5.70  & 5.70  & 5.72  & 119.9  & 3.09  & 0.24  &   473 &  1076 &  1421 &  1852 & 1  \\ 
149764   & 6.95  & 6.88  & 6.93  & 6.93  & 132.6  & 1.63  & 0.11  &   991 &  2253 &  2975 &  3876 & 1  \\ 
149822   & 6.36  & 6.37  & 6.48  & 6.45  & 126.3  & 2.36  & 0.17  &   651 &  1481 &  1955 &  2547 & 1  \\ 
149911   & 6.09  & 5.66  & 5.62  & 5.52  & 113.3  & 3.54  & 0.29  &   390 &   887 &  1172 &  1527 & 1  \\ 
151525   & 5.24  & 5.08  & 5.08  & 5.10  & 120.6  & 4.86  & 0.37  &   302 &   688 &   908 &  1183 & 1  \\ 
151965   & 6.33  & 6.51  & 6.60  & 6.60  & 180.5  & 2.54  & 0.13  &   865 &  1967 &  2596 &  3383 & 1  \\ 
152107   & 4.82  & 4.78  & 4.58  & 4.57  &  55.2  & 2.27  & 0.38  &   296 &   673 &   889 &  1159 & 1  \\ 
153882   & 6.31  & 6.14  & 6.16  & 6.16  & 162.9  & 3.53  & 0.20  &   563 &  1279 &  1689 &  2201 & 1  \\ 
154708   & 8.76  & 8.11  & 7.98  & 7.95  & 148.1  & 1.70  & 0.11  &  1062 &  2415 &  3188 &  4154 & 1  \\ 
157751   & 7.62  & 7.66  & 7.72  & 7.76  & 161.6  & 1.69  & 0.10  &  1167 &  2652 &  3501 &  4562 & 1  \\ 
164258   & 6.37  & 5.93  & 5.89  & 5.89  & 135.3  & 2.85  & 0.20  &   579 &  1317 &  1738 &  2265 & 1  \\ 
165474   & 7.50  &  ...  &  ...  &  ...  & 138.1  & 2.23  & 0.15  &   755 &  1716 &  2265 &  2951 & 1  \\ 
168733   & 5.34  & 5.53  & 5.56  & 5.58  & 170.6  & 4.54  & 0.25  &   458 &  1040 &  1374 &  1790 & 1  \\ 
168856   & 7.05  & 6.59  & 6.61  & 6.55  & 190.8  & 2.38  & 0.12  &   978 &  2223 &  2934 &  3824 & 1  \\ 
170000   & 4.22  & 4.18  & 4.45  & 4.39  &  92.9  & 3.53  & 0.35  &   321 &   729 &   963 &  1254 & 1  \\ 
170397   & 6.03  & 5.94  & 5.96  & 5.93  & 104.8  & 1.83  & 0.16  &   698 &  1587 &  2095 &  2730 & 1  \\ 
171184   & 7.89  & 7.29  & 7.28  & 7.23  & 222.2  & 2.30  & 0.10  &  1179 &  2679 &  3537 &  4609 & 1  \\ 
171586   & 6.46  & 6.23  & 6.31  & 6.29  & 110.3  & 2.14  & 0.18  &   629 &  1431 &  1889 &  2461 & 1  \\ 
172690   & 7.49  & 7.40  & 7.47  & 7.43  & 280.1  & 2.71  & 0.09  &  1262 &  2868 &  3786 &  4933 & 1  \\ 
173650   & 6.52  & 6.38  & 6.43  & 6.40  & 215.5  & 3.92  & 0.17  &   670 &  1523 &  2010 &  2620 & 1  \\ 
175132   & 6.27  & 6.32  & 6.37  & 6.38  & 492.6  & 6.99  & 0.13  &   859 &  1953 &  2578 &  3360 & 1  \\ 
175362   & 5.38  & 5.62  & 5.66  & 5.68  & 131.9  & 2.57  & 0.18  &   626 &  1424 &  1880 &  2450 & 1  \\ 
176196   & 7.51  & 7.56  & 7.62  & 7.58  & 222.2  & 2.68  & 0.11  &  1010 &  2296 &  3031 &  3950 & 1  \\ 
176232   & 5.89  & 5.42  & 5.32  & 5.30  &  78.4  & 2.43  & 0.29  &   393 &   894 &  1180 &  1538 & 1  \\ 
177410   & 6.50  & 6.75  & 6.83  & 6.88  & 213.2  & 2.68  & 0.12  &  1061 &  2413 &  3185 &  4150 & 5  \\ 
179527   & 5.93  & 5.90  & 6.00  & 6.00  & 273.2  & 6.40  & 0.22  &   520 &  1182 &  1561 &  2034 & 1  \\ 
183056   & 5.15  & 5.30  & 5.43  & 5.44  & 216.9  & 5.03  & 0.22  &   526 &  1195 &  1578 &  2056 & 1  \\ 
183339   & 6.58  & 6.88  & 6.99  & 6.97  & 361.0  & 4.80  & 0.12  &   916 &  2083 &  2750 &  3584 & 1  \\ 
183806   & 5.58  & 5.56  & 5.61  & 5.53  & 121.7  & 3.61  & 0.28  &   411 &   934 &  1233 &  1606 & 1  \\ 
184905   & 6.62  & 6.63  & 6.70  & 6.72  & 189.0  & 2.62  & 0.13  &   879 &  1999 &  2639 &  3438 & 1  \\ 
187474   & 5.33  & 5.37  & 5.44  & 5.43  &  92.4  & 2.85  & 0.29  &   394 &   897 &  1184 &  1543 & 1  \\ 
188041   & 5.63  & 5.53  & 5.55  & 5.49  &  80.1  & 2.80  & 0.32  &   349 &   794 &  1048 &  1366 & 1  \\ 
192678   & 7.34  & 7.24  & 7.32  & 7.33  & 198.0  & 2.85  & 0.13  &   848 &  1927 &  2544 &  3315 & 1  \\ 
196178   & 5.77  &  ...  &  ...  &  ...  & 132.3  & 2.80  & 0.20  &   577 &  1311 &  1731 &  2255 & 1  \\ 
196502   & 5.19  & 5.20  & 5.06  & 5.03  & 121.4  & 4.32  & 0.33  &   342 &   778 &  1027 &  1339 & 1  \\ 
199728   & 6.25  & 6.40  & 6.45  & 6.47  & 164.2  & 2.15  & 0.12  &   931 &  2116 &  2794 &  3641 & 1  \\ 
200177   & 7.34  & 7.21  & 7.27  & 7.26  & 159.0  & 1.72  & 0.10  &  1130 &  2568 &  3390 &  4417 & 1  \\ 
201018   & 8.63  &  ...  &  ...  &  ...  & 152.4  & 1.35  & 0.08  &  1373 &  3121 &  4120 &  5369 & 1  \\ 
201601   & 4.68  & 4.28  & 4.18  & 4.10  &  36.3  & 2.07  & 0.53  &   213 &   486 &   641 &   835 & 1  \\ 
203006   & 4.82  & 4.73  & 4.70  & 4.71  &  60.5  & 2.02  & 0.31  &   364 &   829 &  1094 &  1426 & 1  \\ 
204411   & 5.31  & 5.05  & 5.08  & 5.06  & 126.1  & 4.12  & 0.30  &   373 &   849 &  1121 &  1460 & 1  \\ 
205087   & 6.68  & 6.72  & 6.82  & 6.83  & 170.4  & 2.80  & 0.15  &   741 &  1685 &  2224 &  2898 & 1  \\ 
208217   & 7.19  & 6.93  & 6.94  & 6.84  & 139.5  & 2.29  & 0.15  &   741 &  1685 &  2225 &  2899 & 1  \\ 
212385   & 6.84  & 6.67  & 6.69  & 6.63  & 126.3  & 2.03  & 0.15  &   758 &  1724 &  2276 &  2965 & 1  \\ 
215038   & 8.15  & 8.14  & 8.16  & 8.20  & 317.5  & 1.96  & 0.06  &  1979 &  4497 &  5937 &  7736 & 1  \\ 
216018   & 7.62  & 7.34  & 7.37  & 7.31  & 124.5  & 1.53  & 0.11  &   990 &  2252 &  2972 &  3873 & 1  \\ 
217522   & 7.52  & 6.79  & 6.69  & 6.63  &  88.0  & 2.07  & 0.22  &   518 &  1177 &  1554 &  2025 & 1  \\ 
217833   & 6.52  & 6.54  & 6.62  & 6.63  & 260.4  & 3.00  & 0.11  &  1060 &  2409 &  3181 &  4144 & 1  \\ 
220825   & 4.94  & 4.92  & 4.97  & 4.96  &  47.1  & 1.92  & 0.38  &   299 &   680 &   898 &  1170 & 1  \\ 
221006   & 5.68  & 5.92  & 6.03  & 6.00  & 118.5  & 2.23  & 0.18  &   647 &  1472 &  1943 &  2532 & 1  \\ 
221394   & 6.41  & 6.26  & 6.32  & 6.31  & 129.0  & 2.85  & 0.21  &   551 &  1253 &  1654 &  2155 & 1  \\ 
221568   & 7.55  & 7.43  & 7.48  & 7.48  & 217.4  & 3.31  & 0.14  &   801 &  1822 &  2405 &  3133 & 1  \\ 
223640   & 5.18  & 5.42  & 5.51  & 5.50  &  97.8  & 2.57  & 0.24  &   463 &  1053 &  1390 &  1811 & 1  \\ 
224801   & 6.35  & 6.41  & 6.51  & 6.51  & 188.0  & 3.49  & 0.17  &   657 &  1494 &  1972 &  2570 & 1  \\ 
\hline
\multicolumn{13}{c}{HgMn stars}\\ %
\hline
358      & 2.06  & 2.30  & 2.33  & 2.37  &  29.7  & 2.48  & 0.78  &   142 &   323 &   427 &   556 & 6  \\ 
11753    & 5.11  & 5.17  & 5.25  & 5.17  &  94.1  & 9.50  & 0.94  &   407 &   926 &  1222 &  1593 & 7  \\ 
32964    & 5.12  & 5.23  & 5.22  & 5.20  &  94.7  & 2.40  & 0.24  &   597 &  1358 &  1793 &  2336 & 8  \\ 
33904    & 3.29  & 3.53  & 3.59  & 3.57  &  57.0  & 3.39  & 0.55  &   205 &   466 &   615 &   801 & 9  \\ 
34364    & 6.14  & 6.19  & 6.25  & 6.26  & 122.1  & 3.70  & 0.28  &   788 &  1792 &  2366 &  3083 & 10 \\ 
\hline
\end{longtable}
\end{center}
\begin{flushleft}
\medskip
$^{\dagger}$~--~{maximum baseline length needed to fully resolve the star at a given wavelength.}
\par\medskip
\textbf{References:}
(1)~\citet{2006A&A...450..763K}; (2)~\citet{2000AstL...26..177K}; (3)~\citet{2008CoSka..38..245F};
(4)~\citet{2011MNRAS.417..444P}; (5)~\citet{2011MNRAS.417..444P}; (6)~\citet{2002ApJ...575..449A};
(7)~\citet{2012A&A...539A.142M}; (8)~\citet{2011A&A...529A.160M}; (9)~\citet{2011A&A...534L..13K};
(10)~\citet{2010MNRAS.407.2383F}
\end{flushleft}


\begin{thebibliography}{}

\bibitem[Adelman et al.(2002)]{2002ApJ...575..449A} Adelman, S.~J., Gulliver, A.~F., Kochukhov, O.~P., \& Ryabchikova, T.~A.\ 2002, \apj, 575, 449 
\bibitem[Asplund et al.(2009)]{2009ARA&A..47..481A} Asplund, M., Grevesse, N., Sauval, A.~J., \& Scott, P.\ 2009, \araa, 47, 481 
\bibitem[B\'erio et al.(2014)]{berio2014}B\'erio, Ph., Bresson, Y., Clausse, J.-M., et al.\ 2014, SPIE,  Conf. 9146
\bibitem[Boyajian et al.(2013)]{2013ApJ...771...40B} Boyajian, T.~S., von Braun, K., van Belle, G., et al.\ 2013, \apj, 771, 40
\bibitem[Bruntt et al.(2010)]{2010A&A...512A..55B} Bruntt, H., Kervella, P., M{\'e}rand, A., et al.\ 2010, \aap, 512, A55
\bibitem[Bruntt et al.(2008)]{2008MNRAS.386.2039B} Bruntt, H., North, J.~R., Cunha, M., et al.\ 2008, \mnras, 386, 2039
\bibitem[Che et al.(2011)]{2011ApJ...732...68C} Che, X., Monnier, J.~D., Zhao, M., et al.\ 2011, \apj, 732, 68 
\bibitem[Chiavassa et al.(2010)]{2010A&A...511A..51C} Chiavassa, A., Lacour, S., Millour, F., et al.\ 2010, \aap, 511, A51 
\bibitem[Deutsch(1958)]{1958IAUS....6..209D} Deutsch, A.~J.\ 1958, Electromagnetic Phenomena in Cosmical Physics, 6, 209 
\bibitem[Eisenhauer et al.(2008)]{2008SPIE.7013E..69E} Eisenhauer, F., Perrin, G., Brandner, W., et al.\ 2008, \procspie, 7013,  
\bibitem[Folsom et al.(2010)]{2010MNRAS.407.2383F} Folsom, C.~P., Kochukhov, O., Wade, G.~A., Silvester, J., \& Bagnulo, S.\ 2010, \mnras, 407, 2383 
\bibitem[Folsom et al.(2008)]{2008CoSka..38..245F} Folsom, C.~P., Wade, G.~A., Kochukhov, O., et al.\ 2008, Contributions of the Astronomical Observatory Skalnate Pleso, 38, 245 
\bibitem[Freyhammer et al.(2008)]{2008MNRAS.389..441F} Freyhammer, L.~M., Elkin, V.~G., Kurtz, D.~W., Mathys, G., \& Martinez, P.\ 2008, \mnras, 389, 441 
\bibitem[Ghasempour et al.(2012)]{2012AAS...21944613G} Ghasempour, A., Muterspaugh, M., Hutter, D., et al.\ 2012, American Astronomical Society Meeting Abstracts \#219, 219, \#446.13 
\bibitem[Goncharskii et al.(1977)]{1977SvAL....3..147G} Goncharskii, A.~V., Stepanov, V.~V., Kokhlova, V.~L., \& Yagola, A.~G.\ 1977, Soviet Astronomy Letters, 3, 147 
\bibitem[Ireland et al.(2008)]{2008SPIE.7013E..63I} Ireland, M.~J., M{\'e}rand, A., ten Brummelaar, T.~A., et al.\ 2008, \procspie, 7013,  
\bibitem[Khan \& Shulyak(2007)]{2007A&A...469.1083K} Khan, S.~A., \& Shulyak, D.~V.\ 2007, \aap, 469, 1083 
\bibitem[Khokhlova et al.(2000)]{2000AstL...26..177K} Khokhlova, V.~L., Vasilchenko, D.~V., Stepanov, V.~V., \& Romanyuk, I.~I.\ 2000, Astronomy Letters, 26, 177 
\bibitem[Kochukhov et al.(2011)]{2011A&A...534L..13K} Kochukhov, O., Makaganiuk, V., Piskunov, N., et al.\ 2011, \aap, 534, L13 
\bibitem[Kochukhov \& Bagnulo(2006)]{2006A&A...450..763K} Kochukhov, O., \& Bagnulo, S.\ 2006, \aap, 450, 763 
\bibitem[Kochukhov et al.(2005)]{2005A&A...439.1093K} Kochukhov, O., Piskunov, N., Sachkov, M., \& Kudryavtsev, D.\ 2005, \aap, 439, 1093 
\bibitem[Kochukhov et al.(2004)]{2004A&A...424..935K} Kochukhov, O., Drake, N.~A., Piskunov, N., \& de la Reza, R.\ 2004, \aap, 424, 935 
\bibitem[Kochukhov \& Ryabchikova(2001)]{2001A&A...377L..22K} Kochukhov, O., \& Ryabchikova, T.\ 2001, \aap, 377, L22 
\bibitem[Krti{\v c}ka et al.(2012)]{2012A&A...537A..14K} Krti{\v c}ka, J., Mikul{\'a}{\v s}ek, Z., L{\"u}ftinger, T., et al.\ 2012, \aap, 537, A14
\bibitem[Krti{\v c}ka et al.(2009)]{2009A&A...499..567K} Krti{\v c}ka, J., Mikul{\'a}{\v s}ek, Z., Henry, G.~W., et al.\ 2009, \aap, 499, 567 
\bibitem[Krti{\v c}ka et al.(2007)]{2007A&A...470.1089K} Krti{\v c}ka, J., Mikul{\'a}{\v s}ek, Z., Zverko, J., \& {\v Z}i{\v z}{\'n}ovsk{\'y}, J.\ 2007, \aap, 470, 1089 
\bibitem[Kupka et al.(1999)]{1999A&AS..138..119K}Kupka, F., Piskunov, N., Ryabchikova, T. A., Stempels, H. C., \& Weiss, W. W. 1999, \aaps, 138, 119
\bibitem[Le Bouquin et al.(2011)]{2011A&A...535A..67L} Le Bouquin, J.-B., Berger, J.-P., Lazareff, B., et al.\ 2011, \aap, 535, A67 
\bibitem[Le Bouquin et al.(2009)]{2009A&A...496L...1L} Le Bouquin, J.-B., Lacour, S., Renard, S., et al.\ 2009, \aap, 496, L1 
\bibitem[Lemke(1997)]{1997A&AS..122..285L} Lemke, M.\ 1997, \aaps, 122, 285 
\bibitem[Li Causi(2008)]{2008SPIE.7013E.134L} Li Causi, G.\ 2008, \procspie, 7013,
\bibitem[L{\"u}ftinger et al.(2010)]{2010A&A...509A..43L} L{\"u}ftinger, T., Fr{\"o}hlich, H.-E., Weiss, W.~W., et al.\ 2010, \aap, 509, A43 
\bibitem[Lueftinger et al.(2003)]{2003A&A...406.1033L} Lueftinger, T., Kuschnig, R., Piskunov, N.~E., \& Weiss, W.~W.\ 2003, \aap, 406, 1033 
\bibitem[Maestro et al.(2013)]{2013MNRAS.434.1321M} Maestro, V., Che, X., Huber, D., et al.\ 2013, \mnras, 434, 1321
\bibitem[Makaganiuk et al.(2012)]{2012A&A...539A.142M} Makaganiuk, V., Kochukhov, O., Piskunov, N., et al.\ 2012, \aap, 539, A142 
\bibitem[Makaganiuk et al.(2011)]{2011A&A...529A.160M} Makaganiuk, V., Kochukhov, O., Piskunov, N., et al.\ 2011, \aap, 529, A160 
\bibitem[Mashonkina et al.(2009)]{2009A&A...495..297M} Mashonkina, L., Ryabchikova, T., Ryabtsev, A., \& Kildiyarova, R.\ 2009, \aap, 495, 297 
\bibitem[Mazumdar et al.(2009)]{2009A&A...503..521M} Mazumdar, A., M{\'e}rand, A., Demarque, P., et al.\ 2009, \aap, 503, 521 
\bibitem[Michaud(1970)]{1970ApJ...160..641M} Michaud, G.\ 1970, \apj, 160, 641 
\bibitem[Monnier et al.(2012)]{2012ApJ...761L...3M} Monnier, J.~D., Che, X., Zhao, M., et al.\ 2012, \apjl, 761, L3 
\bibitem[Monnier et al.(2007)]{2007Sci...317..342M} Monnier, J.~D., Zhao, M., Pedretti, E., et al.\ 2007, Science, 317, 342
\bibitem[Monnier et al.(2004)]{2004SPIE.5491.1370M} Monnier, J.~D., Berger, J.-P., Millan-Gabet, R., \& ten Brummelaar, T.~A.\ 2004, \procspie, 5491, 1370 
\bibitem[Montarges et al.(2014)]{Montarges}Montarges et al., 2014, in prep.
\bibitem[Mourard et al.(2011)]{2011A&A...531A.110M} Mourard, D., B{\'e}rio, P., Perraut, K., et al.\ 2011, \aap, 531, A110 
\bibitem[Mourard et al.(2009)]{2009A&A...508.1073M} Mourard, D., Clausse, J.~M., Marcotto, A., et al.\ 2009, \aap, 508, 1073 
\bibitem[O'Brien et al.(2011)]{2011ApJ...728..111O} O'Brien, D.~P., McAlister, H.~A., Raghavan, D., et al.\ 2011, \apj, 728, 111 
\bibitem[Ohnaka et al.(2009)]{2009A&A...503..183O} Ohnaka, K., Hofmann, K.-H., Benisty, M., et al.\ 2009, \aap, 503, 183 
\bibitem[Pandey et al.(2011)]{2011MNRAS.417..444P} Pandey, C.~P., Shulyak, D.~V., Ryabchikova, T., \& Kochukhov, O.\ 2011, \mnras, 417, 444 
\bibitem[Petrov et al.(2007)]{2007A&A...464....1P} Petrov, R.~G., Malbet, F., Weigelt, G., et al.\ 2007, \aap, 464, 1 
\bibitem[Piskunov et al.(1995)]{1995A&AS..112..525P}Piskunov, N. E., Kupka, F., Ryabchikova, T. A., Weiss, W. W., \& Jeffery, C. S. 1995, \aaps, 112, 525
\bibitem[Piskunov \& Rice(1993)]{1993PASP..105.1415P} Piskunov, N.~E., \& Rice, J.~B.\ 1993, \pasp, 105, 1415 
\bibitem[Perraut et al.(2013)]{2013A&A...559A..21P} Perraut, K., Borgniet, S., Cunha, M., et al.\ 2013, \aap, 559, A21 
\bibitem[Perraut et al.(2011)]{2011A&A...526A..89P} Perraut, K., Brand{\~a}o, I., Mourard, D., et al.\ 2011, \aap, 526, A89
\bibitem[Ryabchikova et al.(2004)]{2004A&A...423..705R} Ryabchikova, T., Nesvacil, N., Weiss, W.~W., Kochukhov, O., \& St{\"u}tz, C.\ 2004, \aap, 423, 705 
\bibitem[Ryabchikova et al.(1999)]{1999A&A...351..963R} Ryabchikova, T.~A., Malanushenko, V.~P., \& Adelman, S.~J.\ 1999, \aap, 351, 963 
\bibitem[Rousselet-Perraut et al.(2004)]{2004A&A...422..193R} Rousselet-Perraut, K., Stehl{\'e}, C., Lanz, T., et al.\ 2004, \aap, 422, 193 
\bibitem[Scott et al.(2013)]{2013JAI.....240005S} Scott, N.~J., Millan-Gabet, R., Lhom{\'e}, E., et al.\ 2013, Journal of Astronomical Instrumentation , 2, 40005 
\bibitem[Ten Brummelaar et al.(2013)]{2013JAI.....240004T} Ten Brummelaar, T.~A., Sturmann, J., Ridgway, S.~T., et al.\ 2013, Journal of Astronomical Instrumentation , 2, 40004 
\bibitem[Shulyak et al.(2010)]{2010A&A...524A..66S} Shulyak, D., Krti{\v c}ka, J., Mikul{\'a}{\v s}ek, Z., Kochukhov, O., {\ L\"u}ftinger, T.\ 2010, \aap, 524, A66 
\bibitem[Shulyak et al.(2004)]{2004A&A...428..993S} Shulyak, D., Tsymbal, V., Ryabchikova, T., St{\"u}tz, C., \& Weiss, W.~W.\ 2004, \aap, 428, 993 
\bibitem[Vidal et al.(1973)]{1973ApJS...25...37V} Vidal, C.~R., Cooper, J., \& Smith, E.~W.\ 1973, \apjs, 25, 37 
\bibitem[Wade et al.(2000)]{2000MNRAS.313..851W} Wade, G.~A., Donati, J.-F., Landstreet, J.~D., \& Shorlin, S.~L.~S.\ 2000, \mnras, 313, 851 
\bibitem[Zhao et al.(2009)]{2009ApJ...701..209Z} Zhao, M., Monnier, J.~D., Pedretti, E., et al.\ 2009, \apj, 701, 209

\end{thebibliography}
\end{document}